\documentclass[5p]{elsarticle}

\usepackage{amsmath}
\usepackage{amssymb}
\newcommand{\Al}{$^{26}$Al\ }
\newcommand{\about}{$\simeq$}

\newcommand{\Fe}{$^{60}$Fe\ }
\newcommand{\Ni}{$^{56}$Ni\ }

\newcommand{\Msol}{M\ensuremath{_\odot}}

\newcommand{\araa}{Ann.Rev.Astron.\&Astroph. }%
\newcommand{\apj}{Astrophys. J. }%
\newcommand{\apjl}{Astrophys. J. Lett. }%
\newcommand{\apjs}{Astrophys. J. Suppl. Ser. }%
\newcommand{\apss}{Astroph.J.\&Sp.Sci. }%
\newcommand{\aap}{Astron. Astrophys. }%
\newcommand{\aapr}{Astron. Astrophys.~Rev. }%
\newcommand{\aaps}{Astron. Astrophys. Suppl. }%
\newcommand{\aj}{Astron. J. }%
\newcommand{\gca}{Geoch. Cosmoch. Acta}%
\newcommand{\icarus}{Icarus }%
\newcommand{\jcap}{J. Cosmol. Astropart.Phys. }%
\newcommand{\maps}{Met. Plan. Sci. }%
\newcommand{\memsai}{Mem. Soc. Astron. It. }%
\newcommand{\mnras}{Mon. Notices Royal Astron. Soc. }%
\newcommand{\na}{New Astron. }%
\newcommand{\nar}{New Astron. Rev. }%
\newcommand{\prd}{Phys.~Rev.~D }%
\newcommand{\prl}{Phys.~Rev.~Lett. }%
\newcommand{\pasa}{PASA }%
\newcommand{\pasp}{Proc.Astr.Soc.Pac. }%
\newcommand{\rpp}{Rep.Prog.Phys. }%
\newcommand{\solphys}{Sol.~Phys. }%
\newcommand{\sovast}{Soviet~Ast. }%
\newcommand{\ssr}{Space~Sci.~Rev. }%
\newcommand{\nat}{Nature }%
\newcommand{\physrep}{Phys.~Rep. }%
\newcommand{\procspie}{Proc.~SPIE }%

\begin{document}

\title{Cosmic nucleosynthesis: a multi-messenger challenge}

\author{Roland Diehl$^{1,2}$} 
\ead{rod@mpe.mpg.de}
\author{Andreas J.\ Korn$^3$} 
\author{Bruno Leibundgut$^{4,2}$} 
\author{Maria Lugaro$^{5,6,7}$}
\author{Anton Wallner$^{8,9}$}  
\address{$^1$Max Planck Institut f\"ur extraterrestrische Physik, Giessenbachstr.1, D-85748 Garching, Germany}%
\address{$^2$Excellence Clusters \emph{Universe} and \emph{Origins}, Boltzmannstr. 2, D-85748 Garching, Germany}%
\address{$^3$Observational Astrophysics, Department of Physics and Astronomy, Uppsala University, SE-751 05 Uppsala, Sweden}%
\address{$^4$European Southern Observatory, Karl-Schwarzschild-Strasse 2, 85748 Garching, Germany}%
\address{$^5$Konkoly Observatory, Research Centre for Astronomy and Earth Sciences, E\"otv\"os Lor\'and Research Network (ELKH), Konkoly Thege Mikl\'{o}s \'{u}t 15-17, H-1121 Budapest, Hungary}
\address{$^6$ ELTE E\"{o}tv\"{o}s Lor\'and University, Institute of Physics, Budapest 1117, P\'azm\'any P\'eter s\'et\'any 1/A, Hungary}
\address{$^7$ School of Physics and Astronomy, Monash University, VIC 3800, Australia}
\address{$^8$Helmholtz-Zentrum Dresden-Rossendorf, Institute of Ion Beam Physics and Materials Research, 01328 Dresden, Germany}%
\address{$^9$Research School of Physics, Australian National University, Canberra, ACT 2601, Australia}%

\begin{abstract}
The origins of the elements {        and isotopes} of cosmic material is a critical aspect of understanding the evolution of the universe. 
Nucleosynthesis typically requires physical conditions of high temperatures and densities. These are found in the Big Bang, in the interiors of stars, and in explosions with their compressional shocks and high neutrino and neutron fluxes.
Many different tools are available to disentangle the composition of cosmic matter, in material of extraterrestrial origins such as cosmic rays, meteorites, stardust grains, lunar and terrestrial sediments, and through astronomical observations across the electromagnetic spectrum. 
Understanding cosmic abundances and their evolution requires combining such measurements with approaches of astrophysical, nuclear theories and laboratory experiments, and exploiting additional cosmic messengers, such as neutrinos and gravitational waves. Recent years have seen significant progress in almost all these fields; they are presented in this review.

The Sun and the solar system are our reference system for abundances {        of elements and isotopes}. Many direct and indirect methods are employed to establish a refined abundance record from the time when the Sun and the Earth were formed. 
Indications for nucleosynthesis in the local environment when the Sun was formed are derived from meteoritic material and inclusion of radioactive atoms in deep-sea sediments. 
Spectroscopy at many wavelengths and the neutrino flux from the hydrogen fusion processes in the Sun have established a refined model of how the nuclear energy production shapes stars. 
Models are required to explore nuclear fusion of heavier elements. These stellar evolution calculations have been confirmed by observations of nucleosynthesis products in the ejecta of stars and supernovae, as captured by stardust grains and by characteristic lines in spectra seen from these objects. 
One of the successes has been to directly observe $\gamma$~rays from radioactive material synthesised in stellar explosions, which fully support the astrophysical models.
Another has been the observation of radioactive afterglow and characteristic heavy-element spectrum from a neutron-star merger, confirming the neutron rich environments encountered in such rare explosions.
The ejecta material captured by Earth over millions of years in sediments and identified through characteristic radio-isotopes suggests that nearby nucleosynthesis occurred in recent history, with further indications for sites of specific nucleosynthesis.
Together with stardust and diffuse $\gamma$~rays from radioactive ejecta, these help to piece together how cosmic materials are transported in interstellar space and re-cycled into and between generations of stars. 
Our description of cosmic compositional evolution needs such observational support, as it rests on several assumptions that appear challenged by recent recognition of violent events being common during evolution of a galaxy.
This overview presents the flow of cosmic matter and the various sites of nucleosynthesis, as understood from combining many techniques and observations, towards  the current knowledge of how the universe is enriched with elements.
~\\
\\
\end{abstract}
\date{\today}

\maketitle


\section{Introduction}
\label{introduction}
\begin{figure*} 
\centering
\includegraphics[width=1.8\columnwidth,clip]{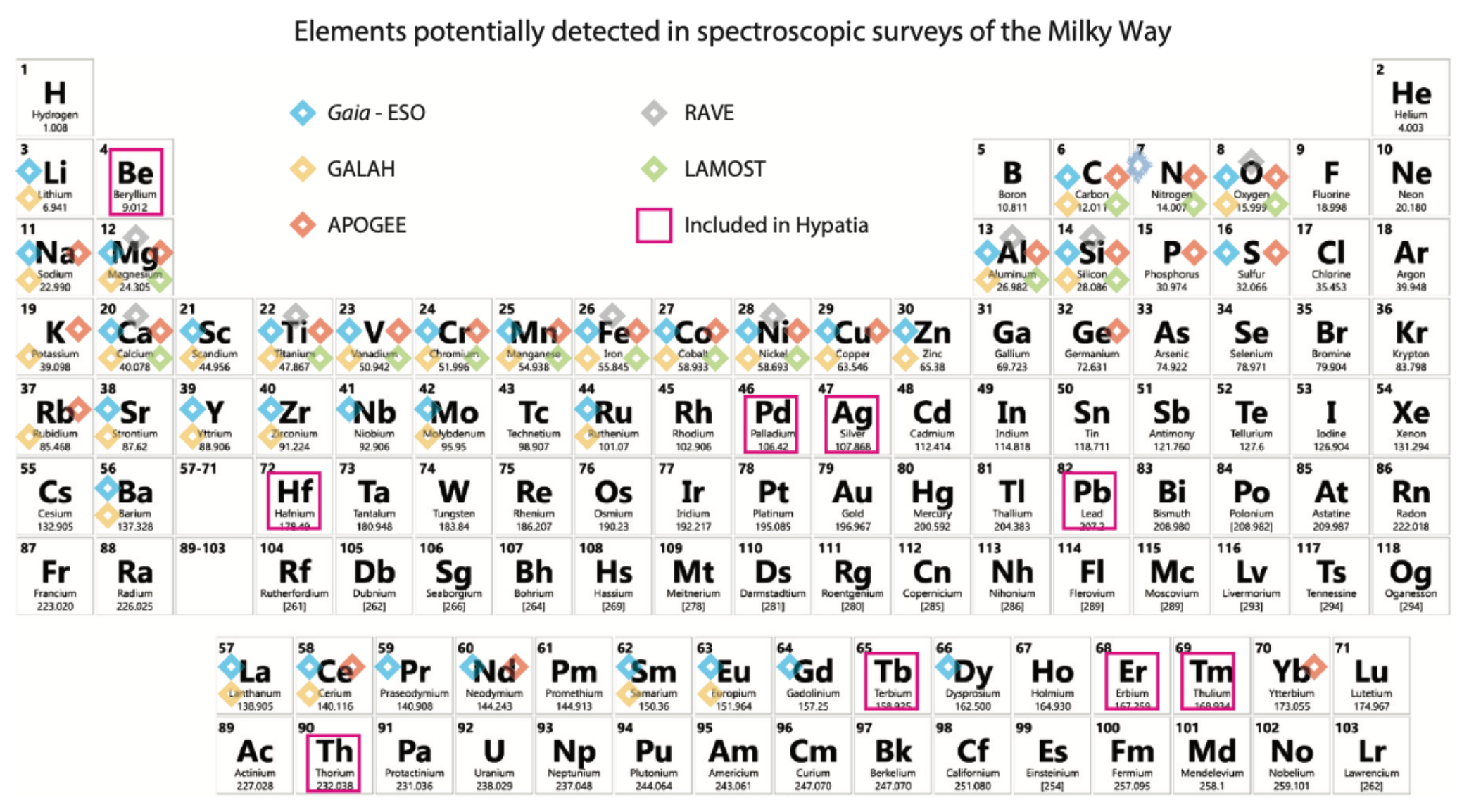}
\caption{The table of elements as currently known. In this representation \citep[from][]{Jofre:2019}, it is indicated for each element if abundance observations from stellar spectroscopy are available.}
\label{fig_stellarAbundElements}       
\end{figure*}

In the decade of the 1860$^{ies}$, a \emph{table of elements} was 
put together by chemists including Dmitri Mendeleev and Lothar Meyer, based on regularities in chemical properties. Mendeleev's 1869 version \citep{Mendeleev:1869} contained 63 elements, and the prediction that many more elements were to be discovered, filling gaps in the regularity scheme. 
Controversies were discussed on how the sequences should properly be arranged in detail, e.g. concerning lanthanides.
The table of elements is shown in Figure~\ref{fig_stellarAbundElements} and currently holds 118 elements, with many stable and also unstable isotopes.

Geochemists such as Frank Wigglesworth Clarke in 1889 and Victor Goldschmidt in the 1930$^{ies}$ \citep{Goldschmidt:1930} developed this early research further towards comparing elemental abundances in different layers of Earth and with meteorites, recognising surprising regularity of abundances. 
The discovery of radioactivity by Henry Becquerel in 1896 had stimulated the work of Pierre and Marie Curie on fission, and it became clear at the turn of the century that atomic nuclei of a kind could be transformed into those of another elemental type. 
In an interesting cosmic viewpoint, Maria Goeppert-Mayer connected radioactivity with nuclear theory \citep{Mayer:1952}: \emph{"It begins to look as if the universe contained a very large number of metastable, "fundamental" particles of which we are able to observe only those which, by some accident, have a low probability of decay"}.

In the 1920$^{ies}$, Cecilia Payne \citep{Payne:1925} analyzed spectra of starlight, to find characteristic absorption lines of chemical elements such as carbon and silicon. In that work, she noticed \emph{``unbelievably-high abundances'' of hydrogen and helium}.
At about the same time, Sir Arthur Eddington made a first astrophysical connection between elemental abundances and the nuclear transformations in stars, stating the hypothesis that the energy source of starlight was the release of some form of subatomic energy \citep{Eddington:1919}.
In the 1950$^{ies}$, Hans E.\ Suess and Harold Urey \citep{Suess:1956} extended the geological aspects to the quest for the cosmic origins of elemental abundances, and also included isotope abundances in their compilation as an essential part of their reasoning. 
Important milestones thereafter were the publication of an insightful theory paper in 1957, the \emph{B$^2$FH paper} \citep{Burbidge:1957}, that named a set of \emph{processes} of cosmic nucleosynthesis \citep{Burbidge:1957}, then the experience of a contemporaneous supernova explosion with 1987A \citep{Arnett:1989a}, which has since been measured with many astronomical instruments as it evolves \citep{McCray:2016}, and lately the celebrated witness of r-process nucleosynthesis in spectra from a kilonova \citep{Smartt:2017,Margutti:2021} associated with signatures of a neutron star merger and formation of a black hole (a gravitational-wave event and a $\gamma$-ray burst) \citep{Abbott:2017g}.
Since more than a century the quest for the origin of the elements and isotopes that make up cosmic material has been one of the main challenges of astrophysics \citep[see][for summaries and milestones of cosmic nucleosynthesis studies and more details]{Burbidge:1957,Clayton:1968,Wallerstein:1997}. 

\begin{figure*} 
\centering
\includegraphics[width=1.8\columnwidth,clip]{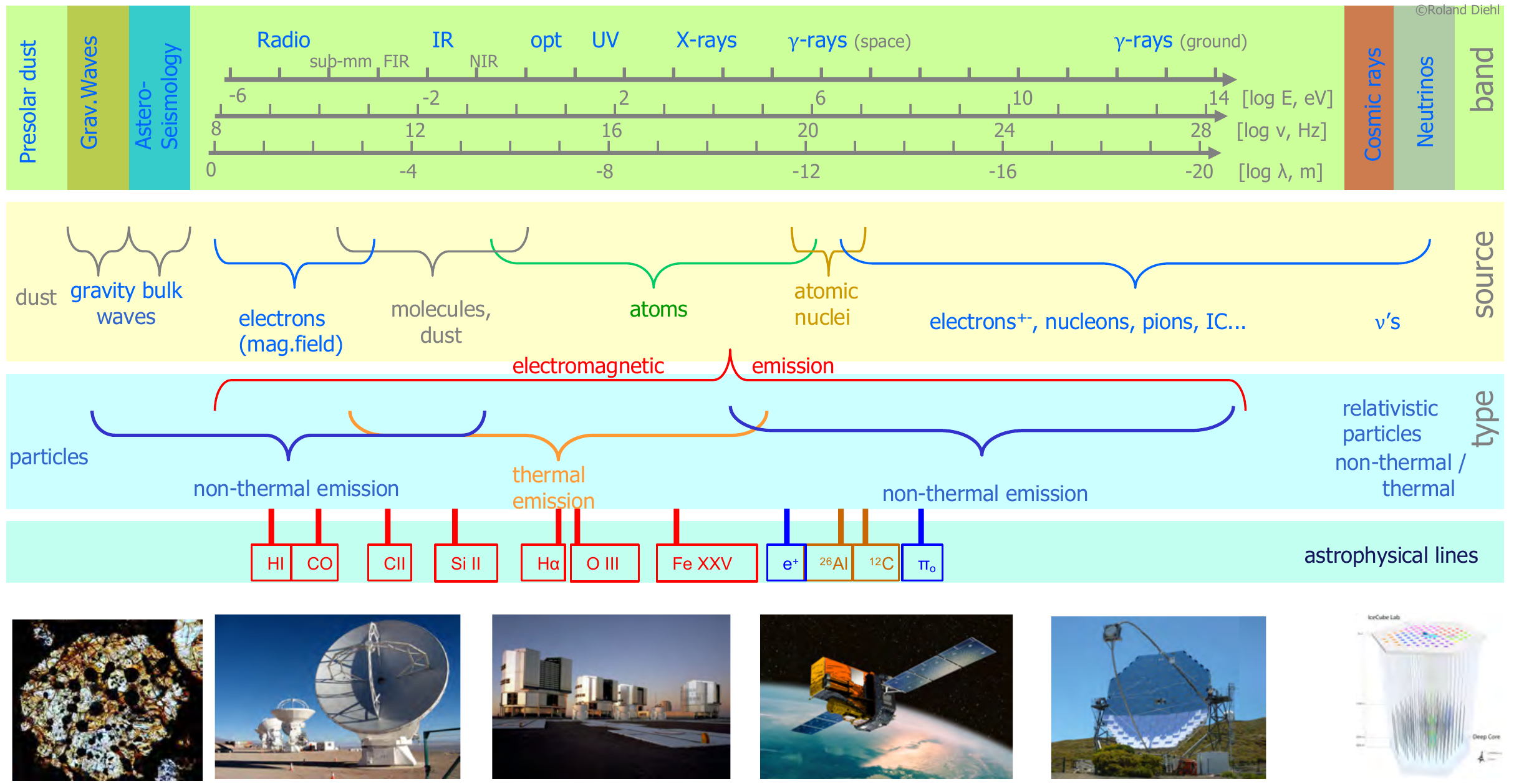}
\caption{The electromagnetic spectrum (top panel) of radiation from radio waves to $\gamma$~rays provides astronomical windows into a variety of cosmic phenomena. Different physical processes are reflected in the origins of emissions within such windows, and are identified below. Spectral lines are most important messengers for nucleosynthesis studies, and are indicated further below. Particles of cosmic origins have been added as messengers on each side of the spectral scale, and also measurements of gravitational waves and stellar oscillations are considered astronomical messengers here. Images in the bottom row show {        instruments used for specific} messengers.}
\label{fig_messengers}       
\end{figure*} 

This science field involves tools from laboratory experiments, astronomical observations of electromagnetic radiation and other cosmic messengers,
a variety of theoretical models, descriptions of nucleosynthesis sources and astrophysical material flows. 
Observations now cover radiation from across the entire electromagnetic spectrum, then other cosmic messengers such as cosmic rays, neutrinos and gravitational waves, and traces of cosmic material transported to Earth by meteorites, or found as collected over time in lunar and geological archives, or captured directly by space missions (Figure~\ref{fig_messengers}). They will all be described in this article. 
 
\begin{figure}[ht] 
\centering
\includegraphics[width=\columnwidth,clip]{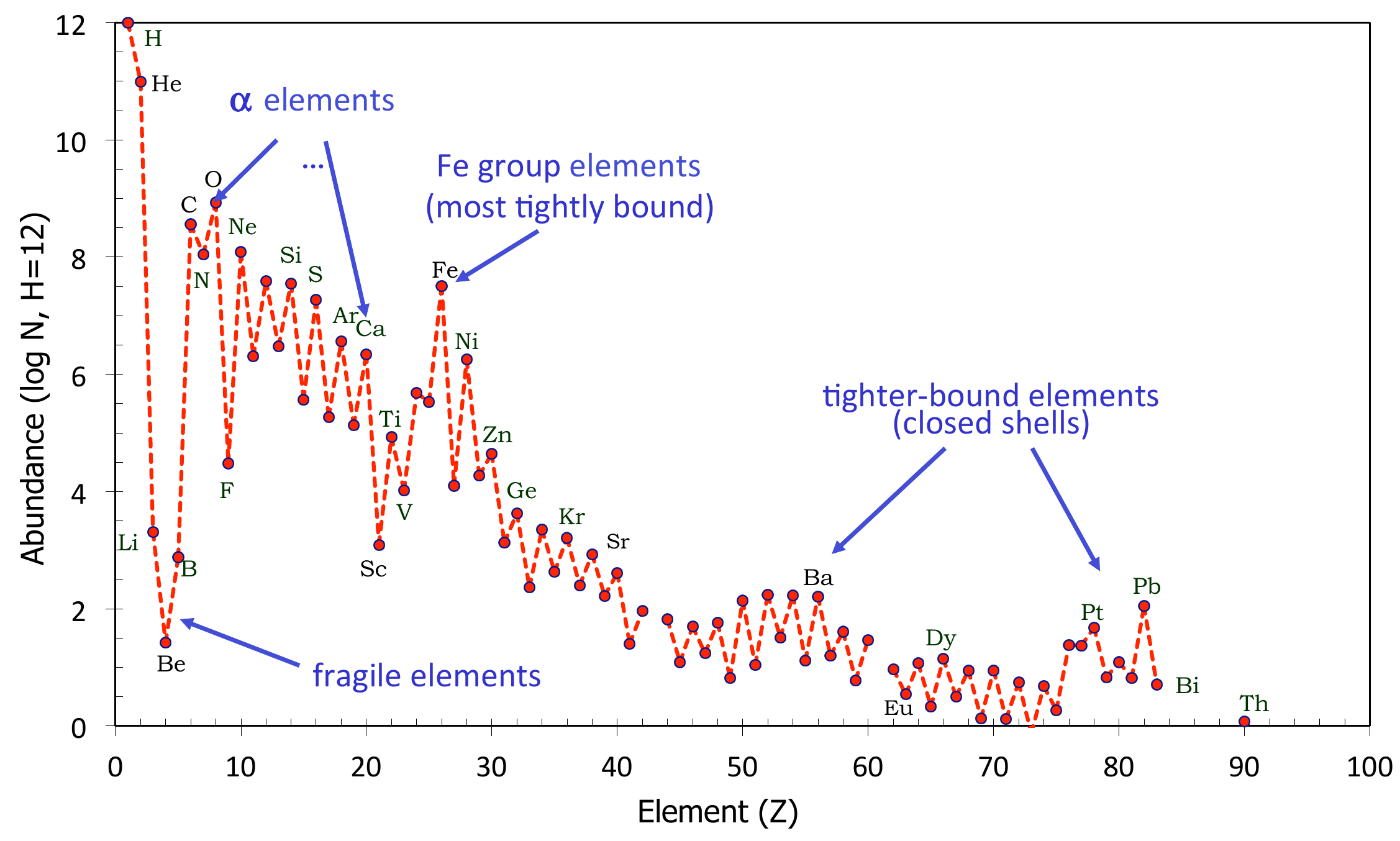}
\caption{The elemental abundances of cosmic gas are best determined for the Sun, and appear to be similar in other regions of the current universe. They are characterised by a dynamic range extending over twelve orders of magnitude, with a dominance of hydrogen and helium, a prominent broad peak around iron-group nuclei with zig-zag abundance patterns bridging the region from light elements to iron, and two characteristic double-peaks across the heavy-element region beyond iron. This plausibly is the result of nuclear-stability physics, as convolved with nuclear-reaction environments for the different cosmic sites where they can occur.}
\label{fig_abundances}       
\end{figure} 

Spectroscopy of starlight was the original tool of astronomy for this field. It had been pioneered by Joseph Fraunhofer's insight that absorption lines against the thermal spectrum of starlight reflect the abundance of specific atomic species in the outer stellar region above the photosphere. 
With refined spectrometry, the standard tools of classical astronomy have built up an impressive wealth of stellar spectra in our Milky Way and galaxies of the Local Group. In rare cases isolated (giant) stars have been observed out to about 4 to 5 Mpc distance.
The observable parts of the electromagnetic spectrum have continuously been expanded over the past century, beginning with radio astronomy, then X-ray and infrared astronomy, and most recently $\gamma$-ray astronomy.
These provided access to abundances of elements and isotopes in interstellar matter, from cold clouds through hot plasma, and with nuclear information independent of thermodynamic environment variables.
Overcoming transparency issues that are characteristic for dense sites of nuclear reactions, neutrinos could be detected from the core of the Sun, and allow for a detailed verification of the reaction chains involved in stellar hydrogen burning.

Photons and neutrinos are not the only cosmic particles that serve as messengers from cosmic nucleosynthesis: Meteorites deliver probes of material from the early history of the solar system 4.5~Gy ago, and can be analysed at great precision in terrestrial laboratories. In this way it has been possible to build a database of isotopic compositions found in a variety of mineral phases,  as solids had formed in the nascent solar system. 
Moreover, tiny stardust grains are embedded in meteorites. Their vastly different isotopic composition identifies them as of extra-solar origin. 
We have dust particles that formed in the immediate vicinity of sites of cosmic nucleosynthesis. But a bias from the need to form dust particles from ambient plasma and gas can be a concern. In particular, explosive nucleosynthesis ashes can also be uncovered in these stardust studies.

The detailed analysis of terrestrial material led geophysicists to speculate about the cosmic origins of elements and their abundances. 
The observation of Tc in stellar atmospheres \citep{Merrill:1952} demonstrated that these origins are not restricted to the distant past, because the element Tc only has one isotope that is radioactive and decays within $\tau=$3.8~My according to  $N(t)=N(0)\cdot e ^{-^{t}/_{\tau}}$. This was reinforced with the detection in the Galactic plane of diffuse $\gamma$-ray emission from radioactive $^{26}$Al \citep{Mahoney:1982}, which has a lifetime of $\tau=$1.04~My. Cosmic nucleosynthesis has been ongoing from the early Galaxy until today.

Cosmic-rays and their nuclear composition near Earth show a striking similarity to the solar abundance, yet with a surprising overabundance for elements lighter than carbon, that hint at interstellar nuclear reactions {         and cosmic-ray spallation} also contributing to the patterns of cosmic abundances.
Most-recently, exploration of gravitational radiation had been established as a cosmic messenger. Together with detectability of $\gamma$-ray bursts, these were key to discover a new type of cosmic explosions, the kilonova phenomenon related to the collision of two neutron stars. 
While it was long believed that ejection of material from such compact-star collisions were unlikely, the follow-up spectroscopy of the first clearly-identified kilonova event led to strong observational support for heavy-element production and their ejection in significant amounts.

\begin{figure}
  \centering
  \includegraphics[width=\columnwidth]{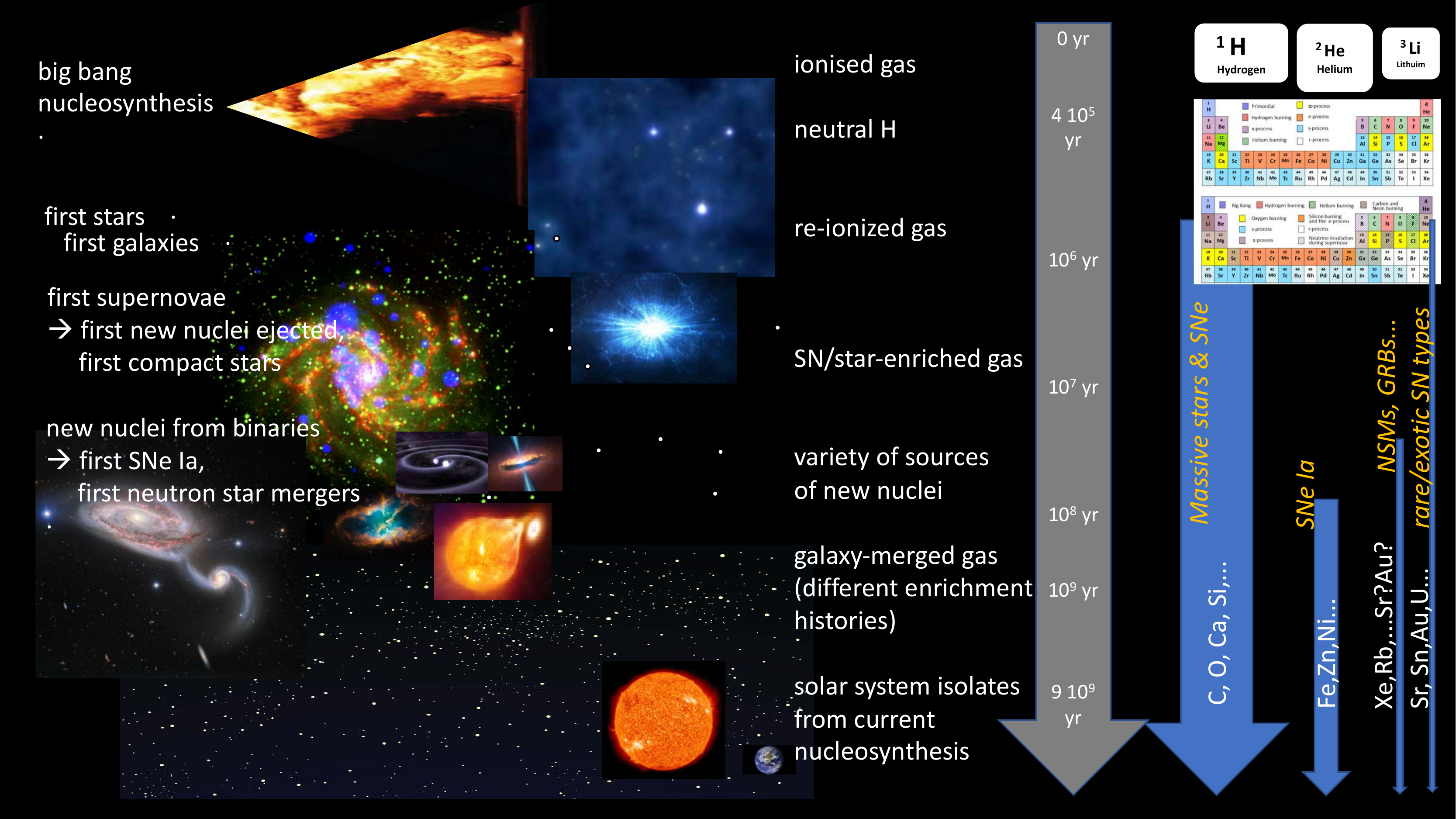}
   \caption{The cosmic enrichment of matter with new nuclei from nucleosynthesis occurs from a variety of sources, and at different times. The y axis of this graph is cosmic time, with the big bang origin at the top, and current time at the bottom. Arrows indicate the epochs where different sources contribute, with arrow width indicating their relative {        importance.}}
  \label{fig_cosmic_evolution_metallicity}
\end{figure}

 Cosmic nucleosynthesis 
 occurs in different sites, where conditions for particular reactions are fulfilled (Table~\ref{tab_nucleosynthesisSites}). 
 This requires a detailed understanding of nuclear reactions as driven by nuclear stability and nucleon binding energies in mostly unstable, nuclear species that can be created in such environments.

\begin{table} 
\begin{center}
\begin{tabular}{|p{2.5cm}|c|c|c|c|}
 \hline
   Site & T & $\rho$ & E & Duration \\
      & [GK] & [g~cm$^{-3}$] & [eV] &   \\
   \hline \hline
   big bang  nucleosynthesis  &   1.0      & 10$^{19}$  & 10$^7$ &  10$^2$ s \\
   \hline
   core  H burning  (Sun)  &   0.02     & 10$^{2}$  & 10$^4$  & {        10$^{10}$ y} \\    
   \hline
  explosive H burning  (nova)  &   0.02     & 10$^{3}$  & 10$^2$  & 10$^{2}$ s \\    
     \hline
   H shell burning  (AGB star)  &   0.1     & 10$^2$  & 10$^2$  & 10$^{8}$ y \\    
      \hline
   core  C burning  (massive star)  &   0.7      & 10$^{5}$  & 10$^6$  & 10$^{3}$ y \\    
   \hline
     explosive burning  (ccSN)  &   2.0      & 10$^{6}$  & 10$^6$  & 1-2 s \\    
      \hline
   explosive burning  (SN Ia)  &   2.0      & 10$^{10}$  & 10$^6$  & 0.5 s \\    
      \hline
   
   explosive n captures  (kilonova)  &   0.02     & 10$^{5}$  & 10$^2$  & 10 s \\    
     \hline
   interstellar spallation    &   0.1      & 10$^{1}$  & 10$^7$ & {        10$^{7}$ y} \\    
      \hline
\end{tabular}
\end{center}
\caption{Examples of the sites of nucleosynthesis and their typical environmental characteristics for nuclear reactions. Given are: typical temperatures in K, typical densities in g~cm$^{-3}$, typical energies E of reacting particles in the Gamov peak in eV, and the characteristic duration of the nucleosynthesis process. Within sites, variations from those typical values may be significant, e.g., few orders of magnitudes in densities, and up to one order of magnitude is temperatures.}
\label{tab_nucleosynthesisSites}
\end{table}

The origin of the hot big bang left little time for nuclear reactions due to the extremely violent expansion, and \emph{big-bang nucleosynthesis (BBN)} products are mainly hydrogen and helium, with some trace amounts of deuterium, lithium, and carbon.
 From this primordial dominance of hydrogen and helium, newly-produced nuclei ejected from nucleosynthesis in stars and explosions  have gradually built up the currently-observed cosmic variety of nuclear species, from carbon through silicon and iron up to lead, platinum, and uranium (see Figures~\ref{fig_abundances} and \ref{fig_cosmic_evolution_metallicity}).
 Identifying and understanding the sources of nucleosynthesis is one aspect: the detailed nuclear reaction paths that   may occur in their specific environments may become obvious, as we are able to constrain environmental parameters. 
 Then, cosmic transports and trajectories of materials are to be understood, to reveal how nuclei once ejected from creation sites may find their way into newly-forming stars and planets or other kinds of observable material such as dust grains and cosmic-ray nuclei.
 We call this process \emph{cosmic chemical evolution}, which combines different disciplines of astrophysics, of nuclear physics, and of analysing data from multiple messengers, 
 but little in terms of chemistry, contrary to what the name might suggest.

In this paper, we aim to trace how the different cosmic messengers as shown in Figure~\ref{fig_messengers} have contributed to the study of cosmic nucleosynthesis, and we also present the results and challenges of each of these messengers.
We begin with a description of our Sun and solar-system material, and how measurements have stimulated the research towards cosmic origins of its materials in terms of nuclear astrophysics. We proceed to address issues of the interiors of stars in general, expanding to massive, rapidly-evolving, and often unstable stars, and to compact stars interacting in binary systems. We then turn to stellar explosions, and how we can understand cosmic extremes of matter and energy density, in terms of how they may contribute to enrich interstellar gas with new nuclei. 
Finally we discuss how this all comes together, in the cosmic cycle of matter and abundance evolution at all scales. This is the final quest, and the goal of chemical-evolution models and their observational tests. Material composition  must be studied throughout the scales and variety of time, space, mass, and density.
We begin with the large-scale picture obtained from cosmological observations, chemical evolution models, and stellar archeology. Then we address in more detail how interstellar gas ends up in stars and how ejecta may be recycled.
We conclude and summarise how different astronomical messengers are woven together within this challenge to understand nuclear transformations and  nucleosynthesis across all times and scales of our cosmos.

\section{Nuclear processes in cosmic environments} 
\label{processes}
\subsubsection*{Basics of nuclear reactions}
Nuclear reactions between two particles are fundamentally described by their \emph{interaction cross section}, which can be understood as their effective size if imagined as colliding hard spheres. 
The typical sizes of atomic nuclei were determined from Rutherford scattering as $10^{-15}$~m, and should scale with the number of nucleons $A$ within each nucleus as $\propto A^{^1 / _3}$.
This is well within the dimensions of the atom of $\sim10^{-11}$~m, 
hence the bare positive charge of the nucleus determines electromagnetic interaction upon encounter.

 All reactions between two nuclei that involve the nuclear force with its short range of $\sim10^{-13}$~m require close encounters, i.e. 'hard' collisions, or  collisions at a small impact parameter. 
Impact parameters can be expressed as angular momentum quantum numbers $l$ of a plane-wave expression in spherical coordinates. Then, the cross section $\sigma$ for the wave component $l$ scales as $\sigma_l = \pi ({r_{l+1}^2}-r_l^2)$, where $r_l = \frac{l \cdot \hbar}{m\cdot v}= l\cdot \lambda_{{B}}$ is the size expressed in units of the \emph{de~Broglie wavelength $\lambda_B$}. This is a purely geometrical aspect of the interaction cross section $\sigma$, and it scales with energy as $\sigma_{\text{geo}}\propto\ ^{1}/_{E}$. 

 One can estimate the cross section in its energetic and geometrical aspects considering that it is required to cross the Coulomb barrier to enter the realm of the nuclear potential. For s-wave scattering, i.e.\ the smallest impact parameter and equivalent to a head-on collision, this re-writes the probability for such a close-enough encounter between nuclei of charges $Z_1$ and $Z_2$ at relative velocity $v$ using the \emph{Sommerfeld parameter} $\eta$ as
 $P=e^{-2\pi\eta}$ with $\eta=2 \pi\cdot \frac{Z_1^2\cdot Z_2^2 \cdot e^2}{\hbar \cdot v}$. 
 This is an aspect of the nuclear reaction cross section between charged particles, called $\sigma_{\text{Coulomb}}$. It is absent for neutron reactions.

\subsubsection*{Nuclear reactions: cosmic reaction specifics} 
The total cross section for a nuclear reaction  drops steeply towards lower energies, which is mostly due to diminished tunneling probability and   the above geometrical aspects. But the reaction cross section can be written as $\sigma (E) = \sigma_{\text{geo}} \cdot \sigma_{\text{Coulomb}} \cdot \sigma_{\text{nuclear}}$. This can be rewritten with the above as  $\sigma (E) = \frac{1}{E} \cdot e^{-2\pi\eta} \cdot S(E)$, with the definition of a term $S$ called the \emph{astrophysical S-factor}.
This is a very useful separation of the components that contribute to the reaction cross section, as it introduces the \emph{astrophysical S factor} to include all nuclear aspects of the reaction.  
In laboratory experiments, a target material is prepared holding one of the reaction partners, while a beam of the other reaction particles is prepared to hit the target at well-defined angle and kinetic energy $E$. This guarantees an optimal preparation of the initial conditions of a reaction between projectile and target, and the measurements of outgoing particles and photons with their directions and momenta then constrains what happened during the reaction.
The astrophysical S factor in its energy dependence then does not extend over many orders of magnitude any more, 
and clear effects of nuclear structure are revealed, e.g. in showing sharp resonances.
Laboratory experiments are combined with nuclear reaction and nuclear force theory
to determine the astrophysical S factors for the reactions of interest; theory is required, because many reactions cannot be measured directly in the laboratory, either because unstable isotopes with short lifetimes are involved, or because required beam energies or target/beam materials cannot be provided. 
Major efforts are being undertaken towards reaching low energies typically found in stars, as typical particle beam energies in the laboratory are $\sim$MeV and above. Underground laboratories are required to suppress cosmic-ray backgrounds.
Exotic reaction partners away from stable isotopes can be provided through isotope separators and sometimes storage rings.
Particular efforts are necessary to measure neutron reactions; neutrons may either be prepared in energy using their time of flight between two detectors, or prepared in a broad range of energies resembling hypothetical cosmic sources. The latter approach results in \emph{Maxwellian-averaged} neutron capture cross sections.

\subsubsection*{Nuclear-reaction environments in cosmic sites}
Atomic nuclei reacting with each other in cosmic sites exist within a plasma of variate composition and thermodynamic state. It is necessary to integrate over energies of the reaction cross section weighing with the particle population. 
For a thermal plasma, energies will be populated according to the Boltzmann distribution, equivalent to a Maxwellian distribution of particle velocities. 
The probabilities for close-enough encounters are obtained by a convolution of such a Maxwellian distribution with the above probability for close encounters between charged particles.  
Quantum tunneling through the remaining Coulomb barrier is essential to describe penetration of reaction partners into the binding force of the strong and short-range nuclear potential for kinetic energies below the repulsive Coulomb energy.
The result of this convolution between an exponentially falling Maxwellian and an exponentially-rising tunneling probability is a peak of almost Gaussian shape that is called the \emph{Gamov peak}. It reveals the energy range where collisions among charged nuclei are close enough for nuclear reactions to occur. Typical energies of the Gamov peak centroid are \about~10~keV for interiors of normal stars and up to \about~1~MeV for explosive environments such as supernovae or type-I X-ray bursts. 

\begin{figure}[ht] 
\centering
\includegraphics[width=\columnwidth,clip]{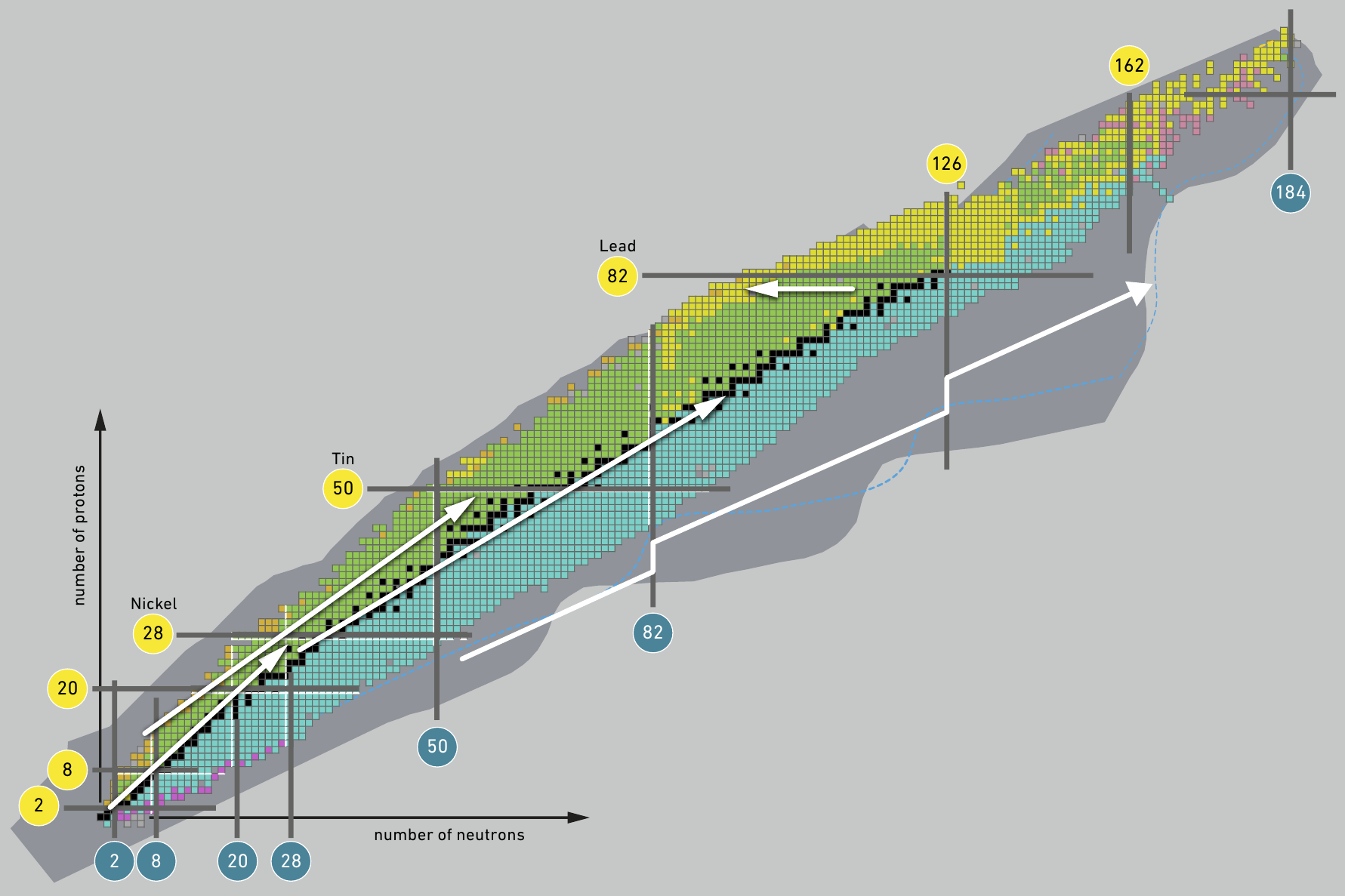}
\caption{The table of isotopes, with the reaction paths of major \emph{processes} of cosmic nucleosynthesis.}
\label{fig_tableOfIsotopes}       
\end{figure} 

Under astrophysical conditions, nuclear reactions lead to a change in the composition. 
Each reaction may either produce or destroy a particular isotope, so that abundances of isotopes of each type are coupled to the full variety of nuclear reactions (charged-particle reactions, neutron reactions, weak reactions such as $\beta$~decays or electron captures). 
Nuclear-reaction network equations need to be solved, often with hundreds of isotopes involved, and the variety of nuclear reactions, including strong and weak processes as well as photon reactions. 

Note that cosmic sites include extremes in terms of explosive or strong-field environments, such as supernovae with strong dynamical evolution in their cores (magnetic-jet supernovae, a-spherical core-collapse supernovae), neutron star collisions, and edges of black-hole accretion disks. Under these conditions, plasma states are dynamic and less well-defined by e.g., a single temperature value or otherwise regular distributions of particle energies. 

This makes it even more challenging to estimate nuclear reaction rates and concepts such as the Gamov peak loose their merits. 
In explosive environments, rapidly changing energy distribution and temperature, and the dynamical motions of plasma regions add to the complexity of nuclear reactions. 
Therefore, care must be taken to applications of simplified nuclear reaction networks with few isotopes, or equilibrium assumptions, and to modelling that ignores the 3D character of the cosmic site of nuclear reactions.  

\subsubsection*{Towards cosmic nucleosynthesis}
 
\begin{figure} 
\centering
\includegraphics[width=\columnwidth]{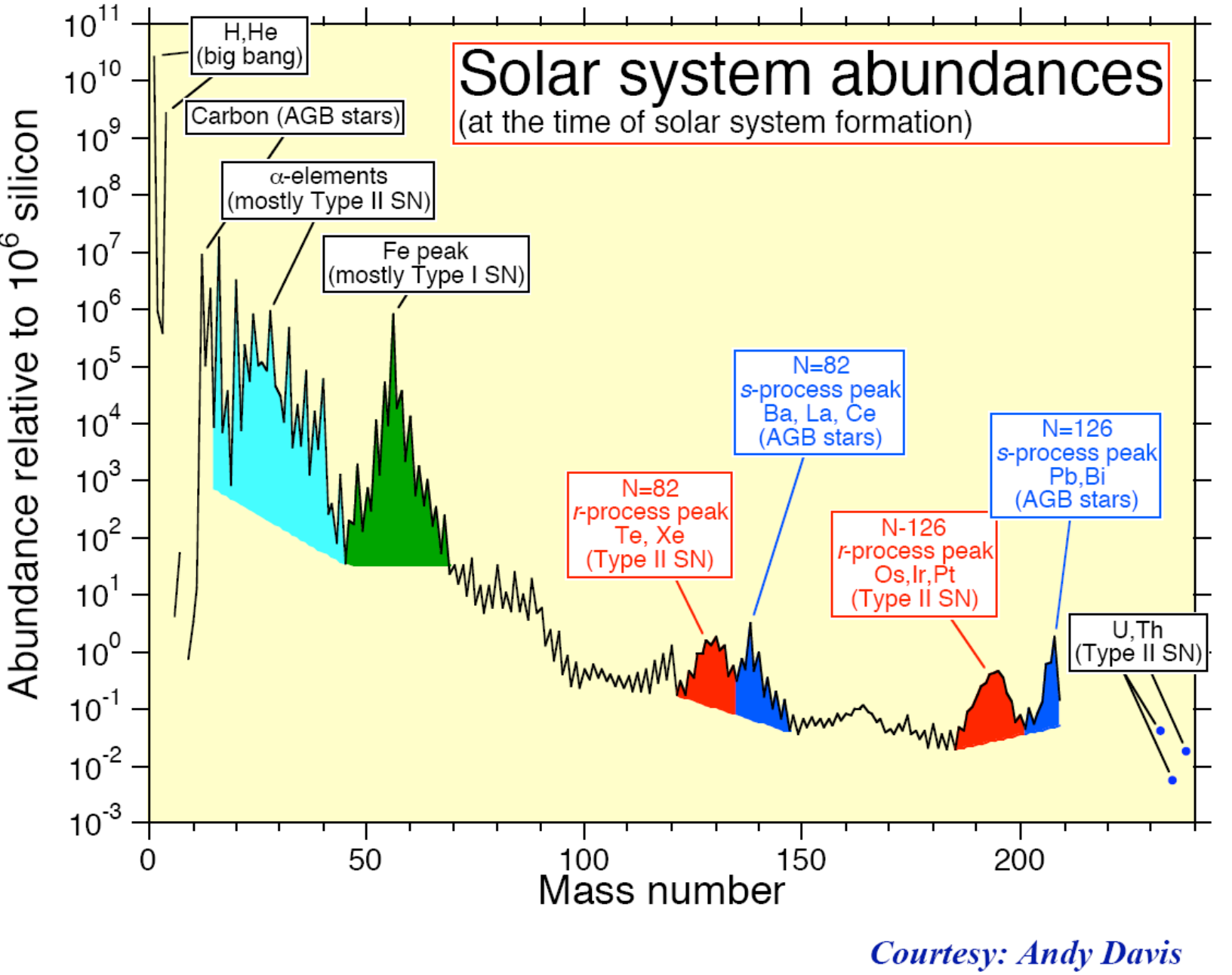}
\caption{The elemental abundances, as commonly attributed to different processes of cosmic nucleosynthesis.  (Figure courtesy Andrew Davis).}
\label{fig_elementOrigins}       
\end{figure} 
 
Astronomical observations have established abundances of elements across a dynamic range of over 12 orders of magnitude, as shown in Figure~\ref{fig_abundances}. This abundance pattern shows several characteristics: The dominance of hydrogen and helium proceeds towards heavier elements with a paucity of nuclei below mass 12, then a plateau of abundances up to the iron group with local maxima for multiple $\alpha$~nuclei, and {        abundances} fall by several orders of magnitude beyond elements heavier than iron.
These signatures are already quite informative, and suggest that aspects of nuclear structure and stability combine with astrophysical conditions for nucleosynthesis in sources of different types.
The strong nuclear binding of $\alpha$~nuclei and iron group nuclei is reflected by the corresponding abundance maxima. 
It also was found that the abundance pattern shown here is fairly typical for objects in the universe at moderately-evolved times, abundances varying among elemental groups, but the general patterns discussed above remain universal.

From remarkably bold extrapolation, several scientist groups including Fred Hoyle, Al Cameron, and the Burbidges \citep{Burbidge:1957} discussed \emph{processes} of cosmic nucleosynthesis, that proceed in different types of cosmic objects.
These processes {        were} identified as: 
\begin{itemize}
\item{} equilibrium nucleosynthesis responsible to produce most-tightly bound isotopes;    
\item{} charged-particle fusion reactions to produce intermediate-mass elements up to the iron group; 
\item{} neutron capture at a slow rate to produce elements heavier than iron along the region of stable nuclei in the chart of isotopes;
\item{} neutron capture at a rapid rate to produce elements heavier than iron all along the chart of isotopes;  
\item{} proton capture or photon reactions to produce elements in the proton-rich region beyond stable nuclei in the chart of isotopes.  
\end{itemize}
{         We add here:
\begin{itemize}
\item{} interstellar cosmic-ray spallation reactions 
\end{itemize}
}
  
\begin{figure} 
\centering
\includegraphics[width=0.8\columnwidth,clip]{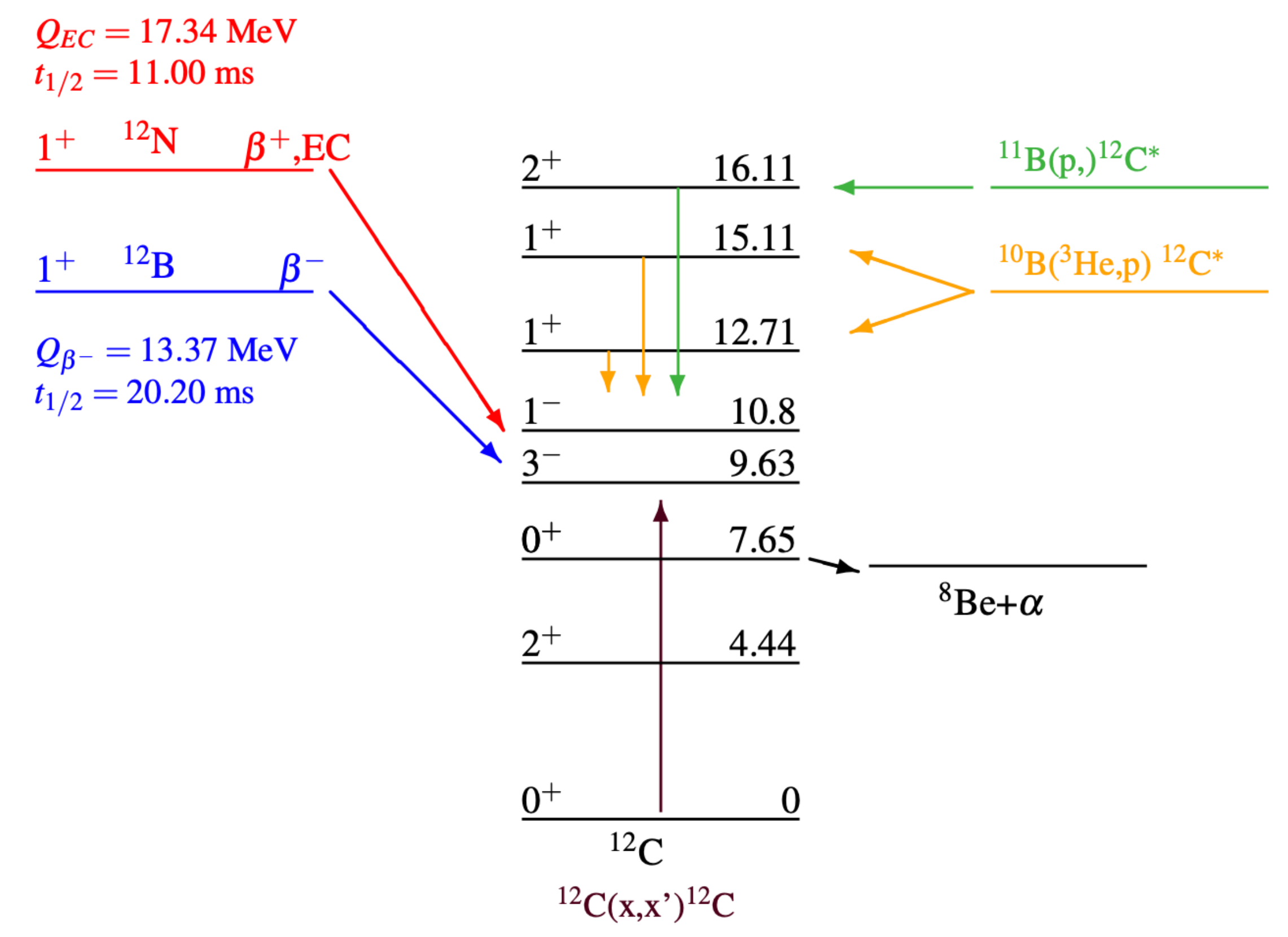}
\caption{The nuclear level scheme of $^{12}$C, with reaction paths of different processes for its production illustrated \citep[from][]{Fynbo:2012}. The level at 7.65 MeV is also called the \emph{Hoyle state}.}
\label{fig_12Clevels}       
\end{figure} 

The above processes are illustrated in the context of the table of isotopes in Figure~\ref{fig_tableOfIsotopes}.
Figure~\ref{fig_elementOrigins} shows how the origins of different elements are commonly understood. 
  
\begin{figure} 
\centering
\includegraphics[width=0.9\columnwidth,clip]{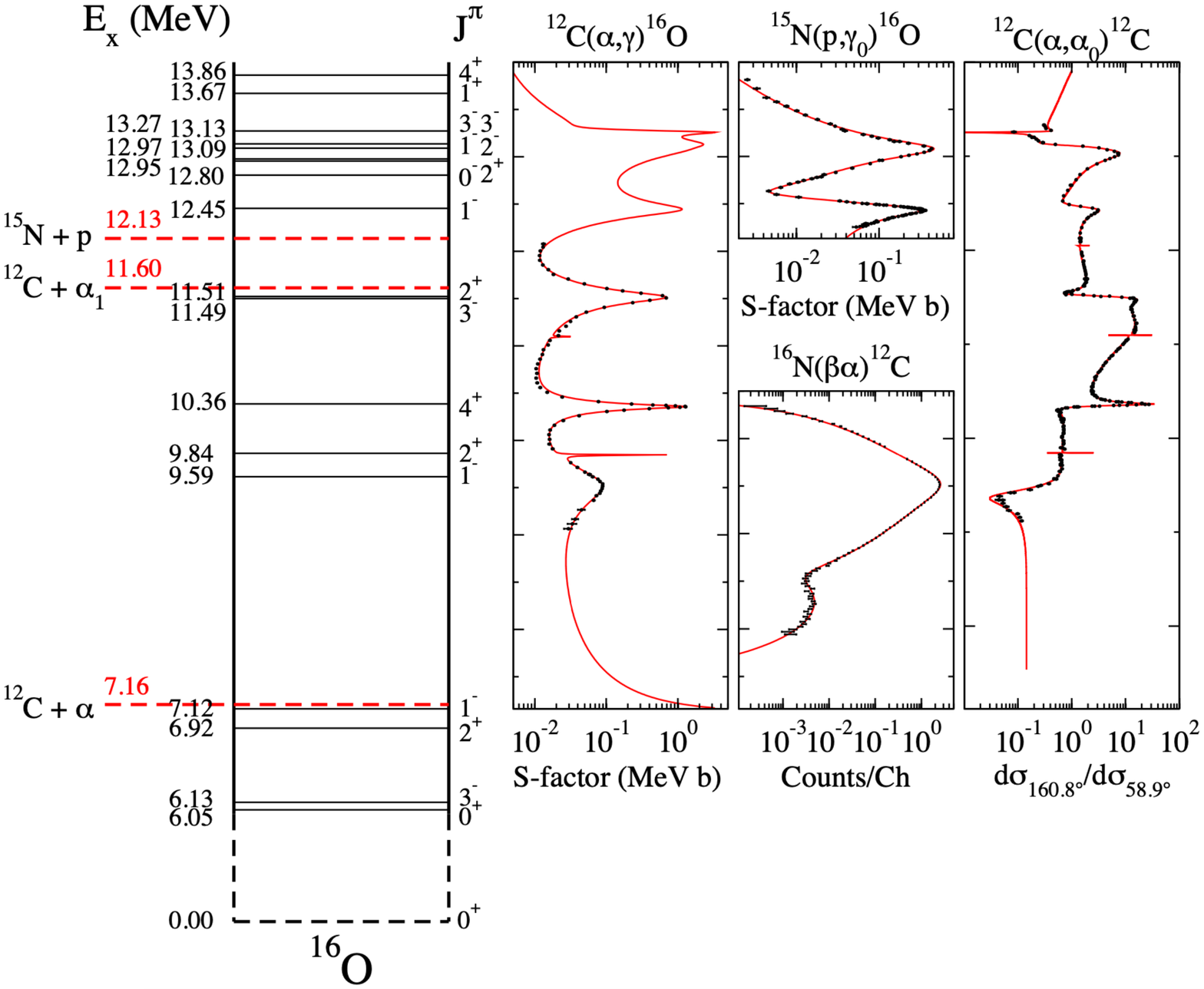}
\caption{The  $^{12}$C($\alpha,\gamma$) reaction, with the level scheme of the $^{16}$O nucleus ({\it left}), and reaction paths of different processes ({\it right}). (From \cite{deBoer:2017}).}
\label{fig_12Cag}       
\end{figure} 

\subsubsection*{Towards the iron group: Charged-particle reactions}
 
\begin{figure} 
\centering
\includegraphics[width=\columnwidth,clip]{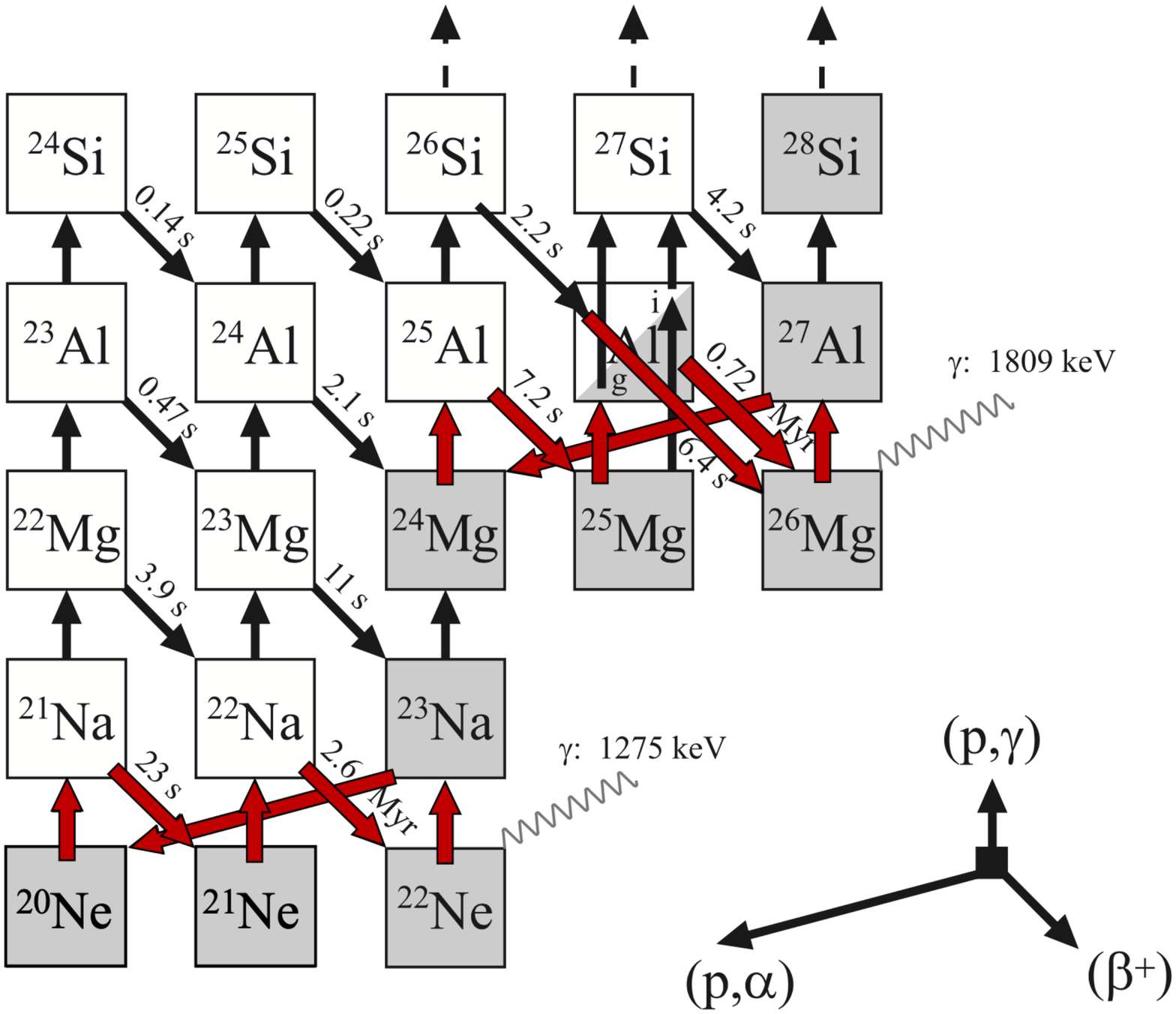}
\caption{The  reaction paths among isotopes in the group of intermediate-mass elements from Na to Si. (From \cite{Laird:2022}).}
\label{fig_Na-Mg-Al-cycle}       
\end{figure} 

A first obstacle to build heavier elements from primordial hydrogen and helium is the lack of a stable nucleus in the mass region between helium and carbon. 
This requires a three-particle reaction to form carbon, the triple-$\alpha$ reaction $^4$He($^4$He,$\gamma$)$^8$Be$\rightarrow$~$^4$He($^8$Be,$\gamma$)$^{12}$C with $^8$Be as intermediate nucleus, 
which has to react with another $\alpha$~particle to form $^{12}$C before $^8$Be decayed. 
The latter reaction requires a resonance in the $^{12}$C nucleus excitation level structure that overlaps with the energy regime that is accessible to the $^4$He$+ ^8$Be~configuration at reaction energies, as shown in Figure~\ref{fig_12Clevels}. This excited state had been predicted by Fred Hoyle before it was confirmed by measurements in nuclear experiments. 
A second obstacle is the reaction variety that can process $^{12}$C into heavier isotopes. Detailed level structures of daughter nuclei are decisive for different reaction channels to open, in the  $^{12}$C($\alpha,\gamma$)$^{16}$O reaction  often called the \emph{holy grail} reaction (see Figure~\ref{fig_12Cag}), as it opens the chain of fusion reactions towards elements heavier than $^{12}$C. 
Note that both of these key reaction rates are subject to significant uncertainty \citep[see][for details on uncertainties]{Kibedi:2020,deBoer:2017}.
   
\begin{figure} 
\centering
\includegraphics[width=0.7\columnwidth,clip]{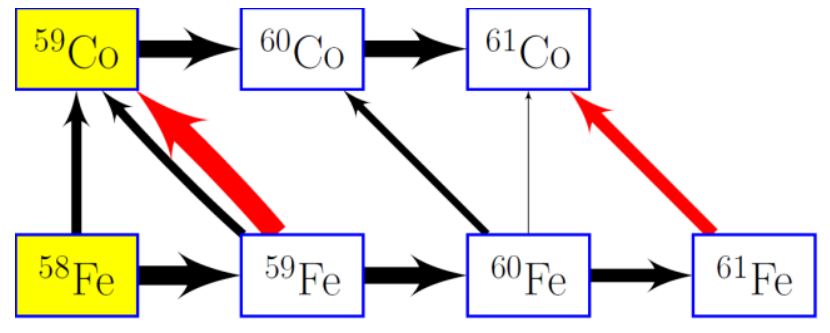}
\caption{The  reaction paths of neutron captures on iron seed nuclei towards production of radioactive $^{60}$Fe \citep[from][]{Diehl:2021b}. Yellow-marked isotopes are stable, and the thickness of the arrows correspond to the intensities of the flows; red and black arrows show $\beta$ decays and nuclear reactions, respectively.}
\label{fig_60Fe-production}       
\end{figure} 

In the realm of intermediate-mass elements, the variety of nuclear reactions includes proton captures $(p,x)$ with $x$ being $p/n/\alpha$/$\gamma$, then $\alpha$~reactions similarly, weak reactions ($\beta$~decays, electron capture),  and photon reactions, all play a role. 
Reaction networks as shown in Figure~\ref{fig_Na-Mg-Al-cycle} are expected to describe nucleosynthesis in environments such as interiors of massive stars.

Towards iron group nucleosynthesis, equilibrium conditions are expected to strongly favour $^{56}$Ni as a product, maximising the binding energy of all nucleons for nuclear burning from symmetric matter (consisting of equal numbers of protons and neutrons).
The abundances of neighbouring isotopes such as $^{57}$Ni and $^{58}$Ni then will characteristically be shaped by the conditions where the reaction equilibrium \emph{freezes out}, as neutrons then can be captured  on $^{56}$Ni as the plasma cools down. 

\subsubsection*{Beyond the iron group: neutron capture reactions}
Neutron capture reactions also can occur in cooler environments further away from equilibrium, and produce heavier isotopes from pre-existing \emph{seed nuclei}. This is called \emph{secondary} nucleosynthesis, as pre-existing seed nuclei from \emph{primary} nucleosynthesis are required, which must have been made before gas condensed to form the star where  such neutron captures subsequently occur. 
Such \emph{s process} nucleosynthesis is, e.g., expected to occur in shell burning regions of massive stars. 
A typical reaction sequence is shown in Figure~\ref{fig_60Fe-production}. 
On the other hand, neutron capture reactions also may occur at opposite extremes, where high neutron fluxes ensure that neutrons   can be captured on unstable isotopes while they have not yet $\beta$-decayed. 
 Neutron capture may continue along the isotope sequence of an element to produce ever neutron richer isotopes, until  an additional neutron cannot be bound any more, and the neutron capture sequence only proceeds after $\beta$~decay of this most neutron-rich isotope; this is referred to as the \emph{r process}.
 
\begin{figure} 
\centering
\includegraphics[width=\columnwidth,clip]{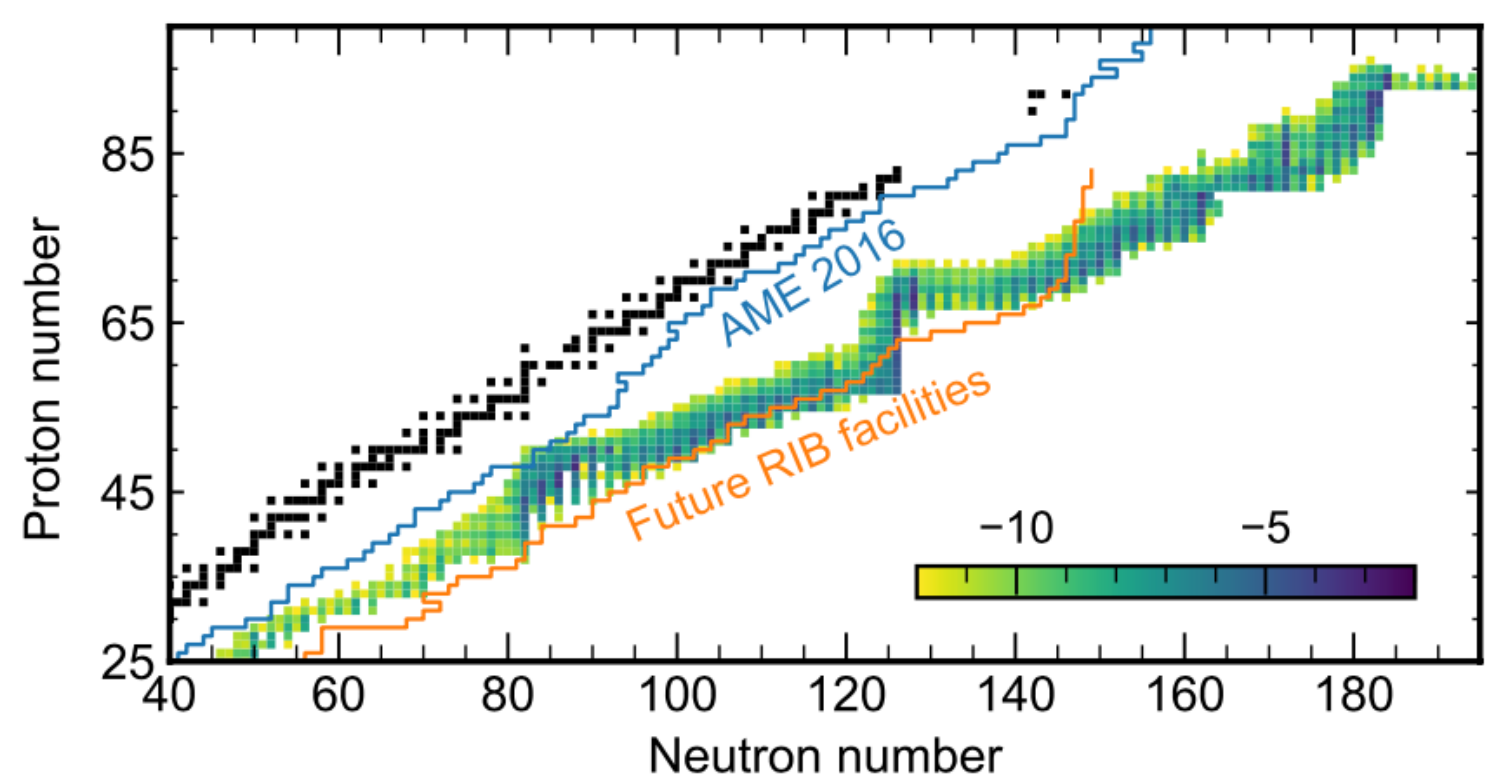}
\caption{The knowledge of nuclear properties as of today (blue line), and as planned to be accessible with future fragments of radioactive ion beam experiments (FRIBs, red line), shown in the diagram of isotopes (cmp. Figure~\ref{fig_tableOfIsotopes}). The reaction path of neutron captures in the r process and including fission recycling are indicated for a specific scenario. (From \cite{Cowan:2021}).}
\label{fig_ToI_rProcess}       
\end{figure}

 The synthesis of elements heavier than iron through neutron capture reactions 
 reveals itself in local maxima at magic numbers of nuclear stability. 
 In the elemental abundance pattern (Figures~\ref{fig_abundances} and \ref{fig_elementOrigins}), these characteristic pairs of local abundance maxima appear around atomic masses 130/140 and 190/215.
The \emph{s~process} of neutron captures, and with time for intermediate $\beta$~decays proceeds along the valley of stable isotopes (example shown in Figure~\ref{fig_60Fe-production}). 
It is thought to occur within stars, as neutron-releasing reactions such as $^{13}$C($\alpha$,n) and $^{22}$Ne($\alpha$,n) occur in He burning shells. 
Neutron-capture cross sections of stable nuclei follow simple rules, so that detailed calculations of s-process nucleosynthesis were considered reliable \citep{Clayton:1968,Schramm:1974,Kappeler:1999,Arnould:2007}.
Subtracting predicted s-process patterns from observed heavy-element abundances, there appears to be a universal pattern, identical in solar and low-metallicity abundance data; this suggests that the cosmic sources of the \emph{r~process} are universal, well regulated and unique \citep{Cowan:1982,Arnould:2007} \citep[see][for a current and refined view]{Cowan:2021}.
Figure~\ref{fig_ToI_rProcess} shows the path of the r process in the table of isotopes, as compared to the location of stable elements, and the regions where our knowledge of nuclear properties ends. 
Environments for the r~process are thought to exist near neutron stars, as they are formed during collapse of a massive star, or as they are destroyed in collisions with other neutron stars or black holes.
From the very nature of such scenarios, a highly-dynamic evolution of environments is to be expected, and explosion kinematics may add complex 3-dimensional plasma flows near compact neutron stars. 
 
\subsubsection*{Nuclear fission}
In the case of a collision of two neutron stars,    
{              one may expect an r~process to occur at an extent that can produce the heaviest elements in the realm of instability towards nuclear fission,
because the collision disrupts the neutron star(s), thus providing an environment which is characterised by extremely neutron-rich nuclei and a large abundance of neutrons.} 
Then, \emph{fission re-cycling} can occur, distributing nucleons back towards lighter isotopes in forms of a variety of fission-product nuclei. 
This re-arranges the seed abundances of the r~process, and thus complicates outcomes of nucleosynthesis.
It is mainly the goal of theory of nuclear stability and of heavy-ion collision experiments to explore the nuclear binding properties of isotopes away from the valley of stability and in the realm where the neutron capture reaction paths of the r~process may fall.
Figure~\ref{fig_tableOfIsotopes} shows the table of isotopes, with arrows indicating the different reaction paths of nucleosynthesis processes. 

\subsubsection*{How to proceed}
Observations play a major role for understanding cosmic nucleosynthesis. 
They are often guided by simulations of the complexities of the cosmic sites; although approximations are necessary herein, such simulations are able to reveal which observational signatures might be teaching us the most about the cosmic-source interior and its dynamical evolution. 
Combining all cosmic messengers leads to a better understanding of cosmic nucleosynthesis and the resulting evolution of the composition of cosmic gas in stars, planets, and interstellar space as well as in space between galaxies.

\section{Understanding the cosmic sources} 
\label{sources}

\subsection{The Big Bang} 
\label{BBN}

Following the hot big-bang origin of the universe, primordial nucleosynthesis begins as the temperature of the early universe has cooled below GeV energies, so that nucleons as particles freeze out of the primordial hot plasma. The abundances of protons and neutrons are then held in a weak equilibrium, as long as the temperature remains above 800~keV, with a number-density ratio 0.17 of neutrons to protons at freeze-out. Free neutrons have a decay lifetime of (880.3 $\pm0.1$) s.
Interactions of protons and neutrons may result in formation (and destruction) of deuterium and He isotopes. But effective nucleosynthesis must wait until the universe cooled down to below the binding energy of deuterium (78~keV), to proceed with the formation of $^3$He as the lightest isotope of helium. 

\begin{figure} 
\centering
\includegraphics[width=0.9\columnwidth,clip]{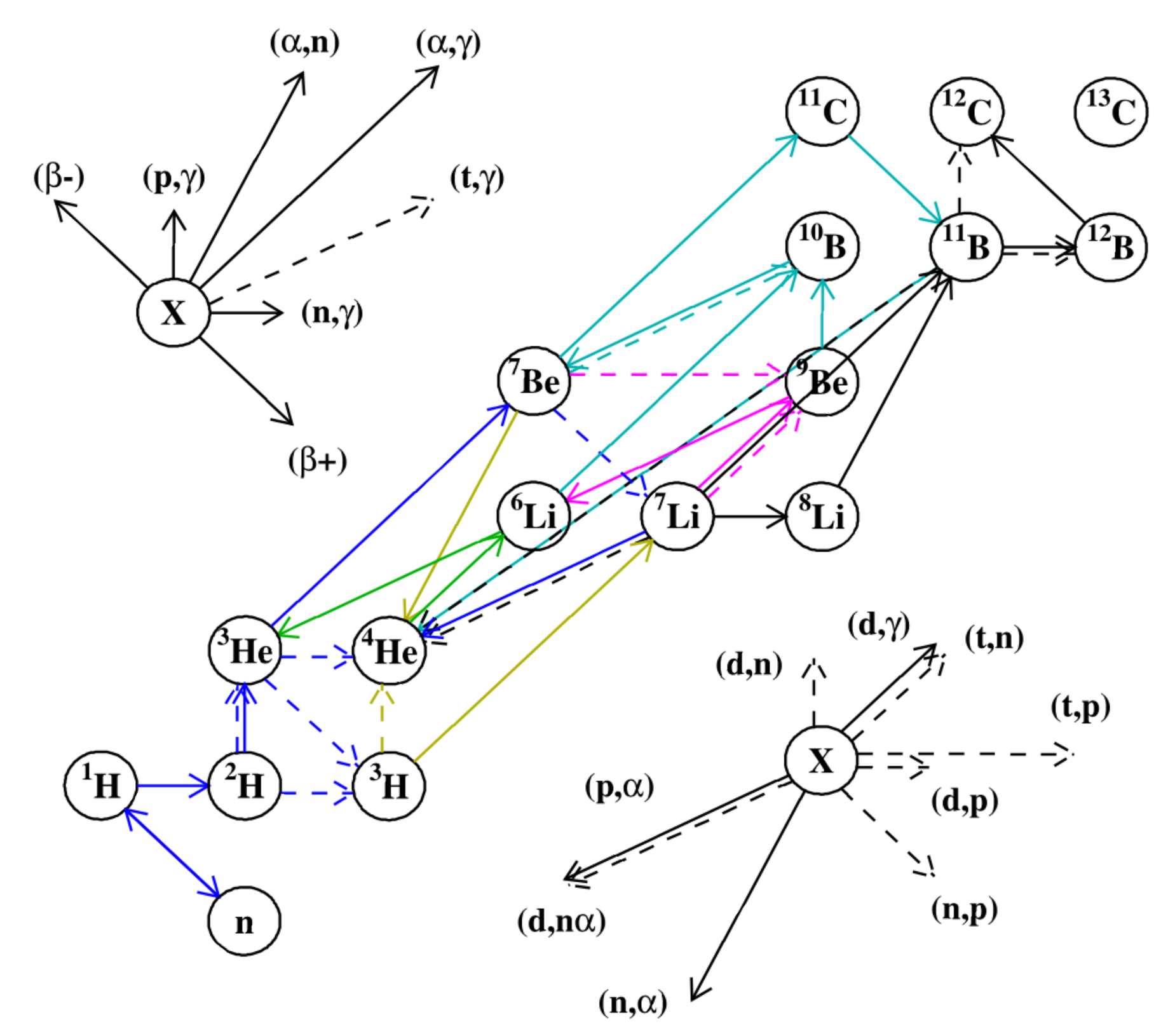}
\caption{The nuclear reactions that are relevant for primordial nucleosynthesis.  \citep[From][see details therein]{Pitrou:2018}.}
\label{fig_BBNreactions}       
\end{figure} 

Figure~\ref{fig_BBNreactions} shows all nuclear reactions that are considered possible and relevant during this period. 
Typical interaction energies are below MeV, which presents the main experimental challenge to measure these reactions in the laboratory and implies the use of low-background environments in underground laboratories \citep{Coc:2015,Pitrou:2021}. 
Beyond He, only trace amounts of Li, Be, B, and C are created in the epoch of big-bang nucleosynthesis. This reflects the non-existence of stable nuclei with masses 5 and 8, so that synthesis of carbon requires the coincident collision of three $^4$He nuclei (\emph{triple alpha reaction}). In the rapidly-expanding universe, this is a rare event.
Below an energy of 10~keV (or about 10$^8$\,K), collision energies are too low for further nucleosynthesis to occur.
This leaves a rather small window for primordial nucleosynthesis, as the universe cools from $\sim$80 to 10~keV. At its beginnings the neutron/proton abundance had fallen to 0.13, from neutron decay.
 
 Nuclear-reaction theory for this reaction network and the rapidly-expanding universe predicts relative abundances for a few key isotopes. 
 These are deuterium $^2$H, the helium isotopes $^3$He and $^4$He, and the lithium isotopes $^6$Li and $^7$Li.

\begin{figure} 
\centering
\includegraphics[width=0.9\columnwidth,clip]{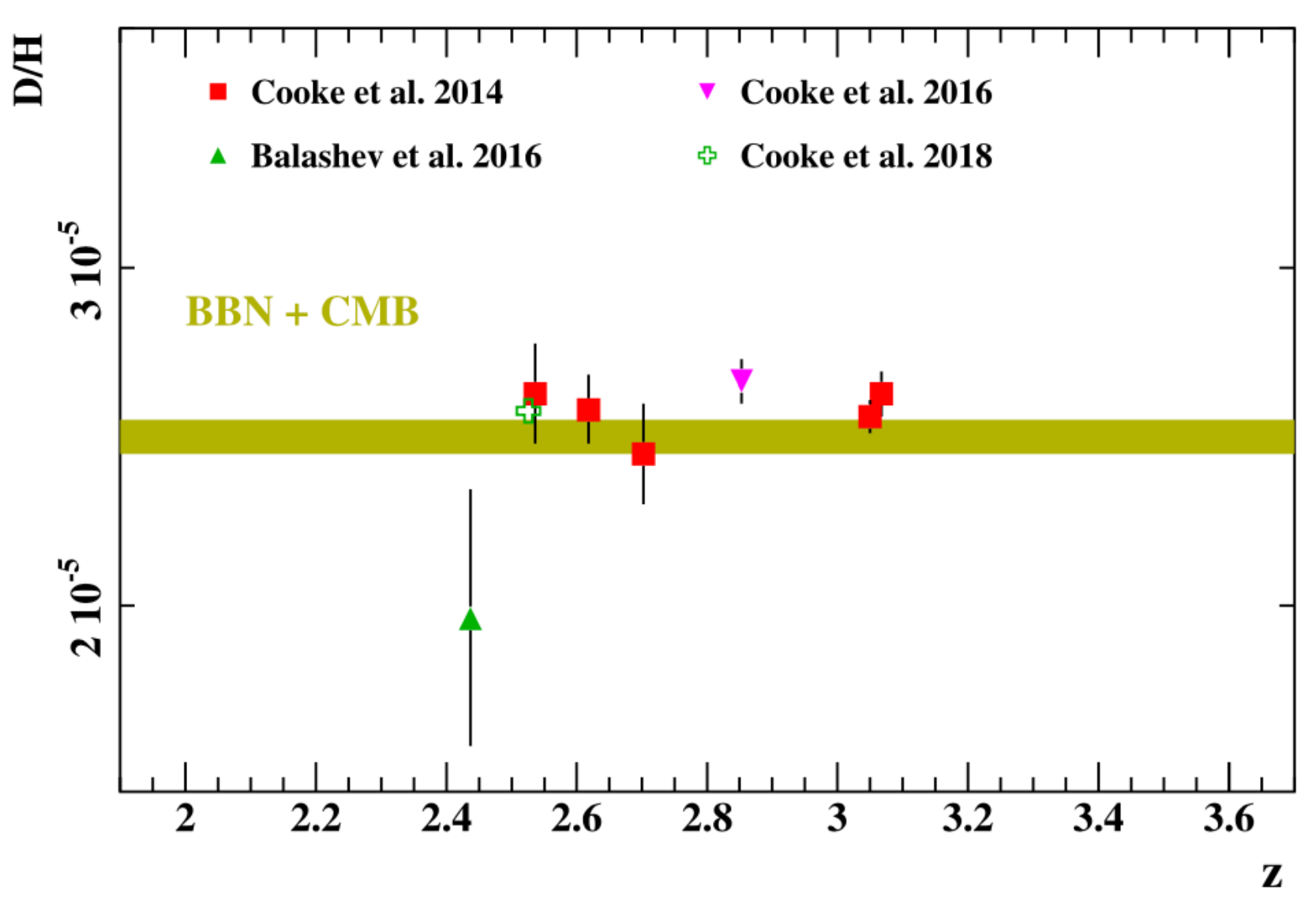}
\includegraphics[width=0.9\columnwidth,clip]{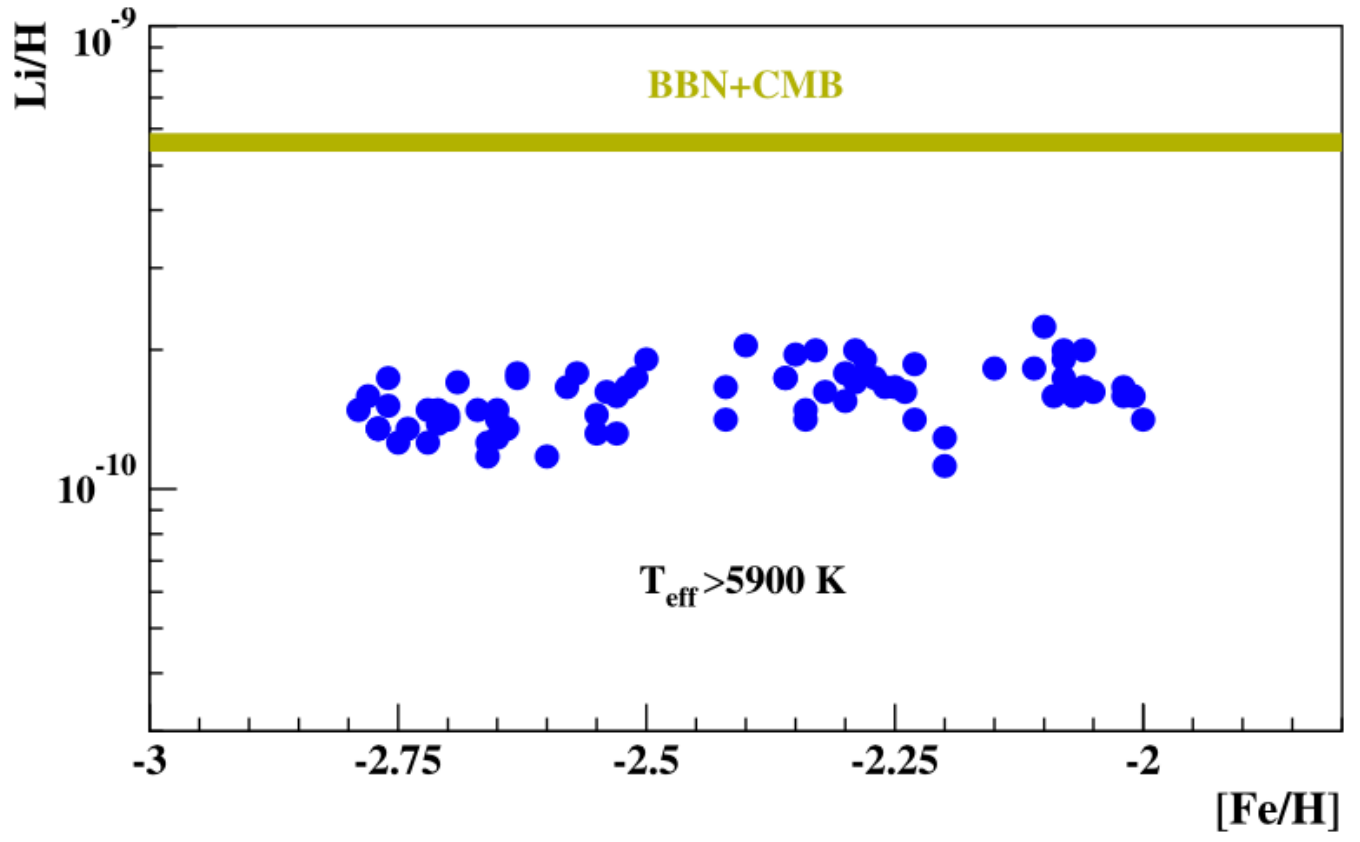}
\caption{Comparison of big-bang nucleosynthesis constraints (horizontal  bands) with cosmic abundances of D/H from high-redshift gas clouds (above) and Li/H from Galactic halo stars (below). \citep[From][ see details therein]{Pitrou:2018}.}
\label{fig_BBNabund}       
\end{figure} 

\begin{figure} 
\centering
\includegraphics[width=0.7\columnwidth,clip]{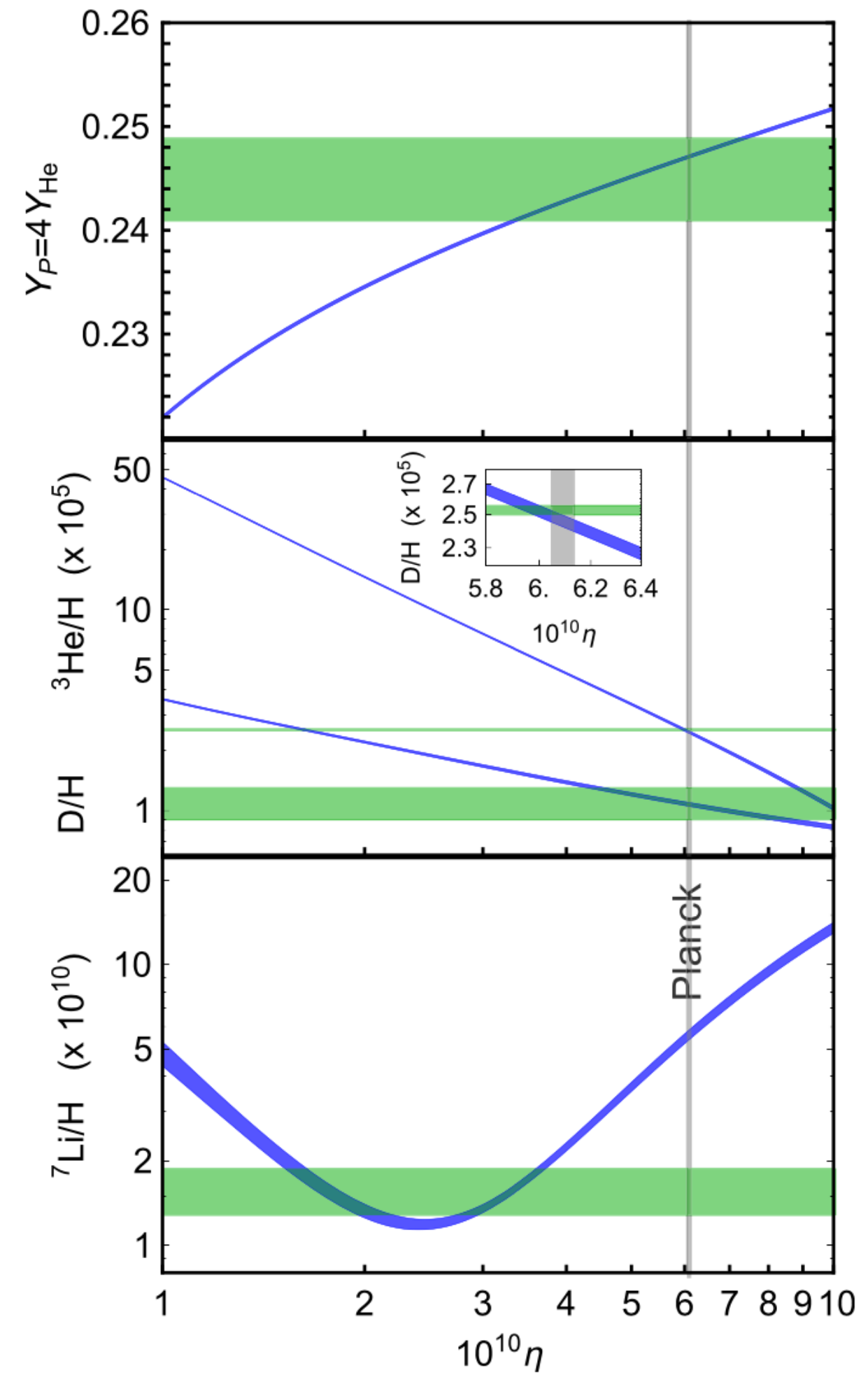}
\caption{The comparisons of observational constraints (horizontal rectangles/lines) to big-bang nucleosynthesis models (blue lines), versus the baryon-to-photon ratio $\eta$. \citep[From][ see details therein]{Pitrou:2018}.}
\label{fig_BBNconstraints}       
\end{figure} 

Deuterium as well as lithium are very weakly bound, and can easily be destroyed in environments that reach temperatures above a few 10$^6$~K. 
The main observational challenge is to find objects that have preserved primordial abundances, and have not been exposed to significant sources of energy. 
Passive material such as cold gas clouds can be identified in quasar spectra. Their redshifts indicate that they are from an early time in the universe. 
The abundances of metals (all elements heavier than helium) grow with time (few very fragile nuclei such as deuterium are the exception to this rule, see below). Low abundance in a well-observable element would be used as a selection criterion for the objects of choice.
Spectral lines from iron are strong even in cold material and guide the search. 
Alternatively, H and He lines could be used, and the requirement for high redshifts of 3 and beyond helps to focus on old and pristine material.
 
Abundances of He have been derived from ionised gas through recombination lines \citep{Cyburt:2016}, from gas clouds in dwarf galaxies at low metallicities. These dwarf galaxies are believed to have had little star forming activity, and hence little ejection of He into their gas reservoir. Models are used to describe line shapes and ratios, such that the thermodynamic conditions of the low-metallicity HII regions can be reproduced. Carefully selecting best-fitting datasets, a metallicity-dependent (using [O/H]) set
 of measurements can be used to extrapolate to zero metallicity, thus determining a best value for primordial $^4$He of $Y_p=(0.2449 \pm 0.0040)$ \citep{Pitrou:2018}. 
 
The deuterium abundance requires selection of passive gas clouds, where absence of any stellar activity suggests that D has not been destroyed substantially. 
The relative abundance D/H thus has been determined to be $(2.527 \pm 0.03) \cdot 10^{-5}$ \citep{Pitrou:2018,Cooke:2018}, from analysis that appeared to show lowest scatter among different objects. Still, such scatter leaves concerns that some systematics from varying D destruction might affect the result. 

The primordial abundance of $^7$Li again can be determined from absorption line spectroscopy of stellar photospheres (as discussed in more detail below for the Sun). 
Care must be taken to account for Li sources other than the big bang nucleosynthesis and Li destruction in stars.
From observed Li abundances versus metallicity, the \emph{Spite plateau} was revealed \citep{Spite:1982a,Spite:1982,Spite:2005}, that suggested one had reached the level of the primordial abundance, while the Li abundances seen in more metal-rich stars included stellar production. 

 With detailed measurements of anisotropies in the cosmic microwave background  \citep{Bennett:2003,Planck-Collaboration:2020}, it became clear that this \emph{Spite plateau} interpretation could not be held up: big-bang nucleosynthesis predictions \citep{Cyburt:2016,Coc:2017} are up to a factor three higher than inferred stellar lithium abundances (see Fig.~\ref{fig_BBNabund}). 
 Refinements of the big-bang nucleosynthesis nuclear reactions over the last decade have increased this \emph{cosmological lithium problem}, rather than helped to diminish it \citep{Coc:2014,Fields:2020}. 
 {        Tracing the cosmic abundance evolution of Li, it is inferred that at most 30\% of solar Li abundance could be attributed to cosmic ray production, while the majority should arise from stellar sources such as novae \citep{Grisoni:2019,Prantzos:2012}. The cosmic-ray production arises from spallation reactions of cosmic rays with interstellar gas, and has been inferred from direct measurements of cosmic rays in near-Earth space \citep{de-Nolfo:2006,Aguilar:2018}. Extrapolating this to low metallicity and primordial Li abundances, we face \emph{the Li problem} of observing less Li than predicted by BBN. Hence, the solution ought to lie in stellar processes or early-Galactic
evolution systematically lowering observable lithium abundances at low
metallicities \citep[see also][]{Prantzos:2017}.
 }
 
 It is not understood how surface lithium could have been reduced in stars on the Spite plateau over the 10$+$ billion years of their existence. The outer convection zones of these stars contain little mass (a few \% of the total mass), which may not be representative of the material the stars were formed from. 
 Refined stellar-structure models including the effects of
 {        pre-main sequence pre-main sequence destruction and re-accretion \citep{Fu:2015},} of stellar-internal atomic diffusion \citep{Michaud:1984}, and of mixing below the outer convection zone \citep{Richard:2005}, obtain surface depletions of order a factor two, partly closing this discrepancy. Given the vast range of length and time scales involved in stellar-evolution calculations, convective-boundary mixing can not be modelled from first principles. It currently needs to be parametrised and constrained observationally. 
 Stars in globular clusters can provide constraints on atomic diffusion and mixing. Ideally, one uses heavy elements observed in stars of different evolutionary phases (and thus differently affected by internal diffusion) to estimate the mixing efficiency; with this model one then makes an inference for the more fragile element lithium. In subgiants of the nearby globular cluster NGC 6397 ([Fe/H]=$-2.1$), a specific lithium signature has been found and interpreted as lithium being mixed up into the atmosphere, as the outer convection zone of stars in this phase of evolution deepens \citep{Korn:2007,Lind:2009,Korn:2020}.  This is the most-direct evidence for lithium diffusion in stars that are located on the Spite plateau. Trends of lithium abundance with the stars' mass lend further support to this semi-quantitative picture \citep{Melendez:2010}, while the partial \emph{meltdown} of the Spite plateau at metallicities below [Fe/H] $\simeq -3$ remains challenging to explain \citep{Sbordone:2010}. A subpopulation of relatively high-mass halo stars has recently been identified which seems to display lithium abundances in agreement with BBN predictions \citep{Gao:2020}.
 
 There were also claims of an observed plateau of $^6$Li abundances for metal-poor stars, alike the Spite plateau of $^7$Li, orders of magnitude in excess of BBN predictions \citep{Asplund:2006}. If confirmed, such a plateau would support discussions towards non-standard big-bang nucleosynthesis \citep{Olive:1993,Jedamzik:2008}. However, the $^6$Li lines are not resolved in stellar spectra due to thermal and convective motions. 
 Rather, they are seen as a line-shape asymmetry in the $^7$Li doublet. Convection has been shown to cause very similar asymmetries in spectral lines. 
 More detailed studies 
 rule out a $^6$Li plateau in Galactic-halo stars at the current level of detectability  \citep{Lind:2013, Wang:2021}.
 
 Adopting predictions of big-bang nucleosynthesis and the baryon-to-photon ratio derived from the Planck-mission data, there are still other possibilities to explain the lithium discrepancy: non-standard big-bang nucleosynthesis (lowering predicted lithium, but leaving in particular deuterium unchanged), a cosmic process globally lowering lithium (possibly in connection with Population III stars, \citep{Piau:2006}) or a better understanding of stellar lithium depletion (diffusion and destruction, before and after the stars reach the main sequence). 
 Just like the abundance of primordial helium can only be correctly inferred by taking into account the stellar contributions to the helium emission lines in HII regions, stellar lithium abundances need to be corrected for effects of stellar evolution. High-redshift deuterium abundances provide the most-direct access to the era of big-bang nucleosynthesis. 
Performing such measurements at even higher redshifts might reveal systematics not captured in the assessment of the few ($<$10) measurements performed to date.
 
\subsection{Stars} 
\label{stars}
 
\begin{figure} 
\centering
\includegraphics[width=0.9\columnwidth,clip]{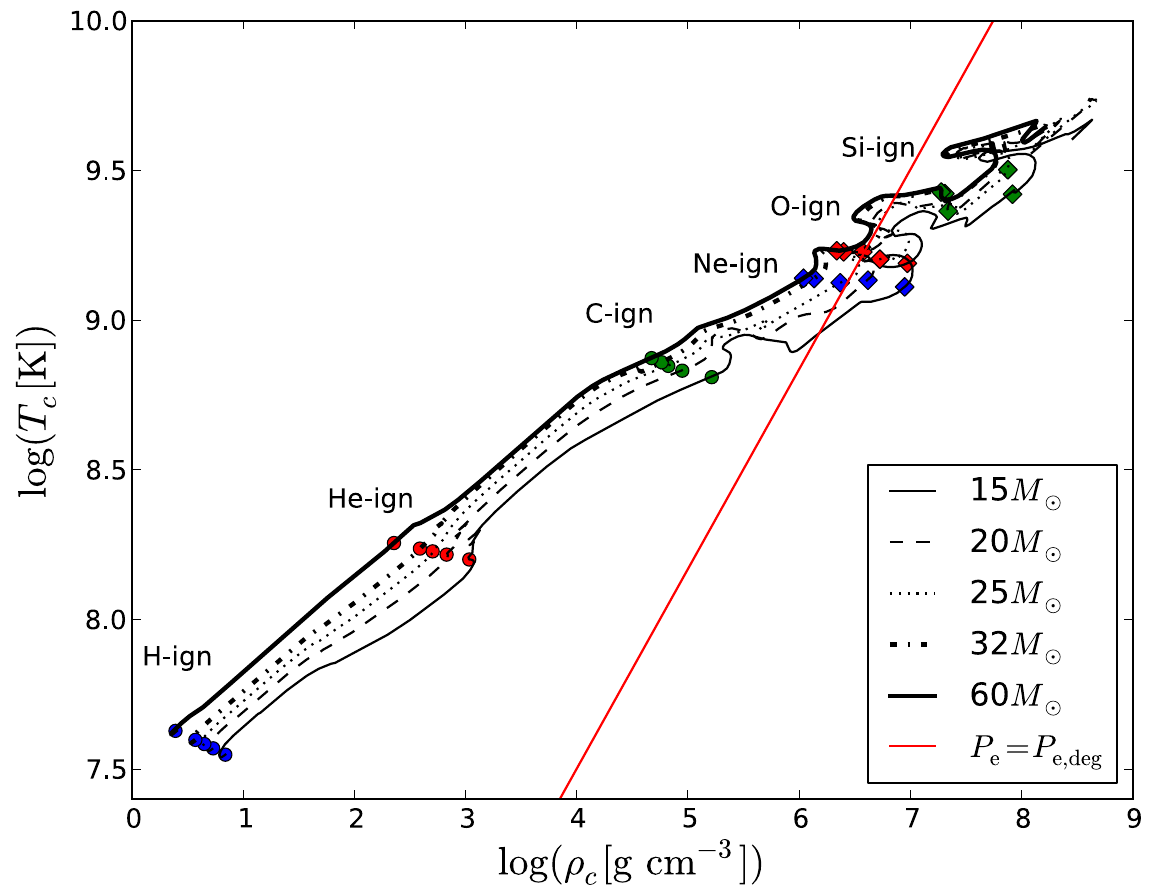}
\caption{The phase diagram for the evolution of stellar cores, through the different burning stages from hydrogen burning until silicon burning before gravitational collapse. The red line shows the separation between thermal and degenerate plasma (towards bottom right), where in the latter the pressure is dominated by Fermi pressure of electrons in degenerate gas. (From \cite{Pignatari:2016}).}
\label{fig_massiveStar_temp_dens}       
\end{figure} 

 Following up on Eddington's hypothesis that subatomic energy is the   source powering stars, stellar models were developed and refined towards higher degrees of realism.
 Today, we understand that stars are assemblies of gas held together by gravity and in hydrostatic equilibrium. 
 Density and temperature are highest in the center, and this is the region where conditions where nuclear reactions take place and hence the release of nuclear binding energy from the transformations of nucleons among different isotopes that are made through nuclear reactions.    
   
Taking this model further through evolution over the nuclear time scale (i.e., when the H fuel is exhausted), further energy releases occur as different reaction thresholds are overcome in a contracting, hence compressionally-heated, star. Stan Woosley called this a \emph{gravitationally-confined nuclear reactor}{        \citep{Woosley:2002}}, a concise description of a star in terms of astrophysics.
Figure~\ref{fig_massiveStar_temp_dens} \citep{Pignatari:2016} shows the phase diagram of massive stars in the temperature-density diagram in log-log units, and illustrates the diagonal path of stars as they evolve through the different core burning stages, from main sequence to silicon burning.   
The more-rapid evolution of later burning stages compared to the main-sequence phase of hydrogen burning is largely due to the increasing involvement of neutrino-emitting reactions and thermal creation of neutrinos, leading to enhanced loss of parts of the nuclear energy that is released. 
The wiggles in the evolutionary path for late stages then reflect deviations from hydrostatic equilibrium, as energy release is stronger and locally disturbs equilibrium conditions.

The theoretical modeling of the hydrostatic evolution of massive stars nowadays is encoded in different simulation tools 
\citep[e.g.][]{Paxton:2011,DeglInnocenti:2008,Maeder:2000}. 
The uncertainties in massive-star evolution are manyfold, including most importantly aspects and drivers of mixing in the stellar envelope and around the different burning shells. 

\subsubsection*{The Sun}
Towards consolidating stellar modelling for an extraordinarily well-observed star,
a specific modelling variant has been developed that most precisely represents our Sun \citep{Jorgensen:2018}: In this case, observations can be a most powerful constraint to validate, or improve, modelling of hydrostatic stars. 
Although photons that reach us from the photosphere of our Sun have lost all specific information about their origins in the energy release from nuclear reactions of hydrogen fusion to helium in the center of the Sun, these photons have been a key to the study of stellar interiors and to cosmic nucleosynthesis.
First generation models for the processes inside the Sun were gradually refined, and have reached impressively detailed treatments of stellar rotation and of convection. Important determinations of uncertainties were achieved \citep{Bahcall:2006},  in particular also bounds for key observational parameters, specifically for the  surface helium abundance, the profile of the sound speed and density versus radius, the extent of the convective zone in the outer envelope, the neutrino fluxes from all burning reactions, and the fractions of nuclear reactions that occur in the CNO cycle or in the three branches of the p-p chains.
These modelling efforts were brought to yet another level of precision \citep{Vinyoles:2017}, when new insights into atomic radiative properties (laid down in consolidated databases \citep{Iglesias:1996,Seaton:2007,Delahaye:2021}) 
could be exploited for precision modelling of the variations of opacities along the solar radius, and after improved nuclear reaction rates had become available. 
In these recent models, specifically two opposing modelling scenarios were evaluated in detail, a \emph{low Z} and a \emph{high Z} model of surface abundances (where Z denotes the mass fraction of all "metals", elements with atomic mass $>$~4).
The solar composition had become a renewed focus, after a sequence of studies since the turn of the century \citep{Asplund:2009} had re-analysed observational values and uncertainties for photospheric abundances, now accounting for atmospheric effects from granulation and 3D characteristics with significantly-improved  atmosphere model,  in addition to also accounting for deviations from local thermodynamic equilibrium and the influence of different atomic data tables for transition probabilities for atomic species.
This new generation of models solves the physically appropriate set of coupled equations (the radiative-transfer equation and the Navier-Stokes equations)
in order to predict the stellar surface flux (or even the specific intensities for spatially resolved observations of the solar surface). This is a numerical and computational challenge. But this approach advances our knowledge beyond the assumptions necessary in one-dimensional models, e.g.\ to model an inherently three-dimensional energy-transport process such as convection, using stability criteria. 
These new models are virtually without free parameters (unlike the 1D models which, e.g., need micro- and macro-turbulence to describe spectral-line broadening).

Hydrodynamic (3D) models have undergone rigorous testing, and generally improve upon all previous semi-empirical or theoretical models. 
Important consistency tests include spectral-line shifts and asymmetries as well as atmospheric velocity fields (as caused by convective motions), the center-to-limb variation of spectral lines, and comparing different diagnostics such as low- vs.\ high-excitation, atomic vs.\ molecular lines, for a specific element (e.g.\ oxygen). 
Thermodynamic equilibrium is a common assumption for gas inside stars, but can only adopted locally in stellar atmospheres (\emph{local thermodynamic equilibrium, LTE}. 
Non-equilibrium (\emph{non-LTE}) line formation plays a key role here, even if departures from LTE are often small in the Sun, and LTE is still assumed to be valid when constructing the atmospheric model. 

However, in the treatment of opacities, i.e.\ how atoms and their absorption lines affect the radiative energy transport, the new models might still be inferior to the classical one-dimensional stellar-atmosphere models: While some classical models treat opacities at 100,000 wavelengths points \citep{Gustafsson:2008}, the 3D models need to group similar opacities into a few opacity bins for computational reasons; comparisons indicate that the changes to the model structures are modest, but efforts need to be made for further improvements. 

\begin{figure} 
\centering
\includegraphics[width=\columnwidth,clip]{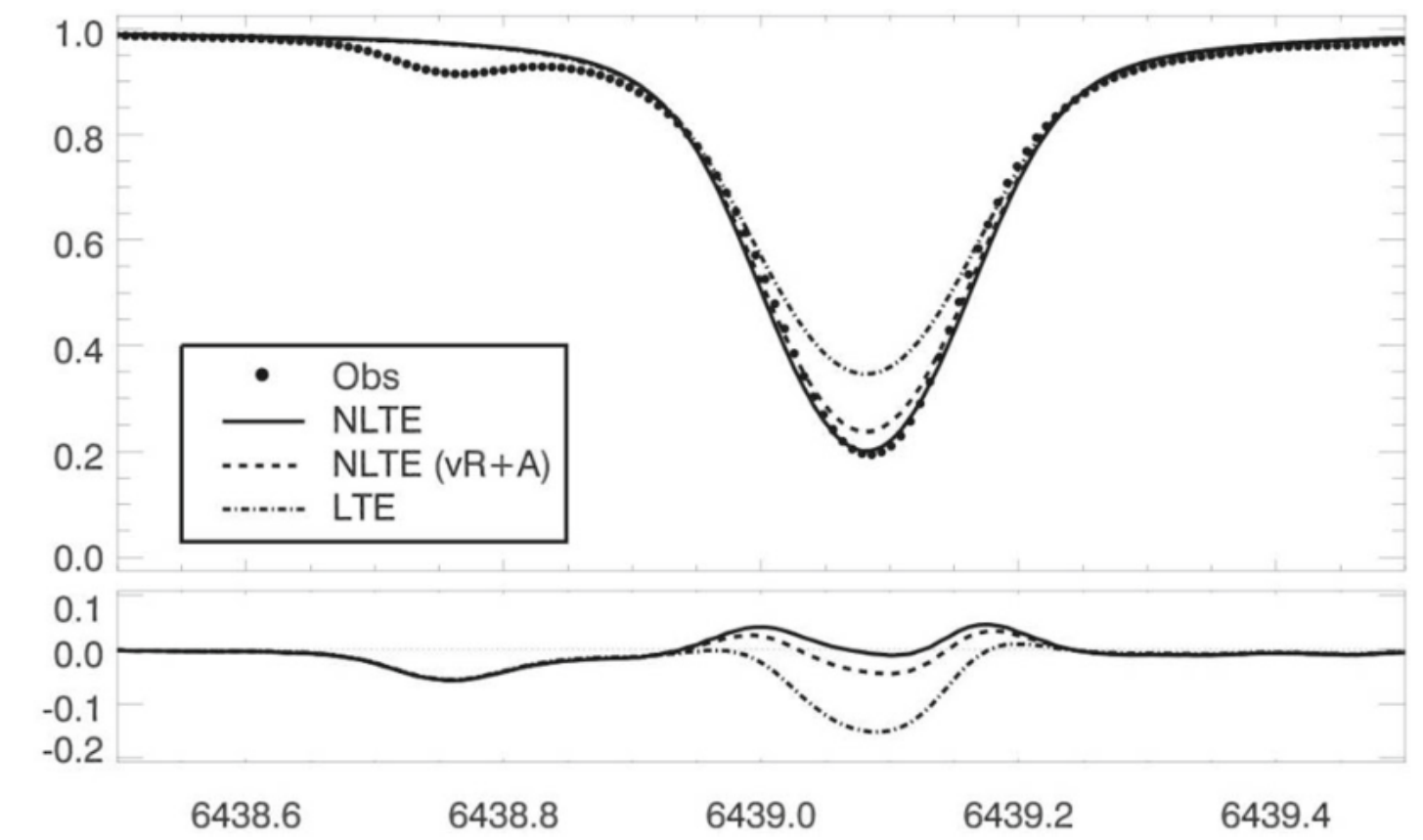}
\caption{The Ca I line at 6439 \AA\ in the solar spectrum is a strong unblended absorption line. It forms over a wide range of photospheric depths. Interpreting the measurement with different models and assumptions about the level populations of Ca atoms and their interactions results in different values for the derived Ca abundance. (From \cite{Osorio:2019}).}
\label{fig_solar_spectrum}       
\end{figure}

Spectroscopy of sunlight shows characteristic features (Figure~\ref{fig_solar_spectrum}), which are caused by selective absorption of the underlying thermal radiation by atoms in the solar photosphere. Absorption lines occur when photons resonate with the atomic excitation levels of an elemental species; therefore, the strength of such absorption lines reflects the abundances of the respective element in the atmosphere of the Sun.
In this way, spectroscopy of solar photons  allows to measure abundances of most elemental species in the Sun, basically unchanged since the formation of the Solar System, about 4.6~Gy ago.  

\begin{figure} 
\centering
\includegraphics[width=\columnwidth,clip]{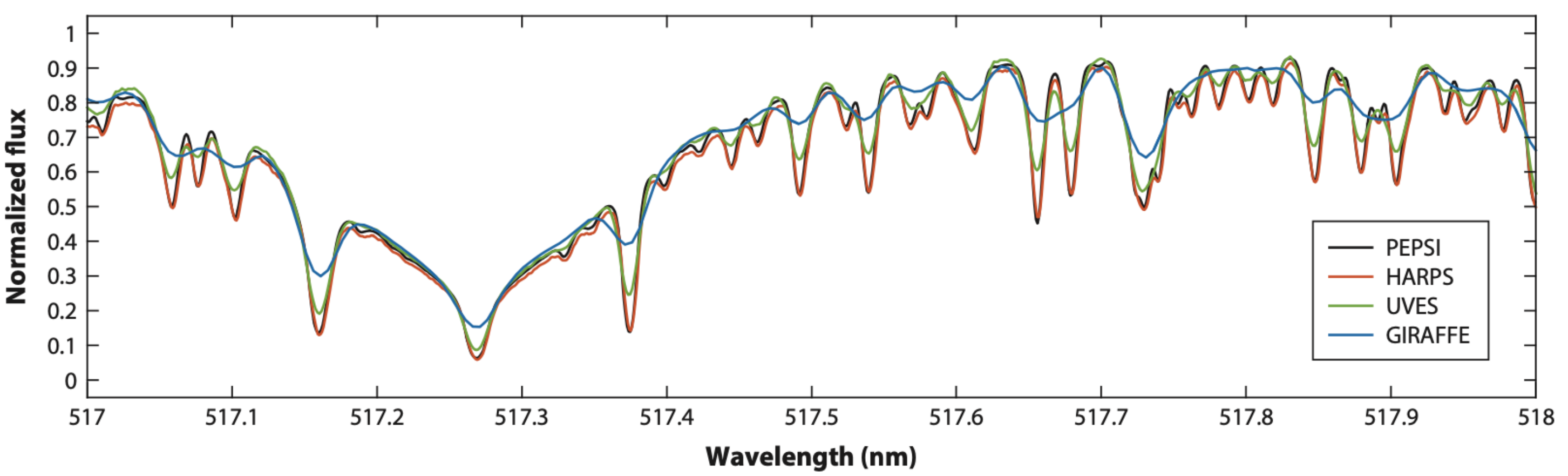}
\caption{The spectrum of the benchmark star $\epsilon$~Eri as recorded with four modern medium-to-high-resolution spectrographs. The signal-to-noise ratio is high facilitating a reliable continuum normalisation. The spectra then mainly differ due to the finite resolution of the spectrograph used. To resolve the individual spectral features requires high resolution.  (From \cite{Jofre:2019}).}
\label{fig_starlight_spectrum}       
\end{figure}

Spectroscopy of starlight in general, and 
telescopes and their spectrographs in particular, have been improved since their beginnings in the 19$^{th}$ century. 
Today's sophisticated instruments show an impressive precision of spectroscopic data (see Figure~\ref{fig_starlight_spectrum}). Unfortunately, exquisite data do not necessarily imply properly-inferred chemical abundances.

\begin{figure} 
\centering
\includegraphics[width=\columnwidth,clip]{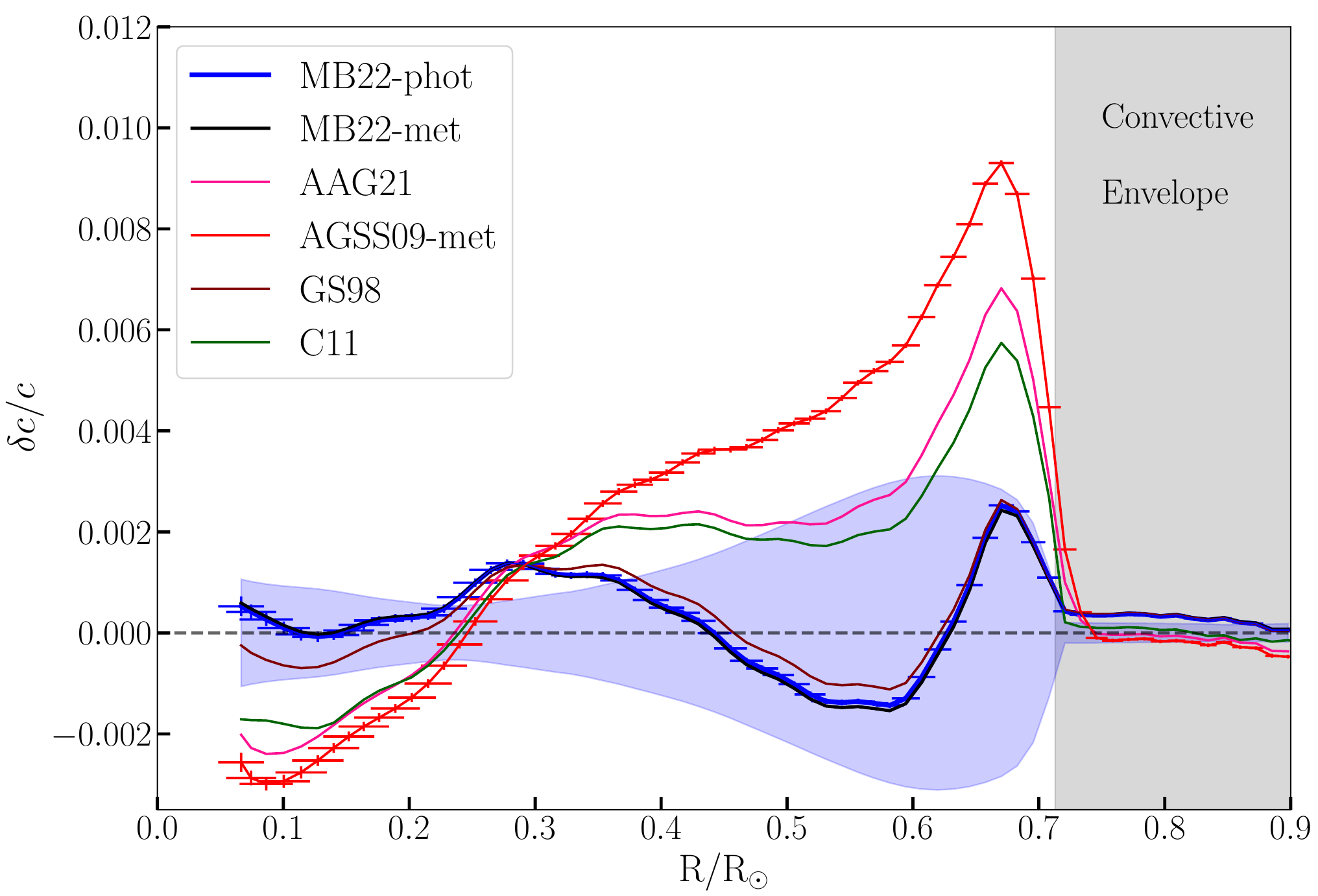}
\caption{The sound speed profile differences between different solar-model predictions and the inversion of the solar-oscillation measurements from heliosesmology. Models have been computed using different solar gas compositions: \citep[][C11]{Caffau:2011}, \citep[][AGSS09-met]{Asplund:2009}, \citep[][GS98]{Grevesse:1998}, \citep[][AG21]{Asplund:2021},  \citep[][MB22-phot and MB22-met]{Magg:2022}. Model uncertainties are represented in shaded blue, while error bars denote uncertainties from helioseismic data and their analysis. (Figure taken from \citep{Magg:2022} as updated from \cite{Vinyoles:2017}).}
\label{fig_solar_soundSpeed}       
\end{figure} 

As one of the main results of the application of hydrodynamical models to the interpretation of solar spectra, CNO abundances in the solar atmosphere have been found lower by almost 30\%,  which results in a solar metallicity Z\,$\simeq$\,0.014 \citep{Asplund:2021}
compared to the earlier agreed and widely used solar metallicity of $\simeq$\,0.019 \citep{Anders:1989}. In particular, the oxygen abundance has decreased from $\log$ (O/H) + 12 = 8.93 $\pm 0.035$ to 8.69 $\pm 0.04$ over the course of the last 30 years. 
Naturally, this transition was not uncontroversial. Mainly two independent teams debated abundance revisions back and forth (\citep{Asplund:2005} vs.\ \citep{Caffau:2008}), but it can now be concluded that the downward revision is basically correct, with remaining differences -- on a much smaller scale than the revision itself -- reflecting different preferences and choices by the spectroscopist teams. As the case of the forbidden oxygen line at 630~nm shows, sometimes the revision of a line-blending contribution makes all the difference \citep{Scott:2009}. 

\begin{figure} 
\centering
\includegraphics[width=\columnwidth,clip]{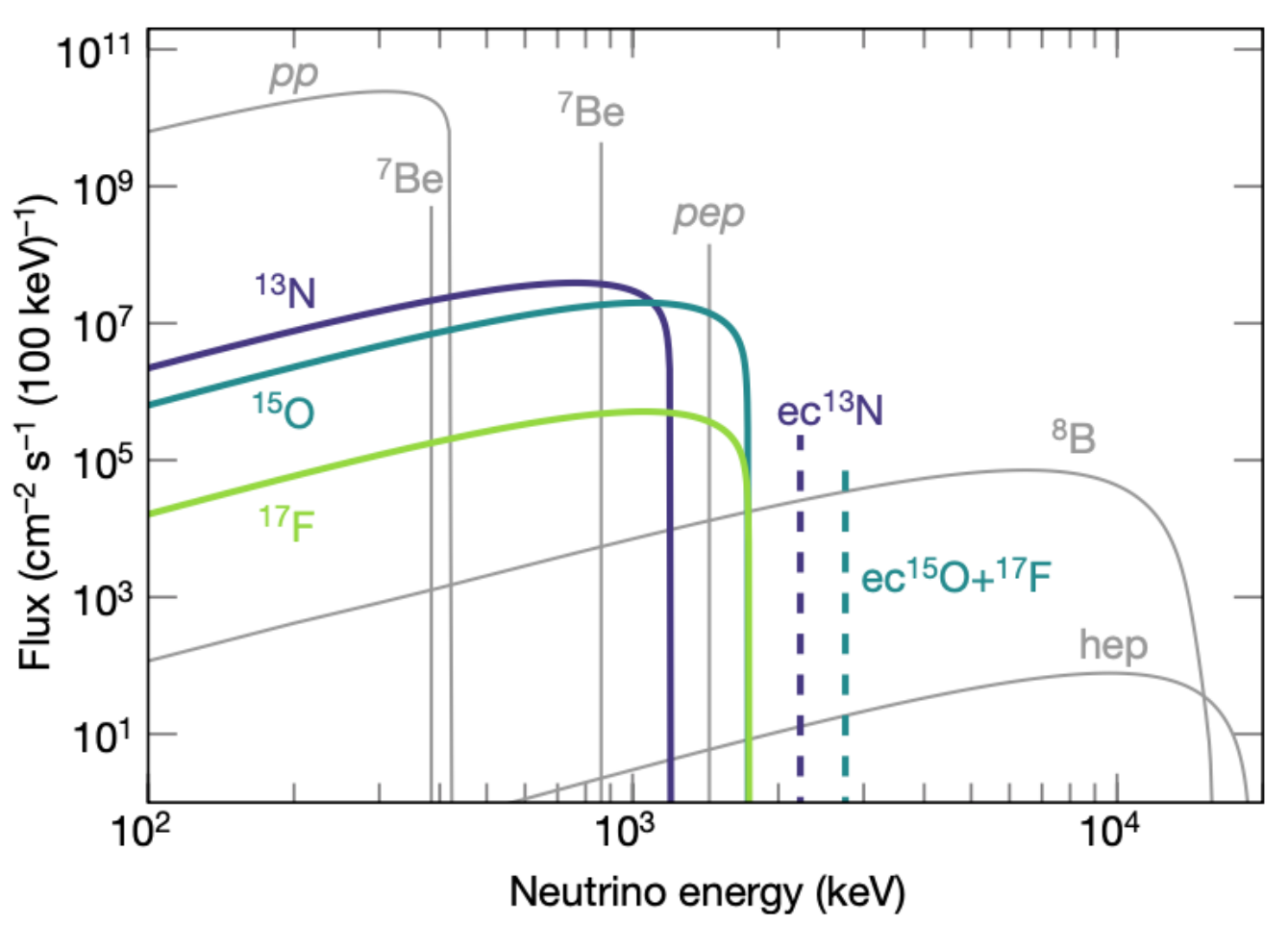}
\caption{The solar neutrino spectrum, identifying the spectra of neutrinos from each of the weak reactions that are part of hydrogen fusion in the Sun due to proton-proton and CNO reaction chains. (From \cite{Borexino-Collaboration:2020}).}
\label{fig_solarNeutrinos}       
\end{figure} 

The quest for a realistic and consistent solar model continues. Comparisons of the solar model with data reveals a significant disagreement of the low-metallicity model (about 3$\sigma$; see Figure~\ref{fig_solar_soundSpeed})
with the sound velocity as inferred from helioseismology measurements around the transition between non-convective and convective energy-transport zones in the Sun.
\emph{Helioseismology} or, more generally, \emph{asteroseismology} is a fairly recent observational discipline for astrophysics, which helps to understand the interior structure of a star, such as the Sun, and its evolution \citep{Aerts:2021,Christensen-Dalsgaard:2021,Basu:2010}. 
The roots of this method lie in the \emph{period-luminosity} relationship for Cepheid stars,  (\emph{Leavitt} law, discovered through Henrietta S. Leavitt's variable-star surveys in the 1900$^s$ \citep{Leavitt:1912}), which enabled the use of Cepheids as distance indicators. 
The physical background are the oscillatory resonances that may occur in a star, as its internal structure is bounded by regions with different stiffness; the laminar/convective boundary in the outer envelope is the most-significant such transition along the solar radius. 
The modes of oscillations caused by such internal structure are propagated to the surface in the forms of luminosity variations at the photosphere, or of gas motions of gas in the upper atmosphere that converts into variations of Doppler shifts. 
Essentially, deviations from sphericity in the baseline hydrostatic model of stellar structure remove degeneracy of oscillatory modes as it would appear in a solid sphere, and result in modes that depend on internal (differential) rotation as well as convective behaviour varying within the star \citep[see][for the method to deconvolve helioseismic observations]{Basu:2010}. 
Gradients in density and pressure are thus related to the sound velocity, and a linear expansion around the standard model for the Sun then relates measured variations to a difference in modelled versus observed sound velocity. This is shown in Figure~\ref{fig_solar_soundSpeed}, as a function of solar radius, highlighting the lower boundary of the Sun's outer convective zone.  
The position of the lower boundary of the outer convection zone depends on the composition of solar gas through opacities caused by metals, as shown in the Figure. Remarkably, the improvements in surface abundance determinations from solar spectroscopy (as described above) increased the mismatch between helioseismic data and the solar model predictions, as shown in Figure~\ref{fig_solar_soundSpeed}.

\begin{figure} 
\centering
\includegraphics[width=1.0\columnwidth,clip]{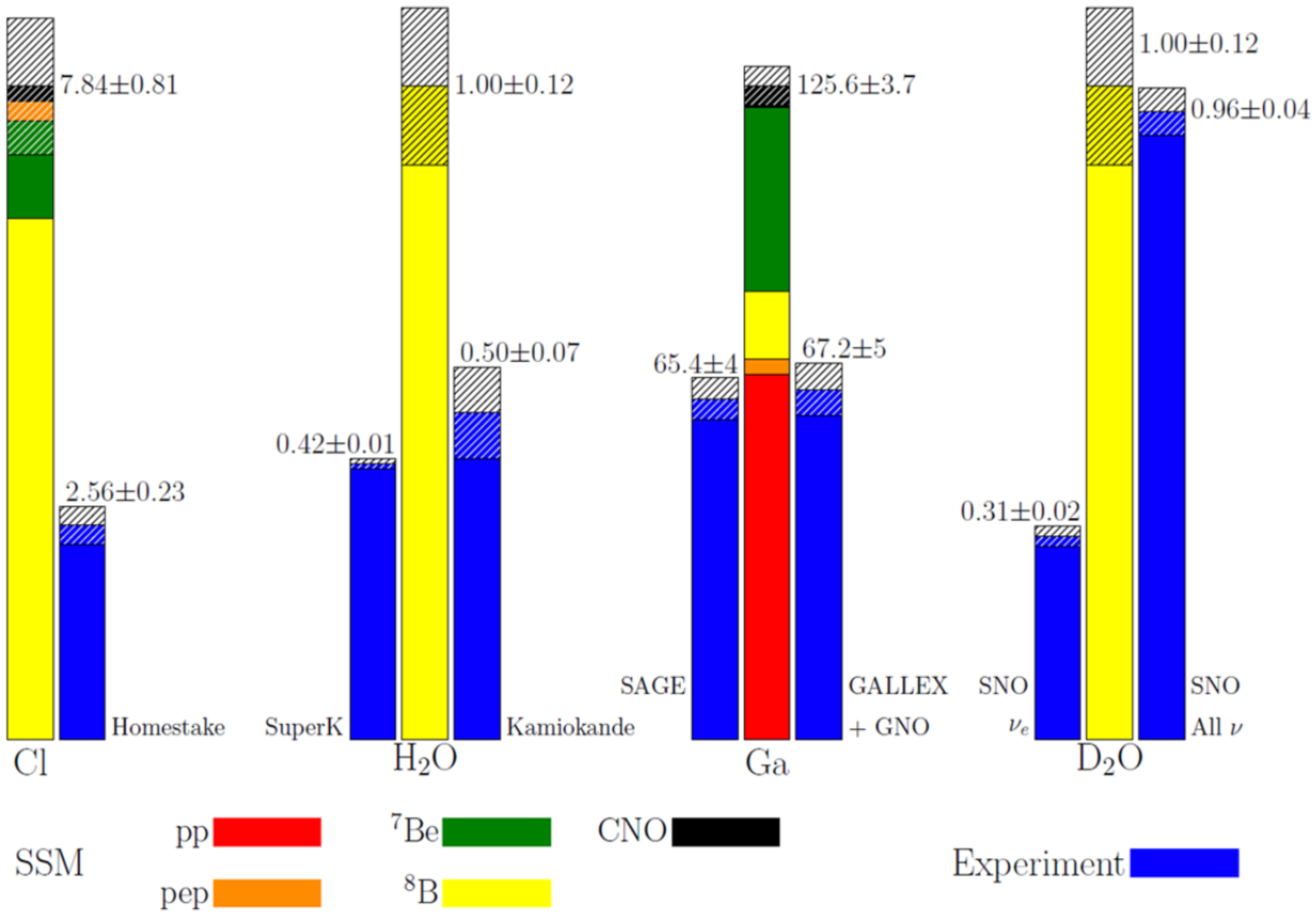}
\includegraphics[width=1.0\columnwidth,clip]{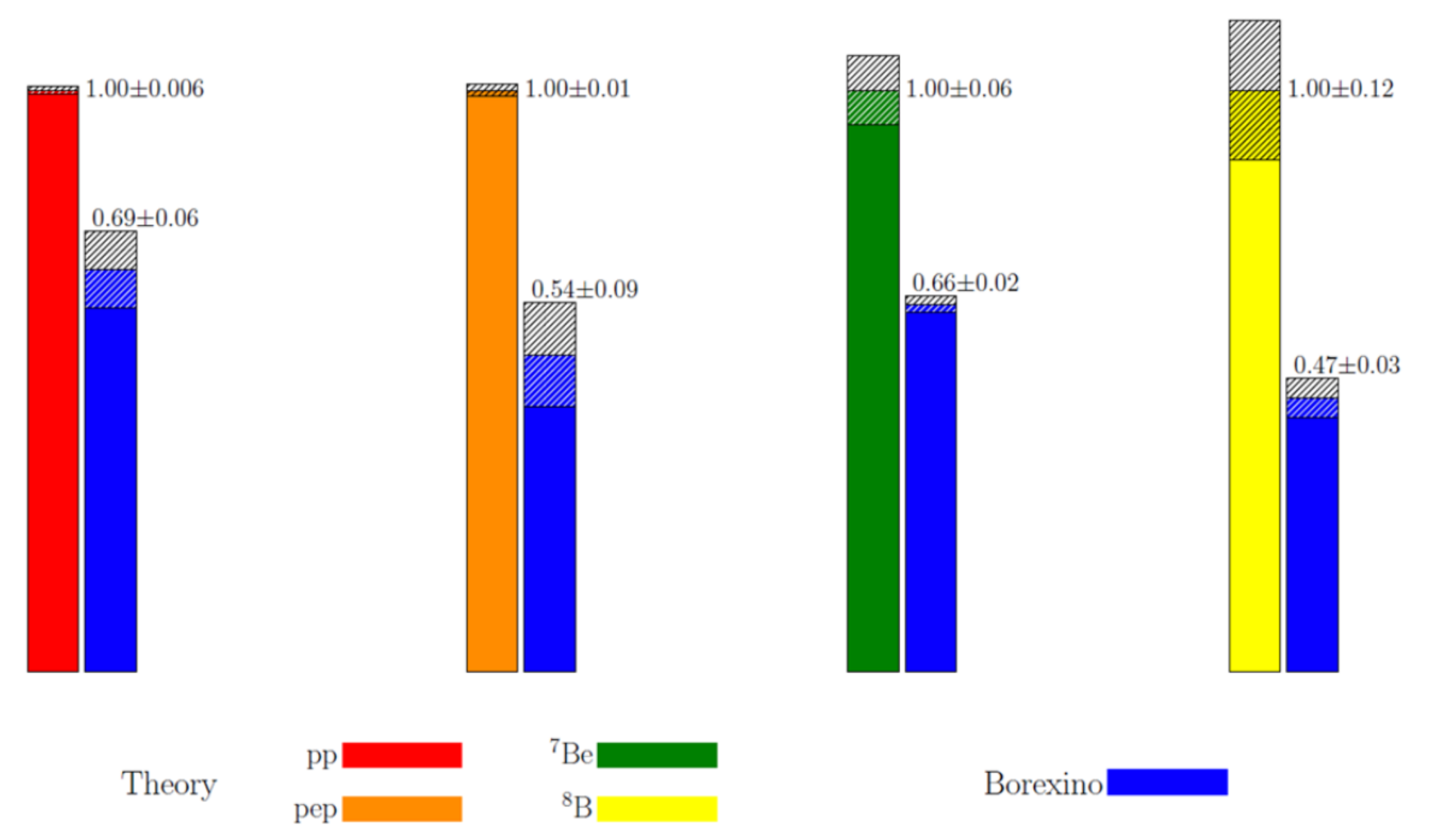}
\caption{Solar neutrino fluxes, measured from different experiments. 
The top panel shows the chemical experiments of the first and second generation, where after a collection time the detector material was analysed for neutrino reactions to have altered the composition creating characteristic isotopes. The second panel below shows results from Borexino for the different hydrogen fusion reaction chains. All experiments record a rate that is substantially below the theoretical predictions; this demonstrated the requirement for neutrino oscillations. (From \cite{Christensen-Dalsgaard:2021}).}
\label{fig_solarNeutrinoFlux}       
\end{figure} 

\begin{figure} 
\centering
\includegraphics[width=\columnwidth,clip]{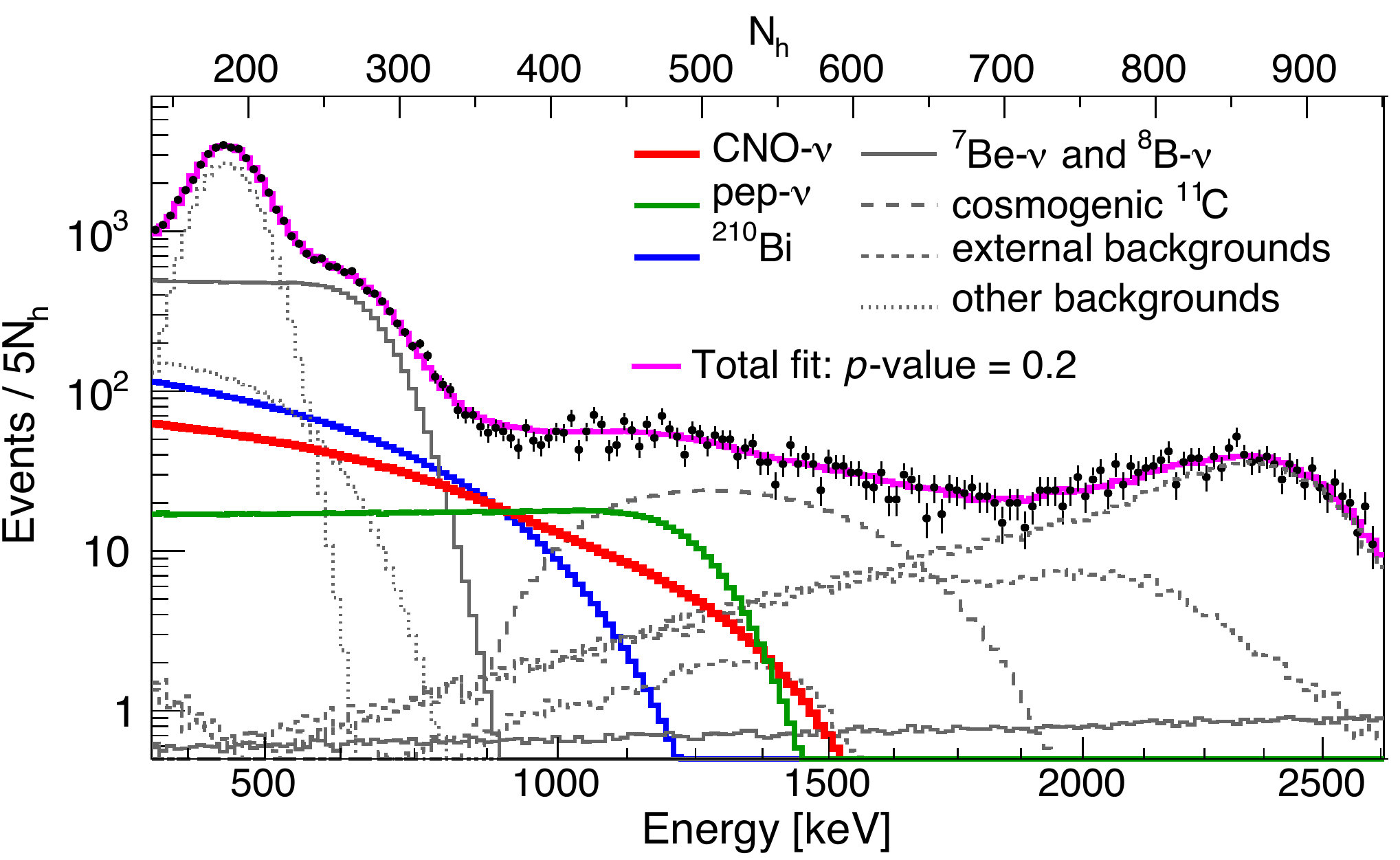}
\caption{The CNO component of solar neutrinos as measured with Borexino. 
The red line shows the CNO neutrino component as fitted in the analysis, the purple line marks the total fit including all background types. The best signal to noise is obtained in the yellow-marked region (see text).
(Updated recently by \citep{Appel:2022} from \citep{Borexino-Collaboration:2020}).}
\label{fig_CNO_Neutrinos}       
\end{figure} 

Observations of stellar interiors are not possible through electromagnetic radiation: The high optical depth of the envelope corresponds to a photon propagation time of 10$^5$y from center to surface in the Sun, due to a large number of individual absorption, re-emission and scattering processes.
Neutrinos, however, have tiny interaction cross sections with matter, and leave the dense core of the Sun. 
The \emph{Borexino} neutrino detector is located underground under more than 1~km of rock in the Gran Sasso mountains of Italy, and is today's most-sensitive detector for neutrinos in the MeV range where nuclear-reaction neutrinos occur. It consists of 278~tons of a hydrocarbon-based liquid scintillator, viewed with 2212 photomultiplier tubes. An outer fiducial volume of the scintillator tank is used as a veto counter to ensure that only scintillation events created from neutrinos inside the inner talk, rather than from any other particles penetrating from the outside, are registered.
With this setup,  the neutrino flux from the dominating proton-proton reaction cycles of hydrogen fusion could be measured at high significance \citep{Bellini:2012,BOREXINO-Collaboration:2014,Borexino-Collaboration:2018}.
Recently, also neutrinos from the \emph{CNO burning} cycle could be discriminated against the background \citep{Appel:2022,Borexino-Collaboration:2020}, as shown in Figure~\ref{fig_CNO_Neutrinos}. Although this cycle only contributes $\sim$1\% of the hydrogen fusion energy in the Sun, it has been part of our solar model, and is now experimentally verified at 5.1$\sigma$ significance to occur  \citep{Borexino-Collaboration:2020}; 
the inferred count rate is 7.2~($+2.9-1.7$) (68\% uncertainties from likelihood analysis) counts per day and 100~tons of scintillator liquid.
The next goal is to advance precision such that a measurement of metallicity in the central regions of the Sun may be achieved through CNO neutrinos.

Another way to determine the abundances of elements and isotopes in solar matter is provided by the laboratory analysis of the chemical and isotopic composition of meteorites, interplanetary dust or cosmic rays. This method analyses sample material in terrestrial laboratories and is free from the potential systematic uncertainties related to models of the solar atmosphere, as needed in spectroscopy of sunlight.

Laboratory analysis of meteorites can provide accurate information on \emph{both} stable elemental \emph{and} isotopic abundances in the presolar nebula. This method analyses  the stable elements and isotopes of meteoritic material. The meteorite studies probe the elemental and isotopic composition at the time the solar system formed 4.6 Gy ago.
In a complementary analysis method, accumulated interplanetary and interstellar dust can be analysed for characteristic radionuclide composition. Interstellar radionuclides if detected before they have decayed away, probes recent nucleosynthesis events. 

Mass spectrometry measures the isotopic composition of physical samples with very high precision. This method is by nature destructive and consumes part of the sample material during the measurement. Conventional mass spectrometry almost exclusively measures stable atoms as isotope ratios. A complementary method is accelerator mass spectrometry (AMS) which also allows to measure signatures of interstellar origin. This method is detailed in chapter 4.

 
While there are also meteorites from Mars, the Moon, and Vesta, most meteorites effectively are the remains of broken-up asteroids \citep{Ciesla:2006a}. 
Iron and stony-iron meteorites (roughly 7\% of all) are understood to be the melted interior cores of their parent asteroid, while the other, stony, meteorites instead represent the asteroid's crust. 
Stony meteorites further divide into chondrites (roughly 86\% of all) and achondrites (roughly 7\%), differentiating if they have \emph{chondrules} (chondrites), or they do not  (achondrites). Chondrules are small ($\sim$ 1 mm sized) spherical inclusions, which are clearly visible in the meteorite's matrix, the finer-grained, often microscopic, mass of material in which the larger grains and structures are embedded. 
Chondrules are believed to have formed very early in the presolar nebula, accreted together, and remained largely unchanged since then \citep{Ciesla:2010}. 
To have a most-useful meteoritic sample of the material that made the Sun, one needs the meteorites that  can be considered to have formed earliest from the pristine gas and have undergone very little chemical modifications after that. 
Carbonaceous chondrites ($\sim$6\% of falls) are therefore considered near-ideal, because they contain a lot of organic compounds, which indicate that they experienced very little heating -- some were never heated above 50~degrees~C. 

By comparing the elemental composition derived from the spectra of sunlight and from the analysis of meteorites, one obtains an extremely close match for elements where bias from systematics are unlikely. The main difference is that meteoritic data are derived from direct laboratory analysis and, although they may still suffer from systematic uncertainties due to the particular sample and its chemistry \citep{Dauphas:2016}, their abundance data are more precise than those from solar spectroscopy \citep{Lodders:2020}. 
It needs to be considered, however, that the abundances of the major elements, such as C, N and O, as well as H and the noble gases, cannot be derived from meteorites because these volatile elements do not fully condense into rocks. 

\begin{figure} 
\centering
\includegraphics[width=\columnwidth,clip]{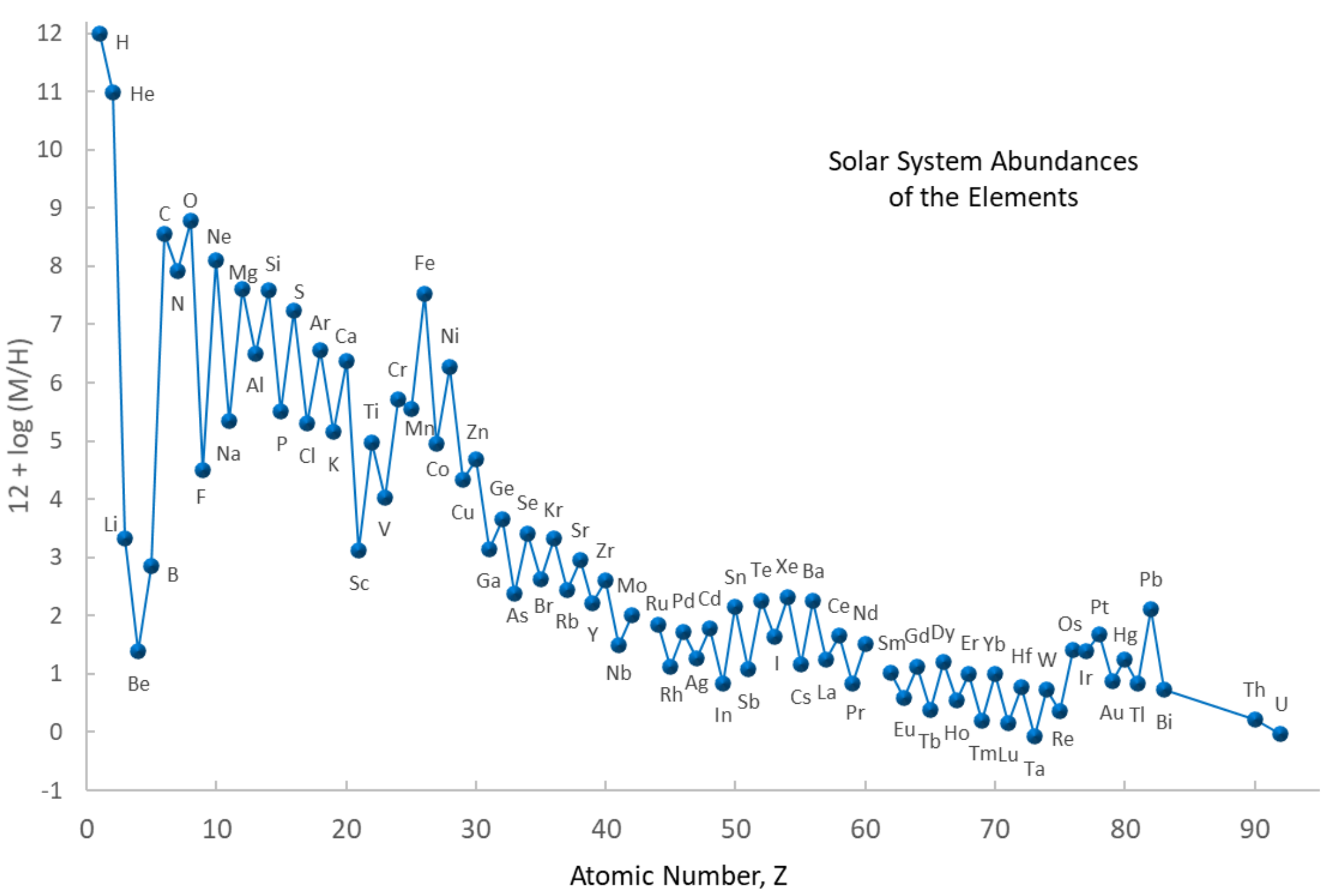}
\caption{
The elemental abundances of solar-system material as obtained from absorption-line spectroscopy of the solar photosphere, supplemented by analysis of the solar-wind, of terrestrial materials, and of material from primitive meteorites that have formed early after the formation of the Sun.
(From  \cite{Lodders:2020}).}
\label{fig_solarAbundances}       
\end{figure}

The first table of elemental and isotopic solar abundances based on meteoritic analysis was published in 1956 \citep{Suess:1956}. Since then, many compilations have been presented and updated, with a recent fully updated discussion of this history, of the analytical methodology, and a revision of solar abundances, as shown in Figure~\ref{fig_solarAbundances} \citep{Lodders:2020}. 

Isotopic abundances are crucial for nuclear astrophysics because they reflect the nuclear processes that produced them (see Section 2 above). 


In spite of these various challenges, present-day chemical abundances in the solar photosphere are nowadays determined at a precision down to 0.03 dex \citep{Asplund:2021,Asplund:2009}. 

\begin{figure} 
\centering
\includegraphics[width=\columnwidth,clip]{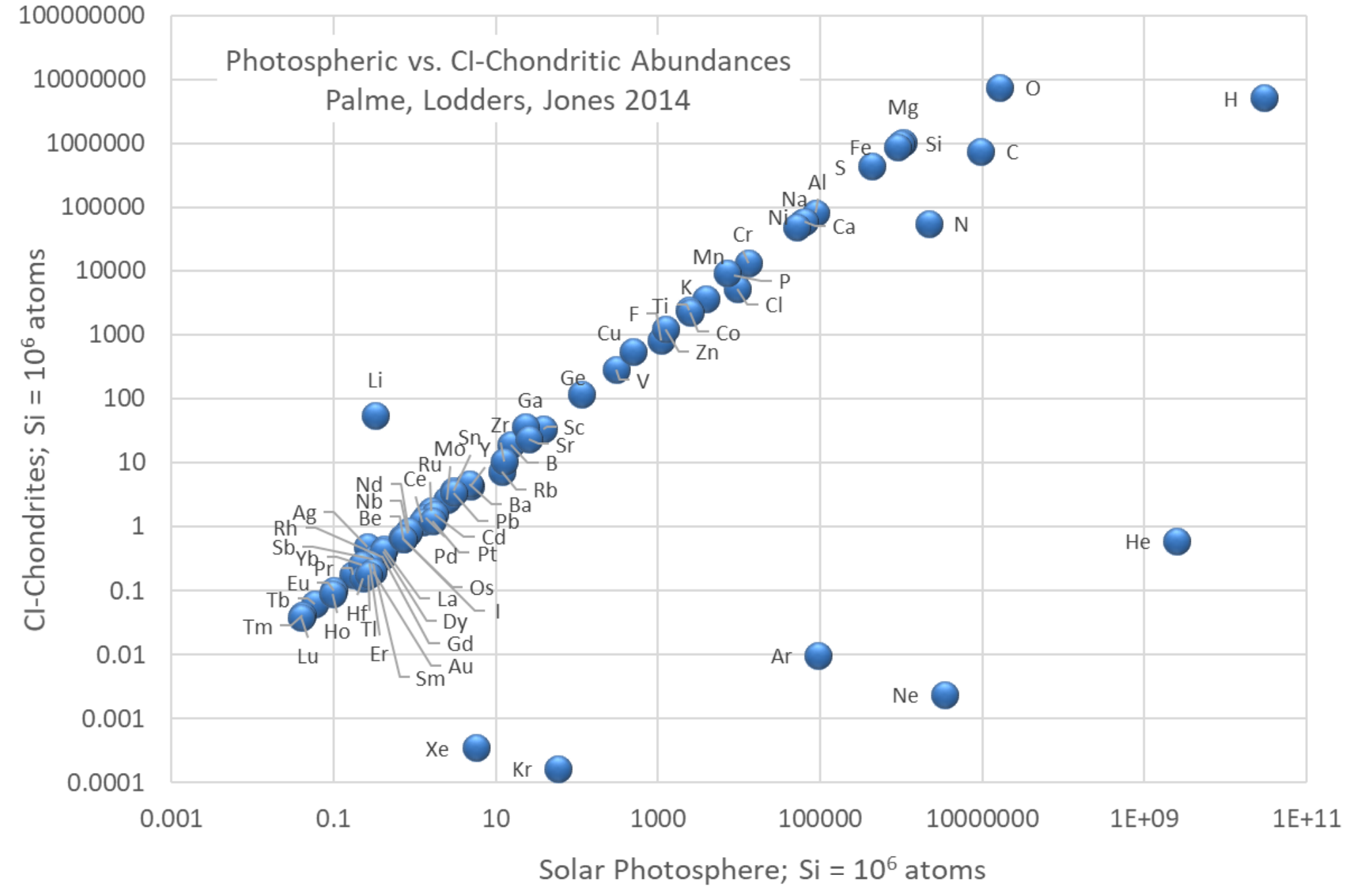}
\caption{The elemental abundances of solar-system material can be obtained from absorption-line spectroscopy of the solar photosphere, and from materials captured and analysed by mass spectrometry from solar-wind samples, from terrestrial materials, and from meteoritic samples. (From \cite{Lodders:2020}).}
\label{fig_solarAbundances_metPhot}       
\end{figure} 

The set of abundances  as shown in Figures~\ref{fig_abundances} and \ref{fig_solarAbundances} are, by necessity, composites from several astronomical messengers.
Photospheric abundances are complemented by solar-wind isotopic abundances, such as the analysis of samples collected with the recent \emph{Genesis} space mission \citep{Heber:2021},  and by the detailed richness of elemental and isotopic abundances extracted at great precision from meteorites.
The solar system provides us with this rich detail of abundances, that serves as a pivotal point for studies of cosmic nucleosynthesis. We take these to be representative of the solar system at birth, 4.56 Gy ago. 
Combining meteoritic and photospheric results, we now have a solid basis of elemental and isotopic abundances, deviations mostly attributed to known biases in the respective messenger \citep{Lodders:2020,Lodders:2021,Asplund:2021} (Figure~\ref{fig_solarAbundances_metPhot}).

\subsubsection*{Beyond the Sun}
Naturally, we are interested to find out how abundances evolved, from solar birth until the present time and of gas in the current interstellar medium in our vicinity, but also for stars other than the Sun, which formed from interstellar gas at their different locations and times.
Spectroscopy as a remote-sensing technique allows astronomy to reach out into the more distant parts of our Galaxy and beyond. 

One would wish to see nucleosynthesis in action, following the increase of surface abundance of an element as it is synthesised inside a star. The long timescales of stellar evolution make this an almost impossible task, and work-arounds have to be found to study nucleosynthesis by approximations. One such \emph{''proxy"} is comparing stars sharing some properties of evolution. Such objects can be binary stars or stellar associations and clusters. The former class of objects has given us important constraints on the s~process in AGB stars \citep[CEMP-s stars, see][]{Beers:2005}, as AGB stars have dynamic complex spectra the modelling of which suffers from large uncertainties and biases; the latter class has been key to piece together a comprehensive picture of stellar evolution, one of the biggest achievements of 20th-century astrophysics. Another proxy is to study the integrated effect of chemical enrichment through generations of stars. The number of generations contributing to the enrichment of a specific star can vary greatly: from potentially one (as discussed in this Section 3) to many thousand (as discussed in Section 4). At the low end we can hope to see the imprints of individual nucleosynthesis events, while at the high end we will measure ensemble-averaged abundances as produced by polluters from a range of stellar masses (i.e.~weighted by the initial mass function). This is the situation in the Galactic thin disk to which the Sun belongs.

The solar neighbourhood has been charted in stars and their respective abundances during the past century, allowing a first discussion of chemical evolution of our nearby part of the universe \citep{Edvardsson:1993,Fuhrmann:1998}.
(This aspect is discussed in more detail in Section~\ref{chemEv} below.) Progress in infrared spectroscopy has allowed the study of stars in farther regions of the disk, towards the Galactic center and in the Galactic bulge \citep{Hayden:2015,Bensby:2017,Zoccali:2018}. 
Abundances have been assembled for many elements (see Figure~\ref{fig_stellarAbundElements}) by modern spectrographs, as exemplified in Figure~\ref{fig_starlight_spectrum}. High spectral resolution is key to isolate, as much as ambient velocities in the light-emitting regions allow, the individual spectral fingerprints of the various atoms and ions.

There are limitations and constraints on how well abundances can be extracted from spectroscopic data. On the one hand, physical details such as the existence or deviations from local thermodynamic equilibrium, 3D motions of gas in the photosphere, and species-dependent photosphere descriptions provide challenges for modelling atmospheres with state-of-the-art tools. Opacities need to be binned or sampled, and ideally departures from local thermodynamic equilibrium considered in an iterative fashion. At present, model atmospheres for solar-type stars rarely consider magnetic fields and are generally constructed assuming local thermodynamic equilibrium. 

On the other hand, radiation transport through the stellar atmosphere should ideally be calculated tracing radiative transfer at the microscopic level, i.e.\ not relying on equilibrium-thermodynamics assumptions which break down in the very layers from which we receive the photons. This requires lots of atomic data: e.g.\ level energies, transition probabilities, photo-ionisation as well as collisional-interaction cross sections. For rarer species, such data are often incomplete or of insufficient accuracy for quantitative spectroscopy, especially in the infrared. Here, astronomy depends heavily on new data provided by laboratory spectroscopists with high-temperature furnaces and theorists performing ab-initio quantum-mechanical calculations 
\citep[e.g.][for recent progress]{Barklem:2016}. 

Classical methods of abundance determinations from spectra, such as techniques based on 1D model atmospheres and equivalent-width measurements, are often too imprecise and depend on questionable assumptions \citep[see][for details on achievable precision of abundance determinations from stellar spectra]{Bergemann:2019}. However, careful differential abundance techniques have revealed interesting differences between the Sun and solar twins. In such analyses, one performs a line-by-line comparison to the solar spectrum and this approach helps to cancel most systematic errors at least for stars very similar to the Sun. One interesting result is the finding that the Sun is not fully representative of local sun-like stars when it comes to the abundances of volatile elements like C, N, O, S and Zn. These are found to be enhanced by $\leq$ 20\% in the Sun \citep{Melendez:2009}. Such departures are a hot topic in contemporary astronomy, as they may reveal planetary signatures \citep{Amarsi:2019}. If this is the case, then we have ever more reason to question the simple-minded picture of stellar atmospheres as perfect time capsules. In reality, the stellar surface layers reflect the sum of all internal and external processes acting on the material since gas accretion in connection with star formation started. Quantifying and correcting for these processes is the forefront of contemporary stellar modelling \citep{Deal:2020,Booth:2020}.

Novel inference techniques which connect stellar observables and stellar parameters have been developed in recent years \citep[e.g.][]{Ness:2015,Ting:2019,Fabbro:2018}. 
Complex algorithms herein train a regression model on previously analysed training sets with relevant stellar labels \citep{Ness:2018}, rather than modelling the matter-light interaction for every object. 
The main advantage of these machine-learning approaches lies in the fact that, after a careful training phase,  large samples of stellar spectra (100,000+)  can be handled very efficiently  \citep[e.g.][]{Buder:2018}.
Variants of these algorithms also make use of genetic hierarchy analyses \citep{Jofre:2019}.

Massive stars are hot enough so that their spectrum reaches higher ionisation levels of gas in their upper photosphere and chromosphere. This results in emission lines from highly-ionised species. 
In particular, it has been a success of high-resolution X-ray spectroscopy to determine abundances of C, N, O, S and Si and to simultaneously constrain the ionisation of the chromospheric plasma. 
This, in turn, constrains the wind and mass-loss properties of massive O and B stars, which are key parameters for their evolution (see above).


\begin{figure} 
\centering
\includegraphics[width=\columnwidth,clip]{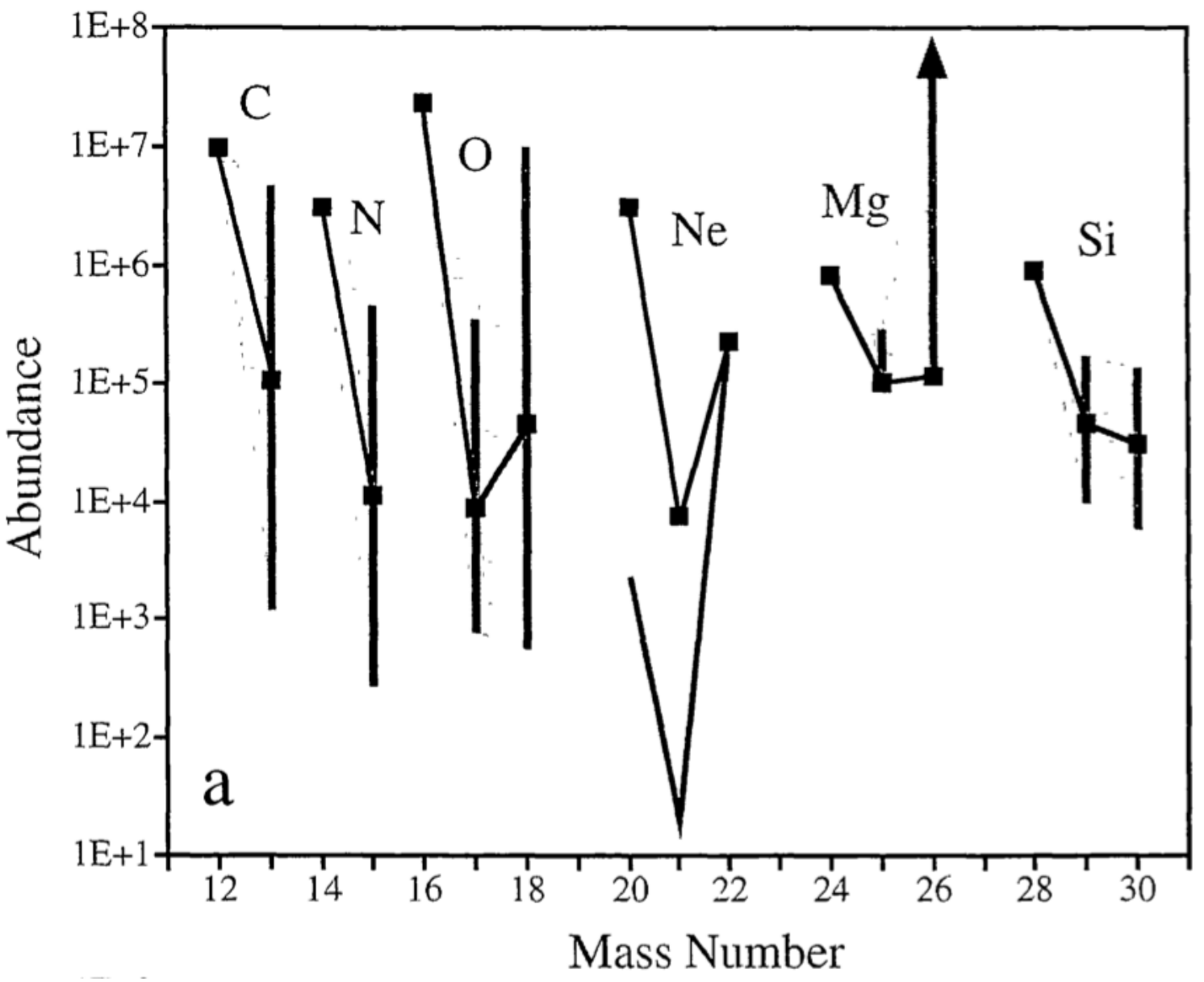}
\caption{The isotope abundances of solar-system material for elements up to Si, with the range of isotopic abundances marked as vertical bars (relative to the most abundant one) for presolar grains found in meteoritic materials. (Figure from \cite{Zinner:1998a}).}
\label{fig_presolarGrainAbundances}       
\end{figure} 

  
Material of extrasolar origins have been collected from meteorites \citep{Nittler:2016,Zinner:2008}, and also directly in space 
\citep{Westphal:2014a} (for the analyses of terrestrial sediments see Section~\ref{recycling_local} below). 
These are often called \emph{stardust} or \emph{pre-solar grains}.
Such pre-solar grains are identified to be of stellar origin due to their very different isotopic composition: while meteoritic materials generally follow the isotopic composition of solar system material to within a few percent, isotopic ratios in stardust grains exceed this range, sometimes even by up to several orders of magnitude. For example, the carbon isotopic ratio $^{12}$C/$^{13}$C in the solar system is 89, while in stardust it can vary from $\sim$2 to $\sim$10,000; the solar  $^{16}$O/$^{18}$O of $\simeq$500 is $>$ 10,000 in some stardust instead -- see Figure~\ref{fig_presolarGrainAbundances} for more examples \citep{Zinner:1998}.
These variations are too large to be attributed to chemical or physical fractionation and could only have been produced by nuclear reactions in stars, therefore, these grains must have formed within  stellar gas. One of the most favourable, and observationally confirmed, location for the formation of dust are the wind-driven, extended envelopes of asymptotic giant branch (AGB) stars, which are the final phases of the evolution of stars of mass between roughly 0.8 and 8 M$_{\odot}$. 
The majority of the known stardust grains came from such stars, and will be discussed below in more detail. Other astrophysical sites where dust is observed to form are the cooled ejecta of core-collapse supernova and nova explosions. Some stardust grains have been attributed to such origins, as they carry the signature of material expected within these ejecta \citep{Nittler:1996,Jose:2004,Hoppe:2019}.

After being ejected into the interstellar medium via stellar winds or explosions, together with most of the gas that formed the star, stardust grains travel through the interstellar medium and are eventually 
incorporated into regions of new star formation, such as the proto-solar nebula.
We find them inside meteorite parent bodies, where they survived until today \citep[see][for a review of the astrophysics of presolar grain studies]{Nittler:2016}. 

In general, stardust grains are made of strongly refractory materials that condense directly from the gas phase at high temperatures (1300-2000 K). The condensation sequence of minerals depends predominantly on the C/O ratio of the gas. If
C/O $<$ 1, all the carbon is locked up in carbon monoxide (CO)
molecules and the condensed minerals are mostly oxides and silicates (like in the Solar System, where C/O $\simeq 0.4$). If C/O $>$ 1, all oxygen is locked up in CO molecules, and carbon compounds can condense, such as graphite and carbides. 
Among the carbon-bearing minerals, stardust silicon carbide (SiC) grains are the most studied, with several thousands of them analysed to date\footnote{Stardust grain data has been collected and can be retrieved from the Presolar Grain Database at presolar.physics.wustl.edu/presolar-grain-database/ \citep{hynes09,stephan20} of the Laboratory for Space Sciences of Washington University in St Louis (USA).}. This is because they are relatively easier to extract and they are also
large enough (from a fraction to a few tens of $\mu$m) to allow the isotopic analysis of single grains.  
Based on their C and Si compositions, SiC grains have been classified into several
populations, the largest of which (\emph{mainstream} SiC)
comprises more than 90\% of the grains. They are believed to originate in C-rich AGB stars of approximately solar metallicity \citep{hoppe97z,lugaro99}. 
The remaining (relatively few) SiC grains are classified into four populations:
A+B grains of unclear origin \citep{amari01AB,liu17JAB,liu17SNAB}, X grains attributed to core-collapse supernovae \citep[e.g][]{Pignatari:2013a,liu18SN}, and Y and Z grains attributed to C-rich AGB stars of metallicity lower than solar \citep{hoppe97z,amari01Y,zinner06}, although there are some inconsistencies with such an interpretation \citep{lewis13,liu19Mo}. Some SiC grains have also been interpreted as material originating from nova explosions \citep{jose16}.


Like the SiC grains, also the graphite grains are large enough to be analysed individually for their isotopic composition. However, their extraction procedure is more complex, and trace elements are present in extremely low abundances. 
About one third of stardust graphite grains have low densities (1.6 - 2.05 g~cm$^{-3}$), and appear to originate from core-collapse supernova explosions \citep{hoppe95,travaglio99,Pignatari:2013a}.
Higher-density grains may originate from a range of stellar environments \citep{Amari:2014}. Silicate and oxide grains are more difficult to locate among the abundant silicates of solar origin (where C/O $<$ 1) that constitute the main part of meteorites. 
Oxide grains comprise mostly corundum and spinel, and also a few hibonites. Only a small fraction of oxide and silicate grains 
are of stellar origin. Special techniques have been developed to recognise them \citep{nguyen07}. Oxide and silicate grains have also been separated into distinct
groups, mostly based on their oxygen isotopic ratios. The composition of most of these grains suggests that they have formed around O-rich AGB stars.

Overall, the vast majority of stardust grains recovered from meteorites originated in AGB stars and have provided us with high-precision constraints on the operation of nuclear reactions in the deep layers of these stars and on the mixing processes that carry processed material from inner to outer layers, where the dust forms. 
Nuclear processes in AGB stars include H and He burning, and the $slow$ neutron-capture process (the $s$ process) that produces roughly half of the elements heavier than Fe in the Universe. 
Mixing processes include (i) the dredge-up of material from the He burning region -- specifically $s$-process elements and C, so that the star can reach C/O$>$1 at the surface and produce carbonaceous dust, (ii) the partial mixing at the border of the H-rich and He-rich regions, leading to the formation of the main source of neutrons ($^{13}$C in the so-called $^{13}$C pocket), and (iii) the convective mixing of H-burning processed material in the more massive AGB stars ($>$ 3 \Msol) in which the base of the convective envelope is hot enough for proton captures to occurs \citep[see][for a review on AGB stars]{Karakas:2014}. 
Stardust mainstream SiC grains carry a clear signature of the $s$ process in the isotopic composition of the elements heavier than Fe, present in trace amounts, which is another strong indication of their origin in AGB stars. The composition of these mainstream grains is in agreement with the composition predicted by models of AGB stars of initial masses between roughly 2 and 4 \Msol\ and metallicity from solar up to roughly twice solar \citep{lugaro03grains,lugaro14Zr,lugaro18grains}.

\begin{figure} 
\centering
\includegraphics[width=\columnwidth,clip]{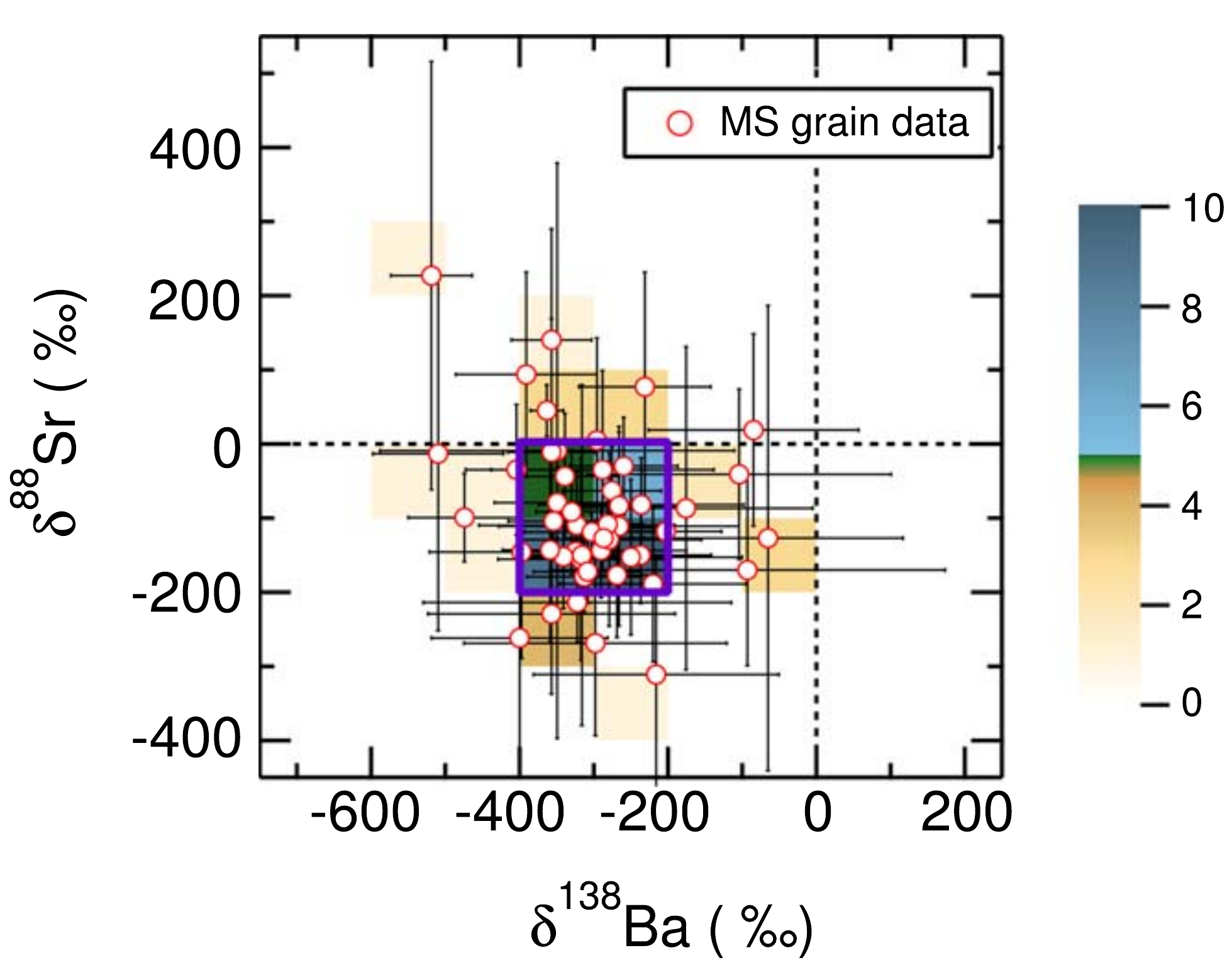}
\caption{High-precision measurements of the $^{88}$Sr/$^{86}$Sr and $^{138}$Ba/$^{136}$Ba isotopic ratios (in $\delta$ notation, i.e., relative to the solar value and multiplied by 1,000) in SiC grains of relatively large size. ($>$ 1~$\mu$m). The color coding on the right corresponds to the number of grains, and shows that the vast majority of the grains have $^{88}$Sr/$^{86}$Sr ratios lower than in solar abundances (i.e., negative $\delta$ values).  (Figure from \cite{Liu:2018}).}
\label{fig_LiuSr}       
\end{figure}

Using these grains it is possible to constrain not only the nucleosynthesis and mixing in AGB stars, but also the dust formation process. For example, the vast majority of relatively large SiC grains measured individually and with high precision, using resonant ionisation mass spectrometry \citep{Liu:2018} 
show $^{88}$Sr/$^{86}$Sr ratios lower than solar (Figure~\ref{fig_LiuSr}). The $s$ process can produce such ratios only together with Ce/Y ratios also lower than solar. 
This is because the $s$~process proceeds from lighter to heavier isotopes, and it is impossible to overproduce (relative to solar) the element Ce (with mass around 140) relative to Y (with mass 89) before Sr at mass 88 is overproduced relative to Sr at mass 86. Ce/Y ratios lower than solar are predominantly observed in Ba stars - the binary stars that accreted $s$-process material from AGB companions - of metallicity higher than solar. Therefore, these large grains must have originated in AGB stars of higher metallicity than solar. This conclusion is complemented by fact that the $^{88}$Sr/$^{86}$Sr ratio measured in smaller grains (this is done for large collections of many grains, since individual measurements are not possible) is on average higher than solar, corresponding to Ce/Y also higher than solar as observed in Ba stars of metallicity lower than solar. Therefore, the data show that the size of the SiC grains increases with the metallicity of the parent star and AGB stars of metallicity higher than solar are predicted to produce larger grains more favourably than AGB stars of metallicity around solar \citep{lugaro20}. Furthermore, this high-metallicity origin for the large SiC grains can explain why the Earth is observed to be more $s$-process enriched than bodies that formed farther away from the Sun \citep{dauphas04}. The larger SiC grains provide us with the best candidates to have carried $s$-process nucleosynthetic signatures in the proto-planetary disk \citep{ek20}. 

\begin{figure} 
\centering
\includegraphics[width=\columnwidth,clip]{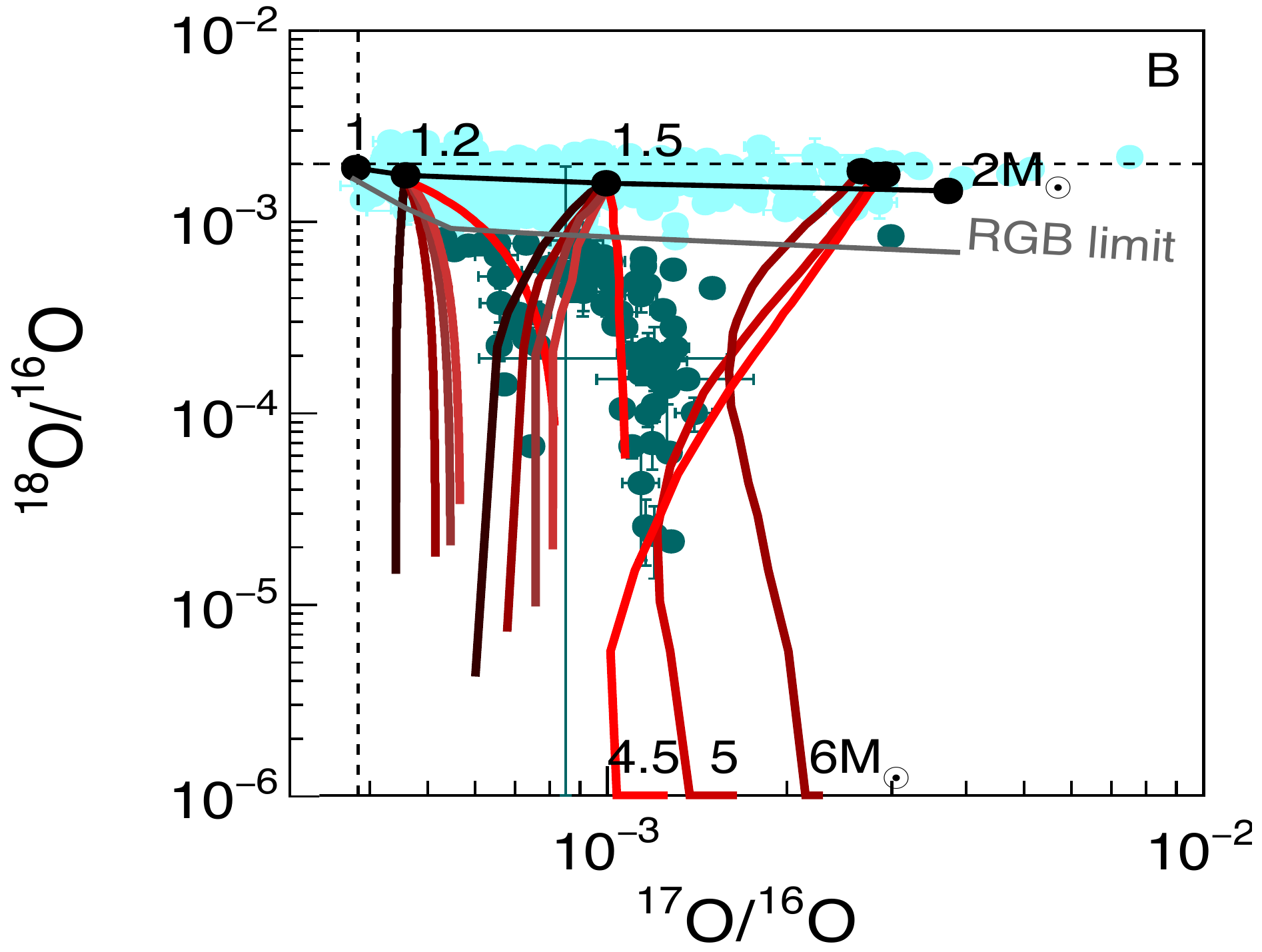}
\caption{The O isotopic ratios measured in stardust oxide grains of Group 1 (cyan points) and Group 2 (green points). Note that the uncertainties of the measurements are typically smaller than the symbol size.
These are compared to predictions from models of AGB stars of mass 1.2 and 1.5 M$_{\odot}$ that remain O-rich because of inefficient third dredge-up, and of mass 4.5, 5, and 6 M$_{\odot}$ that remain O-rich because of hot bottom burning. The 
black and grey solid line and black dots indicate the composition predicted during the red giant branch (RGB, i.e., before the star starts to ascend the AGB phase). The red curves represent the AGB evolution: for the 1.2 and 1.5 M$_{\odot}$ stars they show predictions from models with extra mixing below the formal border of the convective envelope due to magnetic effect for different values of a free parameter representing the depth by the mixing. For the 4.5, 5, and 6 M$_{\odot}$ stars they represent the effect of hot bottom burning on the O isotopic ratios. (From \cite{Palmerini:2021}).}
\label{fig_palmerini}       
\end{figure}

Stardust oxide and silicate grains are also mostly believed to originate from O-rich AGB stars. 
Figure~\ref{fig_palmerini} shows `Group 1 grains', which mostly have excesses in $^{17}$O. These can be explained by the operation of the first and second dredge-up during the red giant phase prior to the AGB phase. However, the origin of Group 2 grains, which also show a strong depletion in $^{18}$O, has so far been elusive; current candidate sites are indicated by the lines in Figure~\ref{fig_palmerini}. 
The first candidates are massive AGB stars ($>$ 4 \Msol), where proton captures destroying $^{18}$O are expected to occur at the base of the convective envelope; this is referred to as \emph{hot bottom burning}. In this case $^{18}$O is predicted to be even more than is observed in the grains, and some few percent dilution with material of solar composition is required to match the grains' data \citep{lugaro17o17}. The other candidate site are low-mass AGB stars ($<$ 1.5 \Msol), if extra mixing processes are assumed to occur below the base of this envelope \citep{Palmerini:2021}. 
It is also possible that Group 2 grains are made of two different populations, with two different origins. 

\begin{figure*} 
\centering
\includegraphics[width=1.45\columnwidth,clip]{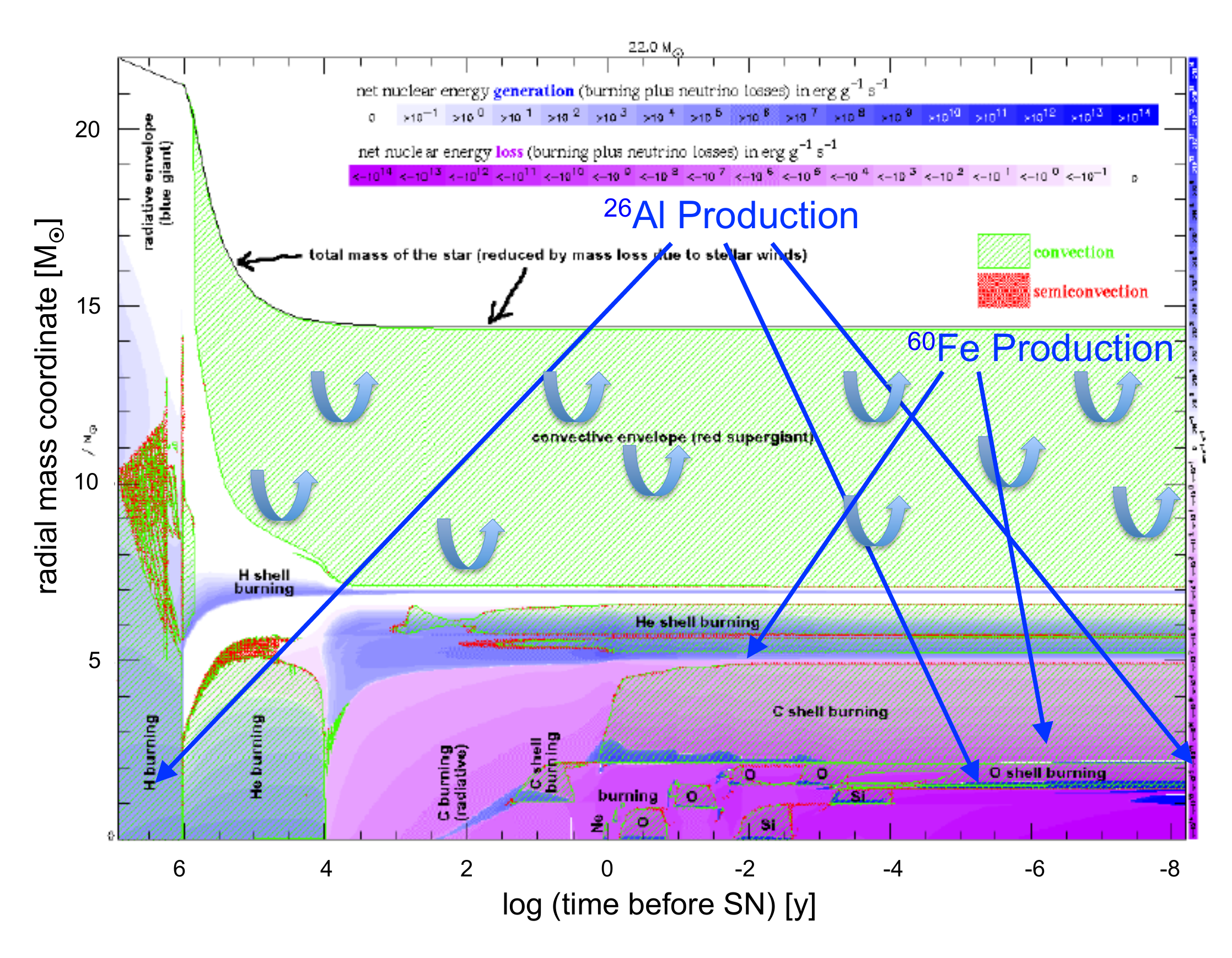}
\caption{This \emph{Kippenhahn diagram} shows the structure and evolution of a massive star, in theoretical modelling \citep{Heger:2002}. The time axis is logarithmic, extending from star birth (left) to core collapse (right), a few million years altogether, and emphasising late evolution towards the supernova from log scaling. The y~axis is a radial coordinate in enclosed mass, from the centre of the star (below) to its surface (above). \Al and \Fe are produced in different locations and at different times, as indicated.}
\label{fig_KippenhahnDiagr_26Al60Fe}       
\end{figure*} 
  
AGB stars as discussed above are of low- and intermediate mass, with masses $\leq$~7-8~\Msol,  and are in their giant phase, burning He and H in outer shells while no core burning of heavier nuclei occurs.  
The structure of more-massive stars becomes increasingly complex during their later evolution, i.e. beyond hydrogen and helium burning stages  (see Figures~\ref{fig_massiveStar_temp_dens} and \ref{fig_KippenhahnDiagr_26Al60Fe}). 
For massive stars, messengers such as presolar grains \citep[e.g.][]{Hoppe:2017}, asteroseismology \citep[e.g.][]{Bowman:2021,Fellay:2021}, and neutrinos \citep[e.g.][]{Yoshida:2019} are becoming available only in single, selective, cases; this does not yet support the detailed exploration of their interiors, unlike what we showed above for the Sun and AGB stars.

But radioactive isotopes produced in the nucleosynthesis reactions within massive stars provide indirect information on nucleosynthesis in these later evolutionary stages. The rapid evolution of massive stars in these late stages with typical time scales of order My and radioactive lifetimes of My, the \Al and \Fe products of massive-star fusion reactions are still alive when they are ejected and their decay in circumstellar space can be observed through $\gamma$-ray telescopes.  

As shown in Figure~\ref{fig_KippenhahnDiagr_26Al60Fe} from our current understanding of massive-star structure and evolution, \Al and \Fe synthesis occurs at different locations within massive stars. 
\Al production in the early main-sequence phase can be mixed into the hydrogen envelope, and may thus be ejected by the strong wind that is typical for such massive stars. Later \Al production is expected from the neon-oxygen shell, and from the explosive nucleosynthesis as the supernova shock proceeds through the star from its central region. Unlike the main-sequence production of $^{26}$Al, these later products are ejected by the supernova itself.
This is also the case for all \Fe synthesised within the star, which is expected to occur in the helium and carbon burning shells; neither of these are coupled to the surface where late mass loss would occur, and are ejected with the supernova.

The origin of these two radioisotopes from the same object, though with some offset in ejection times, implies that the ratio of their amounts can be used as a global test of our understanding of massive-star nucleosynthesis. The detailed location of each massive star cancels out in such a ratio, and cumulative  \Fe to \Al mass ratio for the entire massive-star population of our Galaxy can be measured through integrated diffuse emission in their characteristic decay lines, at 1332 and 1173 keV for $^{60}$Fe, and 1809 keV for $^{26}$Al. 

\begin{figure} 
\centering
\includegraphics[width=\columnwidth,clip]{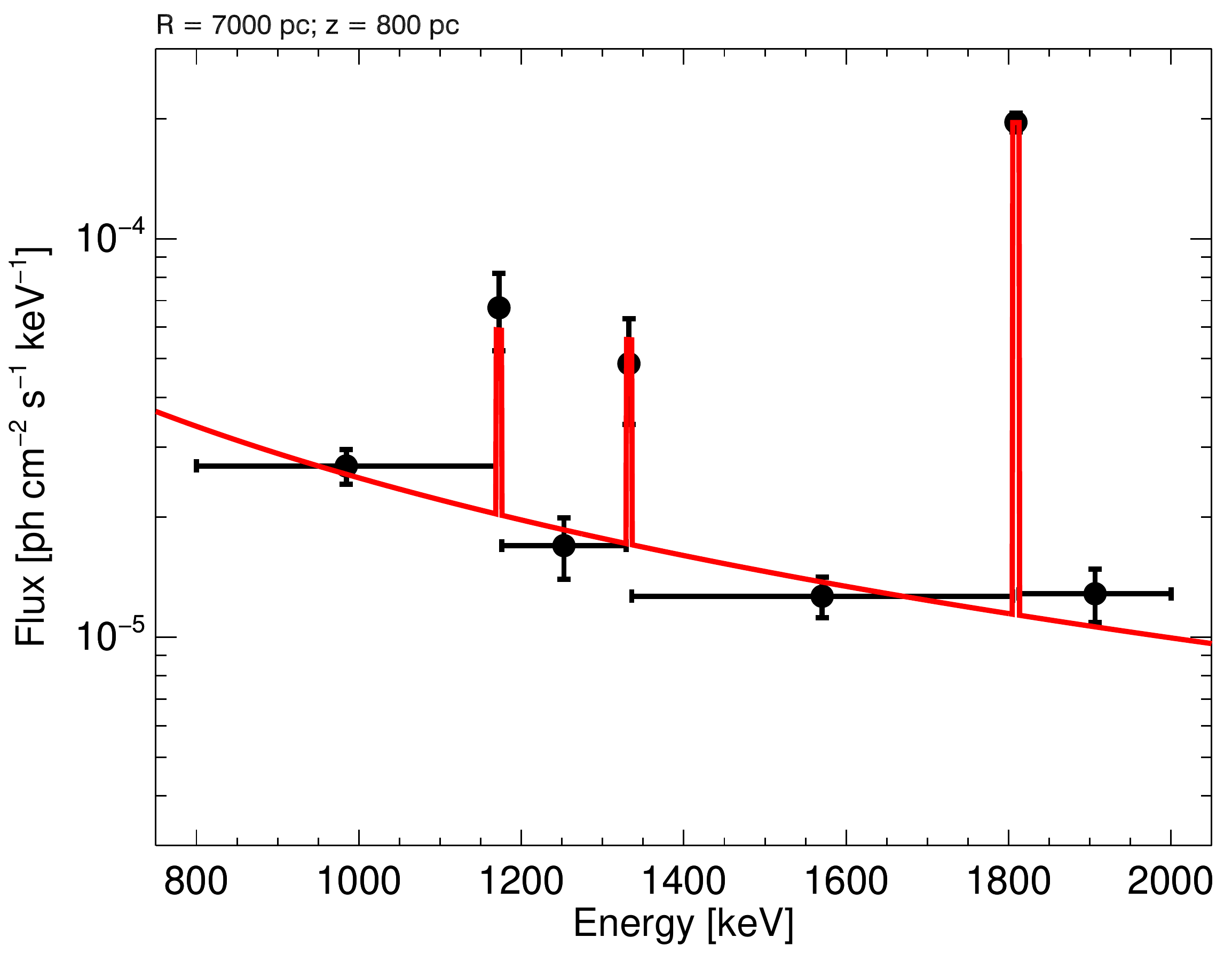}
\caption{The spectrum of diffuse $\gamma$-ray emission from the Galaxy shows the characteristic lines from \Al and \Fe decays, with \Al being several times brighter than $^{60}$Fe \citep{Wang:2020}.}
\label{fig_60Fe26Al-lines}       
\end{figure}

Figure~\ref{fig_60Fe26Al-lines} shows the $\gamma$-ray line intensities as measured. 
From these an observational upper limit on the $^{60}$Fe/$^{26}$Al $\gamma$-ray flux ratio of 0.4 is obtained \citep{Wang:2020}, so that the steady-state mass ratio of $^{60}$Fe/$^{26}$Al is constrained to be below 0.9. 
 
Interestingly, in deep-sea archives on Earth, live \Fe had been discovered about 15 years ago \citep{Knie:2004}. This measurement demonstrates that cosmic material is being deposited on Earth at a small rate. This can be observed through sensitive material analysis, here with accelerator mass spectrometry that can measure particle amounts of \Fe relative to terrestrial Fe as isotope ratios down to 10$^{-17}$. 
The observation of \Fe in layers that can be constrained in age \citep[for details see][]{Wallner:2016} has been taken as evidence that material from supernova activity in the vicinity of the solar system has been collected; stated otherwise, nearby supernova activity can be measured in suitable samples from ocean floors.
(See Section~\ref{recycling_general} and \ref{recycling_local} for more discussion of cosmic material transport).    
If we assume that the solar system vicinity is not too different from a Galactic average as measured in $\gamma$~rays, then also an $^{60}$Fe/\Al ratio constraint from ocean floor analyses should constrain the nucleosynthesis in nearby massive stars, as an ensemble constraint. 
\Al may, however, also be produced in the Earth atmosphere from cosmic-ray spallation reactions, unlike $^{60}$Fe. Therefore, accounting for the rather high contamination, only an upper limit on \Al could be determined from such ocean-floor material analyses \citep{Feige:2018}. This constrains the $^{60}$Fe/\Al ratio for nearby massive stars to a value above 0.2.
Evidently, combined astronomy for this ratio is informative. 
Current models of nucleosynthesis in massive stars and their supernovae predict a yield ratio $^{60}$Fe/\Al of order unity, hence significantly above the region of values as constrained by above $\gamma$-ray measurements. 
The ways out could be diverse, as massive stars and supernovae are not understood in such detail. One of the possibilities discussed is that very massive stars above a mass of about 30~\Msol\ fail to explode as a core-collapse supernova. This would bury their nucleosynthesis products in the black hole remnant, and, as \Fe yields are expected to be much higher than \Al yields for such massive stars, would bring the theoretical value closer to observational constraints.   
    
\subsection{Stellar explosions} 
\label{explosions}   
  
      
      
      
Explosions inside or at the surface of stars occur when energy release from nuclear reactions rises faster than the reaction environment can transport energy away from the reaction site. 
This causes the nuclear-reaction environment (the central region of a star or the burning layer in case of an explosion on the stellar surface) to rapidly expand and absorb energy in expansion work. As a result, an explosion shock wave is created and propagates through the envelope inside-out causing compression and heating. 
This can lead to hot plasma exceeding the thresholds for nuclear reactions. 
The combination of these processes is called \emph{explosive nucleosynthesis} \citep{Arnett:1996,Thielemann:1990}. 
Merging compact stars, or gravitational shredding of stars as they approach a black hole's extreme gravity, are special cases, where also compression and heating is expected to lead to nuclear reactions of stellar material \citep{Metzger:2010,Surman:2006}.
Compact stars that accrete additional material also provide scenarios for thermonuclear explosions, as the accreted material may accumulate and exceed reaction thresholds through higher densities and compressional heating. This leads to nova explosions caused by runaway hydrogen burning on white dwarf surfaces and to type-I X-ray bursts as thermonuclear explosions of accreted helium, followed later under suitable conditions by superbursts caused by explosive carbon burning on the surface of neutron stars \citep{Bildsten:2000}.
 
Explosive nucleosynthesis occurs with rapid dynamical changes of density, pressure, temperature, and composition. Unlike hydrostatic nucleosynthesis the outcomes of explosive nucleosynthesis are the result of a complex interplay of nuclear reactions, their energy release, the local conditions and the hydrodynamic reactions.
The relevance for cosmic abundance evolution hinges on the fraction of nucleosynthesis ashes that is ejected into the circum-source environment, rather than being swallowed by the compact remnant object (neutron star or black hole).

 
  
The nuclear reactions that may occur in such environments can be grouped into their approximate extremes: nuclear statistical equilibrium is likely for extremes of density and temperature, such as in a gravitational collapse of a massive star or explosive runaway of heavy-ion fusion reactions such as $^{12}$C$+^{12}$C~fusion in white dwarf interiors causing a thermonuclear supernova. 
Another extreme is the significant contribution of neutrino reactions, when neutrino intensities become very high and the matter density is high enough to provide an efficient target for neutrino reactions; this is expected in the vicinity of newly-forming neutron stars in the gravitational collapse of the core of a massive star or accretion-induced collapse of a white dwarf to a neutron star in a binary system.
A third extreme is a dominance of neutron reactions in neutron-rich environments, such as is expected from the mergers of neutron stars, when matter of the destroyed neutron star interacts in the strong field and density of a companion neutron star or black hole.
The high-entropy wind that characterises the vicinity of the proto-neutron star in core-collapse supernovae has long been thought to provide such a region with dominance of neutron capture reactions. Other sites of neutron capture element formation are electron-capture supernovae, collapsars, hypernovae and long-duration $\gamma$-ray bursts \citep{Cowan:2021}. 
  
Supernova explosions are the most-prominent events that release nucleosynthesis ashes into their surroundings. We distinguish core-collapse supernovae at the end of stellar evolution of massive stars ($M\geq 8-10$~\Msol) from thermonuclear supernovae caused by the thermonuclear disruption of a white dwarf. Recent reviews on Supernovae include \citep{Arnett:1996a,Branch:2017,Alsabti:2017}. 

Supernovae from massive stars are understood from above-discussed stellar evolution when the end point of nuclear energy release in the hydrostatic evolution with its increasing central compression and heating is reached. This occurs when the central composition is mostly iron, because iron group nuclei have the highest binding energy per nucleon. Further fusion reactions to heavier nuclei, which may occur at such a high temperature and density, are endothermic and will not liberate energy. As a result, the inner energy source of a star can no longer counter gravity and gravitational collapse will occur. Matter that falls onto the central iron core will experience inverse beta decays, as electron pressure increases and favours neutronisation of the central plasma. Intense neutrino emission accompanies this neutronisation. Matter is more compressible towards the densities of nuclear matter as Coulomb repulsion among protons becomes less important. When nuclear density is reached in the centre, further compression is inhibited and in-falling matter now hits a hard wall. This results in a compression shock that is strong enough to decompose in-falling matter at a typical radius of 100~km, while the proto-neutron star in the centre is about 50~km in size. 
Now the environment becomes complex and highly dynamic, as hydrodynamic, magnetic, and nuclear-reaction processes re-distribute internal, kinetic, and thermal energies.
It is likely that nuclear statistical equilibrium is reached in some regions herein, but the intense dynamic evolution results in freeze out of nuclear reactions as materials are redistributed towards an explosion of the collapsed star in a supernova.

\subsubsection*{SN 1987A}

Observations of core-collapse supernovae try to shed light onto these complex interiors. But, as in the case of stars, high opacities of the massive-star envelope prevents direct observations of the collapse. The outgoing shock from the supernova explosion reaches the surface of the star several hours after the implosion and represents the first electromagnetic signal from the supernova. 
Supernova~1987A was discovered in the Large Magellanic Cloud on 23 February 1987 and provided the current generation of scientists with a unique opportunity to witness the disappearance of a massive star and the evolution of the ensuing supernova explosion towards a young supernova remnant, now at an age beyond 35 years \citep{Arnett:1989a, McCray:1993, McCray:2016}.  

\begin{figure} 
\centering 
\includegraphics[width=\columnwidth]{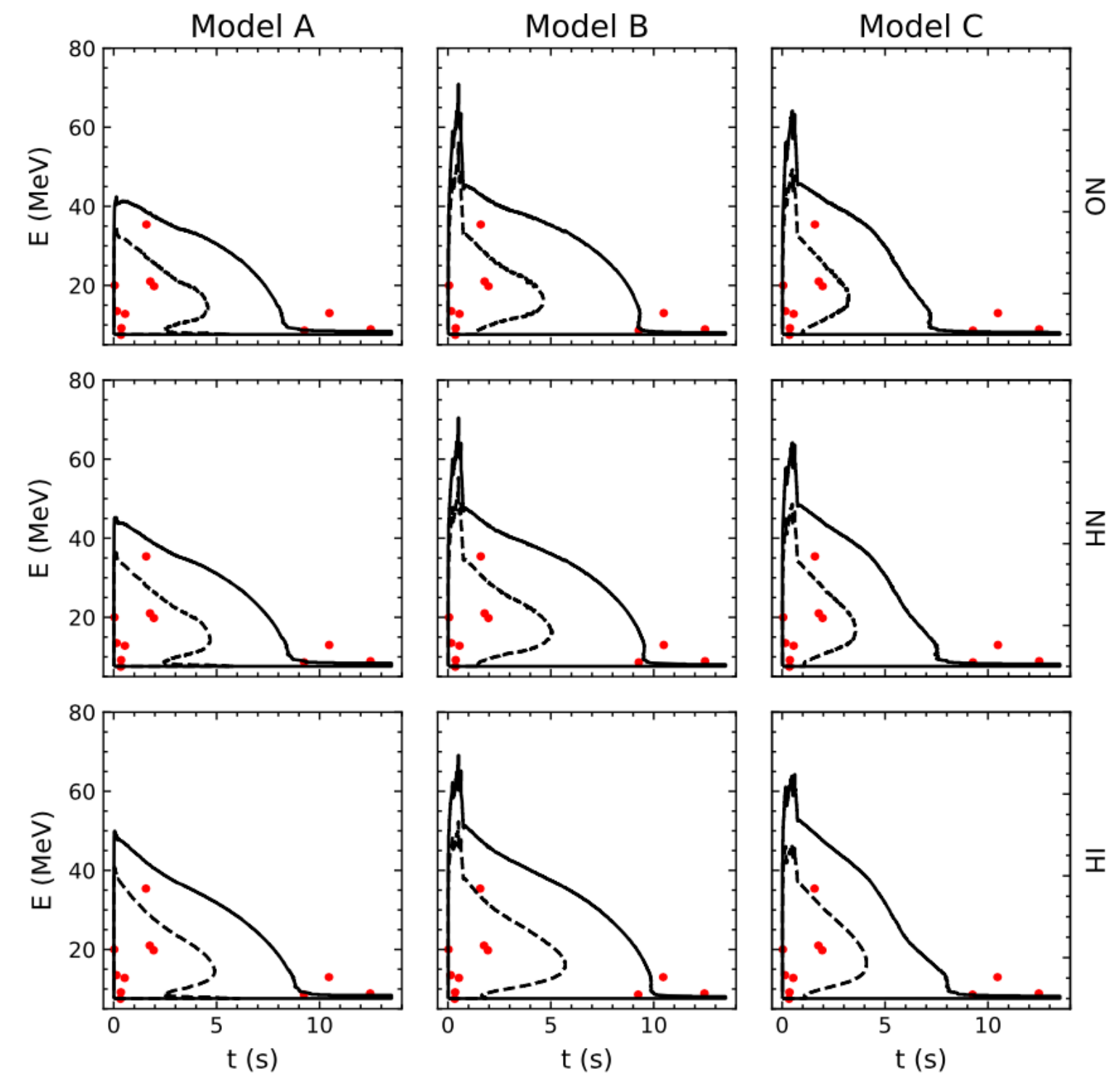}
\caption{A burst of neutrinos has been captured in the IMB neutrino detector (red dots). Shown here are comparisons of three models for SN~1987A neutrino emission and three cases of neutrino oscillation treatment. (From \cite{Olsen:2021}).}
\label{fig_ccSN_neutrinos}
\end{figure} 

The first signal from the exploding SN~1987A was a neutrino burst, that was recorded at several neutrino observatories \citep{Arnett:1989a}. The neutrino burst in itself indicated that the core-collapse picture is correct, and that neutron stars form within stellar explosions (see Fig.~\ref{fig_ccSN_neutrinos}). 
The first optical detection of SN~1987A appeared 3 hours after this neutrino burst.

\begin{figure} 
\centering 
\includegraphics[width=0.9\columnwidth]{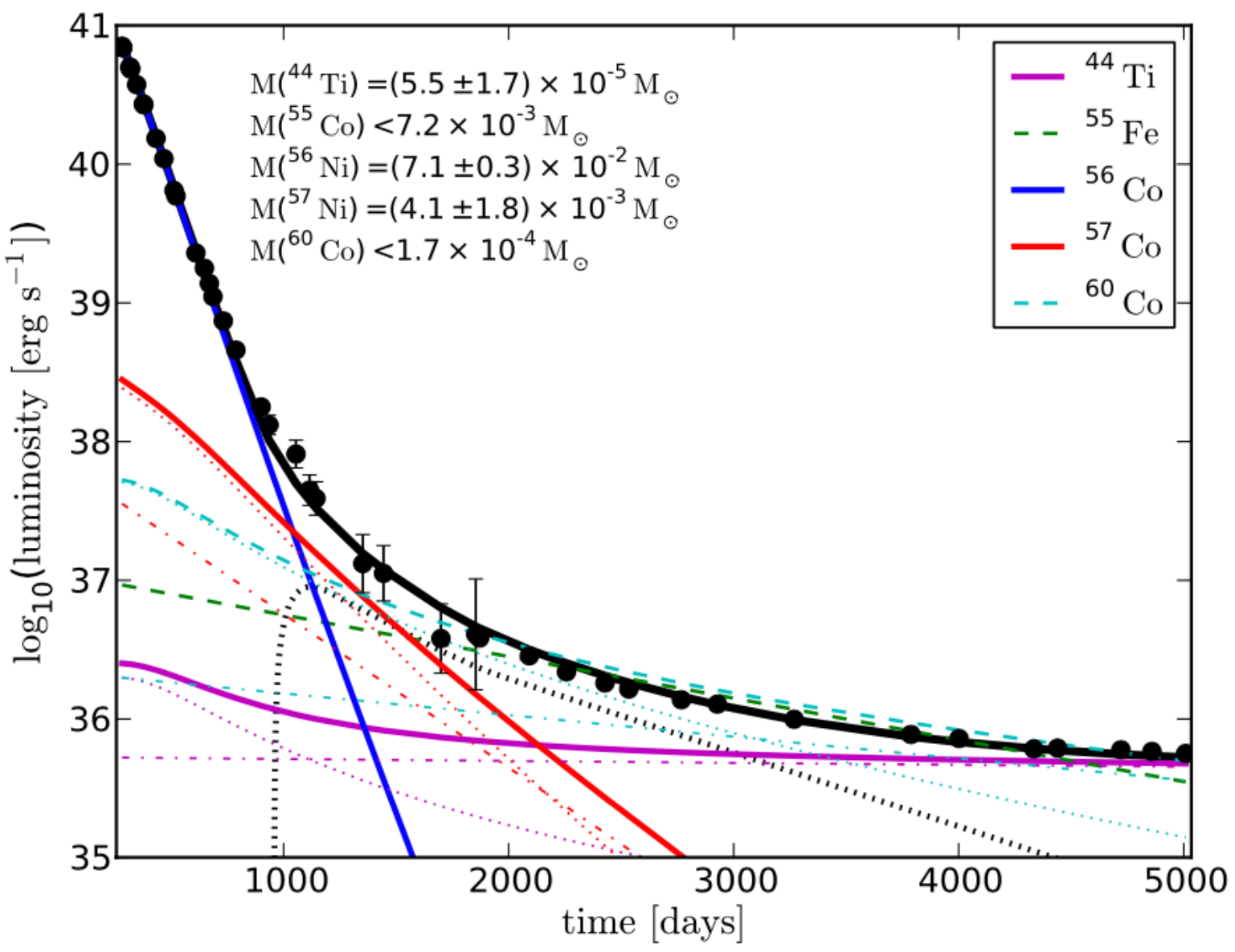}
\caption{Longterm evolution of energy sources in the bolometric light curve of SN~1987A. From \cite{Seitenzahl:2014}.}
\label{fig_SN87A_radioact}
\end{figure}

The first direct indication of radioactive material came from the detection of the 1.238 MeV and 847 keV $\gamma-$lines from the decay of $^{56}$Co  about one hundred days after explosion \citep[e.g.][]{Leising:1990, Tueller:1990}. This was clear evidence that some newly-synthesised material had been transported to the surface of the explosion. 
Elemental-line ratios can also indicate radioactive decays. Since Co decays to Fe, the relevant observed line ratios should change accordingly. The infrared lines of [Co II] 1.547$~\mu$m and [Fe] 1.533$~\mu$m were observed between 250 and 500 days past explosion, and clearly display the expected change in their intensity ratio \citep{Varani:1990}. There was also a first indication of the longer decay time for $^{57}$Co in the line evolution. 
The 67.9 keV and 78.4 keV lines of $^{44}$Ti decay have been observed 25 years after explosion \citep{Grebenev:2012,Boggs:2015}. This means that all predicted radioactive decays have been directly observed in SN~1987A (see Fig.~\ref{fig_SN87A_radioact}). 

The observed Co decay lines contain a further piece of information. Their velocity evolution indicates the distribution of the newly synthesised material within the explosion. In a combination of the direct X-ray decay line evolution with the infrared lines described above, it is possible to deduce a Ni offset from the centre of about 500~km~s$^{-1}$ \citep{Jerkstrand:2020}. This is the signature of an asymmetric explosion.
 
The optical supernova light displays an early cooling of the atmosphere, followed by a re-brightening due to the increasing surface of the explosion. The light curve of SN~1987A was unusual, as it displayed a long rise time of 90 days to maximum followed by a decline for several years. 
The early light curves are dominated by shock energy and recombination of the outer layers in SN~1987A. The radioactive material synthesised in the explosion powered the optical emission after about 100 days. 
The \emph{bolometric} light curve of SN~1987A follows the predictions of fully trapping the energy from $^{56}$Co decay for several hundred days. It can be modulated by changes in the escape fractions due to decreasing densities. While present to some degree in SN~1987A, this was not a dominant effect until about 1000 days after the explosion. 
{        The bolometric light curve started to deviate from the radioactive energy input around 500 days \citep{Bouchet:1991}. At the same time the far-infrared flux started to increase \citep{Wooden:1993}.  This is attributed to dust, formed around this epoch in the ejecta. Another indication of dust formation was a blueshift of several emission lines as the backside of the ejecta was blocked by dust \citep{Lucy:1991}.
The presence of dust was furthermore indicated from mm~wavelength observations after about 1300 days \citep{Bouchet:1991}.
Such formation of dust was another sign of extensive mixing of newly synthesised material to the outer layers of the ejecta and significant cooling. }

At later times the bolometric light curve changed slope, to follow the decay times of other dominating isotopes, first $^{57}$Co and then $^{44}$Ti \citep{Seitenzahl:2014} (Fig.~\ref{fig_SN87A_radioact}). 

\begin{figure} 
\centering 
\includegraphics[width=\columnwidth]{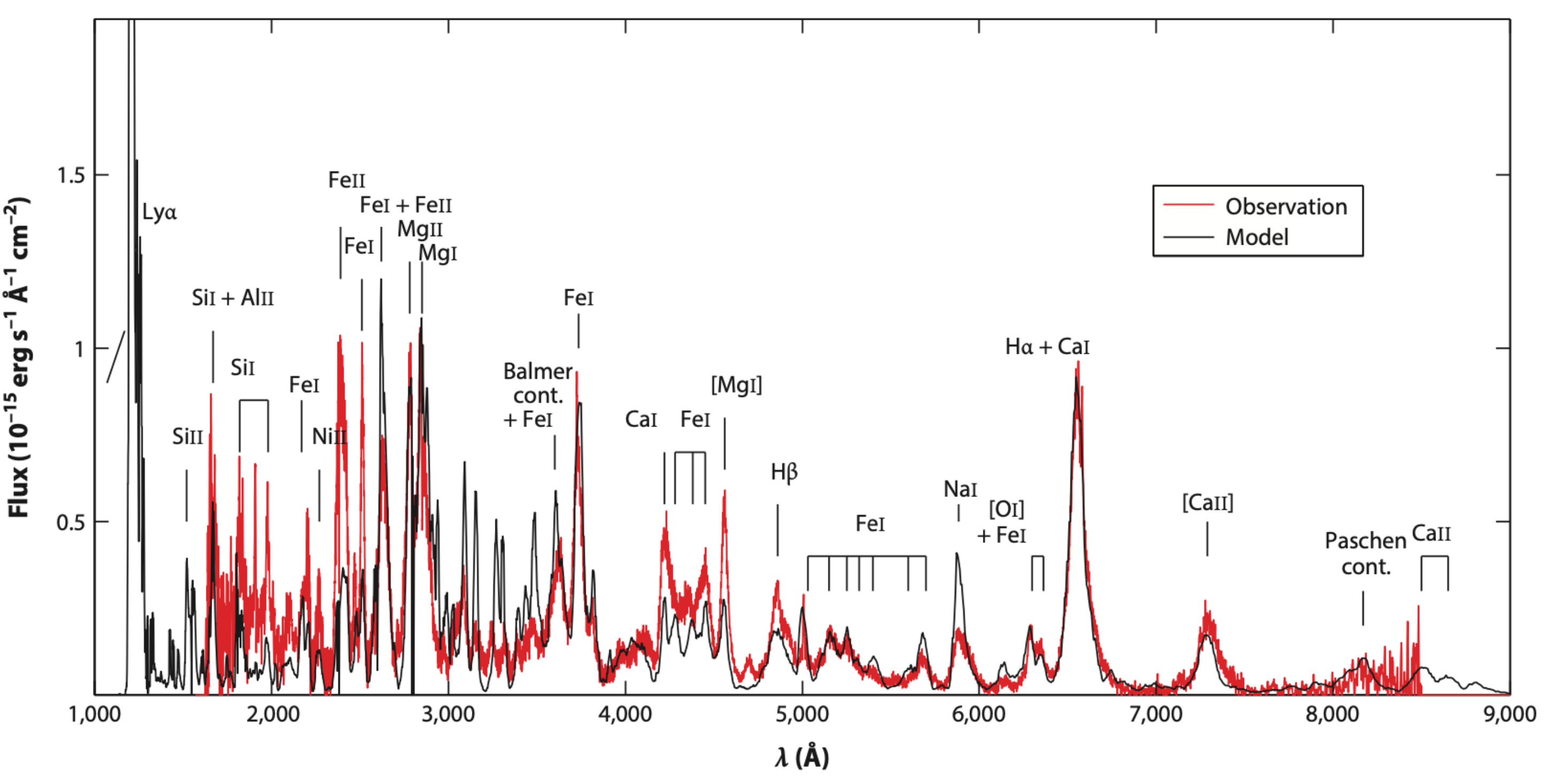}
\caption{Spectrum in the optical/ultraviolet regime from the remnant of SN~1987A, as observed with the Hubble Space Telescope. (From \cite{Jerkstrand:2011}).}
\label{fig_SN87A_spectrumHST}
\end{figure}

Spectral information about nuclear ashes ejected with the supernova are harder to come by, as the explosion results in fully-ionised plasma that gradually neutralises as it cools down. Figure~\ref{fig_SN87A_spectrumHST} shows an optical/UV spectrum collected with the Hubble Space Telescope, indicating a few characteristic lines that convey fresh nucleosynthesis. Their detailed analysis provides evidence of energy input from the $^{44}$Ti decay \citep{Jerkstrand:2011}.

Observations of SN~1987A typically cover several emission sites simultaneously. In addition to the cooling ejecta, SN~1987A is surrounded by an equatorial ring of circum-stellar material, presumably ejected by the progenitor star about 20,000 years prior to explosion \citep{Morris:2007,Podsiadlowski:2017}. After 8 years, the first evidence of interaction of the fastest supernova ejecta with the ring material appeared. 
Strong shocks built up in the ring material and quickly outshone the supernova \citep[e.g.][]{Leibundgut:2003}. 
Separating the various emission sites is not trivial \citep[e.g][]{Kangas:2022}. The inner ejecta are now illuminated by X-rays from the shocks in the equatorial ring, which is visible in the H$\alpha$ lines \citep{Larsson:2011}. The radioactive energy input is observable in other lines, e.g. [Si I]+[Fe II] 1.644~$\mu$m and also He~I 2.05~$\mu$m \citep{Larsson:2016}, which are powered by the $^{44}$Ti decay. 
The inner ejecta display an asymmetric shape. The reason is due to the asymmetric distribution of the radioactive material, which strongly deviates from a spherical distribution \citep{Larsson:2016}. 
Comparisons with explosion models confirm the clearly non-spherical distribution of ejecta material in SN~1987A.  
 
SN~1987A is a rich resource for the understanding of supernova explosions and their interactions with the surrounding medium. It is a backyard example of how stellar material is returned to the interstellar medium and how the kinetic energy is affecting the immediate environment. The transition from supernova to supernova remnant is in full swing and will continue to provide us with insights into the nucleosynthetic products of core-collapse explosions. 

\begin{figure} 
\centering 
\includegraphics[width=0.8\columnwidth]{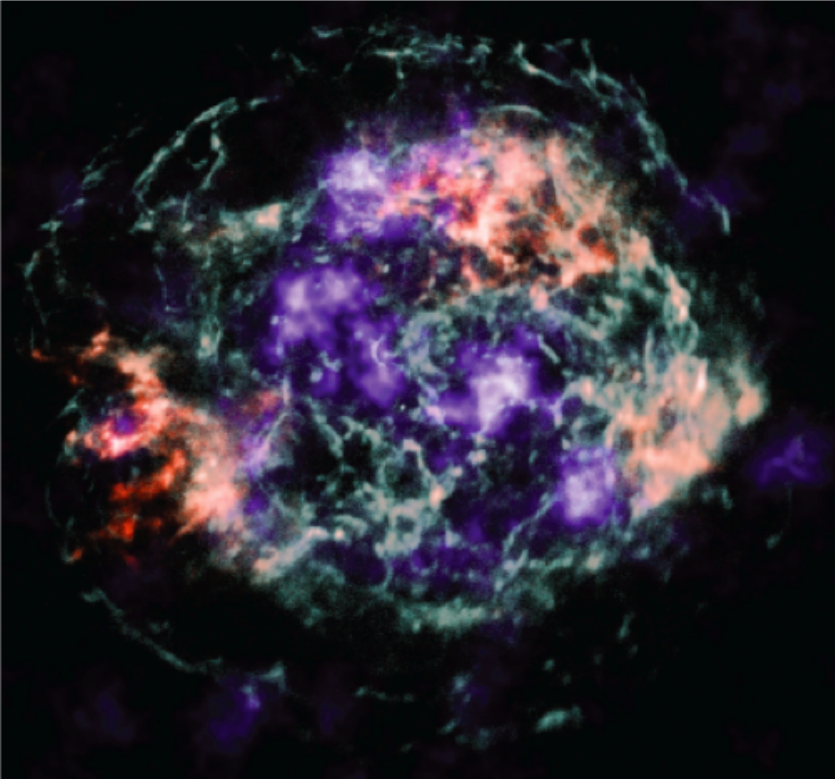}
\caption{Image of the Cas~A supernova remnant from different messengers:  Characteristic lines from $^{44}$Ti decay (blue) reveal the location of inner ejecta through this radioactive nucleosynthesis product, while X-ray line emissions from iron (red) and silicon (green) atoms, that also are emitted from those inner ejecta, show a somewhat different brightness distribution, due to ionisation emphasising the parts of ejecta that have been shocked within the remnant. (From \cite{Grefenstette:2014}).}
\label{fig_CasAimage}
\end{figure} 
 
\subsubsection*{Core-collapse supernovae: Beyond SN~1987A}


$^{44}$Ti as a nucleosynthesis product is also an important energy source of young supernova remnants, as discussed for SN~1987A; due to its 89~yr radioactive lifetime it is expected to contribute significantly to core-collapse supernova remnants at times of order hundreds of years.
The Cas~A supernova remnant at a distance of 3.4~kpc is about 350 years old {        \citep{Fesen:2006}}. 
It was the first source where $^{44}$Ti decay was directly observed through $\gamma$-rays \citep{Iyudin:1994}, later confirmed by several other instruments \citep{The:1996,Rothschild:1999,Vink:2000,Siegert:2015}. 
With its size, it was ideally suited for the NuSTAR imaging hard X-ray telescope \citep{Harrison:2013} to measure its radioactivity $\gamma$-rays from the radioactive $^{44}$Ti that had been ejected with the supernova \citep{Grefenstette:2014}. 
This image (Figure~\ref{fig_CasAimage}) spectacularly shows directly that radioactive ejecta appeared in several clumps, rather than as spherically-symmetric shells. 
After SN~1987A's asphericity indicators (see above), this is another direct demonstration that sphericity is not common in core-collapse supernova explosions.

\begin{figure} 
\centering 
\includegraphics[width=1.0\columnwidth]{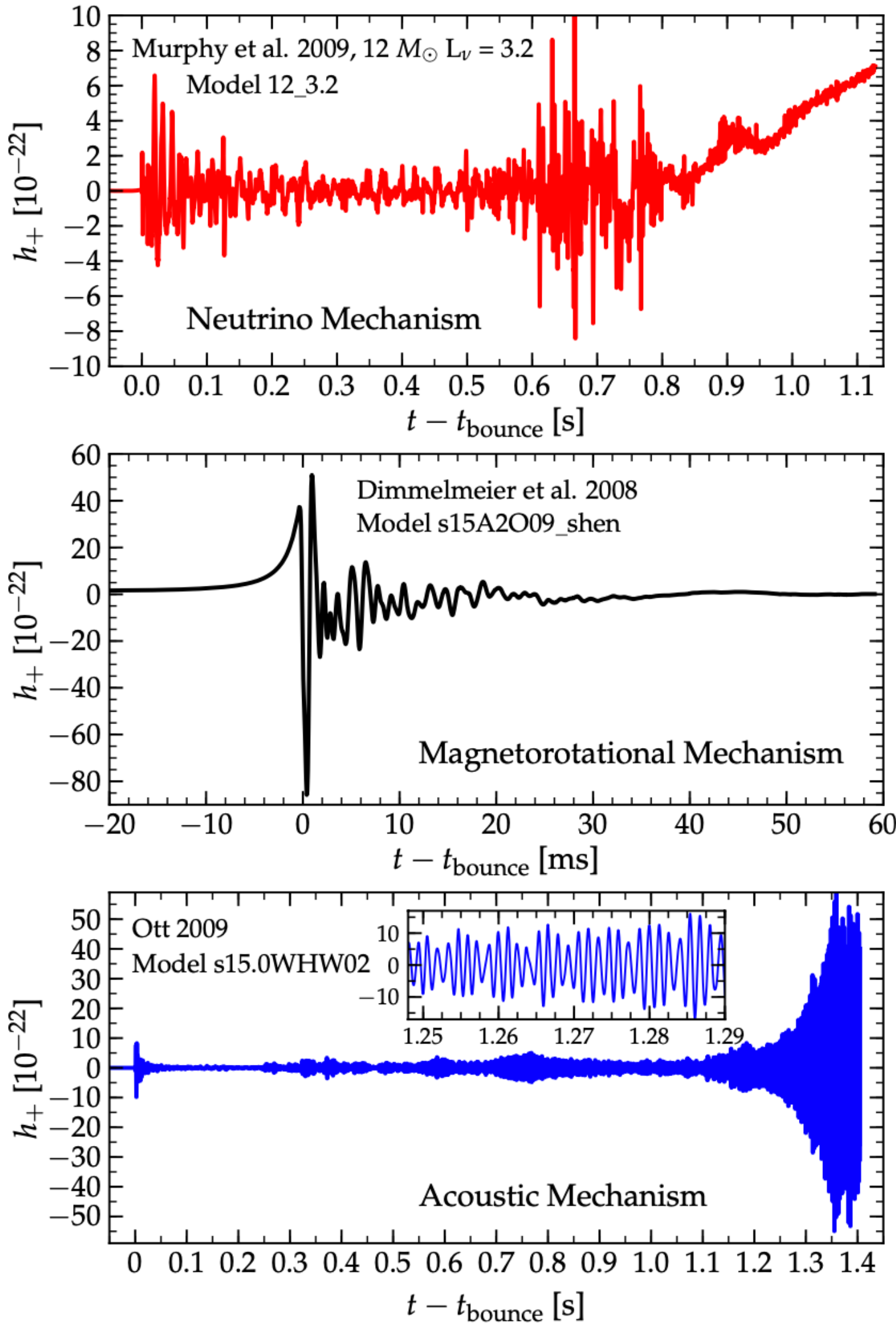}
\caption{Prospects of discriminating different variants of core-collapse supernovae through their signatures in gravitational-waves \citep{Logue:2012}.}
\label{fig_ccSN_gravWav}
\end{figure} 
 
Other messengers may convey even more directly how the explosion develops. 
The rise and spectral evolution of neutrino emission (as discussed above), if measured with better detectors than at the time of SN~1987A, promises a wealth of constraints on how the dynamics of the collapsing core is energised from within \citep{Muller:2019}.
This intense neutrino emission is held responsible for reverting the implosion into an explosion, and constraints on the energetics through neutrino light curve and spectrum will be decisive.
 
Moreover, adding constraints in the future from yet another messenger promises to complement information on the mass distribution and its dynamics during core collapse and early supernova:  gravitational waves will provide a diagnostic on the details of the collapse  \citep{Logue:2012}, and thus help understand the explosive nucleosynthesis from the very inner regions in the core-bounce shock and gain regions. 
Figure~\ref{fig_ccSN_gravWav} illustrates typical gravitational-wave signatures as they are expected from different scenarios of core collapses, from the above-mentioned neutrino-burst initiation, through a magneto-rotational collapse \citep[see also][]{Bisnovatyi-Kogan:1973,Bisnovatyi-Kogan:2018} where in-spiraling magnetic fields are wound up and finally release their energy, to a scenario where the oscillations stimulated by asphericities in the collapse superimpose and steepen up to develop an explosion.
  
\subsubsection*{Searching for sources of the r~process}

\begin{figure} 
\centering 
\includegraphics[width=0.7\columnwidth]{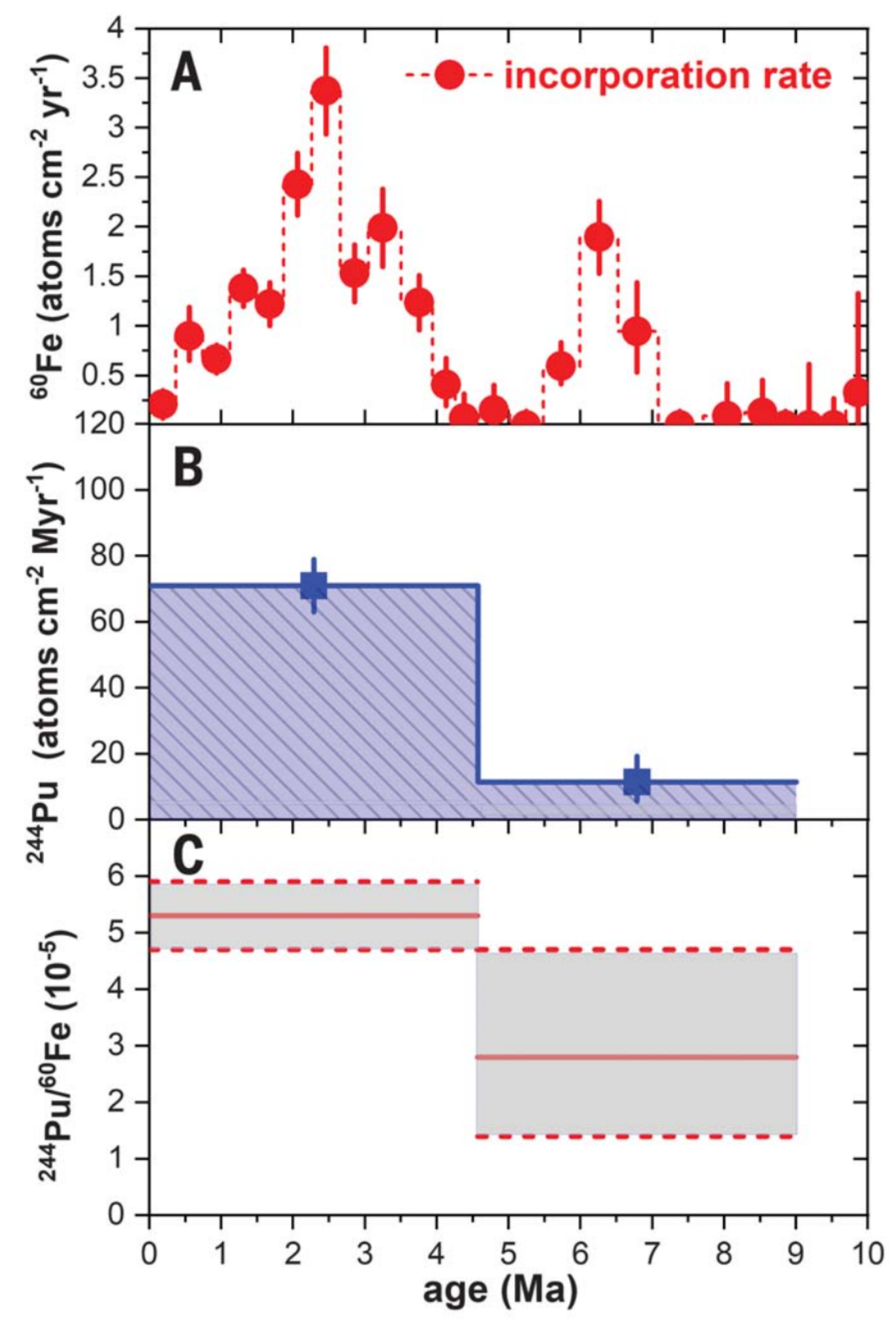}
\caption{Detections of $^{60}$Fe and $^{244}$Pu in the same sediments on ocean floors of the Pacific \citep{Wallner:2021}. The $^{60}$Fe abundance history (\emph{top panel}) shows similarity with the sparse observation history of $^{244}$Pu deposits (\emph{middle panel}). The ratio measured for these two radioactive isotopes (\emph{bottom panel}) is consistent with being constant. This does, however, not necessarily imply the same sources (see text and also Section~\ref{recycling_local}).}
\label{fig_244Pu-60Fe}
\end{figure}

The high-entropy region very close to the proto-neutron star in a core collapse has long been thought to provide an ideal setting for the r-process, i.e. rapid neutron capture reactions on seed nuclei as a process to form elements heavier than iron (see Section 2).
The recent detection of interstellar $^{244}$Pu and $^{60}$Fe in terrestrial ocean-ground sedimentations \citep{Wallner:2021} is a particularly interesting messenger on core-collapse and r-process nuclei, as shown in Figure~\ref{fig_244Pu-60Fe}.
(See Section~\ref{recycling_local} for details on the detection and analysis of radioactive isotopes in terrestrial sedimentation archives).  
As discussed above, core-collapse supernovae are believed to eject $^{60}$Fe. However, from recent simulation and modelling work, they have been questioned to be able to produce sufficient r-process ejecta and the heavy actinides. 
Currently-favored r-process scenarios are rather rare events such as neutron star mergers or a selected subset of supernovae (magnetars, collapsars) \citep{Cowan:2021,Thielemann:2020}. 
Measurable deposition of live $^{244}$Pu, if produced in a rare event,  strongly constrains  time and space of such an event, considering its low probability to occur near the solar system in the recent past. 

A significant influx of interstellar $^{244}$Pu had been observed recently in samples from the deep-sea crusts, in the same samples that also exhibited enhanced $^{60}$Fe values  (Figure~\ref{fig_244Pu-60Fe} from \citep{Wallner:2021}). These are the only two radionuclides with a clear interstellar signal in terrestrial archives. 
Earlier limits  based on no or just one AMS (accelerator mass spectrometry) $^{244}$Pu detection are consistent with these results from improved analysis \citep{Paul:2001,Wallner:2004,Paul:2009,Raisbeck:2007,Wallner:2015}. 
The recent data were obtained from layers that accumulated over 3 to 5 million years, and correspond to an averaged influx rate of only $\sim$10 and $\sim$70 $^{244}$Pu atoms~cm$^{-2}$~Myr$^{-1}$, respectively. 
Two crust layers suggest atom ratios of $^{244}$Pu/$^{60}$Fe between 3$\times$10$^{-5}$ and 5$\times$10$^{-5}$. 

The integrated $^{244}$Pu flux suggests that supernovae may produce some, but not be a major producer, of the heaviest r-process nuclides \citep{Wallner:2015,Wallner:2021,Hotokezaka:2015}.
The concomitant influx of both nuclides in the measured crust sample might relate their sources; but, alternatively, it might be related to how ejecta flow through the interstellar medium on long time scales, and could thus point to a common driver for a simultaneous presence in the terrestrial record. The radioactive decay time of $^{244}$Pu is 30 times longer than that of $^{60}$Fe, so that  $^{244}$Pu would be sampled much further back in time. 
Even if found in the same terrestrial deep-sea layers, $^{244}$Pu could have been produced in older events prior to the $^{60}$Fe-delivering supernovae, if ejecta from such events reached the solar system around the same time as the $^{60}$Fe, leading to a synchronous deposition. 
(Ejecta transport through the interstellar medium and towards the solar system is discussed in more detail below, Section~\ref{recycling_local}.).

\subsubsection*{Supernovae of type Ia}
\begin{figure}
  \centering
  \includegraphics[width=\columnwidth]{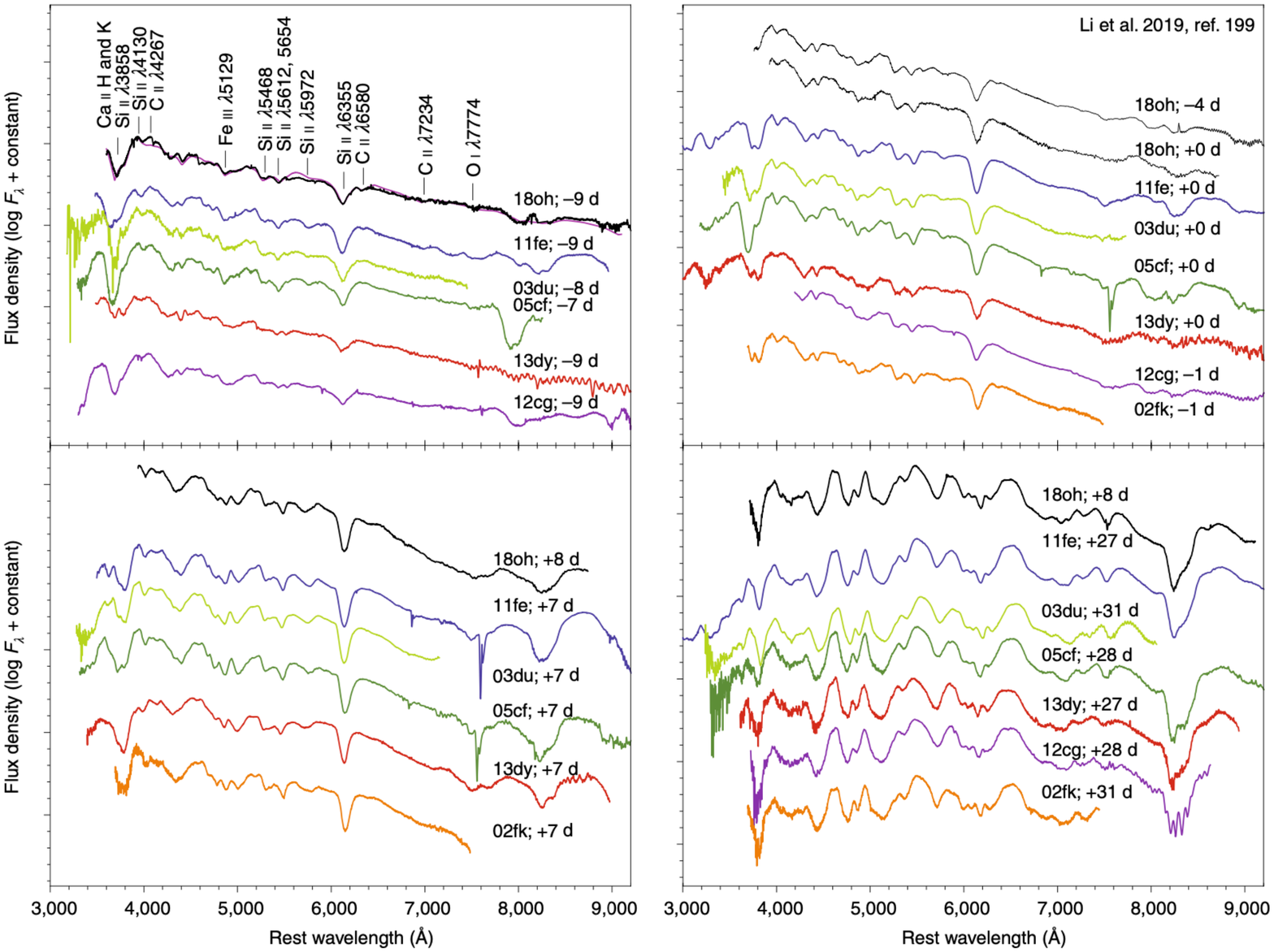}
   \caption{Spectra in the optical window, showing lines from various elements, as they vary before and through the light curve maximum of a few typical supernovae of type Ia.  (From \cite{Jha:2019}).}
  \label{fig_SNIa_maxSpectra}
\end{figure}

\begin{figure}
  \centering
  \includegraphics[width=0.9\columnwidth]{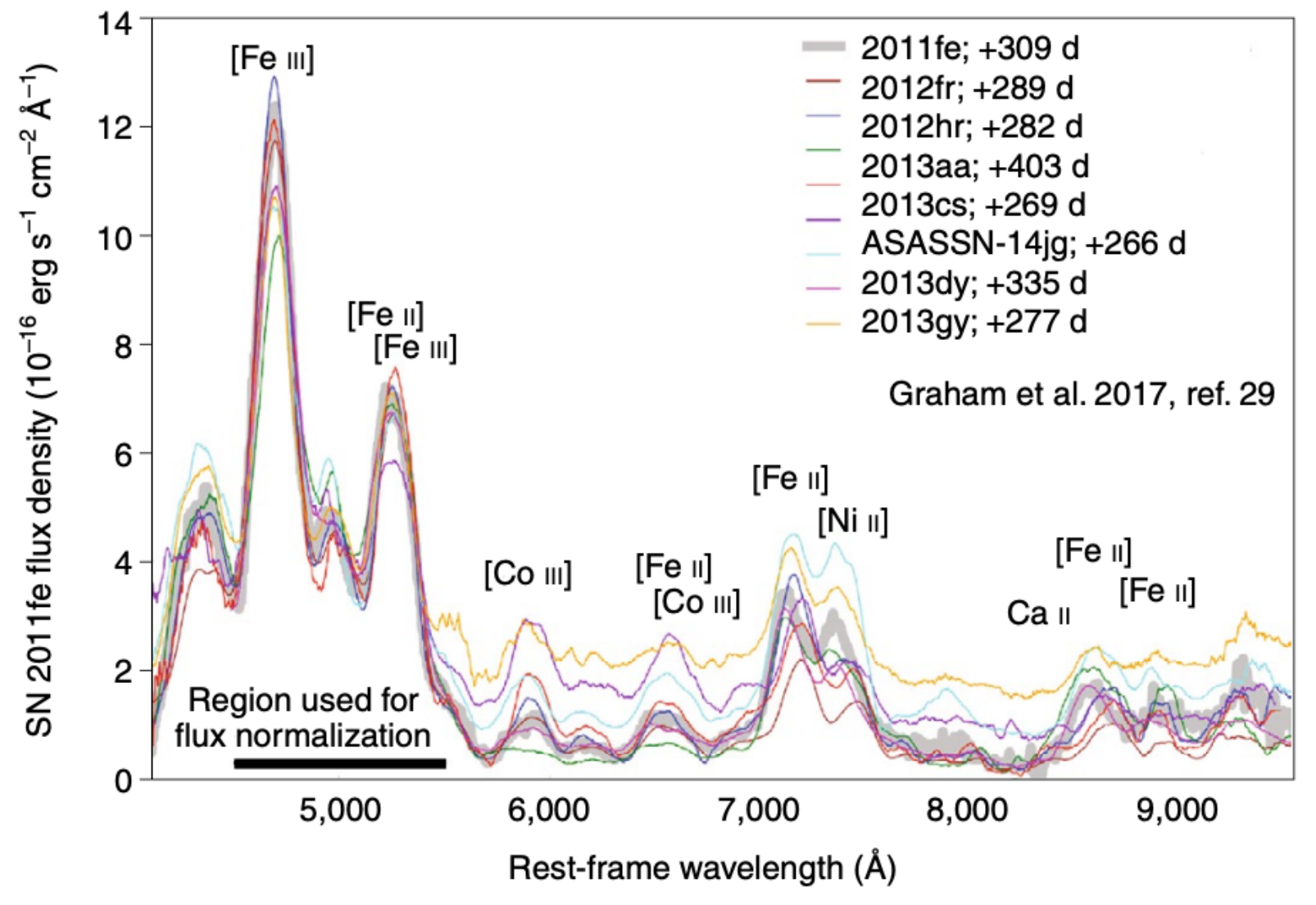}
   \caption{Spectra in the nebular late phase when the supernova is transparent, and lines from the inner regions also can be seen. Lines from various iron group elements provide useful diagnostics of supernovae of type Ia.  (From \cite{Jha:2019}).}
  \label{fig_SNIa_nebularSpectra}
\end{figure}

Thermonuclear explosions are the other physical type of supernova that is rather well studied. They are believed to result from explosive carbon fusion reactions inside a white dwarf star \citep{Seitenzahl:2017}.
These transients are remarkable in that they do not display lines from the most abundant cosmic elements H and He. 
They show a distinct spectral evolution, as measured from absorption spectra during the maximum light phase, with lines of intermediate-mass elements (Fig.~\ref{fig_SNIa_maxSpectra}), changing into emission lines several weeks later. The emission spectrum is dominated by Fe-group elements (Fig.~\ref{fig_SNIa_nebularSpectra}). 

Many uncertainties regarding the exact explosion mechanisms and the progenitor systems of SNe~Ia remain \citep{Jha:2019,Maguire:2017}. But there is general agreement on their nucleosynthesis products  \citep{Seitenzahl:2017}). 
Several observational subtypes of thermonuclear supernovae have been identified \citep{Jha:2017,Taubenberger:2017}. 
The exact mapping of progenitors and explosion mechanisms to these observed types of SNe~Ia remains open. However, it is clear that the explosion of such a small star would cool rapidly, if no extra energy source becomes available. 
Without the energy supplied by radioactive isotopes, type Ia supernovae would have very tight light curves of only a few days and would be very difficult to observe. It is the radioactive decay, which releases the energy over longer time scales and makes SNe~Ia shine for weeks to months. 

The Ni$\rightarrow$Co$\rightarrow$Fe decay chain has been identified as the energy source for SNe~Ia over 50 years ago \citep{Colgate:1969,Clayton:1974,Nadyozhin:1994}. Indirectly, the decay energy could be observed in the \emph{bolometric}, i.e. the integrated electromagnetic spectrum, light curves of the supernovae \citep[e.g.][]{Arnett:1982}. 
Since for most supernovae it is not possible to observe much more than the optical and infrared -- and in recent years also the ultraviolet 
\citep{Brown:2015} -- electromagnetic spectrum of a typical SN~Ia, one often uses the UVOIR (UV-optical-infrared) light curves as a proxy. 

Another attempt to witness the driving energy source was made by analysing the evolution of the emission lines of [Co III] and [Fe~III] in the late spectra (Fig.~\ref{fig_SNIa_nebularSpectra}) out to several hundred days after maximum. Unlike in SN~1987A, where infrared lines were used, here optical lines of a higher excitation were employed. The Fe/Co line ratio increases with time, exactly as expected from the conversion of Co to Fe, providing another piece of observational evidence of the Ni-Co-Fe decay chain acting in SNe~Ia \citep{Kuchner:1994}. 
Arnett \citep{Arnett:1982} presented a connection of the nickel mass and the maximum luminosity of type Ia supernovae through analytic light curves. Using \emph{Arnett's rule} it became obvious that these explosions produce varying amounts of $^{56}$Ni \citep{Contardo:2000}. They are not the accurate 'standard candles' as which they had been promoted before. 

Additional information is encoded in the bolometric light curves: Its shape is dominated by the radioactive-energy input and the increasing escape of $\gamma$-rays due to the decreasing density in the expanding ejecta. 
The escape can be parametrised \citep[see the formalism developed in][]{Jeffery:1999}, which then can be applied to the analysis of observed light curves. This yields an estimate of the relative ejecta masses, assuming that the density structures in SN~Ia explosions to do not vary too strongly. The resulting recognition is that most SNe~Ia appear to be less massive than the maximum-mass limit for a neutron star\footnote{This mass limit arises from the relativistic limit to the energy of electrons as their Fermi energy must rise to support the star through degeneracy pressure. It was discovered by Landau in the 1930$^{ies}$ and is called the \emph{Chandrasekhar mass limit}, attributed to Subrahaman Chandrasekhar from his work on the stability of stars.} \citep{Stritzinger:2006,Scalzo:2014a}. 

The $^{56}$Ni mass in SNe~Ia can be determined also from near-infrared light curves. These IR light curves show a characteristic second maximum after about 20 to 40 days past the B-band maximum. This second maximum is associated with the release of infrared photons due to a decrease in line opacity of iron. The cooling of the supernova leads to the ionisation change from doubly-ionised to singly-ionised iron (from [Fe~III] to [Fe~II]). The removal of the [Fe~III] lines decreases the opacity and temporarily increases the infrared flux \citep{Kasen:2006}. The time of the second maximum correlates with the $^{56}$Ni mass in the supernova \citep{Dhawan:2016}. 
  
Forty years after the direct measurement of $\gamma$~rays of $^{56}$Ni decay had been predicted from supernova models \citep[e.g.][]{Colgate:1969, Clayton:1974}, finally supernova SN~2014J was close enough for current $\gamma$-ray telescopes and the characteristic $^{56}$Co $\gamma$-ray lines could be observed \citep{Churazov:2014}.
In particular,  $\gamma$-ray spectrometry \citep{Diehl:2015,Isern:2016} at adequate spectral resolution contributed to resolve and measure the characteristic lines from the decay chain of $^{56}$Ni. 
This is observational proof of a key ingredient of modelling supernova light from thermonuclear supernovae, from $^{56}$Co radioactivity with its characteristic 111 day decay time. The amount of $^{56}$Ni inferred from the $\gamma$-ray flux of 0.49$\pm$0.09~\Msol\  \citep{Diehl:2015} is in agreement with the amount inferred from the optical brightness of the supernova, as based on the empirical peak-brightness/$^{56}$Ni heating-rate relation discussed above (\emph{Arnett's rule} \citep{Arnett:1982}). 
The $^{56}$Ni mass determination with infrared light curves \citep{Dhawan:2016} (see above) also is in agreement with this direct $\gamma$-ray based $^{56}$Ni mass determination in SN~2014J.
This is reassuring. 

Spectacularly, there had been indications in the time series of $\gamma$-ray spectra from the $^{56}$Co decay lines that individual clumps of $^{56}$Co decay at different bulk velocities appeared at different times, signifying substantial deviations from spherical symmetry of explosion or $^{56}$Co distribution \citep[see][for more detail on the $^{56}$Co $\gamma$-ray signal]{Diehl:2015}.

\begin{figure}
  \centering
  \includegraphics[width=0.8\columnwidth]{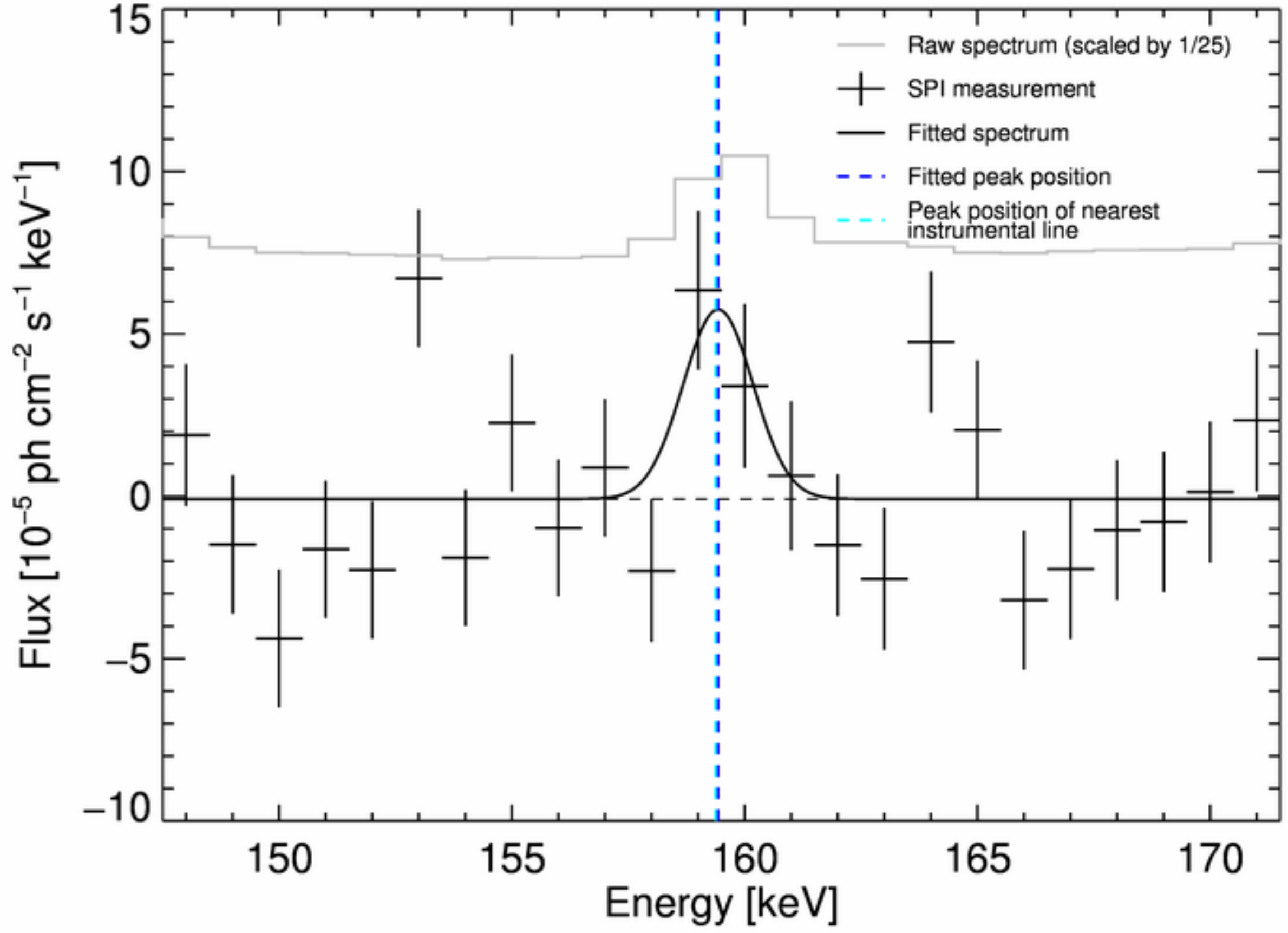}
   \caption{Early $\gamma$-ray spectrum from SN~2014J, finding the characteristic line at 158 keV from \Ni decay. Observed from a three-day interval around day 17.5 after the explosion, this confirms an early visibility of \Ni, probably close to the surface, rather than embedded in the supernova center. The SPI instrumental background is shown as a scaled histogram, showing the SN~2014J line offset from the centroid of a strong background line. The measured intensity corresponds to an initially-synthesised \Ni mass of 0.06 \Msol. (From \cite{Diehl:2015}).}
  \label{fig_SN2014J_SPI-spectrum_158}
\end{figure}

Even more spectacular was the detection of $^{56}$Ni decay lines early-on (Figure~\ref{fig_SN2014J_SPI-spectrum_158}): With a radioactive lifetime of about 9 days, it was believed that $^{56}$Ni would always be embedded so deeply within the supernova's core that even $\gamma$ rays could not leak out before $^{56}$Ni was converted to $^{56}$Co. Only some \emph{He cap} models included the possibility of early $\gamma$-ray emission from $^{56}$Ni decay \citep{The:2014}, as helium deposition on the surface of the white dwarf could cause a helium surface explosion triggering the thermonuclear supernova. Thus, this discovery of early $^{56}$Ni $\gamma$-ray lines was discussed as a support for such a double detonation \citep[see][for more detail]{Diehl:2014}, and double-degenerate progenitor models in general.

\begin{figure}
  \centering
  \includegraphics[width=\columnwidth]{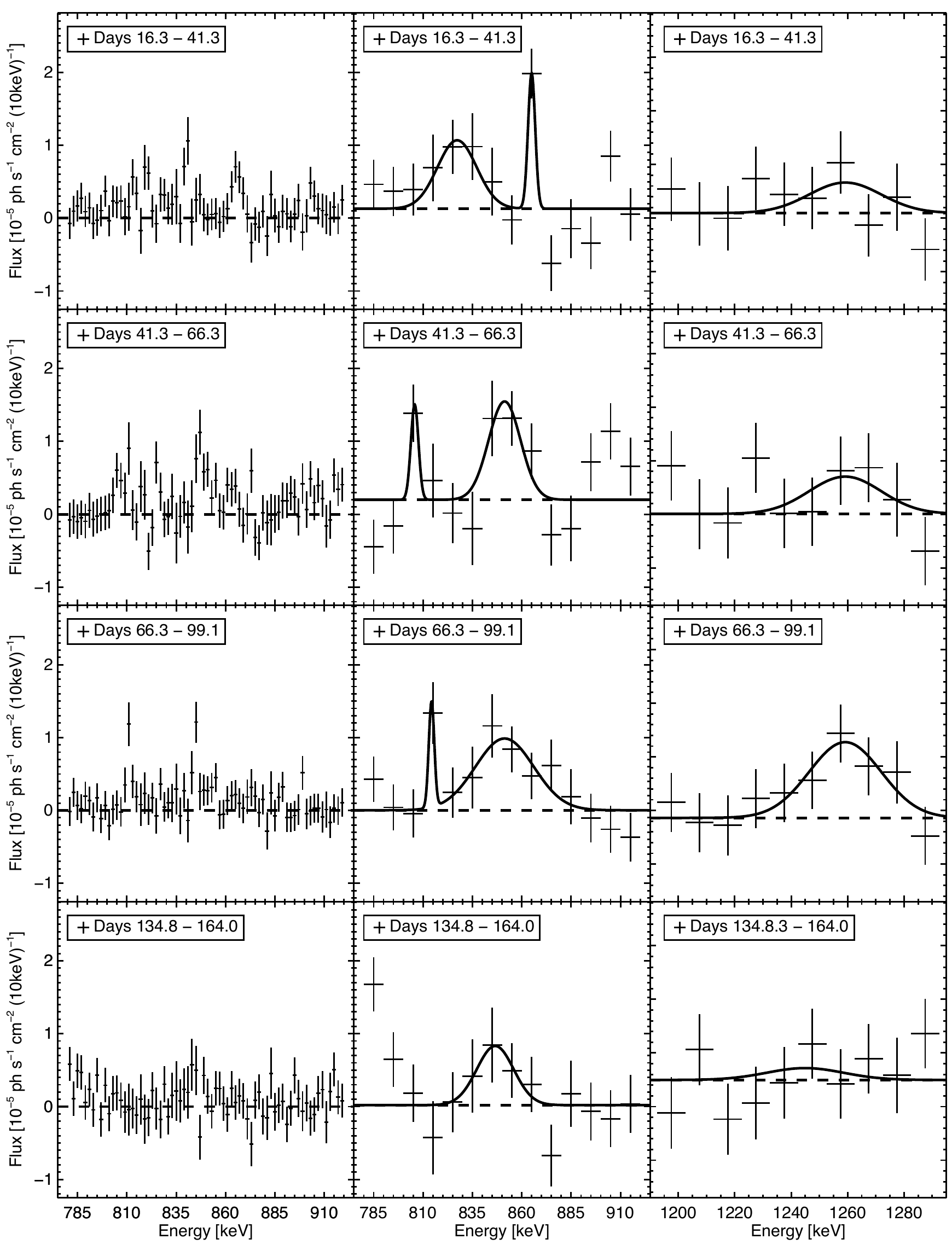}
   \caption{SN~2014J signal intensity variations for the 847 keV line ({\it center}) and  the 1238 keV line ({\it right}) as seen in four epochs of high-resolution $\gamma$-ray observations, in 10 keV energy bins.  Clear and significant emission is seen in the lower energy band ({\it left and center}) through a dominating broad line attributed to 847~keV emission, the emission in the high-energy band in the 1238~keV line appears consistent and weaker, as expected from the branching ratio of 0.68  ({\it right}). 
   For the 847 keV line, in addition a high-spectral resolution analysis is shown at 2 keV energy bin width ({\it left}), confirming an irregular appearance, i.e. not homogeneously in the form of a broad Gaussian.  (From \cite{Diehl:2015}).}
  \label{fig_SN2014J_SPI-spectra-set-847-1238}
\end{figure}

The line emission from decay of $^{56}$Co was traced over 3 months, 
and the line centroid and width could be constrained in their evolution, as shown in Figure~\ref{fig_SN2014J_SPI-spectra-set-847-1238}. Naively, one would have expected a Doppler-broadened Gaussian line to appear and fade in its brightness, with some centroid shift from blue to red as the facing ejecta would shine early and receding ejecta from the distant part of the supernova would add later.
But, as shown in Figure~\ref{fig_SN2014J_SPI-spectra-set-847-1238}, the spectra rather show surprising spikes, which appear to come and go with time. Although statistical noise is substantial, it was asserted that a smooth appearance of the lines as suggested from a spherically-symmetric gradual transparency to centrally located $^{56}$Ni could be excluded \citep{Diehl:2015}. 
Thus, $\gamma$-ray spectroscopy provides an additional indication that non-sphericity may be significant in type Ia explosions, and rather smoothed out in signals based on bolometric light re-radiation from the entire supernova envelope.   
{        The standard M$_{Ch}$ model with its 1D nature appears inadequate to describe these observations.}


Some newly synthesized elements can be tracers of the neutron richness in the explosion sites \citep[e.g][]{Brachwitz:2000} and also the progenitor mass \citep[e.g.][]{Seitenzahl:2017}. The higher central densities in higher-mass progenitors, and the very low densities in mergers on the other hand, can result in the unique production of certain neutron-rich elements. 
Most prominent in the case of SNe~Ia appears to be stable $^{58}$Ni. It has now been identified in a number of late time-spectra \citep{Dhawan:2018, Floers:2018, Floers:2020}. A Ni emission line showing after several hundred days must come from a stable isotope. The comparison with models shows that the most-favoured progenitor white-dwarf mass range is sub-Chandrasekhar \citep{Floers:2020}, {        rather than a M$_{Ch}$ model.}

\begin{figure}
  \centering
  \includegraphics[width=0.8\columnwidth]{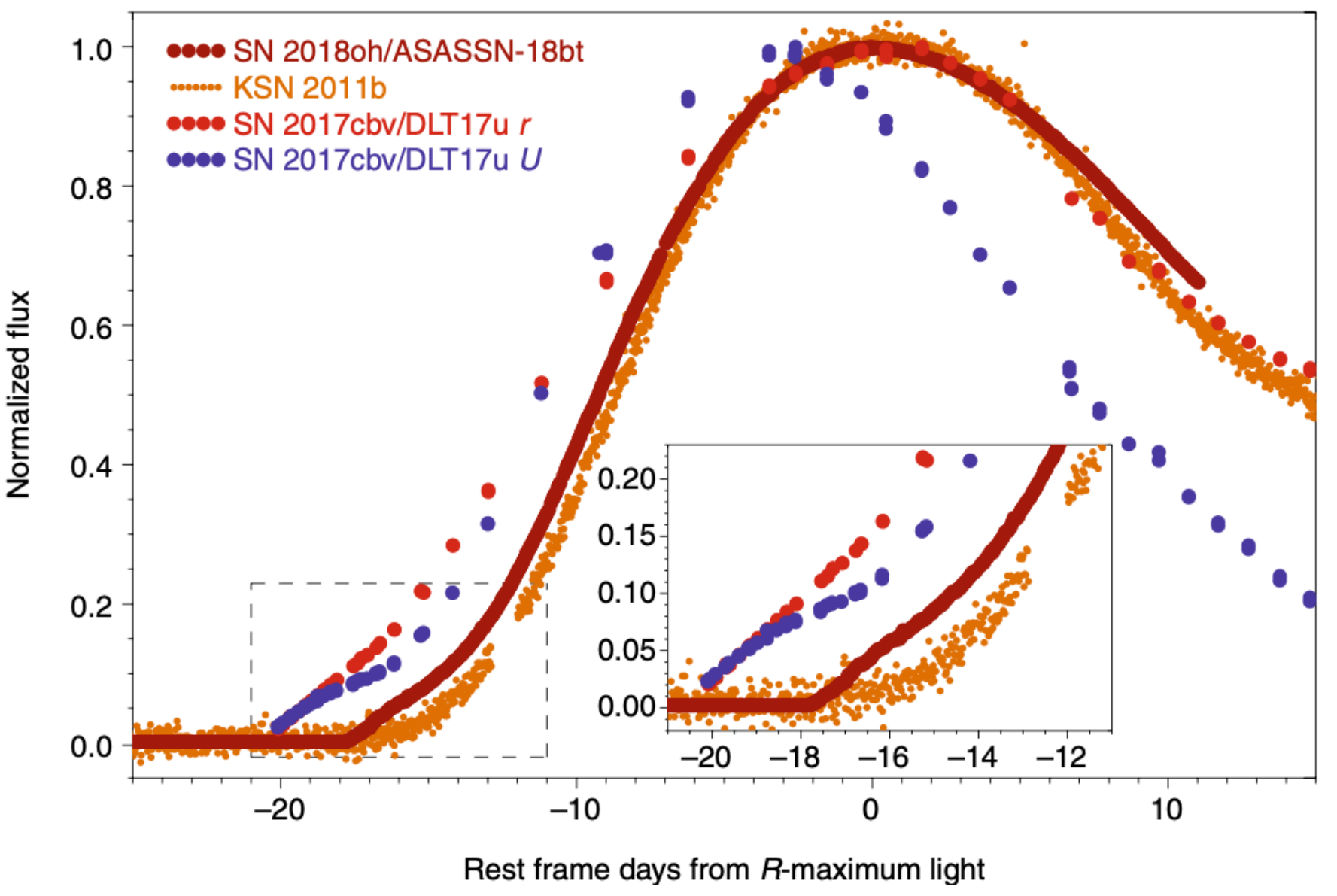}
   \caption{Early light curves in optical/bolometric emission \citep{Jha:2019}, may hold information on how the explosion was started.}
  \label{fig_SNIa_earlyLC}
\end{figure}

It is believed that the very early light curve may be a good diagnostic for the surface composition, potential interaction with circum-stellar material or a companion star. Figure~\ref{fig_SNIa_earlyLC} shows examples of a few supernovae where this early phase could be observed \citep{Jha:2019}. Evidently, different rise behaviour can be seen which may indicate radioactive material close to the surface predicted in some double-detonation models or interaction of the ejecta with a companion star or material lost by the system prior to explosion. The shape of the rising light curves further gives a hint on the extent of mixing of material in the explosion \citep{Noebauer:2017}.

\subsubsection*{Nucleosynthesis clues from supernova remnants}
 

Supernova remnants are formed as a supernova expands into the interstellar medium and the interactions with the surroundings dominate the physical processes \citep[see][for reviews]{Reynolds:2008,Reynolds:2017}.
These interactions with surroundings can lead to illuminations of the supernova ejecta, e.g. through heating and ionisation from the outer discontinuity and the inward-travelling shock wave.

\begin{figure} 
\centering 
\includegraphics[width=0.49\columnwidth]{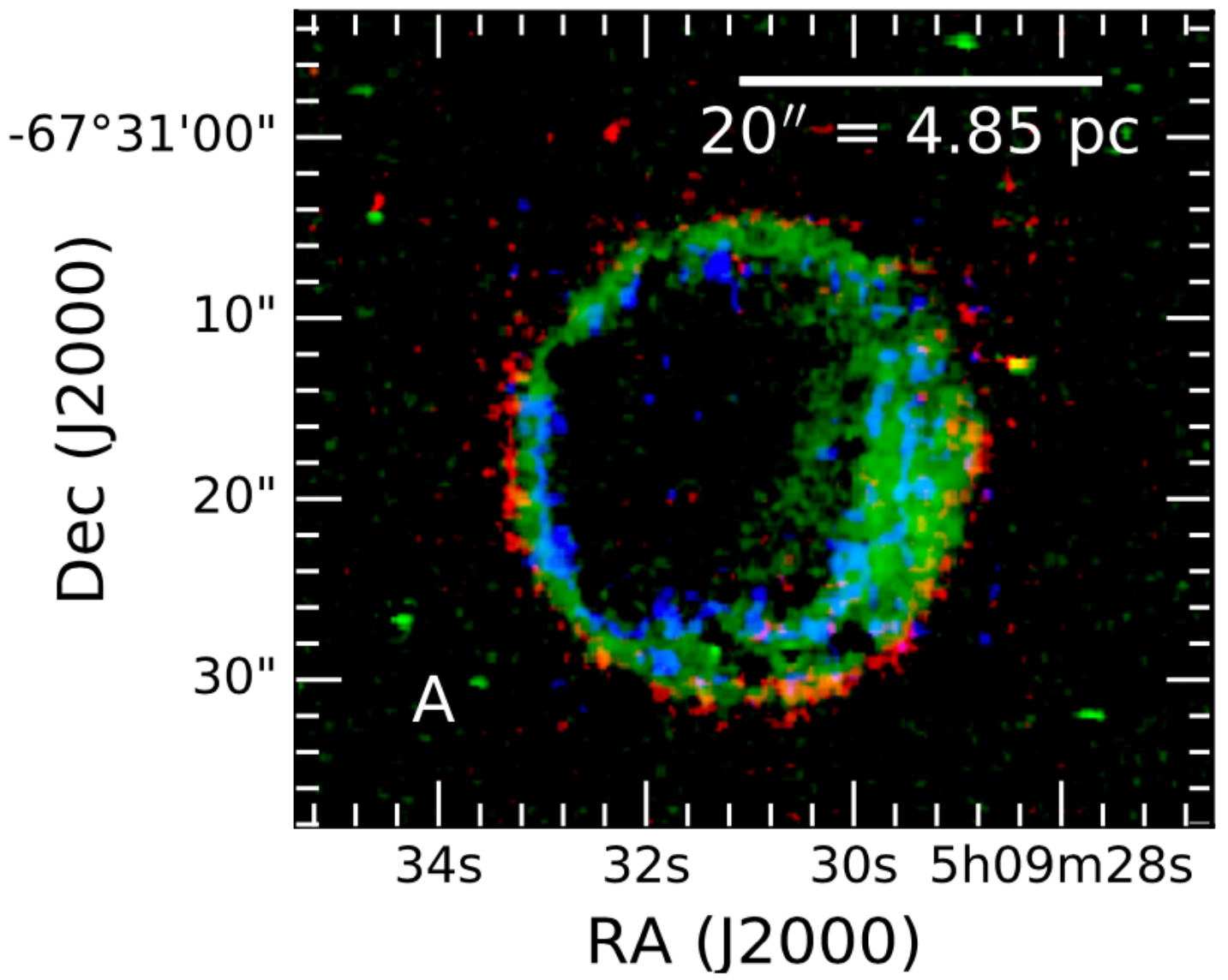}
\includegraphics[width=0.49\columnwidth]{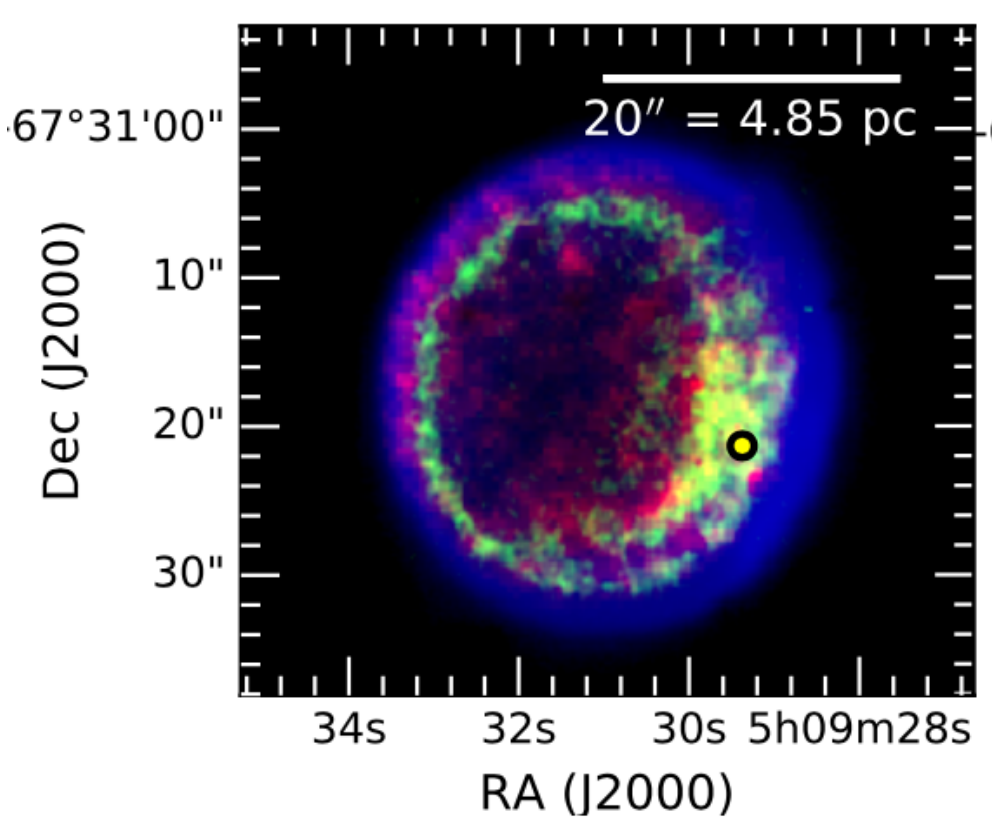}
\caption{{\it Left:} Multi-line image of the young supernova remnant 0509-67.5 in the Large Magellanic Cloud in optical lines from sulfur (red, S~XII) and iron (blue Fe~IX, green Fe~XIV). 
{\it Right:} Multi-messenger image of the same remnant, with X-rays (red; measured with the Chandra satellite), H$\alpha$ (blue), and iron (green; Fe~XIV) (the latter measured with MUSE on the ESO VLT).  \citep[From][]{Seitenzahl:2019}.}
\label{fig_SNR0509}
\end{figure} 

Combining X-ray with optical spectroscopy, the hot plasma in supernova remnants offers an opportunity to obtain a tomographic view of the material composition. Figure~\ref{fig_SNR0509} illustrates recent work
for the type Ia supernova remnant 0509-67.5. Here, material near the reverse shock is seen in different ionisation stages, at the shock in X~rays, and further inside in optical lines from highly-ionised Fe \citep{Seitenzahl:2019}. 
Such studies provide better detail about how atomic-line excitations depend on the kinematics of the ionising reverse shock, avoiding misinterpretations from unknown ionisation levels (as shown in Figure~\ref{fig_CasAimage} for Cas~A).

\begin{figure}
  \centering
  \includegraphics[width=0.8\columnwidth]{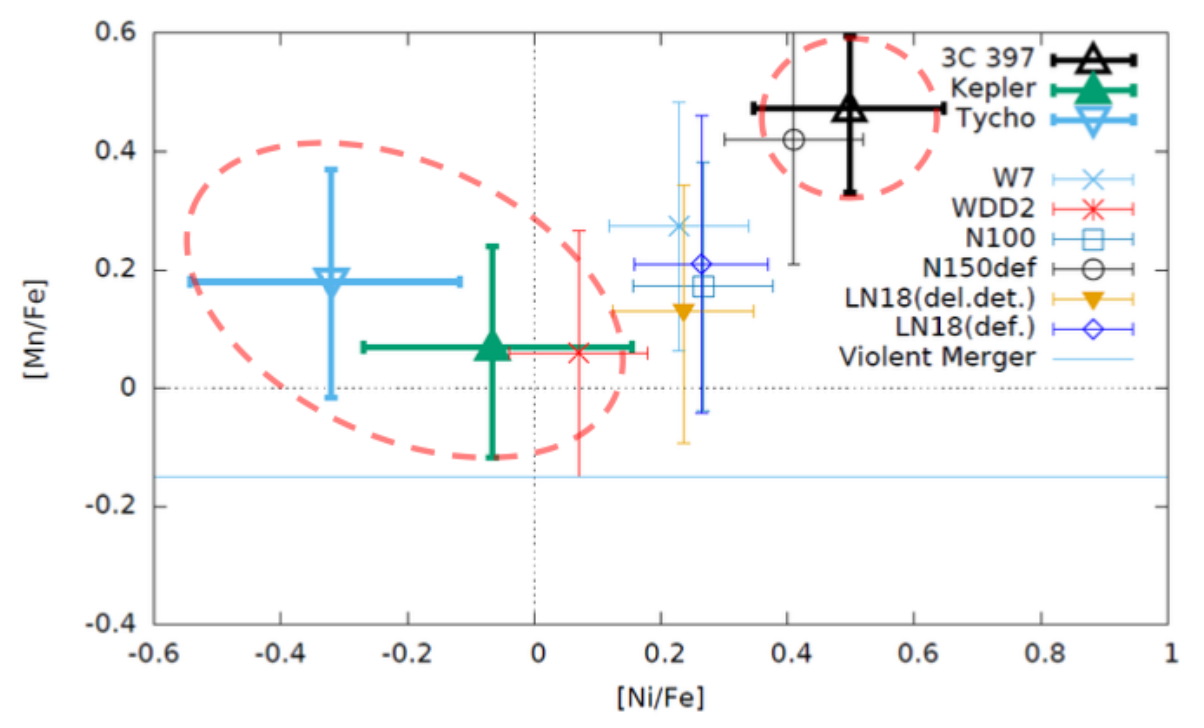}
   \caption{Elemental ratios of Mn and Ni over Fe as extracted from three different supernova remnants and their X-ray spectra, versus calculations from different SNe~Ia models. (From \cite{Mori:2018}).}
  \label{fig_SNIa_Mn-Fe}
\end{figure}

\begin{sloppypar} X-ray spectra from young supernova remnants display emission lines from several species including Mn and Fe, as the reverse shock has ionised inner ejecta and they de-excite \citep{Vink:2012}.
Mn is produced as a decay product from $^{55}$Co through $^{55}$Co$\rightarrow ^{55}$Fe$\rightarrow ^{55}$Mn, 
and has been identified as a tracer for high-mass explosions \citep{Seitenzahl:2013,Seitenzahl:2015}. 
Mn abundances versus metallicity as observed from stellar spectroscopy in stars indicate that some fraction of SNe~Ia explode in high-mass progenitors \citep{Seitenzahl:2013}. 
Analysis of supernova remnants that had been attributed to SNe~Ia explosions \citep{Mori:2018} are inconclusive, however, as different supernova remnants exhibit a variety of Mn/Fe ratios on either side of model calculations (Figure~\ref{fig_SNIa_Mn-Fe}). 
Other explosion models, in particular He-shell detonations of sub-Chandrasekhar mass progenitors, can also create increased amounts of Mn \citep{Lach:2020, Gronow:2021a}. \end{sloppypar} 

$^{55}$Co decay also may be directly measured through an X-ray line at 5.9~keV from the atomic K-shell transition as the $^{55}$Mn daughter de-excites; but this has yet to be observed with future X-ray spectroscopy \citep{Seitenzahl:2015}. 


\begin{figure}
  \centering
  \includegraphics[width=\columnwidth]{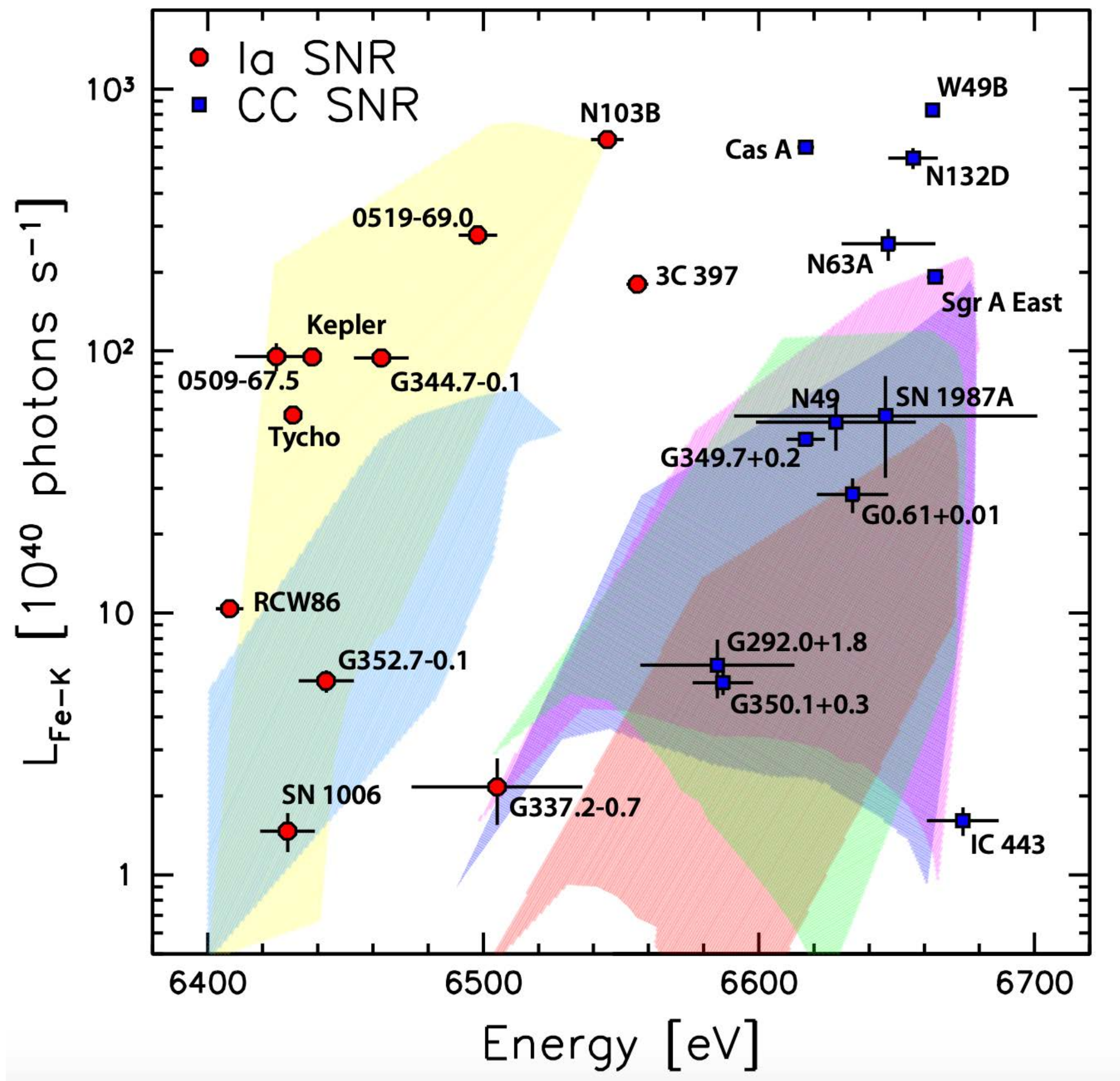}
   \caption{Analysis of X-ray emission from supernova remnants reveals that brightness and line centroid energy of Fe line emission are characteristic for the explosion type. (From  \cite{Patnaude:2015}).}
  \label{fig_X-SNRs}
\end{figure}

Often, supernova remnants that are bright enough for spectroscopy of high-energy emission in X- and $\gamma$-rays cannot be attributed to an explosion type, i.e., thermonuclear or core-collapse. Nevertheless, spectroscopy has provided interesting results.
From a survey for radioactive emission in $^{44}$Ti lines, comparing to expectations from supernova rates, Galactic remnants support our current belief that $^{44}$Ti radioactivity is a characteristic of core-collapse events more than of thermonuclear explosions \citep{Weinberger:2021}. 
Comparing X-ray luminosity in the Fe K$_\alpha$ line to the spectral line energy of the X-ray emission, it could be shown \citep{Patnaude:2015} that X-ray spectroscopy results \citep{Yamaguchi:2014}  cluster to allow discrimination of the remnant origins versus supernova types (Figure~\ref{fig_X-SNRs}). 

\subsubsection*{Supernova dust grains}
As another messenger, a small population of stardust grains show the signatures of explosive nucleosynthesis, and are believed to originate mostly in core-collapse supernovae (and some in novae) \citep{Hoppe:2017,Nittler:1996}. 
Comparison of isotopic details to models of supernova nucleosynthesis is more complex here, compared to the above-discussed AGB-star studies, because it is uncertain how different supernova layers can mix during the explosion and before dust is formed. 
Therefore it is uncertain if the composition of the dust grains forming in the ejecta may reflect either the mixing between inner and outer layers \citep[see, e.g.,][]{travaglio99}, or, for example, nuclear burning at higher energies than those used as standard in  models \citep{Pignatari:2013a},  or the realisation of their nucleosynthetic signatures in different parts of an asymmetric supernova explosion. 
In spite of the intrinsic difficulties of a detailed interpretation of their composition, due to the many uncertainties that affect the modelling of supernova explosions, stardust grains represent a main evidence that supernova stardust survive the explosive shock waves, and is present in the interstellar medium; it even was within the inventory of presolar dust at the time of the formation of the Solar System.

\subsubsection*{Nova explosions}
\label{sec:novae}

Novae are believed to originate from stellar explosions driven by explosive hydrogen burning \citep[see][for recent reviews]{Della-Valle:2020,Chomiuk:2021}. Their contributions to cosmic nucleosynthesis are thought to be dominant for few elements only, including fluorine, sulfur, and lithium, and major contributors to C, N, and O isotopes \citep{Gehrz:1998}.
The hydrogen-burning sequence proceeds on the proton-rich side of the stable isotopes in the nuclide chart (see Figure~\ref{fig_tableOfIsotopes}) up to Calcium, after breaking out of the CNO cycle that dominates H burning initially. 
Many $\beta^+$~unstable isotopes are produced alongside, most of which are quite short-lived, and are expected to produce a large amount of positrons. 
Among observational signatures a bright flash in 511~keV annihilation $\gamma$-rays is predicted to signal the nova explosion, but has not been detected up to now. 
Similarly, a UV/X-ray flash as the shock wave from the explosion reaches the photosphere of the expanding nova, has not yet been detected \citep{Chomiuk:2021}.
Novae are discovered in transient surveys only after reaching sufficient brightness to be recognised as an optical transient, which is typically 1-2 weeks after the thermonuclear runaway occurs \citep{Jose:1998,Hernanz:2002,Jose:2007}. 
The release of nuclear energy leads to expansion of the nova envelope, which, in turn, stops the nuclear runaway due to adiabatic cooling with the expansion, so that radioactive heating of the envelope drives the nova in this expansion phase.

From spectroscopic observations of nova envelopes, large enrichments have been found in elements C, O, Ne, and up to S, confirming the hot-H burning model for nucleosynthesis \citep{Gehrz:1998,Chomiuk:2021}. But spectroscopic identifications are difficult, as the nova envelope expands, producing Doppler shifts and broadenings of characteristic atomic lines.
More recently, lines attributed to $^7$Be have been identified \citep{Molaro:2016,Molaro:2022}, which suggest that novae are significant sources of $^7$Li in the Galaxy.
A search for the corresponding $\gamma$-ray line from the decay of $^7$Be (lifetime 111 days) to $^7$Li, however, could not be found \citep{Siegert:2018,Siegert:2021}.
Similarly, $\gamma$-ray line emission from the decay of $^{22}$Na (lifetime 3.8 y) that is predicted to be produced in particular in novae from O-Ne white dwarfs \citep{Jose:2004} have not been detected \citep{Siegert:2021}.

On the other hand, presolar grains have been found that carry an isotopic signature which is reminiscent of hot hydrogen burning \citep{Jose:2016b}. This interpretation  depends on production of C-rich dust in nova envelopes. Infrared observations of novae have lent support to such C-rich dust grains being present in nova envelopes.

\subsubsection*{Nucleosynthesis in other types of explosions}

Other types of cosmic explosions have been observed. Among them are super-luminous supernovae \citep[e.g.][]{Gal-Yam:2019}, hypernovae and long-duration $\gamma$-ray bursts \citep[e.g.][]{Nomoto:2010a} and kilonovae and short-duration $\gamma$-ray bursts \citep{Abbott:2017, Smartt:2017, Margutti:2021}. All these explosions contribute to the nucleosynthesis of heavy, in particular r-process elements \citep{Cowan:2021}. The exact contributions are difficult to assess as no clear models have been identified for the various explosion types. All these explosions are extremely rare with rates several orders of magnitude lower than supernovae. At the same time, each individual event might produce large masses of heavy elements and even small rates may represent significant contributions. 


\emph{Superluminous supernovae} are a special class of explosions, which emit {        extraordinarily-large} amounts of light. Some of these objects might be powered by the conversion of kinetic energy to radiation by shocks in extended envelopes or circum-stellar material, e.g. the objects displaying narrow emission lines. In general, two types of super-luminous supernovae are distinguished  (analoguous to the regular supernova types): \emph{type I}, which do not display hydrogen lines in their spectra; and \emph{type II}, where prominent hydrogen lines are found \citep{Gal-Yam:2019}. All these supernovae are associated with massive stars and presumably will produce similar nucleosynthesis yields to regular core-collapse supernovae. The energy source is debated, but could come from the braking of highly magnetised neutron stars formed in the collapse -- so called \emph{magnetars} \citep{Woosley:2010}. The exact mechanisms of the energy conversions remain uncertain, but the effects on the nucleosynthesis are expected to be minimal. 

\emph{Long $\gamma$-ray bursts (GRBs)} are highly collimated so that the total (isotropic) energy is several orders of magnitude higher than that of regular core-collapse supernovae.
{        In general, nucleosynthesis reactions are expected to occur in black-hole accretion disks \citep{Mukhopadhyay:2000,Pruet:2003,Surman:2004,Wanajo:2012,Janiuk:2014}, may they arise from neutron star mergers or from collisions of stars with a black-hole companion in binary systems. The latter would produce gravitational-wave events. However, it remains unlikely, and at least unclear, if substantial material would be able to escape into the interstellar medium.}
{         Production of $^{56}$Ni with }masses of several tenths of \Msol\ have been inferred for such explosions \citep{Sollerman:2000a,Woosley:1999,Surman:2014}. In many GRBs a supernova can be observed after the afterglow \citep{Woosley:2006b}. 

Potential r-process element production has been predicted, but the models are very complex and no consensus has emerged so far \citep[see the discussion in][]{Cowan:2021}. 
\emph{Hypernovae} are also often associated with long GRBs, although in these cases the GRB jet is not observed. Hypernovae are characterized by high expansion velocities and are also associated with massive stars, sometimes stripped of their hydrogen envelope \citep{Nomoto:2010}.

\begin{figure}
  \centering
  \includegraphics[width=0.8\columnwidth]{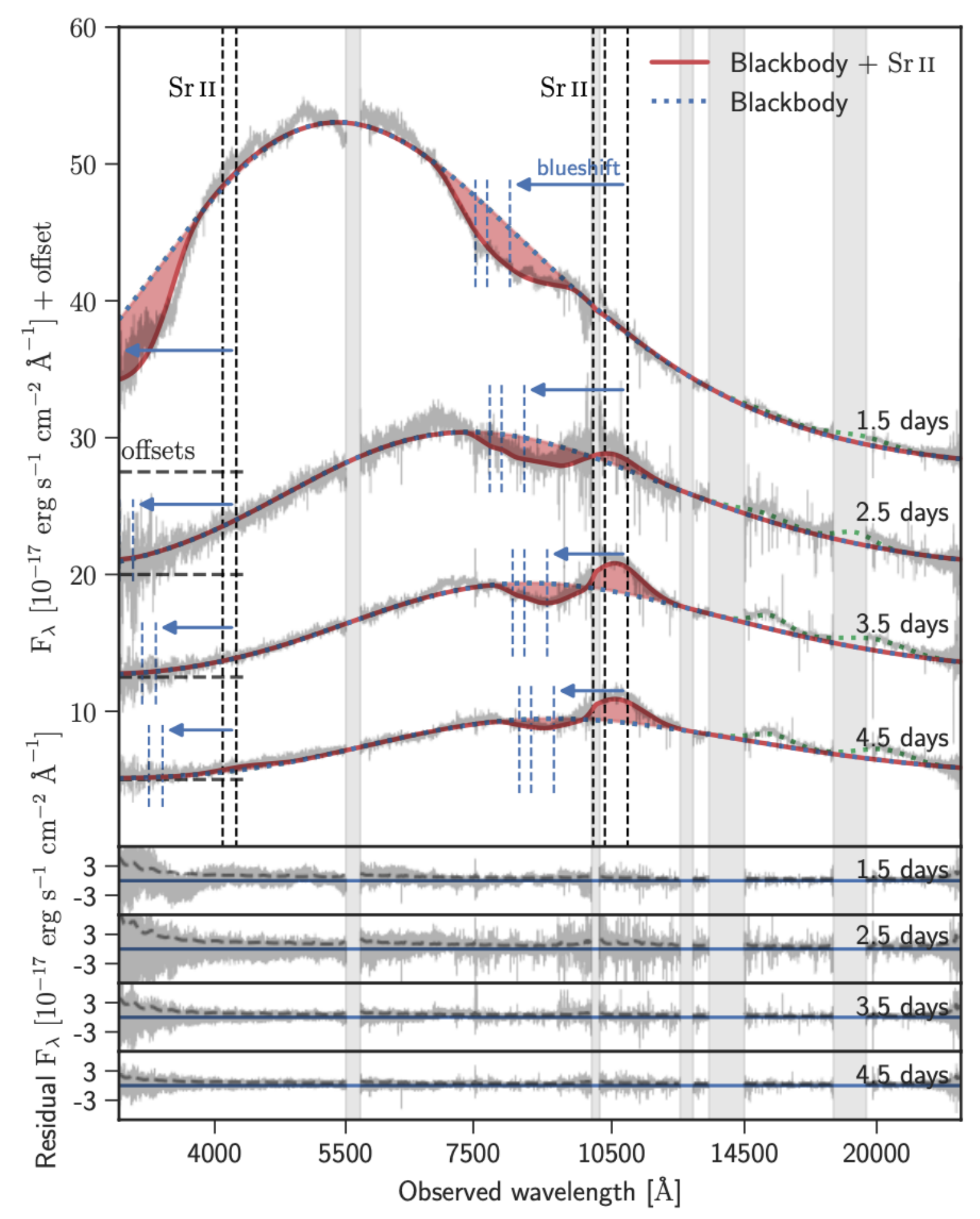}
   \caption{Indication of a spectral line from Sr in the IR spectrum of the kilonova corresponding to gravitational-wave event GW170817.  (From \cite{Watson:2019}).}
  \label{fig_GW170817_Sr}
\end{figure}

\emph{Kilonovae} are associated with short $\gamma$-ray bursts, and the detection of gravitational waves points to the merger of binary neutron stars. 
The recent detection of the electromagnetic counterpart of a gravitational wave source called GW170817 or AT2017gfo demonstrated that such events can be observed \citep{Abbott:2017, Smartt:2017, Pian:2017,Margutti:2021}.
The neutron density in such mergers is naturally extremely high and they had been proposed as potential sites of r-process nucleosynthesis \citep{Rosswog:1999}. 
The kilonova emission that was tracked from the GW170817 source, however, shows a radioactive afterglow \citep{Smartt:2017,Margutti:2021}. Moreover, spectroscopic data supports the synthesis of heavy elements and indicates  elements such as Sr might have been ejected as shown in Figure~\ref{fig_GW170817_Sr}. 
Kilonovae could contribute significantly to the nucleosynthesis of heavy elements, despite their rare occurrences. The rate of kilonovae is estimated to be about 48~Gpc$^{-3}$~y$^{-1}$ \citep{Chruslinska:2018}, translating into $\sim$25 events~My$^{-1}$ for our own Galaxy.
{        But uncertainties are large: a range between $\sim$0.1 events~My$^{-1}$ and  $\sim$100~events~My$^{-1}$ is consistent with a standard model of binary evolution \citep{Chruslinska:2018};  observational uncertainties are even larger, as merely three double neutron stars as candidate sources are known in the Galaxy.} 
The amount of r-process material produced in such mergers can be substantial, but observations of more events are needed to interpret the spectral evolution in terms of astrophysical dynamics and underlying processes. 

Potentially interesting, although rarely-occurring, cosmic nucleosynthesis events also include explosions of very massive ($>100 M_\odot$) progenitors, where the end of stellar evolution might be different: In such \emph{pair instability supernovae} either the progenitor is disrupted entirely or a black hole is formed directly \citep{Heger:2002,Woosley:2007}. No observational evidence of such explosions has been found so far, although there are a number of events among the superluminous supernovae that have been proposed as \emph{pair-instability supernovae} \citep[e.g.][]{Arcavi:2017, Woosley:2018, Sollerman:2019, Gomez:2019}.

\subsubsection*{Specific sources of nucleosynthesis}

Molecular lines can become important for abundance measurements in cold interstellar gas.
Advances in sub-mm spectroscopy, 
and corresponding advances in laboratory studies to identify lines for molecules including radioactive species have been made.
Rotational lines of $^{26}$AlF could be measured from a nova-like source called CK Vul \citep{Kaminski:2018}.
This represents a breakthrough for isotope spectroscopy in molecules due to the spatial resolution, which allowed the direct identification a specific source.
However, it is also clear that molecule production will only occur under very special conditions,  such as in this case. This bias makes it difficult to derive general conclusions on $^{26}$Al sources, like in the cases of meteoritic inclusions. But, learning about specific sources and their nucleosynthesis complements observations of entire source populations, such as in $\gamma$ rays, stardust, and cosmic rays. The sub-mm observations are important for molecule-rich environments such as AGB stars and even proto-planetary discs.

\begin{figure*} 
\centering
\includegraphics[width=1.8\columnwidth,clip]{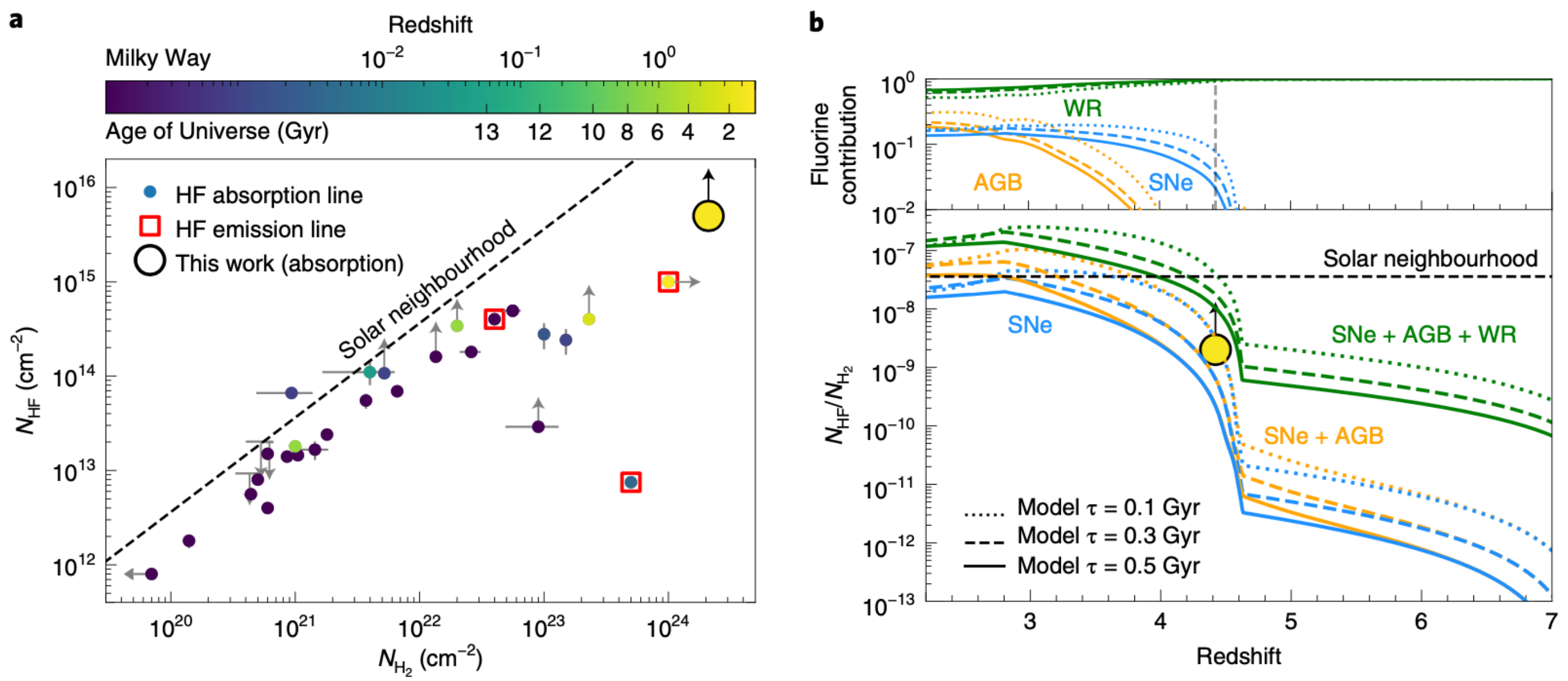}
\caption{The abundance of fluorine in gas clouds of different densities, as measured through the HF molecule and its 1-0 transition, provides a tracer of  nucleosynthesis. 
{\it Left:} HF abundance appeared to decline towards more-massive gas clouds, until the recent measurement showed abundant HF in galaxy NGP-190387. Redshift information allowed to attribute ages to the respective measurements. {\it Right:} The nucleosynthesis of fluorine in our current Galaxy is mostly due to AGB stars. At earlier times, however, supernovae, and in particular Wolf Rayet stars, are much more efficient producers of fluorine, hence from more-massive and faster-evolving stars. (From \cite{Franco:2021}). }
\label{fig_HF}       
\end{figure*} 

There is another application for nucleosynthesis in the distant universe. 
The fluorine-carrying molecule HF shows its lowest-lying excitation line at 1232~GHz and was observed in a lensed galaxy at a redshift of 4.42 \citep{Franco:2021}. 
Fluorine is very reactive and thus capable to split molecular hydrogen to form the HF molecule, which by itself is very stable and accumulates readily as a tracer of fluorine nucleosynthesis.
The galaxy had been identified based on a rising spectrum at 500~$\mu$m wavelength. This indicates a dusty, star-forming galaxy. The C$_I$ line was used to calibrate the molecular hydrogen content.  
As Figure~\ref{fig_HF} shows, in this way we learn that early nucleosynthesis of fluorine is due to very massive stars, while from the current universe, AGB stars are established as dominating fluorine production.

\section{Understanding the cosmic cycle of matter}
\label{cosmicMatterCycle}

\subsection{The cosmic cycle of matter}

\begin{figure}
  \centering
  \includegraphics[width=\columnwidth]{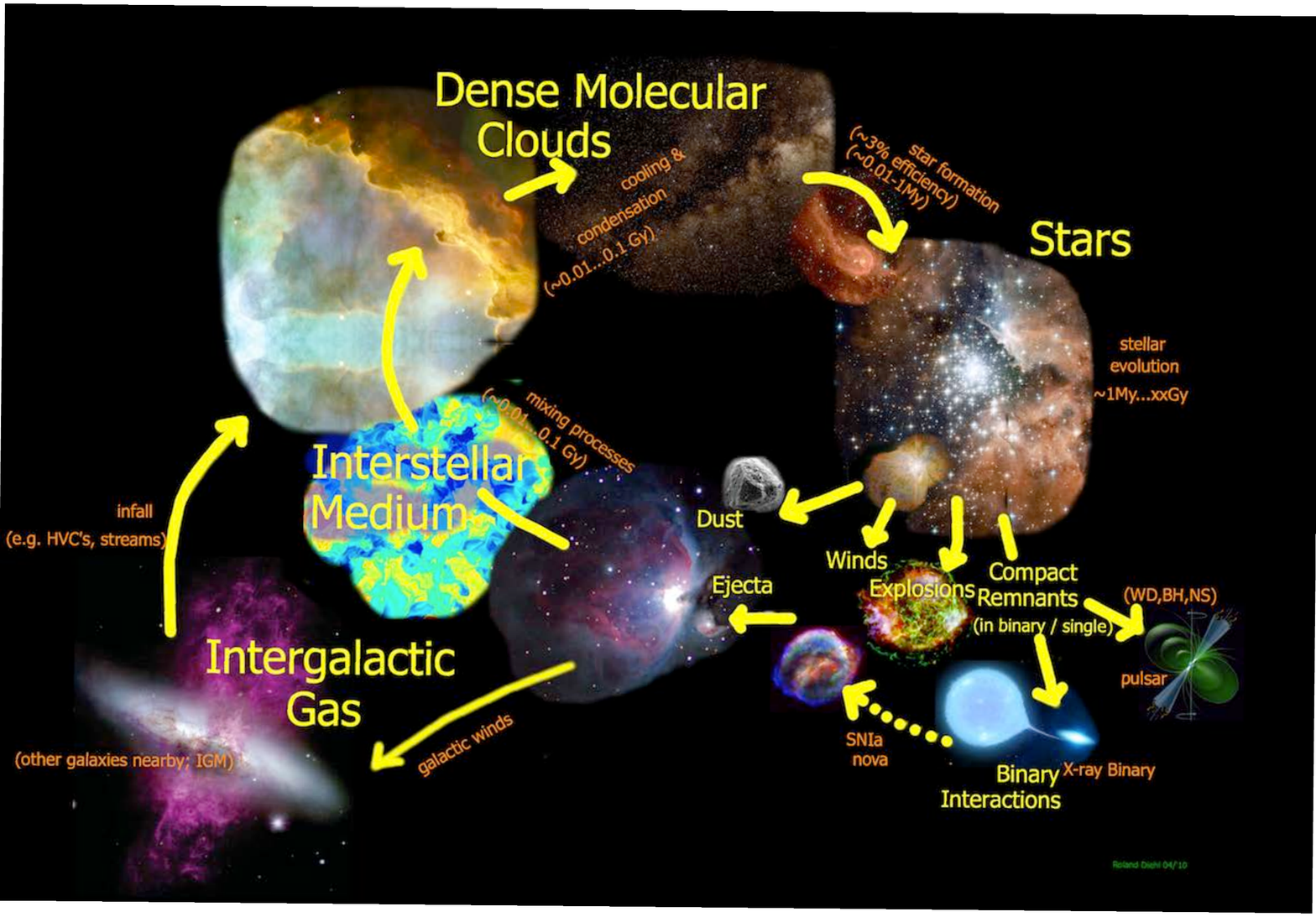}
   \caption{The cycle of matter, with stars forming out of interstellar gas, evolving towards wind and explosions releasing newly-formed isotopes, that are then fed back into the next generation of stars.}
  \label{fig_matter-cycle}
\end{figure}

Once new products of nucleosynthesis have been ejected from stars or explosions, they have to find their way into future generations of stars. 
The cosmic compositional evolution results from the re-cycling of cosmic gas, as shown in Figure~\ref{fig_matter-cycle}, with continuous increase of elemental abundances in every cycle. 
Secondary nucleosynthesis processes have a greater variety of reaction paths to form new nuclei. 
The yields are proportional to the abundances of seed nuclei, hence depend on the nucleosynthetic history of stellar gas, and the re-cycling of nucleosynthesis ashes.
 
The different sources of nucleosynthesis occur with different frequencies at different cosmic times, and in different environments. Figure~\ref{fig_cosmic_evolution_metallicity} shows the cosmic time axis from top (big bang) to bottom (current time), together with the different nucleosynthesis sources. Massive stars appear earliest, and continue to contribute until today, albeit the range of stellar masses that dominates the production changes with increasing metallicity. Lower-mass stars are formed along with the earliest massive stars, but their more modest contributions appear later, due to the longer evolution towards mass-shedding stages. 
Massive stars with their rapid stellar evolution and then their core-collapse supernova explosions are the dominating sources at early times.
The typical environment where nucleosynthesis products are ejected is the turbulent interstellar gas that is energised from stellar winds and explosions.
 
Binary evolution and the accretion of matter onto otherwise inert white dwarf requires more time, typically of order one Gyr \citep{Iben:1984,Maoz:2014}. This implies a new important source of nucleosynthesis appearing with such a delay after the formation of first-generation stars in a galaxy. These are the type Ia supernovae, which are efficient mass ejectors with abundant product nuclei up to the iron group. Their contribution can be easily identified in histories of observed abundances relative to the abundance of iron. Therefore, the abundance of iron also serves as the most widely used cosmic clock, and would be strictly proportional to time if supernovae of type Ia occurred regularly and their ejected materials would be well mixed into observable cosmic gas, such as in stellar photospheres. 
Due to the delay time of such binary evolution, supernovae of type Ia are expected to occur in an environment typically different from massive stars, as during the binary evolution the interstellar gas may have cooled and be dispersed, plus the binary may have moved away from denser regions and other stars due to even small kicks received when the binary system formed.
Also, the evolution of binaries towards these supernovae is independent of the star-formation activity within the galaxy, and thus continues even in epochs of galaxy evolution where feedback and other processes may have increased, or extinguished, formation of stars.
Therefore, enrichments from type Ia supernovae occurs in all types of galaxies, even \emph{elliptical} galaxies that have ceased forming stars. 

An example of using Galactic chemical evolution studies to constrain binary evolution addresses different subtypes of supernova type Ia progenitors (see above):
The ratio of Mn and Zn abundances in the Milky Way shows a flat evolution compared to the Fe abundance. This implies that most type Ia SNe in the Galaxy are well below a M$_{Ch}$.
Rather, a dominance of the double-degenerate progenitor models (75\% contribution) is suggested.
Thus, abundance archeology includes messages about supernova explosion types and binary evolution towards these.

Similarly, the evolution of binaries including a neutron star towards becoming a source of nucleosynthesis takes time; the binary orbit shrinks gradually due to the energy loss from gravitational radiation.  Estimates of such evolutionary times are uncertain, but are $\geq$1~Gy. Therefore, the disconnection to the massive-star forming environment and appearance of nucleosynthesis ashes in a more remote and different environment, as for type I supernovae, holds even more for kilonovae and the ashes ejected from neutron star mergers.

Tracing the re-cycling efficiency of matter from the nucleosynthesis sources is a challenging task for astronomy in the attempt to describe the abundance evolution on all scales. 

\begin{figure*}[ht] 
\centering
\includegraphics[width=2.0\columnwidth,clip]{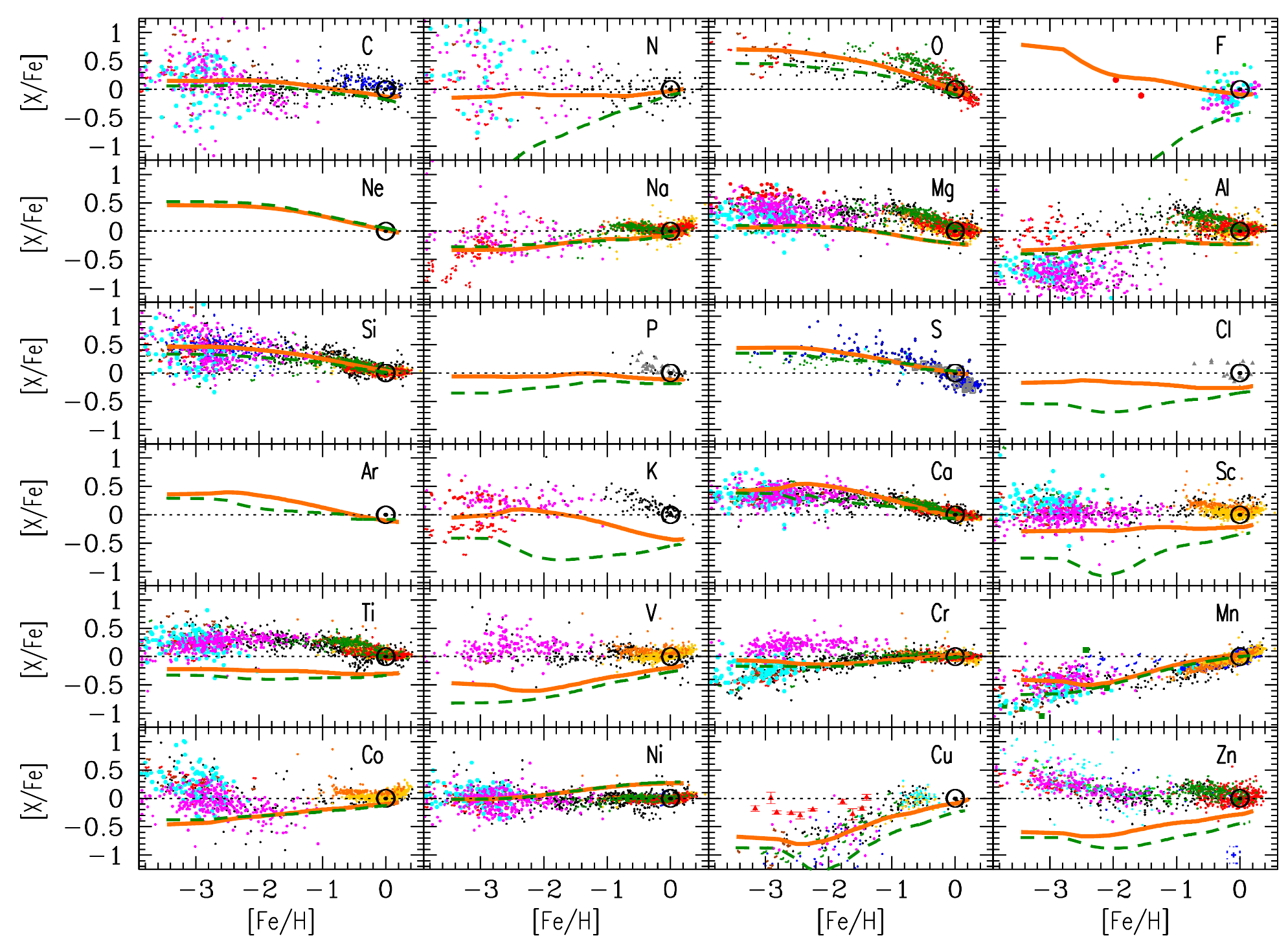}
\caption{The elemental abundances for elements up to Fe, as observed in stars (dots), and as modelled in current treatments of Galactic compositional evolution (lines). (From \cite{Prantzos:2018}).  }
\label{fig_abundanceGal}       
\end{figure*} 

\subsection{Cosmic compositional evolution} 
\label{chemEv}

%
%
%

\subsubsection*{The solar neighborhood and the Galaxy}
The solar neighbourhood has been studied using hundreds of normal stars like our Sun \citep[e.g.][]{Edvardsson:1993,Fuhrmann:2004,Reddy:2006,Adibekyan:2013,Bensby:2014}. The solar neighbourhood is dominated by stars attributed to the \emph{thin disk} of our Galaxy, with a small fraction of \emph{thick-disk} and \emph{halo} stars. These global, Galactic, populations have been distinguished through their mean metallicities, their $\alpha$-to-iron ratios, their kinematics, and their ages \citep{Gilmore:1989}. From abundance ratios such as [Mg/Fe], it can be inferred that both the halo and the thick disk formed on short timescales: their elevated [Mg/Fe] ratios reflect an earlier time, before   supernovae of type Ia (producing copious amounts of iron, but no $\alpha$ elements; see Section 3.3)  bring these ratios down towards solar-like values as seen in the thin disk \citep{Tinsley:1980}. 

More generally, the successive build-up of elemental abundances over the course of the last 10 Gyr can be studied using disk stars \citep[e.g.][]{Magrini:2018}.
This is done employing descriptions, models, and simulations of \emph{chemical evolution}, developed after pioneering work \citep{Tinsley:1972,Tinsley:1974,Tinsley:1980} in semi-analytical \citep[e.g.][]{Matteucci:1989a,Boissier:1999,Chiappini:2003,Prantzos:2018}, numerical-cosmological \citep[e.g.][]{Gibson:2003,Kobayashi:2011,Cote:2019a,Kobayashi:2020a}, and hybrid \citep[e.g.][]{Minchev:2013} forms. 
Figure~\ref{fig_abundanceGal} shows an example of elemental-abundance data from stars, as compared to recent models of chemical evolution \citep{Prantzos:2018}.
We expect that improvements in stellar abundance precisions (from improved models of the stellar-atmosphere physics; see above), from nucleosynthesis yields of the various source types, and from chemical-evolution modelling (inhomogeneous evolution, treatment of ejecta recycling from specific source types) will lead to improvements in future comparisons of observations and models. 
Such modelling includes explaining abundance gradients with Galacto-centric radius, the effects of radial migration of stars, and inhomogeneities from spiral arms and star-forming regions along these, and analysis of nucleosynthesis yields for the different sources (discussed in Section 3) and their parameters.
Recent progress and challenges are met, as star clustering and the impact of binary evolution on nucleosynthesis yields are being addressed \citep[e.g.][]{Izzard:2013,Mennekens:2013,Mandel:2016,Song:2016}.

A new population of stars in the Galactic halo was discovered with the astrometric space mission Gaia, and named \emph{Gaia Enceladus} \citep{Babusiaux:2018,Vincenzo:2019}.
These stars show lower $\alpha$-to-iron ratios compared to other halo stars of the same metallicity. This points towards an origin in a smaller galaxy that merged with our Galaxy some~10~Gyrs ago, and in which star formation had progressed at a lower pace. Other such events have been inferred from identifying stellar streams \citep{Ibata:2021}, but Gaia Enceladus likely represents the \emph{last major merger}, which is believed to be connected to the formation of the Galaxy's \emph{thick disk} \citep{Helmi:2018}.

Studying stellar systems in the Galactic halo has become scientific routine since the advent of multi-object spectrographs at 8m telescopes. Two classes of stellar systems have received particular attention: globular clusters and dwarf galaxies. The detailed study of elemental anti-correlations in Galactic globular clusters has revealed a remarkable complexity of systems which not even 50 years ago were considered the best realizations of single stellar populations, i.e.\ stars born from one and the same gas cloud at the same time. Practically all globular clusters are now known to contain multiple stellar populations \citep{Gratton:2019} and seem unique in their abundance fingerprints \citep{Milone:2017}. The variations are inherent to the cluster environment and not an effect of stellar internal processes, as even unevolved stars show abundance variations \citep{Gratton:2001}. A coherent picture that explains all observed abundance variations (involving elements ranging from He to K, and Fe-peak elements in some cases) has not emerged yet \citep{Bastian:2018}. The cores of asymptotic giant branch stars reach temperatures sufficient to produce some of the observed elemental variations, but not all. The mechanical winds of fast-rotating massive stars are other contenders, as are super-massive stars. Isotopic ratios are important constraints, but with few exceptions have not received adequate attention \citep{Yong:2003}. Observations are very challenging, however. Explaining globular-cluster compositions is one of the central riddles of 21-century nuclear astrophysics.

In recent years, astero-seismology has added to the domain of Galactic archeology, through provision of significantly-improved age estimates for stars \citep{Spitoni:2019,Spitoni:2020,Silva-Aguirre:2018}. 
This is a key step to disentangle stellar groups of different origins, as the degeneracy between age effects and different star forming histories is lifted.

Abundances can be determined from emission lines of hot and ionised gas \citep{Crowther:2007}. Planetary nebulae have received a prominent role herein \citep{Garcia-Hernandez:2016,Stanghellini:2018}. These are the ionized cirum-stellar gas regions around low- and intermediate-mass stars, as the central star transitions from the AGB phase to being a white-dwarf remnant which strongly ionises its surroundings. The gas is shed off outer shells of the central star \citep{Gesicki:2003}.
Planetary nebulae have provided an independent set of stellar abundances \citep{Kingsburgh:1994}. 
Beyond tracing abundances and chemical evolution across the Galaxy \citep[e.g.][]{Stanghellini:2018}, these data also provide constraints on AGB stars \citep{Karakas:2014} (see Section 3).
Some tension exists between results from collisionally excited lines versus those from recombination lines for these regions around the central ionising star \citep{Garcia-Rojas:2022}.

Compared to this, the \emph{nucleocosmochronology} based on radioactive isotopes and their information regarding the age of stars and the Galaxy are less precise, though important as an entirely dfferent messenger \citep{Tinsley:1975,Meyer:2000}.
The method of age determination is similar to age-dating on earth using the $^{14}$C/$^{12}$C ratio and the fact that $^{12}$C is stable while $^{14}$C has a radioactive lifetime of 8270~yr. 
Assuming deposition of C isotopes at material condensation with the solar-system isotopic abundance ratio of $^{14}$C/$^{12}$C, a measurement of the current isotopic ratio determines the age, 
Cosmo-clocks are isotopes with radioactive decay times beyond $\sim$10$^8$y, such as $^{238}$U (6.49$\times$10$^9$ yr), $^{235}$U (1.01$\times$10$^9$ yr), $^{232}$Th (20.2$\times$10$^9$ yr), or $^{187}$Rh (72.1$\times$10$^9$ yr). 
These all are expected to be produced from the r~process (see Section 2). Metal-poor stars have shown abundance patterns that resemble the solar system r-process abundance pattern, except at a much lower abundance level due to the overall lower metallicity \citep{Cowan:2006}. 
Thus, if one can predict the production ratio of these chronometers to a stable reference isotope (or element) that is well measured and known, then application of the radioactive-decay law results in an age constraint.
A commonly-used stable reference for the r-process products is the abundance of Eu; for the $^{187}$Rh isotope measurements in meteorites, the $^{187}$Os abundance provides the reference.

One must adopt an enrichment model for reference. As an example, one may assume the r~process contributing linearly with time since the early Galaxy; another extreme would be to adopt single events. Depending on these assumptions, the radioactive age constraints provide upper or lower limits. The adopted enrichment history, however, appears as a second-order correction, which reflects the astrophysical usefulness of the radioactive age method \citep{Meyer:2000}.

Using meteoritic measurements of the abundances as reference at the time of solar system formation, 
one obtains a lower limit for first r-process nucleosynthesis in our Galaxy of larger than 8.7~Gy \citep{Meyer:2000}. 
Ages for r-process-enhanced metal-poor stars can be determined assuming a single source for the nucleosynthesis. 
The results are 15.6$\pm$4.6~Gy for CS22892-052 and  12.5$\pm$3~Gy for CS31082-001, two of the best-studied halo stars \citep{Thielemann:2002b}. 

Nuclear properties (the decay may be different from above values in a stellar environment), production ratios (for the r-process sources that are fundamentally unknown, and nucleosynthesis unknowns of the r~process including fission cycling and nuclear stability models \citep[see Section 2 and][]{Thielemann:2002b}, and the various uncertainties of stellar photosphere abundance determinations contribute to the uncertainties. 
It is important to have this age determination method from nucleocosmochronometers that is fundamentally independent of metallicity indicators and chemical evolution models.


The presolar dust grains of type SiC from meteoritic material and discussed above (Section~\ref{stars}) allow us to investigate the chemical evolution of the Galaxy. This is achieved by exploiting specific isotopic compositions of the intermediate-mass elements, such as Si and Ti, that are not affected by nucleosynthesis in the parent AGB stars, but reflect the initial composition of the star. These elements provide an independent constraint on the age-metallicity relationship of stars in the solar neighbourhood at around the time of the formation of the Sun. 
Si shows a trend with the grain size. From the distribution of isotopic ratios, when compared to models of the chemical evolution of the Galaxy, one infers that the grains must have originated in stars with on average higher metallicities than observed in the solar neighbourhood \citep{lugaro99,lewis13}. These correspond to a high-metallicity stellar population that may have been identified recently  \citep{nissen20}.

\subsubsection*{The high-redshift universe}
Towards larger distances, absorption line spectroscopy becomes more challenging, as objects get fainter. Stars are no longer observable individually (except in rare cases supported by gravitational lensing), but integrated stellar populations can be sampled in observations.  
At high redshifts ground-based observatories gain access to UV lines. These provide important information on the composition of distant gas clouds. 

\begin{figure} 
\centering
\includegraphics[width=0.7\columnwidth,clip]{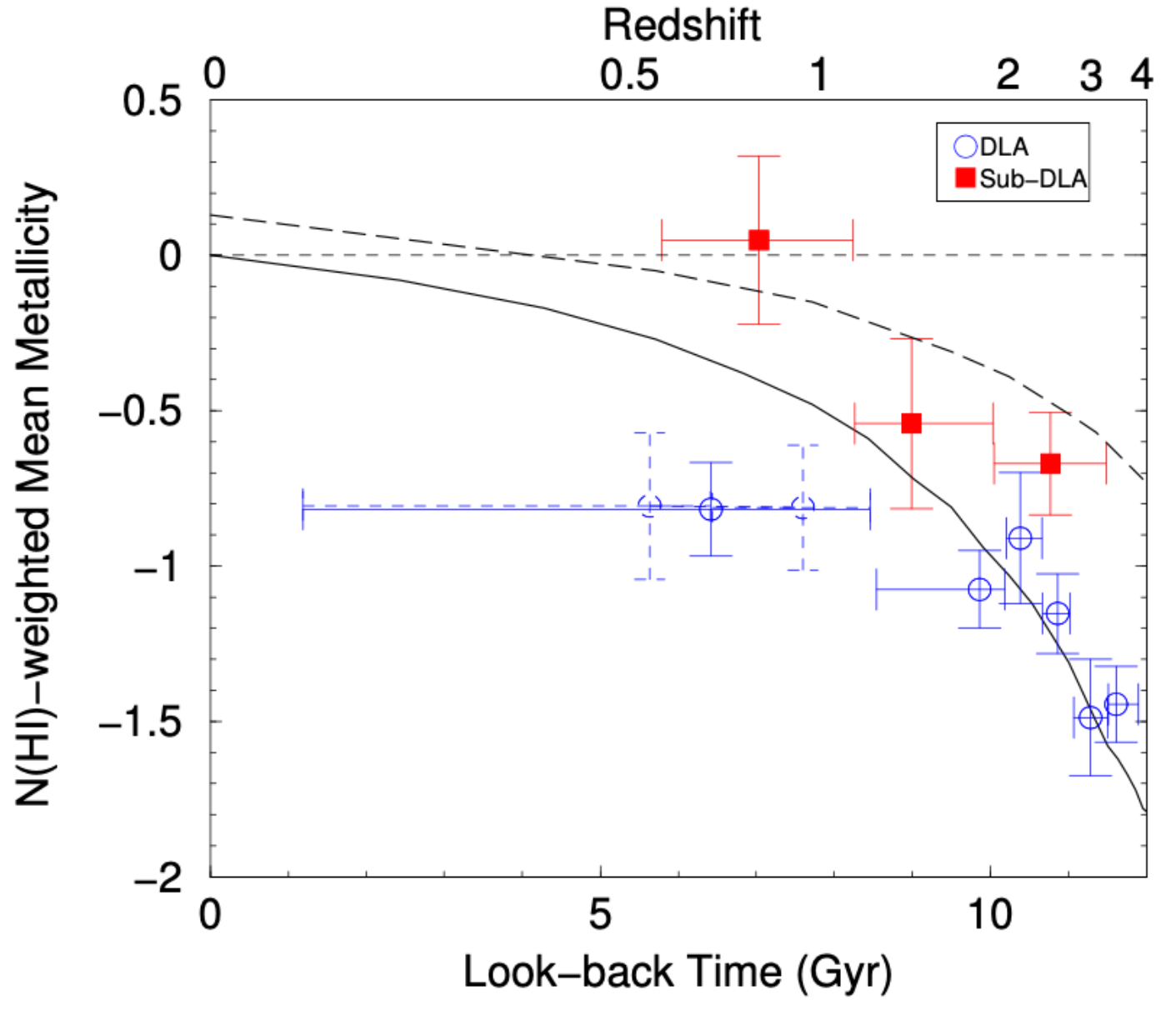}
\caption{The abundance of metals in  the early (distant) universe can be studied in gas clouds {        from absorption lines. Clouds must be carefully selected to be plausibly shaped by }  star formation {         (see text)}.  (From \cite{Kulkarni:2007a}). }
\label{fig_DLAs}       
\end{figure} 

The most prominent line accessible at redshifts larger than about 1.9 is Hydrogen Lyman-$\alpha$ at a rest wavelength of 121.6~nm. This is the transition from the first excited level into the ground-state of the hydrogen atom. In emission, this line indicates warmer to hot regions and points towards star formation activity, while in absorption, it is formed by the integrated spectrum of the (older) stellar population. The Lyman-$\alpha$ emission line strength is often used as an indicator for star formation. 
High-redshift quasars and $\gamma$-ray bursts show absorption lines shortward of Lyman~$\alpha$ from the source. These \emph{Lyman-$\alpha$ forest} lines come from absorbing gas clouds on the near side of the quasars or $\gamma$-ray bursts \citep[see][for a review]{Peroux:2020}.
High-density clouds, corresponding to sight lines through galaxy disks and a column density of $log \, N_H\geq 20.3$ display saturated absorption lines, and are called \emph{Damped Lyman-$\alpha$ Absorbers (DLA)}. They show characteristic metal lines, and provide a way to measure metallicity at high redshifts. 
Several complications have to be considered. Since typically lines of highly ionised atoms are observed, ionisation corrections need to be applied to derive abundances. Then, some elements may deplete onto dust grains and are not visible in the gas abundances any longer. Both these effects require corrections, which can carry significant uncertainties. Nevertheless, a clear trend has been observed of decreasing metallicity with increasing redshift ($0<z<5$), i.e. the gas has been enriched with cosmic time (the past 9~Gyr are covered) \citep{Peroux:2020}.
The effect is more pronounced in sub-DLA systems (with column densities $log \, N_H\geq 19.0$). Lower density clouds are more strongly affected by local star formation, which leads to a stronger enrichment with time (Figure~\ref{fig_DLAs}, \citep{Kulkarni:2007a}).

\subsection{Recycling of ejecta on the galactic scale} 
\label{recycling_general}
%
%
%
%

\subsubsection*{Dwarf galaxies and Pop III stars}

Significant amounts of matter flow 
in and out of galaxies. 
The kinetic energy imparted onto interstellar gas through stellar activity locally overcomes the gravitational potential of a galaxy to escape from that galaxy. 
Vice versa, the gravitational potential of a galaxy captures material from the intergalactic medium and feeds low-metallicity gas into a galaxy during, and even at late times, of its evolution.

Dwarf galaxies are small galaxies with stellar content less than 10$^8$~\Msol  \citep{Tolstoy:2009}. 
Studying chemical evolution in dwarf galaxies opens a window into  matter recycling that is very different from our Galaxy,  gravitational potential and star formation activity being important conditions 
\citep{Chiappini:2003a,Venn:2004,Venn:2007,Annibali:2022}.
This has been demonstrated from stellar photospheric abundances of several elements. These show, e.g., that the characteristic kink in [X/Fe] evolution versus [Fe/H] sets in at times that are characteristic for the evolutionary timescale of each such galaxy with its amount of star-forming gas \citep{Annibali:2022}. 
Few, if not single, events of nucleosynthetic enrichment are believed to  occur in dwarf galaxies. 
Spectroscopy of a handful of stars in the dwarf galaxy \emph{Reticulum II}  shows prominent enrichment with r-process material that seems incompatible with normal chemical evolution over longer times and may result from the enrichment by a single r-process event \citep{Ji:2016a,Ji:2018,Frebel:2005}. 
{         In another case, a single hypernova is deemed responsible for the observed abundance signature in a galaxy with extremely-low metallicity \citep{Skuladottir:2021}.
Such hints are important; interpretations should be made with care, as they rely on spectra of a few single stars, as well as on models for typical/exceptional source yields (Section 3) and matter recycling (Section 4) in such small galaxies.}



\begin{figure}
  \centering
  \includegraphics[width=\columnwidth,clip]{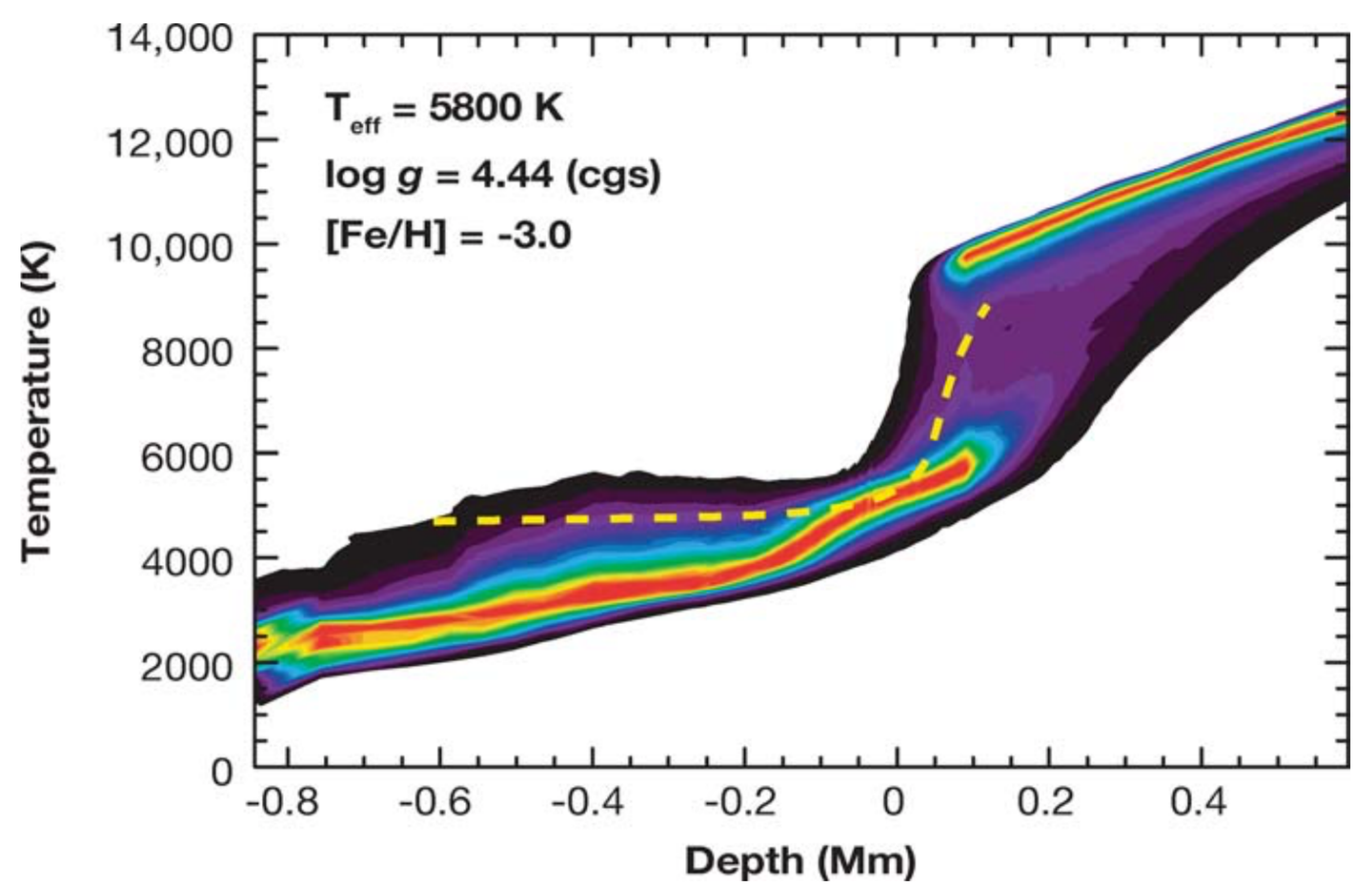}
   \caption{Distribution of temperatures as a function of geometrical depth for a hydrodynamical model atmosphere of a metal-poor sun-like star. The dashed yellow line shows a corresponding 1D model. The hydrodynamical model has a steeper average temperature gradient which leads to modified Saha-Boltzmann level populations of lines forming in the upper photosphere. (From  \cite{Asplund:2005a}).}
  \label{fig_T-tau}
\end{figure}

The search for metal-free \emph{Population III} stars in the Galactic halo provides hints for interesting enrichment histories which differ fundamentally from that of the present-day Galaxy. 
Individual stars so metal-poor (one millionth solar or below), that their iron abundance becomes hard to infer from even the best spectra have recently been identified \citep{Nordlander:2019}. While being so poor in iron content, they display large relative over-abundances of C, N or O, and are called \emph{carbon-enhanced metal-poor stars} (CEMPs) \citep[see][for a review]{Beers:2005}. 
Stars with more normal compositions are also seen at the lowest metallicities, but seem to be the exception \citep{Caffau:2017}. 
Such ultra metal-poor stars may potentially have formed from the ejecta of a single supernova in the early universe.
Thus they may open a window to study the early universe in our current Galactic halo (\emph{Galactic archaeology}). 
(As an aside, it is worth mentioning that there are prospects for discovering individual Population III stars in the distant early universe as gravitationally lensed objects with the James Webb space telescope as it begins its mission in 2022 \citep{Vanzella:2020}).
Constraints on the various hypothesised supernova models for the earliest stars (hypernovae, pair instability supernovae, supermassive star progenitors) and their parameters (mass, explosion energy, mixing and fallback etc.) are rather indirect and based on just a handful of elemental abundances. 
 The most iron-poor star observed to date, SMSS 0313-6708, has been studied with the best available methods. Departures from both hydrostatic structure and local thermodynamic equilibrium are large in such stars (see Fig.~\ref{fig_T-tau}), so that abundances from classical models are often biased by up to a factor of 10. Even with the best modelling, constraints on the supernova progenitor mass are weak: 10 M$_\odot < $M$_{\rm SN} < $60 M$_\odot$.  Explosion energies are inferred to be below 2.5 B (10$^{51}$~erg) \citep{Nordlander:2017}, which excludes hypernovae and pair-instability supernovae as progenitors in this particular case. For another star, SMSS~J200322.54-114203.3, with a strong enhancement in r-process elements, the best abundance-pattern fit is obtained for a 25~\Msol\ magneto-rotational hypernova \citep{Yong:2021}. It thus seems that neutron-star mergers do not exclusively produce r-process elements in the early Galaxy \citep{Cote:2019}. The delay time of contributions from neutron-star mergers, even if poorly known, make this plausible.

\subsubsection*{Radioactive afterglows}

\begin{figure}  
\centering
\includegraphics[width=\columnwidth]{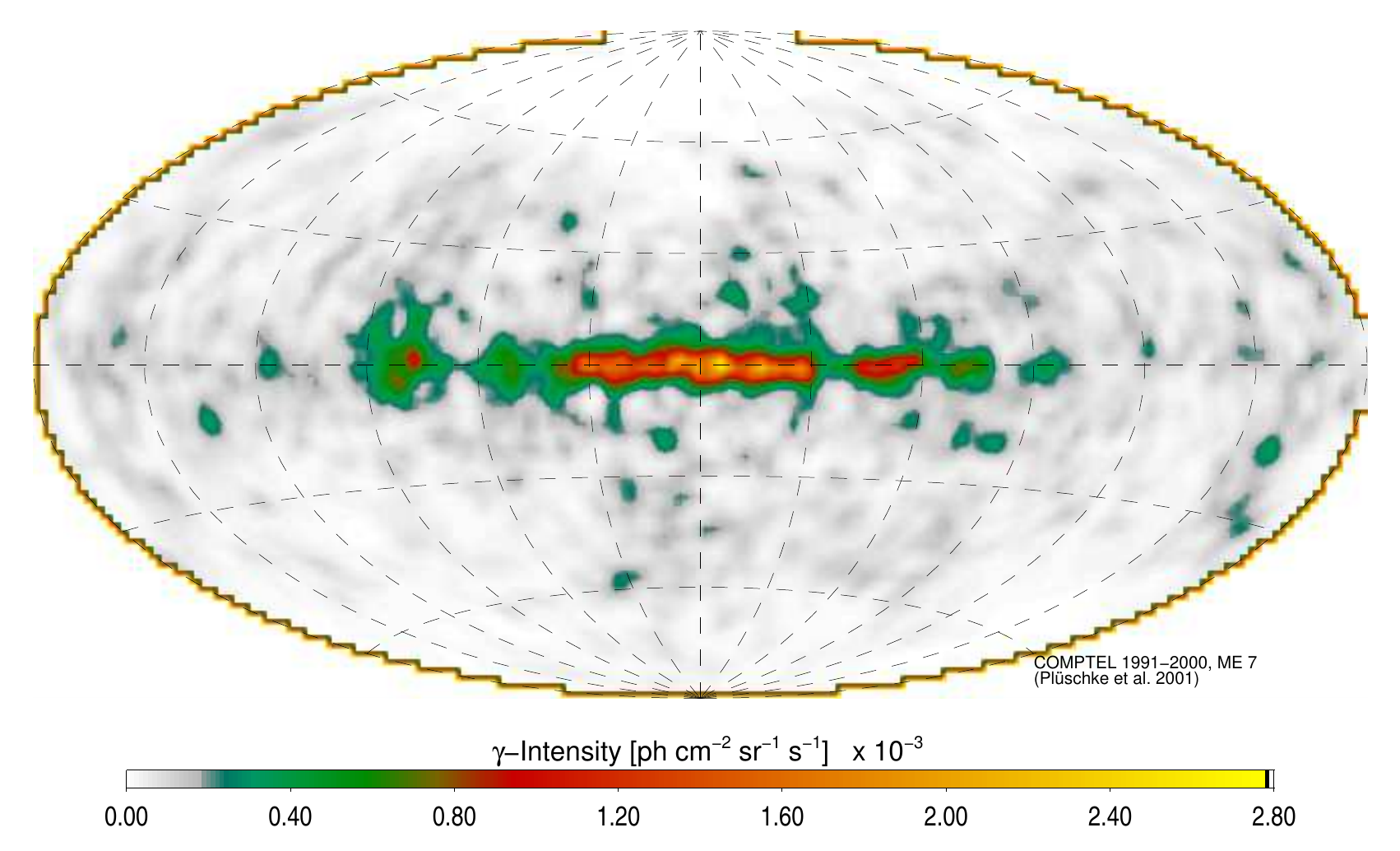}
\caption{The all-sky emission of $\gamma$~rays from radioactive decay of $^{26}$Al.  This image \citep[from][]{Pluschke:2001c} was obtained using a maximum-entropy regularization together with the maximum-likelihood method to iteratively fit a best image to the measured photon events.}
\label{fig_Al26map}
\end{figure}   

\begin{figure}  
\centering
\includegraphics[width=0.6\columnwidth]{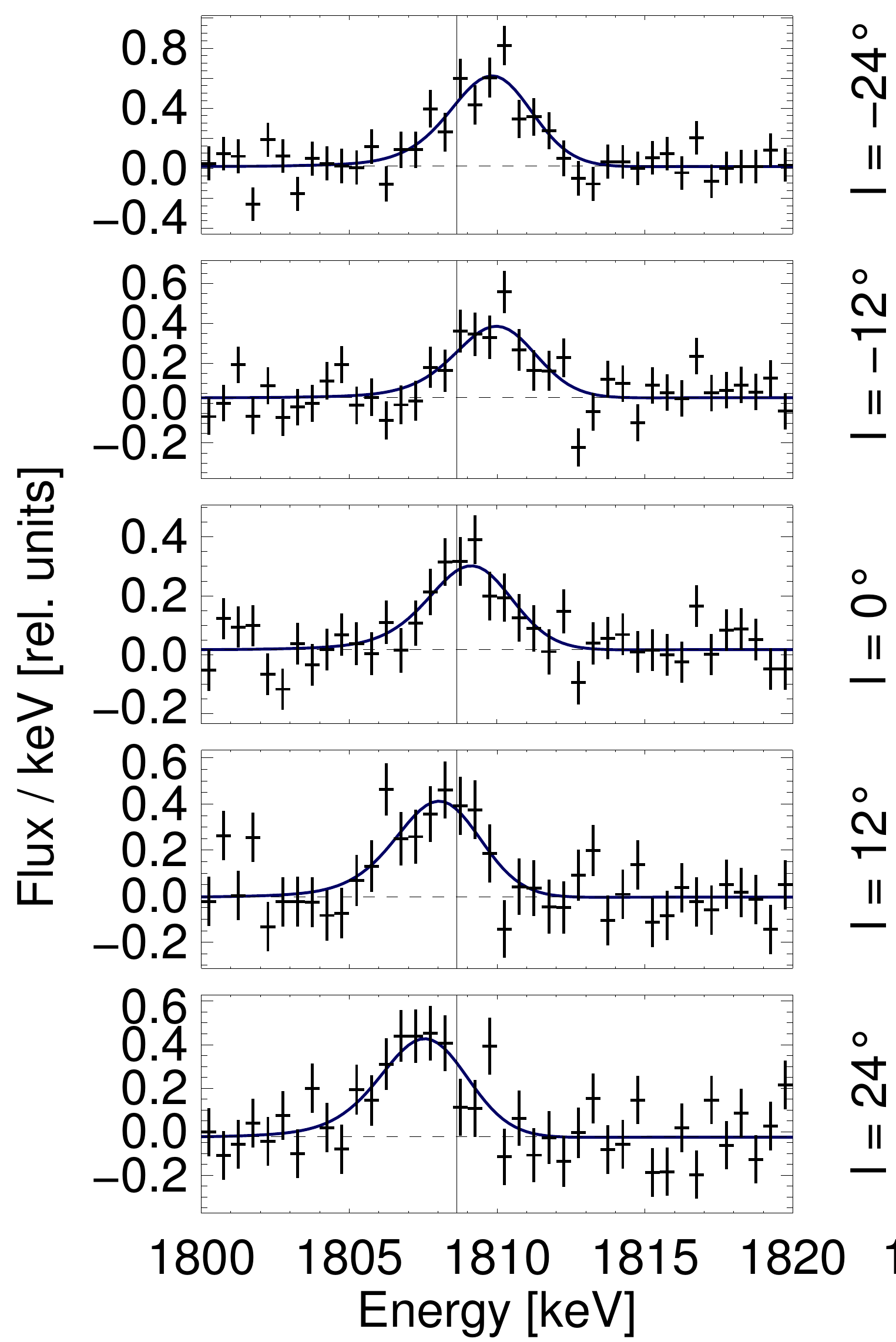}
\caption{The \Al line as seen towards different directions (in Galactic longitude) \citep{Kretschmer:2013}. This demonstrates kinematic line shifts from the Doppler effect, due to large-scale Galactic rotation. }
\label{fig_Al26_longitudes}
\end{figure}   
 
Radioactive isotopes with a lifetime approaching the typical recycling time scale of 10$^7$ to 10$^8$ years can offer a way to \emph{trace} the flow of nucleosynthesis ejecta directly. 
\Al with its lifetime of 1.04$\times$10$^6$~y is on the short side, yet longer-lived than the times over which individual supernova remnants can be traced (up to 10$^5$ years at most). 

\begin{figure}[t]  
\centering 
\includegraphics[width=\columnwidth]{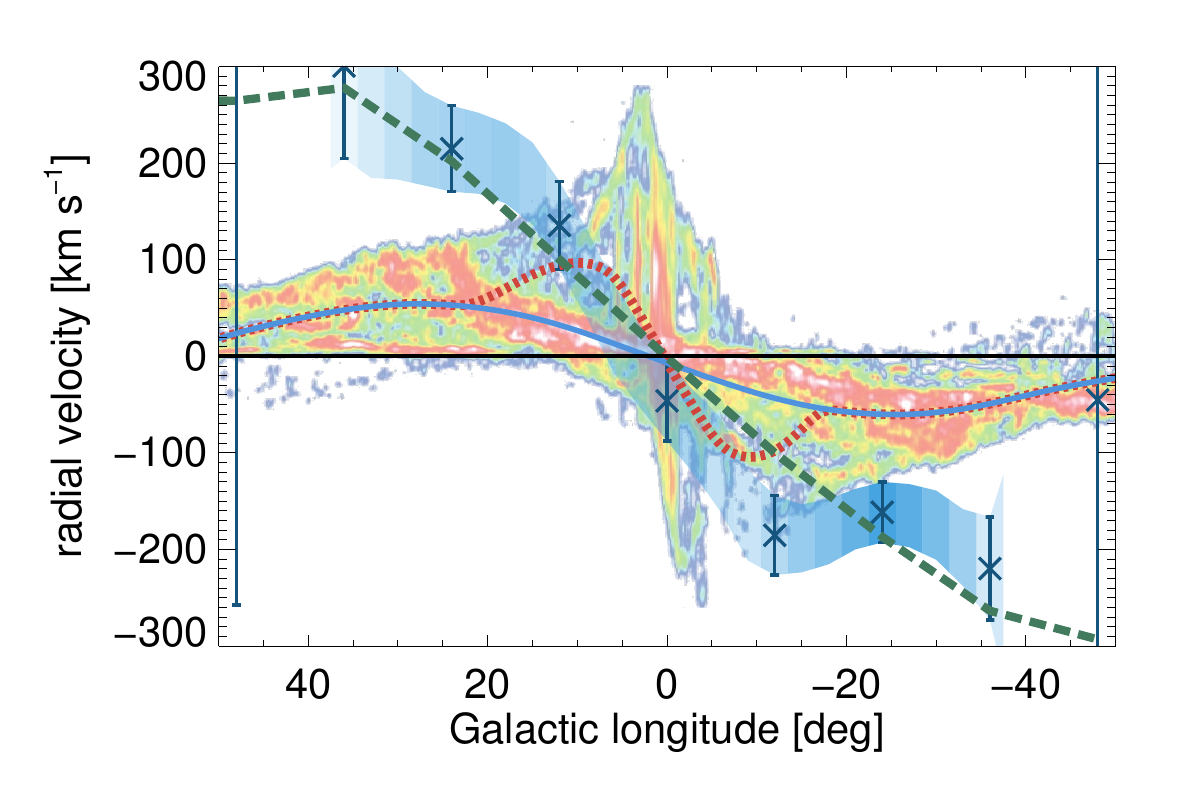}
\caption{The line-of-sight velocity shifts seen in the \Al\ line versus Galactic longitude, compared to measurements for molecular gas. (From  \cite{Kretschmer:2013}). }
\label{fig:al_long-velocity} 
\end{figure}

The diffuse emission from \Al decay (Figure~\ref{fig_Al26map}) had been measured in the first sky survey in $\gamma$~rays towards the turn of the century. 
Apparently, \Al is of large-scale, galactic, origins. A main supporting argument is the observation of the characteristic signature of large-scale Galactic rotation in spectroscopy of the $\gamma$-ray line from \Al decay.
Already in 1995, a balloon experiment had reported an indication that the \Al~line was significantly broadened to 6.4~keV \citep{Naya:1996}, from a measurement with their high-resolution Ge detectors. 
More recent high-quality spectroscopic data obtain a resolution of 3~keV at the energy of the \Al line (1809~keV) for multi-year data accumulation; this corresponds to a Doppler velocity shift resolution of about 100 km~s$^{-1}$ for bright source regions \citep{Kretschmer:2013}.
The characteristic signature of large-scale Galactic rotation, clearly indicated in early results \citep{Diehl:2006d}, is a blue shift when viewing towards the fourth quadrant (objects on Galactic orbits approaching) and a red shift when viewing towards the first quadrant (receding objects, on average).
The consolidation with more years of exposure is shown in Figs.~\ref{fig_Al26_longitudes} and \ref{fig:al_long-velocity}).
The observed $\gamma$-ray flux could be translated into a total emitting mass of $^{26}$Al, using geometrical models of how sources are distributed within the Galaxy, such as double-exponential disks and spiral-arm models \citep[see][for details]{Diehl:2006d}. 
The total mass of Al in the Galaxy is constrained to be between 1.8 and 2 ~\Msol\   \citep{Diehl:2006d,Pleintinger:2020}. 

This observed mass of a single unstable isotope in the current Galaxy's interstellar medium can be converted into a rate of core-collapse supernovae. 
The rate is an important ingredient to understand the dynamical state of interstellar medium as supernova explosions drive turbulence within the interstellar medium \citep{Krumholz:2018,Koo:2020}.
Using \Al yields from theoretical models of massive stars and their supernovae,   a core-collapse supernova rate of 1.9$\pm$1.1 events per century for our Galaxy was derived \citep{Diehl:2006d}; this value was updated with new measurements and model yields to 1.4$\pm$1.1~century$^{-1}$ \citep{Diehl:2018,Pleintinger:2020}. 
At face value, the 294 supernova remnants observed in the Galaxy through their radio emission \citep{Green:2019} are in a tension with this value. With the $\gamma$-ray based supernova rate, the radio count would suggest a maximum sampling age of 21,000~y, while significantly larger ages have been inferred for some of the radio supernova remnants. This is evidence that the supernova remnant appearance strongly depends on their surroundings.

The observed \Al velocity from large-scale Galactic rotation \citep{Kretschmer:2013} with the velocity of molecular gas exhibits a puzzling discrepancy (Fig.~\ref{fig:al_long-velocity}).
If interpreted as kinematic Doppler shifts of astrophysical origin, the \Al line translates into a motion of 540~km~s$^{-1}$ \citep{Naya:1996}. 
Considering the $1.04~10^6$~y decay time of $^{26}$Al, such a large velocity observed for averaged interstellar decay of \Al would naively translate into kpc-sized cavities around \Al~sources, so that velocities at the time of ejection would be maintained during the radioactive lifetime. An alternative hypothesis is that major fractions of  \Al condensed onto grains, which would maintain ballistic trajectories in the tenuous interstellar medium \citep{Chen:1997,Sturner:1999}. 
Although formation of dust has been found to occur in supernovae (see discussion of SN1987A above), it is unclear if the reverse shock of the remnant would not rather destroy most of it within an \Al lifetime.

\begin{figure}[t]  
\centering 
  \includegraphics[width=0.65\linewidth]{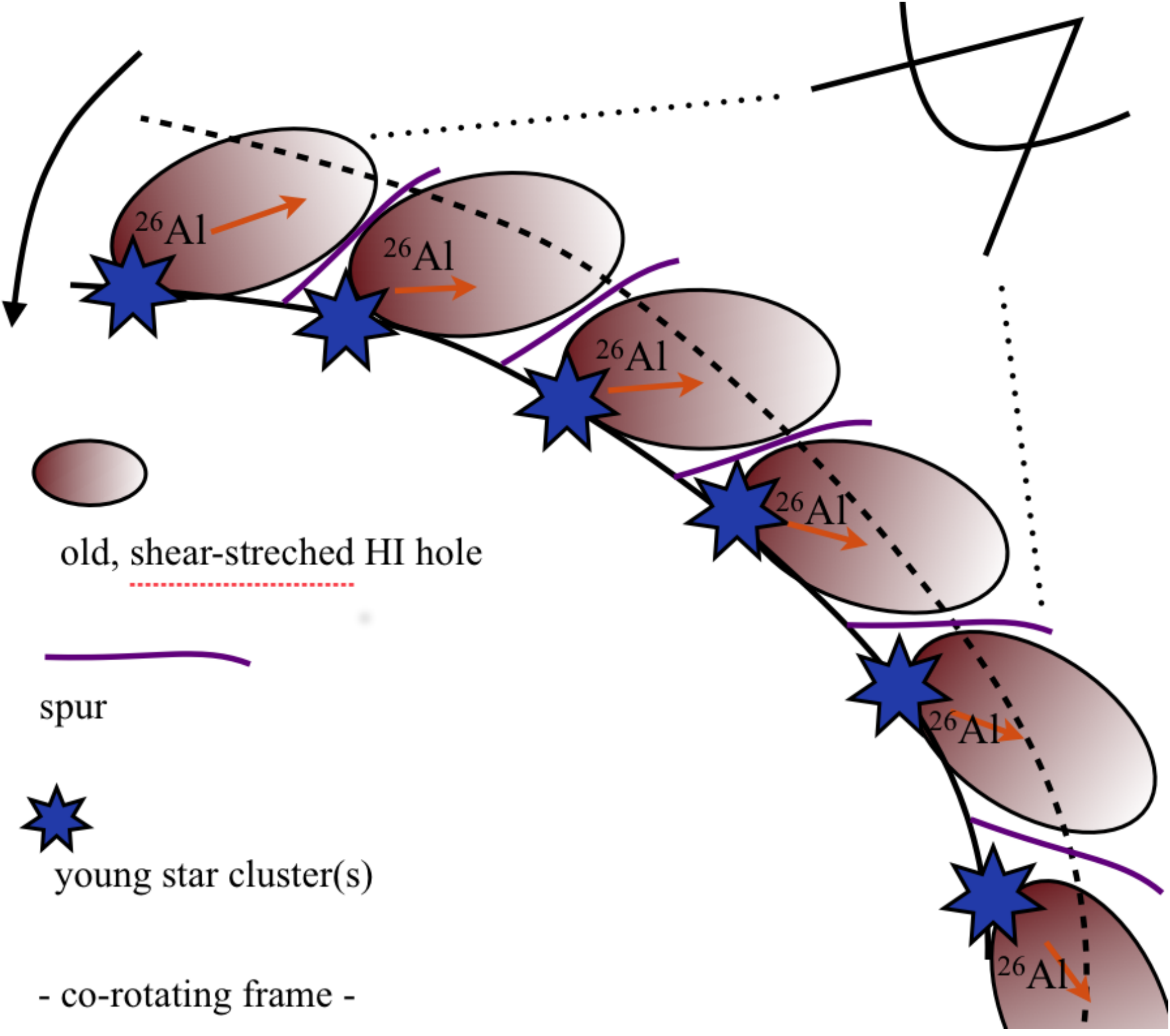}
\caption{A model for the different longitude-velocity signature of $^{26}$Al, assuming \Al\ blown into inter-arm cavities at the leading side of spiral arms.  (From \cite{Krause:2015}). }
\label{fig:spiralarmbubbles} 
\end{figure}   

The velocities seen for \Al throughout the plane of the Galaxy   \citep{Diehl:2006d,Kretschmer:2013} exceed the velocities measured for  molecular clouds, stars, and (most precisely measured velocities) maser sources by typically as much as 200~km~s$^{-1}$ (see Fig.~\ref{fig:al_long-velocity}). 
This high apparent bulk motion of decaying $^{26}$Al means that its velocities remain higher than the velocities within typical interstellar gas for 10$^6$ years. Additionally, there is a bias for this excess average velocity in the direction of Galactic rotation. 
This has been interpreted as $^{26}$Al decay occurring preferentially within large cavities (superbubbles), which are elongated into the direction of large-scale Galactic rotation (Fig.~\ref{fig:spiralarmbubbles}). 
If such cavities are interpreted as resulting from the early onset of stellar winds in massive-star groups, they characterise the source surroundings at times when stellar evolution terminates in  core-collapse supernovae. Such  wind-blown superbubbles around massive-star groups plausibly extend further in space in forward directions and away from spiral arms (that host the sources), as has been seen in images of interstellar gas from other galaxies  \citep[][and references therein]{Schinnerer:2019}.  
Such superbubbles can extend up to kpc \citep{Krause:2015} \citep[see also][]{Rodgers-Lee:2019,Krause:2021,Nath:2020}, which allows \Al to move at speeds of the sound velocity within superbubbles. 
The dynamics of such superbubbles into the Galactic halo above the disk are unclear; this is perpendicular to the line of sight, so that measurements of \Al line Doppler velocities cannot provide an answer \citep[see][for a discussion of outflows into the halo]{Krause:2021}.
Nevertheless, these measurements underline the important consequences of massive-star clustering in shaping the interstellar medium, with connections to the astrophysics of stellar feedback in general \citep[see][for recent theoretical considerations]{Krause:2020,Chevance:2022}.

\begin{figure} 
\centering
\includegraphics[width=\columnwidth,clip]{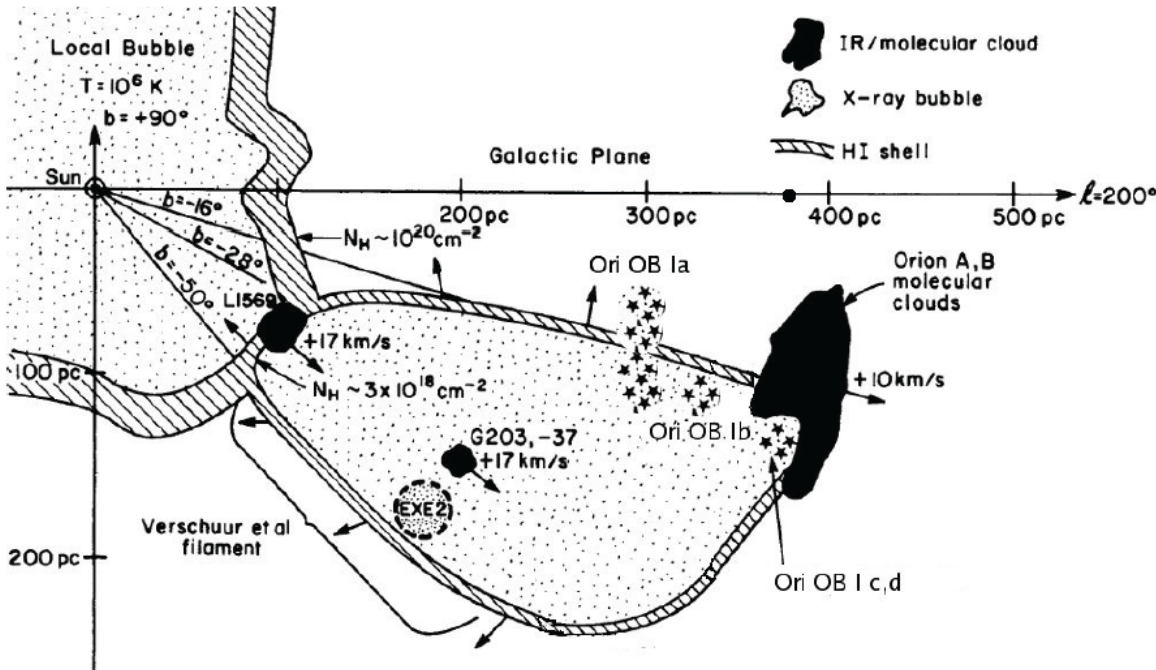}
\caption{The Orion region is characterised by two massive molecular clouds at about 450~pc distance, with recent massive-star formation preferentially on its near side (star symbols), and a large cavity bounded by denser HI emission, which extends from the clouds towards the direction of the Sun. ( \citep[Updated from][]{Burrows:1993} \citep[by][]{Fierlinger:2016}). }
\label{fig_Ori-Eri}       
\end{figure}

One particular such superbubble has been identified near the solar system and towards the Orion region: 
The Eridanus cavity has been recognised in diffuse X-ray emission \citep{Burrows:1993}, its boundaries are delineated in HI radio emission \citep{Heiles:1999}, and $\gamma$~rays from \Al have been detected \citep{Siegert:2017a}. 
Figure~\ref{fig_Ori-Eri} illustrates the geometry of the objects in this region, showing the stellar associations of the Orion massive star region. 

\subsection{Local constraints on matter cycling}
\label{recycling_local}

Our solar system is exposed to a influx of cosmic gas and dust particles. We can capture such influx either directly through detectors in near-earth space, or analyse what has been captured and deposited onto Earth and Moon. 

\subsubsection*{Cosmic rays}
\emph{Galactic cosmic rays} with their predominant component of atomic nuclei provide a sample of matter from outside the Solar system. 
After almost a century of research, the origins of cosmic rays, their acceleration, and propagation
in the Galaxy, remain open questions of current astrophysics \citep[see][for a recent review]{Tatischeff:2018b}. 
Shock acceleration is believed to be the main acceleration mechanism and regions of massive-star nucleosynthesis with wind and supernova shocks play a main role \citep[e.g.][]{Berezhko:1996,Bykov:2001,Ellison:2007,Parizot:2004}. 
The connection of cosmic ray studies to nucleosynthesis has been considered plausible \citep[e.g.][]{Biermann:2010}. 
Cosmic-ray material is thought to originate from winds of Wolf-Rayet stars, for example, but accelerated on time scales longer than $\sim$10$^5$ years  \citep{Binns:2001,Wiedenbeck:1999,Biermann:2018}. 
Key issues herein are related to the spatial and temporal scales of various processes, such as acceleration in one event or in a series of events, and the propagation and confinement of cosmic rays in the complex magnetic-field configurations of interstellar medium in the Galaxy.
In spite of these uncertainties, the connection of cosmic ray origins to nucleosynthesis sources such as massive-star clusters and their supernova remnants has been considered \citep[e.g.][]{Biermann:2010,Parizot:2004}. 

Observations {        of cosmic rays in its forms of positrons, electrons, protons, heavy nuclei, but also antiprotons and anti-helium up to multi-TeV energies are possible in near-Earth space (the Alpha Magnetic Spectrometer (AMS-02) \citep{Aguilar:2018} is a precision particle physics detector on the International Space Station, capable to measure velocity, momentum, charge and rigidity of traversing particles) and on ground for highest energies. Results} related to nucleosynthesis concentrate on isotopes that either are directly attributed to a nuclear fusion process, such as $^{59}$Ni or $^{60}$Fe, or to isotopes that are likely products of spallation reactions of cosmic rays in the interstellar gas \citep{Yanasak:2001,Wiedenbeck:2007}. 
We expect a bias to such measurements from how the solar system's magnetic field and the solar wind redirect cosmic gas and dust flows. This get weaker with increasing energy, so that the high-energy part of cosmic rays, above a \emph{cutoff rigidity} of $\sim$30~GeV per nucleon, would be unaffected, while at energies accessible to mass spectrometers in space ($\sim$200~MeV per nucleon), the particle spectrum still rises with energy due to solar modulation \citep{Wiedenbeck:2007}. 

The radioactive isotope $^{60}$Fe can only be a primary product of stellar nucleosynthesis. The abundance of heavier cosmic-ray nuclei is insufficient to produce $^{60}$Fe by fragmentation as a secondary product. 
It is the only primary cosmic-ray radioactive isotope with atomic number Z~$\leq$~30  decaying slowly enough to potentially survive
the time interval between nucleosynthesis, cosmic-ray acceleration, and detection near Earth.

The Cosmic Ray Isotope Spectrometer (CRIS) aboard NASA's Advanced Composition
Explorer (ACE) provided  a $^{60}$Fe detection after 17 mission years \citep{Binns:2016}.
Models of supernova nucleosynthesis  \citep{Woosley:2007b,Sukhbold:2016,Limongi:2018} suggest a small production ratio relative to $^{56}$Fe making its detection difficult with present-day instruments.
Fifteen CRIS events were attributed to $^{60}$Fe nuclei, with 2.95$\times$10$^5$ $^{56}$Fe nuclei.
This detection of $^{60}$Fe is the first observation of a primary cosmic-ray clock. 
The inferred ratio of  $^{60}$Fe/$^{56}$Fe is (4.6~$\pm$~1.7)$\times$10$^{-5}$ at the detector entrance \citep{Binns:2016}.

This can be converted to the ratio at the acceleration site only through an understanding of cosmic-ray propagation in the Galaxy. 
In this analysis  \citep{Binns:2016}, a simple \emph{leaky box model} was adopted and adjusted to match the  observations of $\beta$-decay secondaries $^{10}$Be, $^{26}$Al, $^{36}$Cl, $^{54}$Mn measured with the same instrument. These elements are thought to be produced by spallation along the cosmic-ray trajectory \citep{Yanasak:2001}.
All four radioactivities are fit by this leaky box model with a leakage parameter of an \emph{escape time} of 15.0~$\pm$~1.6 Myr.
The resulting ratio inferred for the acceleration site is  $^{60}$Fe/$^{56}$Fe$=$ (7.5$\pm$ 2.9)$\times$10$^{-5}$  \citep{Binns:2016}.
More-complex disk/halo cosmic-ray diffusion models lead to a slightly lower ratio at the source of (6.9$\pm$2.6)$\times$10$^{-5}$ \citep{Morlino:2020,Morlino:2021}.

The production ratio of the two Fe isotopes at the nucleosynthesis source will be different. While the nuclei are accelerated radioactive decay  converts some $^{60}$Fe to $^{60}$Ni. Some mixing of ejecta with ambient interstellar gas may occur. 
If cosmic-ray acceleration occurs in superbubbles generated by massive stars in a group, the shocks in such a region accelerate a mixture of material from previous supernova explosions, winds and any material left over in the bubble since the onset of star formation.
This scenario is, however, debated \citep[see][for controversial views]{Prantzos:2012,Tatischeff:2018} .

In the framework of a simple propagation model,  the distance from which the cosmic rays originate was estimated to be below 1~kpc \citep{Binns:2016}.
Such a volume could include more than twenty OB associations and stellar sub-groups, forming superbubbles favouring cosmic-ray acceleration \citep{Nath:2020,Nath:2020a}. 
In any case, it suggests that the solar vicinity has been active in synthesis of Fe group isotopes and their acceleration in the recent past.

\subsubsection*{Meteorites}

\emph{Meteorites} (as already discussed above for the determination of solar abundances and for stellar nucleosynthesis studies from embedded stardust) are of great use for the study of cosmic nucleosynthesis. 
Meteorite parent bodies are asteroids that formed within the presolar nebula, i.e. the gas that decoupled from the interstellar medium and formed the solar system. Asteroids and comets are the remains of the so-called planetesimals: solid objects, with a radius between roughly 1 and 100 km, that formed roughly 4.5 billion years ago. 
Most of the asteroids are located in the main asteroid belt, a region between the orbits of Mars and Jupiter. Comets, on the other hand, are on more elliptical or irregular orbits. 
Meteorites themselves are fragments of asteroids that eventually collided with Earth. If they were large enough, some of their material survived the intense heating-up as they pass through the atmosphere (thus creating the \emph{meteor} phenomenon). The remains of a meteorite then land on Earth. If found and recovered, these can be analysed in the laboratory, and we can measure the composition of Solar-System bodies other than our planet Earth. 
As mentioned in Section.~\ref{stars}, meteorites attributed to the most pristine and unaltered asteroids can be used to determine the elemental composition of the Sun. 

The overall isotopic composition of meteorites is relatively homogeneous, as the Solar System matter was overall heated and homogenised at the start. However, meteorites do carry isotopic anomalies that are directly related to stellar nucleosynthesis. There are three types of anomalies. The most evident is that carried by the stardust grains already discussed in Section~\ref{stars}. As mentioned there, isotopic ratios in stardust can vary by orders of magnitude and directly reflect the composition of their individual parent star. 

\begin{figure} 
\centering
	\includegraphics[width=0.8\columnwidth]{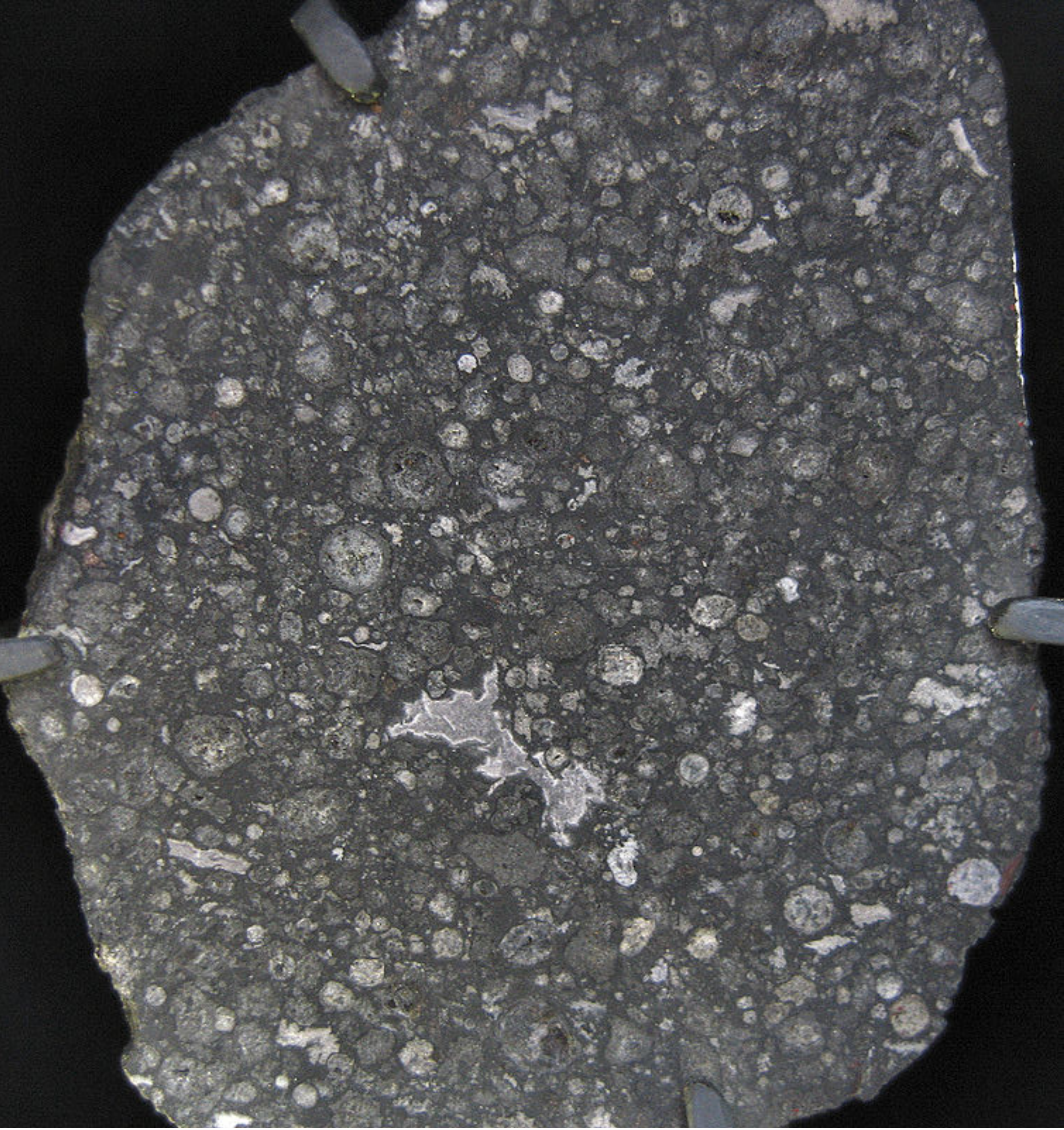}
	\caption{A slice from the Allende meteorite, which formed 4.5 Gyr ago at the beginning of the Solar System. The elongated white shape almost in the centre of the slice is a CAI, size of the order of cm. The round droplets of previously melted material are the \emph{chondrules} of size mm to cm. These inclusions are embedded int he darker matrix, made of microscopic minerals (including stardust), not visible to the naked eye. (Image from Wikipedia, licences under Creative Commons).} 
	\label{fig_meteorite}
\end{figure} 

The second type of nucleosynthetic signatures found in meteorites are variations in isotopic ratios of much smaller extent than in stardust. They are usually reported as parts per ten thousand ($\epsilon$) or parts per million ($\mu$), relative to the bulk solar isotopic ratios \citep[see, e.g.][for a review]{Dauphas:2016}. 
Such variations have been found first in some types of Calcium-Aluminium-rich Inclusions (CAIs; see Figure~\ref{fig_meteorite}). These are considered samples of the oldest material of the Solar System, and their age is in fact considered to be the age of the Sun, 4.6 Gyr. 
{         Meteorites are often named after their finding sites, such as \emph{Allende} (Figure~\ref{fig_meteorite}; Mexico) or \emph{Murchison} (Australia). These two have been particularly massive, and have been precious resources for laboratory mass spectroscopy.}
Some of these minerals contain isotopic variations with $\mu$-values up to $10^2-10^3$, in particular for neutron-rich isotopes such as $^{48}$Ca, $^{50}$Ti, and $^{54}$Cr. 
These variations are interpreted as the leftover of anomalies in the original presolar stardust, which did not completely homogenise before the formation of the CAIs. 
For example, stardust chromite grains carrying $^{54}$Cr excesses have been discovered \cite{Nittler:2018,denhartogh22}.
Other anomalies in stable isotopes, of smaller magnitude than those seen in CAIs (order of $\mu$) but with data covering more elements and isotopes, are found also in bulk meteoritic rocks. 
They appear to vary according to the location where the parent body of the meteorite formed, i.e. its distance from the Sun. Variations in the isotopic ratios affected by the $s$ process have been observed in many elements (from Mo and Ru to Pd, e.g., \cite{dauphas04,ek20}). 
As mentioned in Section~\ref{stars}, these were most likely carried by stardust SiC grains from AGB stars
\citep[see also][]{stephan21}. 
Another crucial observation of the past decade is the presence of a dichotomy in the composition of the Solar System material, whereby bodies that formed further away from the Sun (referred as CC) have a different composition than bodies that formed closer to the Sun (referred as NC), and this range is well separated  \citep[see][for a review]{kleine20}. 
One proposed explanation is that material could not mix due the early formation of Jupiter \citep{warren11,nanne19,brasser20}. 
The origin of the nucleosynthesis signature and why the infall material varied is, however, still unexplained, although explosive nucleosynthesis needs to be invoked due to $^{58}$Ni excesses \cite{nanne19}. 

In summary, stable isotope anomalies can shed light on the formation of our Solar System, such as the composition and timing of the infalling material, the processes that affected the distribution of dust in the 
protoplanetary disk, and the timescales of planet formation. For some of these anomalies we have not yet clearly identified the original stardust carrier nor the astrophysical production site. Since the measured variations are so small, they need to be corrected by effects due to mass-dependent fractionation of isotopes \cite{Dauphas:2016}, and the effect of cosmic rays which are all of the same order of the nucleosynthetic variations. 

Finally, the third type of direct nucleosynthetic signature found in meteorites and their inclusions are excesses in the daughter isotopes of radioactive nuclei. 
These correlated excesses are clear signature that radioactive isotopes were incorporated into solids in the early Solar System, while they were still alive with half lives of the order of 1 to 100 Myr \citep[for reviews see][]{Dauphas:2011,Lugaro:2018}. 
More than a dozen such isotopes have been confirmed in the early Solar System, and for another half dozen isotopes upper limits of their abundance are available. These isotopes range from the relatively light $^{26}$Al, which was also responsible for the heating and thermal evolution of the planetesimals \citep{lichtenberg16b}, to $^{129}$I and $^{247}$Cm, whose abundances are direct evidence of the last $r$-process event to have contributed matter to the Solar System \cite{Cote:2021}. While a self-consistent picture of the origin of these nuclei is still missing, current investigations shed light not only on the nature of nucleosynthesis sites in the Galaxy, but also on the environment of the Sun's birth \citep{adams10}, and on the availability of water in terrestrial exoplanets \cite{ciesla15,Lichtenberg:2019}. 
\citep[See][for some recent efforts towards a self-consistent solution for the origin of all radioactive isotopes in the early Solar System]{Lugaro:2022}.

  


\subsubsection*{Ejecta deposits in the Solar System}

Extrasolar gas and dust may reach the surfaces of Earth and Moon. Interstellar dust grains may be particularly efficient vehicles to bring ejecta into the solar system. 
Earth and Moon act as a \emph{collection area} that captures such interstellar material. 
The \emph{uptake} efficiency and the efficiency of a transport into sedimentations are complex problems. 


Is there any change for deposition on Earth and direct detection of interstellar nuclides in geological archives? Can traces of freshly synthesised nuclides be found in the terrestrial record before they decay away? Such signatures would represent direct messengers of recent nucleosynthesis events.

These ideas were first discussed for the longer-lived actinides $^{244}$Pu (80.6 My) and $^{247}$Cm (15.7 My) as messengers for nearby supernova explosions imprinted into the lunar soil 
\citep{Fields:1970}. First measurements in the late 60$^{ies}$ and early 70$^{ies}$ could not find any evidence and the focus changed to a potential steady state influx from the ISM of such nuclides on Earth \citep{Cowan:1972,Sakamoto:1974}.

In the mid 90$^{ies}$ of the last century  \emph{geological isotope anomalies} in terrestrial archives were proposed as potential sites of nucleosynthetic products \citep{Ellis:1996}. 
The proposal was to search in ice cores and deep-sea sediments for direct deposition of supernova debris, but also for indirect signatures such as an increased production of cosmogenic nuclides (e.g. $^{10}$Be) in the Earth atmosphere due to an enhanced cosmic ray intensity caused by a nearby supernova. Such work was also motivated by 35 and 60 kyr-old $^{10}$Be anomalies that had been observed in the Vostok Antarctic ice cores. 
Several specific radio-isotopes had been suggested: $^{10}$Be, $^{26}$Al, $^{36}$Cl, $^{53}$Mn, $^{60}$Fe, and $^{59}$Ni, and also the longer-lived $^{129}$I, $^{146}$Sm, and $^{244}$Pu \citep{Ellis:1996}. 

When approaching the solar system, charged particles will be deflected by the magnetic field of the Sun within the heliosphere. Also, they will have to overcome the pressure of the solar wind \citep{Fields:2005,Athanassiadou:2011,Fry:2016,Fry:2020}. 
This appears possible only for larger dust grains. Interstellar dust grains have been observed by instruments on the Ulysses, Galileo and Cassini space missions \citep{Altobelli:2003,Altobelli:2005,Grun:1993,Frisch:2009,Mann:2010,Westphal:2014a} for distances from the Sun between 0.4 and $>$5 AU. 
These dust grains were found to follow the flow velocity of the local interstellar medium (25~km~s$^{-1}$). 
Freshly-produced radionuclides, if condensed into dust particles, can penetrate deep into the solar system, and eventually get incorporated in terrestrial archives or deposited on the lunar surface \citep{Ellis:1996,Korschinek:1996,Paul:2001}. 

In the complex path of isotopes from nucleosynthesis to deposits in terrestrial archives, several factors need to be taken into account when relating any observed particle deposition in terrestrial archives to an interstellar flux, These  conversion factors can vary from some 10\%  to less than permil, and need to be determined through independent methods:

\begin{itemize}
\item{} \emph{The radionuclides of interest need to be condensed into dust particles}.

Infrared observations suggest that dust grains in SN shocks survive for sizes $>$0.2~$\mu$m and might permeate the interstellar medium. In particular, Al, Fe and Pu are refractory elements and it is expected that a significant fraction of them will be condensed into dust particles within a short time after the explosion. The fraction of such elements bound in dust in the interstellar medium over longer time periods has been estimated to be between 10 and 100\%. Infrared depletion measurements also suggest a value close to unity.

\item{} \emph{Only a fraction will penetrate into the solar system}

The direct detections of interstellar dust particles by satellites \citep{Altobelli:2005,Grun:1993a} show a mass filtering tendency towards particles sizes $>$0.2~$\mu$m near Earth orbit. As only larger dust grains will survive a supernova shock, there seems to be no additional size filtering and deflection of interstellar dust by the solar system. Cassini measurements report an interstellar dust flux and mass distribution which suggests that 3-9\%  of the dust component of the interstellar medium that is intercepted by the solar system would be found at Earth orbit.

\item{} \emph{Only a fraction may be (or will remain) incorporated in terrestrial archives}.

Eventually, these dust grains become gravitationally bound to Earth. They will ablate during their passage through the atmosphere and the interstellar nuclides will be released from the dust grain, to rain down onto the Earth surface. 
They might be deposited on land, e.g., captured in snow and finally trapped in ice, or on the surface of the ocean. There they will be transferred to the ocean floor by scavenging processes. 
These processes usually happen at time scales much shorter than the influx time. Typical residence times in ocean could be a few 100 years for Fe or Pu. Once incorporated, the time distribution in terrestrial archives should reflect the time-dependent particle flux from the interstellar medium into the solar system.
\end{itemize}

Stable isotopes that come from space in the same way will not be distinguishable from their terrestrial counterparts.

Geological archives such as ice cores or deep-sea sediments and crusts grow over long time.  Individual layers therein will keep their integrity, so that time information can be extracted. 
Growing over millions of years, these archives thus continuously accumulate material including extraterrestrial particles, deposited within well-defined layers. 


We expect deep-sea sediments to incorporate essentially all the radionuclides discussed above from the water column. This is due to the high particle reactivity, which results in an estimated 100\% transfer efficiency into the sediments. 
But for the cases of ocean \emph{crusts and nodules}, this efficiency may be substantially lower than 100\% \citep{Wallner:2004}. 

$^{60}$Fe incorporation efficiencies into crusts were found to lie between 7 and 17\%, and for nodules of order of a few~\%. Similar incorporation efficiencies were found for Pu \citep{Wallner:2015, Wallner:2016, Wallner:2021}.

\begin{figure} 
\centering
	\includegraphics[width=\columnwidth]{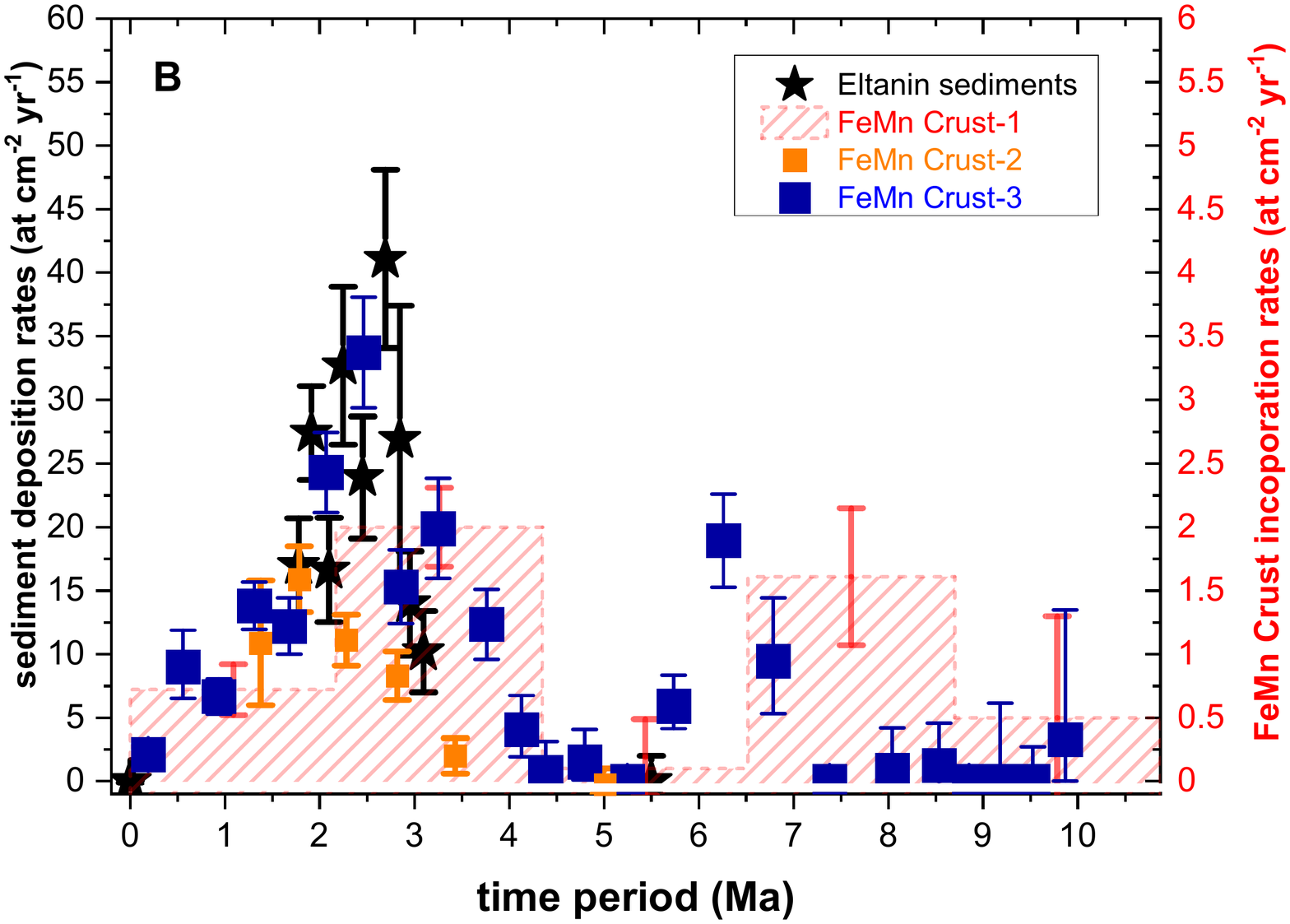}
	\caption{\Fe detections in a variate set of sediments, versus sediment age \citep[updated from][]{Wallner:2016}. An enhanced exposure of Earth to cosmic \Fe influx is seen around 3~Myr ago, and a second influx period is indicated earlier. }
	\label{fig_60Fe-sediments}
\end{figure} 

Archives can be considerably different in their accumulation or growth rates. For example, ice core layers of cm length can represent time periods of years, depending strongly on the depth, while deep-sea sediments show sedimentation rates of some mm to cm per 1,000 years. 
Deep-sea nodules and deep-sea crusts grow much slower at rates of some mm per million years. 
Therefore, the concentrations of any extraterrestrial nuclide in the matrix of a terrestrial archive will vary correspondingly. $^{60}$Fe detection in ice cores requires hundreds of kg of material to collect a sufficient number of atoms, while for sediments this can be a few gram, and for crusts even a few mg might be sufficient. The terrestrial archives can be dated via terrestrial $^{10}$Be (t$_{1/2}$=1.387 My) which is produced by cosmic-ray spallation of nitrogen and oxygen in the Earth atmosphere and incorporated into the crust at constant rates \citep{Segl:1984,Feige:2018,Lachner:2020}.


The time resolution of an archive scales with the growth rates; for crusts, thin layers of 0.5 to 1 mm (this is the limit in precision) correspond to time resolutions of a few 100,000 years per layer while sediments allow time resolutions of a few 1000 years per individual sample. 

Lunar material is a different archive as interstellar dust particles are simply implanted into the lunar surface. The top few cm of lunar surface are constantly reworked due to (solar system) particle bombardment (micrometeorites, meteorites) and no simple age can be associated. Some exposure ages of lunar regolith may extend to several 100 million years of potential integration time of interstellar particle flux. Comparatively high interstellar particle concentrations can be expected due to the thin layer of reworking depth. Clearly, lunar material is the most difficult sample material to acquire.

The expected low concentrations of interstellar particles in terrestrial or lunar archives require the most sensitive detection techniques and only direct atom-counting techniques can be applied. While originally mass spectrometric techniques had been tested in the 70$^{ies}$ of the last century to search for interstellar messengers \citep{Fields:1970,Hoffman:1971}, first in lunar and later in terrestrial samples, only accelerator mass spectrometry (AMS) achieved the required sensitivity. In AMS a particle accelerator is sandwiched between two mass spectrometers. Such a setup reduces the measurement background efficiently.

Only a few dedicated facilities have become sensitive enough to detect these extremely small interstellar signals. 
The pioneering study of a ferromanganese (FeMn) crust from the Pacific Ocean in 1999 at TU Munich resulted in the first detection of interstellar $^{60}$Fe in a terrestrial archive \citep{Knie:1999}. With improved time resolution, they identified enhanced $^{60}$Fe concentrations peaking within layers dated to an age of about 2.5 million years before present \citep{Knie:2004}.

Since then, interstellar signatures were investigated in 9 different deep-sea FeMn-crust samples \citep{Knie:1999,Knie:2004,Fitoussi:2008,Wallner:2004,Wallner:2015,Wallner:2016,Ludwig:2016,Korschinek:2020,Wallner:2021}, in 2 deep-sea nodules \citep{Wallner:2016}, in 8 different deep-sea sediment cores \citep{Fitoussi:2008,Wallner:2016,Ludwig:2016,Feige:2018,Wallner:2020,Wallner:2021}, in Antarctic snow \citep{Koll:2020}, and in lunar soil samples from three Apollo missions \citep{Fimiani:2016}. Consistently, enhancements in $^{60}$Fe concentrations have been detected.
This $^{60}$Fe signal is interpreted as a signature from nearby supernovae, because terrestrial production of $^{60}$Fe is negligible and any interplanetary influx would be much lower than their measured signal \citep{Fitoussi:2008,Wallner:2016}.
Note that in general, crust data show incorporation rates that are about 10 times lower compared to sediments; this reflects the lower incorporation efficiency (10 to 20\%) into crusts. 

The time profile of $^{60}$Fe deposition rates into deep-sea sediments and deep-sea crusts is shown in Figure~\ref{fig_60Fe-sediments}. 
A general time profile is seen in all archives: it peaks at $\sim$2.5 million years before present and continues until recent times, in particular throughout the whole Holocene until today, as evidenced in deep-sea sediments \citep{Wallner:2020} and Antarctic snow \citep{Koll:2019}, respectively. Between 4 and 5.5 million years ago no significant influx had been observed, but between 6 and 7 million years a second influx period has been clearly detected \citep{Wallner:2016,Wallner:2021}. 
The broad peak observed for the younger scenario indicates continued influx over more than 1.5 million years, and lasts much longer than a simple transient passage of supernova ejecta would suggest.
Note that the $^{60}$Fe nuclei detected in the current-day cosmic ray particle spectrum \citep{Binns:2016} may have different origins than $^{60}$Fe found in the deeper layers of the terrestrial and lunar archive. 

Indication for interstellar $^{53}$Mn influx was reported for the same time-period between 2 and 3~My before present where $^{60}$Fe influx is seen  \citep{Korschinek:2020}. 
Assuming the same fate for $^{53}$Mn nuclei as for $^{60}$Fe, from production in a massive star until deposition in a terrestrial deep-sea crust, these data represent an atom ratio $^{53}$Mn/$^{60}$Fe of about 14. 
This is consistent with ratios calculated by models for a supernova with a 11\textendash25~M$_{\odot}$ progenitor mass and solar metallicity \citep{Korschinek:2020}.

$^{26}$Al in interstellar space is another isotope attributed to massive stars (as discussed above). 
Therefore the ratio $^{60}$Fe/$^{26}$Al in terrestrial archives could constrain nucleosynthesis yields as well. 
Measurements of $^{26}$Al in the same deep-sea sediments showing clearly enhanced $^{60}$Fe signals did not find any significant $^{26}$Al excess above a dominating cosmogenic terrestrial component. 
Nevertheless, these data provide a lower limit of $^{60}$Fe/$^{26}$Al of (0.18+0.15-0.08) \citep{Feige:2018}. This is consistent with the observed ratio from galactic $\gamma$-rays (see Section 3.2).

The interstellar medium should also include  longer-lived radionuclides, such as $^{244}$Pu (80.6 My). We refer to our discussion above of the implication of detections of $^{244}$Pu together with $^{60}$Fe in the same terrestrial sediment archive, with respect to nearby sources of the r~process. 
The connection of the solar vicinity to the Local Bubble and nucleosynthesis sources within or nearby has been addressed by numerous theoretical and observational studies  \citep[more details can be found e.g. in][]{Frisch:2017,Cox:1982,Smith:2001,Frisch:2009,Breitschwerdt:2016,Zucker:2022}.

In summary, direct detection of radionuclides in deep-sea archives complements the direct observations of their radioactive decay in the wider interstellar space.  These data, however, will only reflect the local interstellar environment, because influx is only possible for dust particles that originate from sources within a few 100 pc distance to Earth.
Terrestrial archives, on the other hand, can provide well-defined time information, which links to production or deposition time, and can thus collect interstellar signatures over tens of million years.


\section{Summary and Conclusions}
\label{conclusions}

\begin{table*} 
\begin{center}
\begin{tabular}{|p{6cm}|p{8cm}|p{2.5cm}|}
 \hline
  \centerline{Messenger} &  \centerline{Astrophysical Goals} & References \\
   \hline \hline
    starlight spectroscopy                                       &    elemental abundances at time of starbirth;         & \citep{Payne:1928,Caffau:2014,Jofre:2019} \\
   - absorption lines   in  UV/opt/IR                            & cosmic abundance evolution;                                 &    \citep{Buder:2021}                                                                  \\
                                                                               & mixing \& burning processes in stellar interiors;    &                                                                      \\
                                                                               & nucleocosmochronology    &  \citep{Meyer:2000,Thielemann:2002b} \\
       - emission lines        in  X/UV/opt/IR                                           &   hot massive stars and winds & \citep{Crowther:2007} \\
                          &  wind interactions; binaries  & \citep{Pittard:2010} \\
                          &  planetary-nebula abundances  & \citep{Garcia-Hernandez:2016} \\
   \hline
   interstellar emission spectroscopy                       & stellar and nucleosynthesis activity                         &  \\
   - UV/opt/IR                                                           & in regions of galaxies;                                              &  \\
   - radio                                          & abundances in specific sources/ regions             &  \citep{Smith:2010,Kaminski:2018,Tang:2019} \\
   - X rays                                                                & abundances in hot plasma (supernova remnants)     &  \citep{Vink:2012} \\
                                                                                &  cosmic abundance evolution (intergalactic medium) &      \\
   - $\gamma$ rays                                                  & abundances of specific radioactive isotopes             & \citep{Diehl:2013}     \\
                                                                               & (short-lived: explosive nucleosynthesis;   & \citep{Diehl:2015,Jose:2007}\\
                                                                               & long-lived: massive-star late shell burning phases;     &                                                                      \\      
                                                                               & propagation of nucleosynthesis ejecta) & \citep{Krause:2014} \\ 
   \hline 
  spectrometry of material samples                & isotopic composition of matter samples:   & \\
  - cosmic rays                                                        & abundances in accelerated interstellar matter            & \citep{Wiedenbeck:2001a} \\
  - stardust                                                              & abundances in ejecta of specific sources         & \citep{Clayton:2004,Zinner:2008,Westphal:2014} \\
                                                                               & (AGB stars: s process, hot-bottom burning; &  \citep{Karakas:2014} \\
                                                                               &  supernovae: explosive nucleosynthesis;   &  \citep{Hoppe:2017,Leung:2018,Thielemann:2018b}\\
                                                                               & novae: hot hydrogen burning) & \citep{Jose:2016b}\\
  - meteorites                                                          & abundances in the early solar system                         & \citep{Dauphas:2016,Kleine:2020} \\
  - sediments (oceanfloor; earth \& moon     & abundances in inflows of cosmic dust in recent history                         & \citep{Ellis:1996,Wallner:2021} \\
  \quad surface)  & & \\
  \hline
  asteroseismology                                                 & precision test of stellar (solar) model   with inclusion of abundances    and nucleosynthesis    & \citep{Christensen-Dalsgaard:2021,Silva-Aguirre:2018} \\
   \hline
   neutrinos   &     &  \\
   - low energy  &   H burning reactions in the Sun & \citep{Borexino-Collaboration:2020,Borexino-Collaboration:2018,BOREXINO-Collaboration:2014} \\
   - MeV energies  &   URCA reactions and $\nu$ cooling in massive stars & \citep{Bahcall:2006} \\
   - $\sim$10-50 MeV  &   proto-neutron star cooling (explosive nucleosynthesis) & \citep{Muller:2019} \\
        \hline 
  gravitational waves                                              & core-collapse supernova dynamics (explosive nucleosynthesis);                   &   \citep{Logue:2012}     \\ 
                                                                               & neutron-star collisions (explosive nucleosynthesis and r process) &    \citep{Thielemann:2017}   \\
                                                                           \hline
\end{tabular}
\end{center}
\caption{Messengers of cosmic nucleosynthesis (1$^{st}$ column), with their main goals (2$^{nd}$ column) within the study of cosmic nucleosynthesis. Key references are given in the 3$^{rd}$ column.}
\label{tab_nucleosynthesisMessengers}
\end{table*}
%

Our knowledge of cosmic nucleosynthesis has been shaped by different astronomical messengers and methods: 
The information from electromagnetic radiation, from material samples, from neutrinos, and recently from gravitational waves have all been incorporated, in order to understand the cosmic sites that provide the environment of nuclear fusion reactions, and how these may co-act to produce the cosmic composition of isotopes and elements  in its impressive variety as seen in the current universe.
Fundamental nuclear reaction theory within the extremes of cosmic environments had to be obtained from nuclear experiments and theory. This also works in the opposite direction, as cosmic isotopic abundances can teach us about the complexities of reaction networks among nuclei in plasma conditions that may not be tested by terrestrial experiments.

The foundation of cosmic elemental abundances has been laid by spectroscopy of light from a variety of cosmic sources, as matter imprints characteristic absorption signals.
Spectra have complex appearances, as many elements with their different abundances and ionisation states lead to the superposition of many lines. Sophisticated data analysis tools
can provide satisfactory identifications of elements and determinations of their abundances. 
Using such tools, a substantial library of elemental abundances has been assembled for a great variety of stars in the Milky Way and in nearby galaxies.
These have led to a basic understanding of the processes of cosmic nucleosynthesis and to a fundamental understanding of stellar evolution.

Stellar explosions and their observations add essentials to our understanding of nucleosynthesis. 
A concept of cosmic nucleosynthesis has been applied with success to describe the observed abundance evolution. 
Additional astronomical messengers and their astrophysical exploitation add to our understanding of cosmic nucleosynthesis.

Neutrino measurements 
have confirmed the hydrogen fusion reactions in the Sun. This is the most-direct messenger from nuclear fusion reactions as they happen inside stars.
Less directly, specific isotopes produced by nuclear reactions among heavier nuclei have been made accessible to measurements, through dedicated spectroscopy, such as C and O isotopes, 
and widely through dust grains that have formed in the circumstellar regions, preserving isotopic abundances of stellar ejecta to analyse them in mass spectrometers in terrestrial laboratories. 
These grains have been brought to Earth via meteorites. 
But information of their nucleosynthetic sources has been lost in their trajectory before incorporation into meteorites, and must be obtained indirectly. 
Pattern recognition of isotopic abundances in terrestrial laboratories has led to inferred origins of these dust grains 
in terms of stars and stellar explosions. 
From these measurements we learn about mixing of stellar plasma throughout the star, and specifically in the boundary regions between shells that either differ greatly in their composition or in their hydrostatic stability. 
Measurements in photospheric spectra, and in presolar dust grains, constrain the s~process of neutron capture reactions in stars. 
Indirectly, we also learn about neutron-releasing reactions in stars and explosions.

Astero-seismology adds a diagnostic of interior structure within stars, from our Sun to red giants.
For the Sun, this has enabled a precision test of the solar model, which led to interesting efforts to consolidate observed solar abundances because of an apparent inconsistency.
For other stars, seismic age determinations are much more precise than the main-sequence turn-off ages derived for stellar groups. 

Supernova explosions, with matter diluting as the expansion proceeds, reveal a three-dimensional view of the composition of the exploding ejecta as the photosphere effectively travels to regions further inside with time. In addition to light curves, the spectra from such \emph{tomography} of supernovae have helped to constrain the physical processes and nuclear reactions in the pre-supernova star and in the explosion.  
Essential additional information is provided by direct observation through characteristic $\gamma$~rays of radioactive isotopes that have been produced in the explosion. On the one hand, they create an energy source from within, and the progressively leaking $\gamma$~rays provide information about the expanding envelope. On the other hand, they reflect conditions for their nucleosynthesis at the time of the explosion and measure these through their yields more directly than the light curves.
Neutrinos from neutronisation of matter and cooling of the proto-neutron star are a direct probe of the processes that launch the supernova explosion. After SN~1987A's first signal, better detectors have been built and await the next core-collapse supernova in our Galaxy, to provide neutrino light curves and spectra. 
Gravitational-wave astronomy adds an important diagnostic of the central matter distribution as the star launches a supernova. An explosion event in the nearby universe is to be awaited and will lead to a leap in physical models of supernova explosions.

Tracing the flow of nucleosynthesis ejecta, radioactive by-products have been key to direct measurements due to their intrinsic clock. 
Cosmochronology had been initiated from abundance measurements of fissioning radioactive isotopes such as uranium and thorium. 
On the shorter time scale, abundant isotopes $^{60}$Fe and $^{26}$Al, but also $^{244}$Pu, have proven useful to constrain nucleosynthesis event rates.
In situ transport measurements with $\gamma$~rays have complemented transport history constraints from abundance measurements of deposits on/in solar-system bodies.  
Galactic archeology based on stellar photospheric abundances of elements for stars of the variety of cosmic ages is complemented by specific constraints on trajectories of cosmic material. These will have to be combined into gas transport models that make up the stage after nucleosynthesis events and before star formation. For a complete description of abundance evolution, also the process of star formation itself will have to be sufficiently well understood. 
{        The efficiency of recycling ejecta can be expected to vary for different sources of nucleosynthesis, because their environments likely are different, ranging from dense massive-star forming regions to dilute parts of a galaxy's halo. Therefore, measurements are important which can trace ejecta flows in these cases, and future models of chemical evolution may resolve their inherent degeneracy of model yields and recycling efficiency.}

The detailed study of cosmic nucleosynthesis in its current stage is characterised by combining {        the variety of }these different astronomical messengers {         (see Table 2)}. 
{         Contributions from different sources of nucleosynthesis are accumulated at different times, from different source types, and into different environments (see Figure~\ref{fig_cosmic_evolution_metallicity}).}
In this way, we advance beyond description of specific abundance data, refining a model that implements astrophysical understandings and assumptions. The aim is towards a consistent description of observables from multiple messengers within a coherent astrophysical model that minimises priors and assumptions. 
Advances in the radiation and materials messengers, including neutrinos and gravitational waves, are all contributing to this process.
This challenge is characteristic for complex interconnected systems. 
Once a satisfactory description is obtained in terms of physics, the origins of the isotopes that make up our world can be claimed to be understood. 

%
\section*{Acknowledgments}
This study was supported by the Deutsche Forschungsgemeinschaft (DFG, German Research Foundation) under its Excellence Strategy, the Munich Clusters of Excellence \emph{Origin and Structure of the Universe} and \emph{Origins} (EXC-2094-390783311), by the EU through COST action ChETEC CA160117, and by ERC via CoG-2016 RADIOSTAR (Grant Agreement 724560).{         We appreciate being able to re-use several figures as referenced, per copyright agreements, and thank both referees for their constructive comments.}


\begin{thebibliography}{100}

\bibitem{Mendeleev:1869}
D.~{Mendeleev}, On the relationship of the properties of the elements to their
  atomic weights, Zhurnal Russkoe Fiziko-Khimicheskoe Obshchestvo 1 (1869)
  60--77.

\bibitem{Jofre:2019}
P.~{Jofr{\'e}}, U.~{Heiter}, C.~{Soubiran}, {Accuracy and Precision of
  Industrial Stellar Abundances}, \araa 57 (2019) 571--616.

\bibitem{Goldschmidt:1930}
V.~M. {Goldschmidt}, {Geochemische Verteilungsgesetze und kosmische
  H{\"a}ufigkeit der Elemente}, Naturwissenschaften 18~(47-49) (1930)
  999--1013.

\bibitem{Mayer:1952}
M.~G. {Mayer}, {Radioactivity and Nuclear Theory}, Annual Review of Physical
  Chemistry 3 (1952) 19--38.

\bibitem{Payne:1925}
C.~H. {Payne}, {Astrophysical Data Bearing on the Relative Abundance of the
  Elements}, Proceedings of the National Academy of Science 11~(3) (1925)
  192--198.

\bibitem{Eddington:1919}
A.~S. {Eddington}, {The sources of stellar energy}, The Observatory 42 (1919)
  371--376.

\bibitem{Suess:1956}
H.~E. {Suess}, H.~C. {Urey}, {Abundances of the Elements}, Reviews of Modern
  Physics 28~(1) (1956) 53--74.

\bibitem{Burbidge:1957}
E.~M. {Burbidge}, G.~R. {Burbidge}, W.~A. {Fowler}, F.~{Hoyle}, {Synthesis of
  the Elements in Stars}, Reviews of Modern Physics 29 (1957) 547--650.

\bibitem{Arnett:1989a}
W.~D. {Arnett}, J.~N. {Bahcall}, R.~P. {Kirshner}, S.~E. {Woosley}, {Supernova
  1987A}, \araa 27 (1989) 629--700.

\bibitem{McCray:2016}
R.~{McCray}, C.~{Fransson}, {The Remnant of Supernova 1987A}, \araa 54 (2016)
  19--52.

\bibitem{Smartt:2017}
S.~J. {Smartt}, T.-W. {Chen}, A.~{Jerkstrand}, M.~{Coughlin}, E.~{Kankare},
{\it et al.},
 {A kilonova as
  the electromagnetic counterpart to a gravitational-wave source}, \nat 551
  (2017) 75--79.

\bibitem{Margutti:2021}
R.~{Margutti}, R.~{Chornock}, {First Multimessenger Observations of a Neutron
  Star Merger}, \araa  59 (2021), 155.

\bibitem{Abbott:2017g}
B.~P. {Abbott}, R.~{Abbott}, T.~D. {Abbott}, F.~{Acernese}, K.~{Ackley},
{\it et al.},
{\it et al.},
  {Gravitational Waves and Gamma-Rays from a Binary Neutron Star Merger:
  GW170817 and GRB 170817A}, \apjl 848~(2) (2017) L13.

\bibitem{Clayton:1968}
D.~D. {Clayton}, {Principles of stellar evolution and nucleosynthesis}, New
  York: McGraw-Hill (1968).

\bibitem{Wallerstein:1997}
G.~{Wallerstein}, I.~J. {Iben}, P.~{Parker}, A.~M. {Boesgaard}, G.~M. {Hale},
{\it et al.},
{Synthesis of the elements in stars: forty years of progress},
  Reviews of Modern Physics 69 (1997) 995--1084.

\bibitem{Merrill:1952}
P.~W. {Merrill}, {Spectroscopic Observations of Stars of Class S}, \apj 116
  (1952) 21.

\bibitem{Mahoney:1982}
W.~A. {Mahoney}, J.~C. {Ling}, A.~S. {Jacobson}, R.~E. {Lingenfelter}, {Diffuse
  galactic gamma-ray line emission from nucleosynthetic Fe-60, Al-26, and Na-22
  - Preliminary limits from HEAO 3}, \apj 262 (1982) 742.

\bibitem{Fynbo:2012}
H.~{Fynbo}, {New experimental results for the triple-alpha reaction}, in:
  J.~{Lattanzio}, A.~{Karakas}, M.~{Lugaro}, G.~{Dracoulis} (Eds.), Nuclei in
  the Cosmos (NIC XII), (2012) 11.

\bibitem{deBoer:2017}
R.~J. {deBoer}, J.~{G{\"o}rres}, M.~{Wiescher}, R.~E. {Azuma}, A.~{Best}, {\it et al.}, {The $^{12}$C({$\alpha$} ,{$\gamma$}
  )$^{16}$O reaction and its implications for stellar helium burning}, Reviews
  of Modern Physics 89~(3) (2017) 035007.

\bibitem{Laird:2022}
A.~{Laird}, {\it et al.}, Modelling of the 26al abundance in stars and progress
  on the relevant nuclear reaction rates, J. Phys. G. (2022) {\it in review}.

\bibitem{Kibedi:2020}
T.~{Kib{\'e}di}, B.~{Alshahrani}, A.~E. {Stuchbery}, A.~C. {Larsen},
  A.~{G{\"o}rgen}, {\it et al.}: 
   {Radiative Width of the Hoyle State from
  {\ensuremath{\gamma}} -Ray Spectroscopy}, \prl 125~(18) (2020) 182701.

\bibitem{Diehl:2021b}
R.~{Diehl}, M.~{Lugaro}, A.~{Heger}, A.~{Sieverding}, X.~{Tang}, 
{\it et al.:}
 {The
  Radioactive Nuclei $^{\textbf{26}}$Al and $^{\textbf{60}}$Fe in the Cosmos
  and in the Solar System}, Publications of the Astronomical Society of
  Australia 38 (2021) e062.

\bibitem{Schramm:1974}
D.~N. {Schramm}, B.~M. {Tinsley}, {On the origin and evolution of s-process
  elements}, \apj 193 (1974) 151--155.

\bibitem{Kappeler:1999}
F.~{Kappeler}, {The origin of the heavy elements: The s process}, Progress in
  Particle and Nuclear Physics 43 (1999) 419--483.

\bibitem{Arnould:2007}
M.~{Arnould}, S.~{Goriely}, K.~{Takahashi}, {The r-process of stellar
  nucleosynthesis: Astrophysics and nuclear physics achievements and
  mysteries}, \physrep 450 (2007) 97--213.

\bibitem{Cowan:1982}
J.~J. {Cowan}, A.~G.~W. {Cameron}, J.~W. {Truran}, {The thermal runaway
  r-process}, \apj 252 (1982) 348--355.

\bibitem{Cowan:2021}
J.~J. {Cowan}, C.~{Sneden}, J.~E. {Lawler}, A.~{Aprahamian}, M.~{Wiescher},
  K.~{Langanke}, G.~{Mart{\'\i}nez-Pinedo}, F.-K. {Thielemann}, {Origin of the
  heaviest elements: The rapid neutron-capture process}, Reviews of Modern
  Physics 93~(1) (2021) 015002.

\bibitem{Pitrou:2018}
C.~{Pitrou}, A.~{Coc}, J.-P. {Uzan}, E.~{Vangioni}, {Precision big bang
  nucleosynthesis with improved Helium-4 predictions}, \physrep 754 (2018)
  1--66.

\bibitem{Coc:2015}
A.~{Coc}, P.~{Petitjean}, J.-P. {Uzan}, E.~{Vangioni}, P.~{Descouvemont},
  C.~{Iliadis}, R.~{Longland}, {New reaction rates for improved primordial D /H
  calculation and the cosmic evolution of deuterium}, \prd 92~(12) (2015)
  123526.

\bibitem{Pitrou:2021}
C.~{Pitrou}, A.~{Coc}, J.-P. {Uzan}, E.~{Vangioni}: {Resolving conclusions
  about the early Universe requires accurate nuclear measurements}, Nature
  Reviews Physics 3~(4) (2021) 231--232.

\bibitem{Cyburt:2016}
R.~H. {Cyburt}, B.~D. {Fields}, K.~A. {Olive}, T.-H. {Yeh}: {Big bang
  nucleosynthesis: Present status}, Reviews of Modern Physics 88~(1) (2016)
  015004.

\bibitem{Cooke:2018}
R.~J. {Cooke}, M.~{Fumagalli}: {Measurement of the primordial helium abundance
  from the intergalactic medium}, Nature Astronomy 2 (2018) 957--961.

\bibitem{Spite:1982a}
M.~{Spite}, F.~{Spite}: {Lithium abundance at the formation of the Galaxy},
  \nat 297~(5866) (1982) 483--485.

\bibitem{Spite:1982}
F.~{Spite}, M.~{Spite}, {Abundance of lithium in unevolved stars and old disk
  stars : Interpretation and consequences.}, \aap 115 (1982) 357--366.

\bibitem{Spite:2005}
M.~{Spite}, R.~{Cayrel}, B.~{Plez}, V.~{Hill}, F.~{Spite}, {\it et al.}:
{First stars VI - Abundances of
  C, N, O, Li, and mixing in extremely metal-poor giants. Galactic evolution of
  the light elements}, \aap 430 (2005) 655--668.

\bibitem{Bennett:2003}
C.~L. {Bennett}, M.~{Halpern}, G.~{Hinshaw}, N.~{Jarosik}, A.~{Kogut},
{\it et al.},
{First-Year Wilkinson Microwave Anisotropy Probe (WMAP)
  Observations: Preliminary Maps and Basic Results}, \apjs 148 (2003) 1--27.

\bibitem{Planck-Collaboration:2020}
{Planck Collaboration}, N.~{Aghanim}, Y.~{Akrami}, M.~{Ashdown}, J.~{Aumont},
  C.~{Baccigalupi}, {\it et al.},
  {Planck 2018 results.
  VI. Cosmological parameters}, \aap 641 (2020) A6.

\bibitem{Coc:2017}
A.~{Coc}, E.~{Vangioni}, {Primordial nucleosynthesis}, International Journal of
  Modern Physics E 26~(8) (2017) 1741002.

\bibitem{Coc:2014}
A.~{Coc}, J.-P. {Uzan}, E.~{Vangioni}, {Standard big bang nucleosynthesis and
  primordial CNO abundances after Planck}, \jcap 2014~(10) (2014) 050.

\bibitem{Fields:2020}
B.~D. {Fields}, K.~A. {Olive}, T.-H. {Yeh}, C.~{Young}, {Big-Bang
  Nucleosynthesis after Planck}, \jcap 2020~(3) (2020) 010.

\bibitem{Grisoni:2019}
V.~{Grisoni}, F.~{Matteucci}, D.~{Romano}, X.~{Fu}, {Evolution of lithium in
  the Milky Way halo, discs, and bulge}, \mnras 489~(3) (2019) 3539--3546.

\bibitem{Prantzos:2012}
N.~{Prantzos}, {On the origin and composition of Galactic cosmic rays}, \aap
  538 (2012) A80.

\bibitem{de-Nolfo:2006}
G.~A. {de Nolfo}, I.~V. {Moskalenko}, W.~R. {Binns}, E.~R. {Christian}, A.~C.
  {Cummings}, {\it et al.},
  {Observations of
  the Li, Be, and B isotopes and constraints on cosmic-ray propagation},
  Advances in Space Research 38~(7) (2006) 1558--1564.

\bibitem{Aguilar:2018}
M.~{Aguilar}, L.~{Ali Cavasonza}, G.~{Ambrosi}, L.~{Arruda}, N.~{Attig},
{\it et al.}, {AMS
  Collaboration}, {Observation of New Properties of Secondary Cosmic Rays
  Lithium, Beryllium, and Boron by the Alpha Magnetic Spectrometer on the
  International Space Station}, \prl 120~(2) (2018) 021101.

\bibitem{Prantzos:2017}
N.~{Prantzos}, P.~{de Laverny}, G.~{Guiglion}, A.~{Recio-Blanco}, C.~C.
  {Worley}, {The AMBRE project: a study of Li evolution in the Galactic thin
  and thick discs}, \aap 606 (2017) A132.

\bibitem{Fu:2015}
X.~{Fu}, A.~{Bressan}, P.~{Molaro}, P.~{Marigo}, {Lithium evolution in
  metal-poor stars: from pre-main sequence to the Spite plateau}, \mnras
  452~(3) (2015) 3256--3265.
  
\bibitem{Michaud:1984}
G.~{Michaud}, G.~{Fontaine}, G.~{Beaudet}, {The lithium abundance - Constraints
  on stellar evolution}, \apj 282 (1984) 206--213.

\bibitem{Richard:2005}
O.~{Richard}, G.~{Michaud}, J.~{Richer}, {Implications of WMAP Observations on
  Li Abundance and Stellar Evolution Models}, \apj 619~(1) (2005) 538--548.

\bibitem{Korn:2007}
A.~J. {Korn}, F.~{Grundahl}, O.~{Richard}, L.~{Mashonkina}, P.~S. {Barklem},
  R.~{Collet}, B.~{Gustafsson}, N.~{Piskunov}, {Atomic Diffusion and Mixing in
  Old Stars. I. Very Large Telescope FLAMES-UVES Observations of Stars in NGC
  6397}, \apj 671~(1) (2007) 402--419.

\bibitem{Lind:2009}
K.~{Lind}, F.~{Primas}, C.~{Charbonnel}, F.~{Grundahl}, M.~{Asplund},
  {Signatures of intrinsic Li depletion and Li-Na anti-correlation in the
  metal-poor globular cluster NGC 6397}, \aap 503~(2) (2009) 545--557.

\bibitem{Korn:2020}
A.~J. {Korn}, {How stars in globular clusters reveal the depletion of the Spite
  plateau of lithium.}, \memsai 91 (2020) 105.

\bibitem{Melendez:2010}
J.~{Mel{\'e}ndez}, L.~{Casagrande}, I.~{Ram{\'\i}rez}, M.~{Asplund}, W.~J.
  {Schuster}, {Observational evidence for a broken Li Spite plateau and
  mass-dependent Li depletion}, \aap 515 (2010) L3.

\bibitem{Sbordone:2010}
L.~{Sbordone}, P.~{Bonifacio}, E.~{Caffau}, H.-G. {Ludwig}, N.~T. {Behara},
{\it et al.}:
  {The metal-poor end of the Spite plateau. I. Stellar parameters,
  metallicities, and lithium abundances}, \aap 522 (2010) A26.

\bibitem{Gao:2020}
X.~{Gao}, K.~{Lind}, A.~M. {Amarsi}, S.~{Buder}, J.~{Bland-Hawthorn}, S.~W.
  {Campbell}, {\it et al.}, {GALAH
  Collaboration}, {The GALAH survey: a new constraint on cosmological lithium
  and Galactic lithium evolution from warm dwarf stars}, \mnras 497~(1) (2020)
  L30--L34.

\bibitem{Asplund:2006}
M.~{Asplund}, N.~{Grevesse}, A.~{Jacques Sauval}, {The solar chemical
  composition}, Nuclear Physics A 777 (2006) 1--4.

\bibitem{Olive:1993}
K.~A. {Olive}, {Big Bang Nucleosynthesis: Non-Standard Models}, in: C.~W.
  {Akerlof}, M.~A. {Srednicki} (Eds.), Texas/PASCOS '92: Relativistic
  Astrophysics and Particle Cosmology, Vol. 688  (1993) 769.

\bibitem{Jedamzik:2008}
K.~{Jedamzik}: {Cosmic Li6 and Li7 problems and BBN with long-lived charged
  massive particles}, \prd 77~(6) (2008) 063524.

\bibitem{Lind:2013}
K.~{Lind}, J.~{Melendez}, M.~{Asplund}, R.~{Collet}, Z.~{Magic}: {The lithium
  isotopic ratio in very metal-poor stars}, \aap 554 (2013) A96.

\bibitem{Wang:2021}
E.~X. {Wang}, T.~{Nordlander}, M.~{Asplund}, A.~M. {Amarsi}, K.~{Lind},
  Y.~{Zhou}: {3D NLTE spectral line formation of lithium in late-type stars},
  \mnras 500~(2) (2021) 2159--2176.

\bibitem{Piau:2006}
L.~{Piau}, T.~C. {Beers}, D.~S. {Balsara}, T.~{Sivarani}, J.~W. {Truran}, J.~W.
  {Ferguson}: {From First Stars to the Spite Plateau: A Possible Reconciliation
  of Halo Stars Observations with Predictions from Big Bang Nucleosynthesis},
  \apj 653~(1) (2006) 300--315.

\bibitem{Pignatari:2016}
M.~{Pignatari}, F.~{Herwig}, R.~{Hirschi}, M.~{Bennett}, G.~{Rockefeller},
 {\it et al.}: {NuGrid Stellar
  Data Set. I.Stellar Yields from H to Bi for Stars with Metallicities Z = 0.02
  and Z = 0.01}, \apjs 225 (2016) 24.

\bibitem{Woosley:2002}
S.~E. {Woosley}, A.~{Heger}, T.~A. {Weaver}: {The evolution and explosion of
  massive stars}, Reviews of Modern Physics 74~(4) (2002) 1015--1071.

\bibitem{Paxton:2011}
B.~{Paxton}, L.~{Bildsten}, A.~{Dotter}, F.~{Herwig}, P.~{Lesaffre},
  F.~{Timmes}: {Modules for Experiments in Stellar Astrophysics (MESA)}, \apjs
  192~(1) (2011) 3.

\bibitem{DeglInnocenti:2008}
S.~{Degl'Innocenti}, P.~G. {Prada Moroni}, M.~{Marconi}, A.~{Ruoppo}: {The
  FRANEC stellar evolutionary code}, \apss 316~(1-4) (2008) 25--30.

\bibitem{Maeder:2000}
A.~{Maeder}, G.~{Meynet}, {The Evolution of Rotating Stars}, \araa 38 (2000)
  143--190.

\bibitem{Jorgensen:2018}
A.~C.~S. {J{\o}rgensen}, J.~R. {Mosumgaard}, A.~{Weiss}, V.~{Silva Aguirre},
  J.~{Christensen-Dalsgaard}, {Coupling 1D stellar evolution with
  3D-hydrodynamical simulations on the fly - I. A new standard solar model},
  \mnras 481~(1) (2018) L35--L39.

\bibitem{Bahcall:2006}
J.~N. {Bahcall}, A.~M. {Serenelli}, S.~{Basu}, {10,000 Standard Solar Models: A
  Monte Carlo Simulation}, \apjs 165 (2006) 400--431.

\bibitem{Vinyoles:2017}
N.~{Vinyoles}, A.~M. {Serenelli}, F.~L. {Villante}, S.~{Basu},
  J.~{Bergstr{\"o}m}, M.~C. {Gonzalez-Garcia}, M.~{Maltoni},
  C.~{Pe{\~n}a-Garay}, N.~{Song}, {A New Generation of Standard Solar Models},
  \apj 835 (2017) 202.

\bibitem{Iglesias:1996}
C.~A. {Iglesias}, F.~J. {Rogers}, {Updated Opal Opacities}, \apj 464 (1996)
  943.

\bibitem{Seaton:2007}
M.~J. {Seaton}, {Updated Opacity Project radiative accelerations}, \mnras
  382~(1) (2007) 245--250.

\bibitem{Delahaye:2021}
F.~{Delahaye}, C.~P. {Ballance}, R.~T. {Smyth}, N.~R. {Badnell}, {Quantitative
  comparison of opacities calculated using the R-matrix and distorted-wave
  methods: Fe XVII}, \mnras 508~(1) (2021) 421--432.

\bibitem{Asplund:2009}
M.~{Asplund}, N.~{Grevesse}, A.~J. {Sauval}, P.~{Scott}, {The Chemical
  Composition of the Sun}, \araa 47 (2009) 481--522.

\bibitem{Gustafsson:2008}
B.~{Gustafsson}, B.~{Edvardsson}, K.~{Eriksson}, U.~G. {J{\o}rgensen},
  {\r{A}}.~{Nordlund}, B.~{Plez}, {A grid of MARCS model atmospheres for
  late-type stars. I. Methods and general properties}, \aap 486~(3) (2008)
  951--970.

\bibitem{Caffau:2011}
E.~{Caffau}, H.~G. {Ludwig}, M.~{Steffen}, B.~{Freytag}, P.~{Bonifacio}, {Solar
  Chemical Abundances Determined with a CO5BOLD 3D Model Atmosphere}, \solphys
  268~(2) (2011) 255--269.

\bibitem{Grevesse:1998}
N.~{Grevesse}, A.~J. {Sauval}, {Standard Solar Composition}, \ssr 85 (1998)
  161--174.

\bibitem{Asplund:2021}
M.~{Asplund}, A.~M. {Amarsi}, N.~{Grevesse}, {The chemical make-up of the Sun:
  A 2020 vision}, \aap 653 (2021) A141.

\bibitem{Magg:2022}
E.~{Magg}, M.~{Bergemann}, A.~{Serenelli}, M.~{Bautista}, B.~{Plez},
{\it et al.}: {Observational
  constraints on the origin of the elements. IV. Standard composition of the
  Sun}, \aap 661 (2022) A140.

\bibitem{Anders:1989}
E.~{Anders}, N.~{Grevesse}, {Abundances of the elements - Meteoritic and
  solar}, \gca 53 (1989) 197--214.

\bibitem{Asplund:2005}
M.~{Asplund}, N.~{Grevesse}, A.~J. {Sauval}: {The Solar Chemical Composition},
  in: I.~{Barnes}, Thomas~G., F.~N. {Bash} (Eds.), Cosmic Abundances as Records
  of Stellar Evolution and Nucleosynthesis, Vol. 336 of Astronomical Society of
  the Pacific Conference Series (2005) 25.

\bibitem{Caffau:2008}
E.~{Caffau}, H.~G. {Ludwig}, M.~{Steffen}, T.~R. {Ayres}, P.~{Bonifacio},
  R.~{Cayrel}, B.~{Freytag}, B.~{Plez}, {The photospheric solar oxygen project.
  I. Abundance analysis of atomic lines and influence of atmospheric models},
  \aap 488~(3) (2008) 1031--1046.

\bibitem{Scott:2009}
P.~{Scott}, M.~{Asplund}, N.~{Grevesse}, A.~J. {Sauval}, {On the Solar Nickel
  and Oxygen Abundances}, \apjl 691~(2) (2009) L119--L122.

\bibitem{Aerts:2021}
C.~{Aerts}, {Probing the interior physics of stars through asteroseismology},
  Reviews of Modern Physics 93~(1) (2021) 015001.

\bibitem{Christensen-Dalsgaard:2021}
J.~{Christensen-Dalsgaard}, {Solar structure and evolution}, Living Reviews in
  Solar Physics 18~(1) (2021) 2.

\bibitem{Basu:2010}
S.~{Basu}, {Helioseismology as a diagnostic of the solar interior}, \apss
  328~(1-2) (2010) 43--50.

\bibitem{Leavitt:1912}
H.~S. {Leavitt}, E.~C. {Pickering}, {Periods of 25 Variable Stars in the Small
  Magellanic Cloud.}, Harvard College Observatory Circular 173 (1912) 1--3.

\bibitem{Borexino-Collaboration:2020}
M.~{Borexino Collaboration}, Agostini, K.~{Altenm{\"u}ller}, S.~{Appel},
  V.~{Atroshchenko}, Z.~{Bagdasarian}, {\it et al.}, {Experimental evidence of neutrinos
  produced in the CNO fusion cycle in the Sun}, \nat 587~(7835) (2020)
  577--582.

\bibitem{Bellini:2012}
G.~{Bellini}, J.~{Benziger}, D.~{Bick}, S.~{Bonetti}, G.~{Bonfini}, {\it et al.}, {First Evidence of pep Solar Neutrinos by
  Direct Detection in Borexino}, Physical Review Letters 108~(5) (2012) 051302.

\bibitem{BOREXINO-Collaboration:2014}
{BOREXINO Collaboration}, G.~{Bellini}, J.~{Benziger}, D.~{Bick}, G.~{Bonfini},
  D.~{Bravo}, {\it et al.},
  {Neutrinos from the primary proton-proton fusion process in the Sun}, \nat
  512 (2014) 383--386.

\bibitem{Borexino-Collaboration:2018}
{Borexino Collaboration}, M.~{Agostini}, K.~{Altenm{\"u}ller}, S.~{Appel},
  V.~{Atroshchenko}, Z.~{Bagdasarian}, {\it et al.}, {Comprehensive measurement of pp-chain solar neutrinos}, \nat
  562~(7728) (2018) 505--510.

\bibitem{Appel:2022}
S.~{Appel}, Z.~{Bagdasarian}, D.~{Basilico}, G.~{Bellini}, J.~{Benziger},
 {\it et al.}: {Improved measurement of solar neutrinos from the
  Carbon-Nitrogen-Oxygen cycle by Borexino and its implications for the
  Standard Solar Model}, arXiv e-prints (2022) arXiv:2205.15975

\bibitem{Ciesla:2006a}
F.~J. {Ciesla}, S.~B. {Charnley}, {The Physics and Chemistry of Nebular
  Evolution}, in: D.~S. {Lauretta}, H.~Y. {McSween} (Eds.), Meteorites and the
  Early Solar System II (2006) 209.

\bibitem{Ciesla:2010}
F.~J. {Ciesla}, {The distributions and ages of refractory objects in the solar
  nebula}, \icarus 208~(1) (2010) 455--467.

\bibitem{Dauphas:2016}
N.~{Dauphas}, E.~A. {Schauble}: {Mass Fractionation Laws, Mass-Independent
  Effects, and Isotopic Anomalies}, Annual Review of Earth and Planetary
  Sciences 44 (2016) 709--783.

\bibitem{Lodders:2020}
K.~{Lodders}, {Solar Elemental Abundances}, in: The Oxford Research
  Encyclopedia of Planetary Science, Oxford University Press (2020), arxiv:1912.00844.

\bibitem{Osorio:2019}
Y.~{Osorio}, K.~{Lind}, P.~S. {Barklem}, C.~{Allende Prieto}, O.~{Zatsarinny}:
  {Ca line formation in late-type stellar atmospheres. I. The model atom}, \aap
  623 (2019) A103.

\bibitem{Heber:2021}
V.~S. {Heber}, K.~D. {McKeegan}, R.~C.~J. {Steele}, A.~J.~G. {Jurewicz}, K.~D.
  {Rieck}, Y.~{Guan}, R.~{Wieler}, D.~S. {Burnett}, {Elemental Abundances of
  Major Elements in the Solar Wind as Measured in Genesis Targets and
  Implications on Solar Wind Fractionation}, \apj 907~(1) (2021) 15.

\bibitem{Lodders:2021}
K.~{Lodders}: {Relative Atomic Solar System Abundances, Mass Fractions, and
  Atomic Masses of the Elements and Their Isotopes, Composition of the Solar
  Photosphere, and Compositions of the Major Chondritic Meteorite Groups}, \ssr
  217~(3) (2021) 44.

\bibitem{Beers:2005}
T.~C. {Beers}, N.~{Christlieb}: {The Discovery and Analysis of Very Metal-Poor
  Stars in the Galaxy}, \araa 43 (2005) 531--580.

\bibitem{Edvardsson:1993}
B.~{Edvardsson}, J.~{Andersen}, B.~{Gustafsson}, D.~L. {Lambert}, P.~E.
  {Nissen}, J.~{Tomkin}: {The Chemical Evolution of the Galactic Disk - Part
  Two - Observational Data}, \aaps 102 (1993) 603.

\bibitem{Fuhrmann:1998}
K.~{Fuhrmann}: {Nearby stars of the Galactic disk and halo}, \aap 338 (1998)
  161--183.

\bibitem{Hayden:2015}
M.~R. {Hayden}, J.~{Bovy}, J.~A. {Holtzman}, D.~L. {Nidever}, J.~C. {Bird},
 {\it et al.}: {Chemical Cartography with APOGEE: Metallicity Distribution
  Functions and the Chemical Structure of the Milky Way Disk}, \apj 808~(2)
  (2015) 132.

\bibitem{Bensby:2017}
T.~{Bensby}, S.~{Feltzing}, A.~{Gould}, J.~C. {Yee}, J.~A. {Johnson},
 {\it et al.}:
  {Chemical evolution of the Galactic bulge as traced by microlensed dwarf and
  subgiant stars. VI. Age and abundance structure of the stellar populations in
  the central sub-kpc of the Milky Way}, \aap 605 (2017) A89.

\bibitem{Zoccali:2018}
M.~{Zoccali}, E.~{Valenti}, O.~A. {Gonzalez}: {Weighing the two stellar
  components of the Galactic bulge}, \aap 618 (2018) A147.

\bibitem{Barklem:2016}
P.~S. {Barklem}, {Accurate abundance analysis of late-type stars: advances in
  atomic physics}, \aapr 24~(1) (2016) 9.

\bibitem{Bergemann:2019}
M.~{Bergemann}, A.~J. {Gallagher}, P.~{Eitner}, M.~{Bautista}, R.~{Collet},
{\it et al.}:
{Observational constraints on the origin of the
  elements. I. 3D NLTE formation of Mn lines in late-type stars}, \aap 631
  (2019) A80.

\bibitem{Melendez:2009}
J.~{Mel{\'e}ndez}, M.~{Asplund}, B.~{Gustafsson}, D.~{Yong}, {The Peculiar
  Solar Composition and Its Possible Relation to Planet Formation}, \apjl
  704~(1) (2009) L66--L70.

\bibitem{Amarsi:2019}
A.~M. {Amarsi}, P.~E. {Nissen}, {\'A}.~{Sk{\'u}lad{\'o}ttir}, {Carbon, oxygen,
  and iron abundances in disk and halo stars. Implications of 3D non-LTE
  spectral line formation}, \aap 630 (2019) A104.

\bibitem{Deal:2020}
M.~{Deal}, M.~J. {Goupil}, J.~P. {Marques}, D.~R. {Reese}, Y.~{Lebreton},
  {Chemical mixing in low mass stars. I. Rotation against atomic diffusion
  including radiative acceleration}, \aap 633 (2020) A23.

\bibitem{Booth:2020}
R.~A. {Booth}, J.~E. {Owen}, {Fingerprints of giant planets in the composition
  of solar twins}, \mnras 493~(4) (2020) 5079--5088.

\bibitem{Ness:2015}
M.~{Ness}, D.~W. {Hogg}, H.~W. {Rix}, A.~Y.~Q. {Ho}, G.~{Zasowski}, {The
  Cannon: A data-driven approach to Stellar Label Determination}, \apj 808~(1)
  (2015) 16.

\bibitem{Ting:2019}
Y.-S. {Ting}, C.~{Conroy}, H.-W. {Rix}, P.~{Cargile}, {The Payne:
  Self-consistent ab initio Fitting of Stellar Spectra}, \apj 879~(2) (2019)
  69.

\bibitem{Fabbro:2018}
S.~{Fabbro}, K.~A. {Venn}, T.~{O'Briain}, S.~{Bialek}, C.~L. {Kielty},
  F.~{Jahandar}, S.~{Monty}, {An application of deep learning in the analysis
  of stellar spectra}, \mnras 475~(3) (2018) 2978--2993.

\bibitem{Ness:2018}
M.~{Ness}, {The Data-Driven Approach to Spectroscopic Analyses}, \pasa 35
  (2018) e003.

\bibitem{Buder:2018}
S.~{Buder}, M.~{Asplund}, L.~{Duong}, J.~{Kos}, K.~{Lind}, {\it et al.}, {Galah
  Collaboration}, {The GALAH Survey: second data release}, \mnras 478~(4)
  (2018) 4513--4552.

\bibitem{Zinner:1998a}
E.~{Zinner}, {Leonard Award Address--Trends in the study of presolar dust
  grains from primitive meteorites}, \maps 33~(4) (1998) 549--564.

\bibitem{Nittler:2016}
L.~R. {Nittler}, F.~{Ciesla}, {Astrophysics with Extraterrestrial Materials},
  \araa 54 (2016) 53--93.

\bibitem{Zinner:2008}
E.~{Zinner}, {Stardust in the Laboratory}, \pasa 25 (2008) 7--17.

\bibitem{Westphal:2014a}
A.~J. {Westphal}, R.~M. {Stroud}, H.~A. {Bechtel}, F.~E. {Brenker}, A.~L.
  {Butterworth}, {\it et al.},
  {Evidence for interstellar origin of seven dust particles collected by the
  Stardust spacecraft}, Science 345~(6198) (2014) 786--791.

\bibitem{Zinner:1998}
E.~{Zinner}, {Stellar Nucleosynthesis and the Isotopic Composition of Presolar
  Grains from Primitive Meteorites}, Annual Review of Earth and Planetary
  Sciences 26 (1998) 147--188.

\bibitem{Nittler:1996}
L.~R. {Nittler}, S.~{Amari}, E.~{Zinner}, S.~E. {Woosley}, R.~S. {Lewis},
  {Extinct 44Ti in Presolar Graphite and SiC: Proof of a Supernova Origin},
  \apjl 462 (1996) L31.

\bibitem{Jose:2004}
J.~{Jos{\'e}}, M.~{Hernanz}, S.~{Amari}, K.~{Lodders}, E.~{Zinner}, {The
  Imprint of Nova Nucleosynthesis in Presolar Grains}, \apj 612 (2004)
  414--428.

\bibitem{Hoppe:2019}
P.~{Hoppe}, R.~J. {Stancliffe}, M.~{Pignatari}, S.~{Amari}, {Isotopic
  Signatures of Supernova Nucleosynthesis in Presolar Silicon Carbide Grains of
  Type AB with Supersolar $^{14}$N/$^{15}$N Ratios}, \apj 887~(1) (2019) 8.

\bibitem{hynes09}
K.~M. {Hynes}, F.~{Gyngard}, {The Presolar Grain Database:
  http://presolar.wustl.edu/\~{}pgd}, in: Lunar and Planetary Science
  Conference, Vol.~40 of Lunar and Planetary Science Conference (2009) 1198.

\bibitem{stephan20}
T.~{Stephan}, M.~{Bose}, A.~{Boujibar}, A.~M. {Davis}, C.~J. {Dory},
 {\it et al.}, {The Presolar Grain Database Reloaded - Silicon
  Carbide}, in: 51st Annual Lunar and Planetary Science Conference, Lunar and
  Planetary Science Conference (2020) 2140.

\bibitem{hoppe97z}
P.~{Hoppe}, P.~{Annen}, R.~{Strebel}, P.~{Eberhardt}, R.~{Gallino},
  M.~{Lugaro}, S.~{Amari}, R.~S. {Lewis}, {Meteoritic Silicon Carbide Grains
  with Unusual Si Isotopic Compositions: Evidence for an Origin in Low-Mass,
  Low-Metallicity Asymptotic Giant Branch Stars}, \apjl 487~(1) (1997)
  L101--L104.

\bibitem{lugaro99}
M.~{Lugaro}, E.~{Zinner}, R.~{Gallino}, S.~{Amari}, {Si Isotopic Ratios in
  Mainstream Presolar SIC Grains Revisited} 527 (1999) 369--394.

\bibitem{amari01AB}
S.~{Amari}, L.~R. {Nittler}, E.~{Zinner}, K.~{Lodders}, R.~S. {Lewis},
  {Presolar SiC Grains of Type A and B: Their Isotopic Compositions and Stellar
  Origins} 559 (2001) 463--483.

\bibitem{liu17JAB}
N.~{Liu}, T.~{Stephan}, P.~{Boehnke}, L.~R. {Nittler}, C.~M. {O'D. Alexander},
  J.~{Wang}, A.~M. {Davis}, R.~{Trappitsch}, M.~J. {Pellin}, {J-type Carbon
  Stars: A Dominant Source of $^{14}$N-rich Presolar SiC Grains of Type AB},
  \apjl 844~(1) (2017) L12.

\bibitem{liu17SNAB}
N.~{Liu}, L.~R. {Nittler}, M.~{Pignatari}, C.~M. {O'D. Alexander}, J.~{Wang},
  {Stellar Origin of $^{15}$N-rich Presolar SiC Grains of Type AB: Supernovae
  with Explosive Hydrogen Burning}, \apjl 842~(1) (2017) L1.

\bibitem{Pignatari:2013a}
M.~{Pignatari}, M.~{Wiescher}, F.~X. {Timmes}, R.~J. {de Boer}, F.-K.
  {Thielemann}, C.~{Fryer}, A.~{Heger}, F.~{Herwig}, R.~{Hirschi}, {Production
  of Carbon-rich Presolar Grains from Massive Stars}, \apjl 767 (2013) L22.

\bibitem{liu18SN}
N.~{Liu}, L.~R. {Nittler}, C.~M.~O.~D. {Alexander}, J.~{Wang}, {Late formation
  of silicon carbide in type II supernovae}, {\it Science Advances} 4~(1)
  (2018) eaao1054.

\bibitem{amari01Y}
S.~{Amari}, L.~R. {Nittler}, E.~{Zinner}, R.~{Gallino}, M.~{Lugaro}, R.~S.
  {Lewis}, {Presolar SIC Grains of Type Y: Origin from Low-Metallicity
  Asymptotic Giant Branch Stars} 546 (2001) 248--266.

\bibitem{zinner06}
E.~{Zinner}, L.~R. {Nittler}, R.~{Gallino}, A.~I. {Karakas}, M.~{Lugaro},
  O.~{Straniero}, J.~C. {Lattanzio}, {Silicon and Carbon Isotopic Ratios in AGB
  Stars: SiC Grain Data, Models, and the Galactic Evolution of the Si
  Isotopes}, \apj 650 (2006) 350--373.

\bibitem{lewis13}
K.~M. {Lewis}, M.~{Lugaro}, B.~K. {Gibson}, K.~{Pilkington}, {Decoding the
  message from meteoritic stardust silicon carbide grains}, \apjl 768 (2013)
  L19.

\bibitem{liu19Mo}
N.~{Liu}, T.~{Stephan}, S.~{Cristallo}, R.~{Gallino}, P.~{Boehnke}, L.~R.
  {Nittler}, C.~M. {O'D. Alexander}, A.~M. {Davis}, R.~{Trappitsch}, M.~J.
  {Pellin}, I.~{Dillmann}, {Presolar Silicon Carbide Grains of Types Y and Z:
  Their Molybdenum Isotopic Compositions and Stellar Origins}, \apj 881~(1)
  (2019) 28.

\bibitem{jose16}
J.~{Jos{\'e}}, G.~M. {Halabi}, M.~F. {El Eid}, {Synthesis of C-rich dust in CO
  nova outbursts}, \aap 593 (2016) A54.

\bibitem{hoppe95}
P.~{Hoppe}, S.~{Amari}, E.~{Zinner}, R.~S. {Lewis}, {Isotopic compositions of
  C, N, O, Mg, and Si, trace element abundances, and morphologies of single
  circumstellar graphite grains in four density fractions from the Murchison
  meteorite}, \gca 59~(19) (1995) 4029--4056.

\bibitem{travaglio99}
C.~{Travaglio}, R.~{Gallino}, S.~{Amari}, E.~{Zinner}, S.~{Woosley}, R.~S.
  {Lewis}, {Low-Density Graphite Grains and Mixing in Type II Supernovae}, \apj
  510 (1999) 325--354.

\bibitem{Amari:2014}
S.~{Amari}, E.~{Zinner}, R.~{Gallino}, {Presolar graphite from the Murchison
  meteorite: An isotopic study}, \gca 133 (2014) 479--522.

\bibitem{nguyen07}
A.~N. {Nguyen}, F.~J. {Stadermann}, E.~{Zinner}, R.~M. {Stroud}, C.~M.~O.
  {Alexander}, L.~R. {Nittler}, {Characterization of Presolar Silicate and
  Oxide Grains in Primitive Carbonaceous Chondrites}, \apj 656~(2) (2007)
  1223--1240.

\bibitem{Karakas:2014}
A.~I. {Karakas}, J.~C. {Lattanzio}, {The Dawes Review 2: Nucleosynthesis and
  Stellar Yields of Low- and Intermediate-Mass Single Stars}, \pasa 31 (2014)
  e030.

\bibitem{lugaro03grains}
M.~{Lugaro}, A.~M. {Davis}, R.~{Gallino}, M.~J. {Pellin}, O.~{Straniero},
  F.~{K{\"a}ppeler}, {Isotopic Compositions of Strontium, Zirconium,
  Molybdenum, and Barium in Single Presolar SiC Grains and Asymptotic Giant
  Branch Stars} 593 (2003) 486--508.

\bibitem{lugaro14Zr}
M.~{Lugaro}, G.~{Tagliente}, A.~I. {Karakas}, P.~M. {Milazzo},
  F.~{K{\"a}ppeler}, A.~M. {Davis}, M.~R. {Savina}, {The impact of updated Zr
  neutron-capture cross sections and new asymptotic giant branch models on our
  understanding of the s process and the origin of stardust} 780 (2014) 95.

\bibitem{lugaro18grains}
M.~{Lugaro}, A.~I. {Karakas}, M.~{Pet{\H{o}}}, E.~{Plachy}, {Do meteoritic
  silicon carbide grains originate from asymptotic giant branch stars of
  super-solar metallicity?}, \gca 221 (2018) 6--20.

\bibitem{Liu:2018}
N.~{Liu}, R.~{Gallino}, S.~{Cristallo}, S.~{Bisterzo}, A.~M. {Davis},
  R.~{Trappitsch}, L.~R. {Nittler}, {New Constraints on the Major Neutron
  Source in Low-mass AGB Stars}, \apj 865~(2) (2018) 112.

\bibitem{lugaro20}
M.~{Lugaro}, B.~{Cseh}, B.~{Vil{\'a}gos}, A.~I. {Karakas}, P.~{Ventura},
  {\it et al.},
  {Origin of Large Meteoritic SiC Stardust Grains in Metal-rich AGB
  Stars}, \apj 898~(2) (2020) 96.

\bibitem{dauphas04}
N.~{Dauphas}, A.~M. {Davis}, B.~{Marty}, L.~{Reisberg}, {The cosmic
  molybdenum-ruthenium isotope correlation}, {\it Earth Planet. Sci. Lett.}
  226~(3-4) (2004) 465--475.

\bibitem{ek20}
M.~{Ek}, A.~C. {Hunt}, M.~{Lugaro}, M.~{Sch{\"o}nb{\"a}chler}, {The origin of
  s-process isotope heterogeneity in the solar protoplanetary disk}, {\it
  Nature Astronomy} 4 (2020) 273--281.

\bibitem{lugaro17o17}
M.~{Lugaro}, A.~I. {Karakas}, C.~G. {Bruno}, M.~{Aliotta}, L.~R. {Nittler},
 {\it et al.},
 {Origin of meteoritic
  stardust unveiled by a revised proton-capture rate of $^{17}$O}, {\it Nature
  Astronomy} 1 (2017) 0027.

\bibitem{Palmerini:2021}
S.~{Palmerini}, S.~{Cristallo}, L.~{Piersanti}, D.~{Vescovi}, M.~{Busso},
  {Group II Oxide Grains: How Massive Are Their AGB Star Progenitors?},
  Universe 7~(6) (2021) 175.

\bibitem{Heger:2002}
A.~{Heger}, S.~E. {Woosley}, {The Nucleosynthetic Signature of Population III},
  \apj 567 (2002) 532--543.

\bibitem{Hoppe:2017}
P.~{Hoppe}, {Stardust from Supernovae and Its Isotopes}, in: A.~W. {Alsabti},
  P.~{Murdin} (Eds.), Handbook of Supernovae (2017) 2473.

\bibitem{Bowman:2021}
D.~M. {Bowman}, {Asteroseismology of massive stars: new insights of stellar
  interiors from their pulsations}, in: OBA Stars: Variability and Magnetic
  Fields (2021) 27.

\bibitem{Fellay:2021}
L.~{Fellay}, G.~{Buldgen}, P.~{Eggenberger}, S.~{Khan}, S.~J.~A.~J. {Salmon},
  A.~{Miglio}, J.~{Montalb{\'a}n}, {Asteroseismology of evolved stars to
  constrain the internal transport of angular momentum. IV. Internal rotation
  of Kepler-56 from an MCMC analysis of the rotational splittings}, \aap 654
  (2021) A133.

\bibitem{Yoshida:2019}
T.~{Yoshida}, K.~{Takahashi}, H.~{Umeda}, K.~{Ishidoshiro}, {Neutrinos from
  Presupernova Stars}, in: Nuclei in the Cosmos XV, Vol. 219 (2019)
  157--161.

\bibitem{Wang:2020}
W.~{Wang}, T.~{Siegert}, Z.~G. {Dai}, R.~{Diehl}, J.~{Greiner}, A.~{Heger},
  M.~{Krause}, M.~{Lang}, M.~M.~M. {Pleintinger}, X.~L. {Zhang}, {Gamma-Ray
  Emission of $^{60}$Fe and $^{26}$Al Radioactivity in Our Galaxy}, \apj
  889~(2) (2020) 169.

\bibitem{Knie:2004}
K.~{Knie}, G.~{Korschinek}, T.~{Faestermann}, E.~A. {Dorfi}, G.~{Rugel},
  A.~{Wallner}: {$^{60}$Fe Anomaly in a Deep-Sea Manganese Crust and
  Implications for a Nearby Supernova Source}, Physical Review Letters 93~(17)
  (2004) 171103.

\bibitem{Wallner:2016}
A.~{Wallner}, J.~{Feige}, N.~{Kinoshita}, M.~{Paul}, L.~K. {Fifield},
 {\it et al.}:
  {Recent near-Earth supernovae probed by global deposition of interstellar
  radioactive $^{60}$Fe}, \nat 532 (2016) 69--72.

\bibitem{Feige:2018}
J.~{Feige}, A.~{Wallner}, R.~{Altmeyer}, L.~K. {Fifield}, R.~{Golser},
  S.~{Merchel}, G.~{Rugel}, P.~{Steier}, S.~G. {Tims}, S.~R. {Winkler}: {Limits
  on Supernova-Associated <mml:mmultiscripts>Fe 60
  </mml:mmultiscripts>/<mml:mmultiscripts>Al 26 </mml:mmultiscripts>
  Nucleosynthesis Ratios from Accelerator Mass Spectrometry Measurements of
  Deep-Sea Sediments}, \prl 121~(22) (2018) 221103.

\bibitem{Arnett:1996}
D.~{Arnett}: {2D Simulations of Supernovae}, in: IAU Colloq. 145: Supernovae
  and Supernova Remnants (1996) 91.

\bibitem{Thielemann:1990}
F.~{Thielemann}, M.~{Hashimoto}, K.~{Nomoto}, {Explosive nucleosynthesis in SN
  1987A. II - Composition, radioactivities, and the neutron star mass}, \apj
  349 (1990) 222--240.

\bibitem{Metzger:2010}
B.~D. {Metzger}, G.~{Mart{\'{\i}}nez-Pinedo}, S.~{Darbha}, E.~{Quataert},
  A.~{Arcones}, D.~{Kasen}, R.~{Thomas}, P.~{Nugent}, I.~V. {Panov}, N.~T.
  {Zinner}, {Electromagnetic counterparts of compact object mergers powered by
  the radioactive decay of r-process nuclei}, \mnras 406 (2010) 2650--2662.

\bibitem{Surman:2006}
R.~{Surman}, G.~C. {McLaughlin}, W.~R. {Hix}, {Nucleosynthesis in the Outflow
  from Gamma-Ray Burst Accretion Disks}, \apj 643 (2006) 1057--1064.

\bibitem{Bildsten:2000}
L.~{Bildsten}, {Theory and observations of Type I X-Ray bursts from neutron
  stars}, in: S.~S. {Holt}, W.~W. {Zhang} (Eds.), American Institute of Physics
  Conference Series, Vol. 522 of American Institute of Physics Conference
  Series (2000) 359--369.

\bibitem{Arnett:1996a}
D.~{Arnett}, {Supernovae and nucleosynthesis. An investigation of the history
  of matter, from the Big Bang to the present}, Princeton series in
  astrophysics, Princeton University Press, New Jersey, USA (1996).

\bibitem{Branch:2017}
D.~{Branch}, J.~C. {Wheeler}, {Supernova Explosions}, Springer Astronomy and Astrophysics Library (2017).

\bibitem{Alsabti:2017}
A.~W. {Alsabti}, P.~{Murdin}, {Handbook of Supernovae} (2017).

\bibitem{McCray:1993}
R.~{McCray}, {Supernova 1987A revisited.}, \araa 31 (1993) 175--216.

\bibitem{Olsen:2021}
J.~{Olsen}, Y.-Z. {Qian}, {Comparison of simulated neutrino emission models
  with data on Supernova 1987A}, \prd 104~(12) (2021) 123020.
  
\bibitem{Leising:1990}
M.~D. {Leising}, G.~H. {Share}, {The gamma-ray light curves of SN 1987A}, \apj
  357 (1990) 638--648.

\bibitem{Tueller:1990}
J.~{Tueller}, S.~{Barthelmy}, N.~{Gehrels}, B.~J. {Teegarden}, M.~{Leventhal},
  C.~J. {MacCallum}, {Observations of gamma-ray line profiles from SN 1987A},
  \apjl 351 (1990) L41--L44.

\bibitem{Varani:1990}
G.~F. {Varani}, W.~P.~S. {Meikle}, J.~{Spyromilio}, D.~A. {Allen}, {Direct
  observation of radioactive cobalt decay in supernova 1987A.}, \mnras 245
  (1990) 570.

\bibitem{Grebenev:2012}
S.~A. {Grebenev}, A.~A. {Lutovinov}, S.~{Tsygankov}, C.~{Winkler}, Hard-x-ray
  emission lines from the decay of $^{44}$ti in the remnant of supernova 1987a,
  Nature, Vol. 490, 7420, (2012) 373-375.

\bibitem{Boggs:2015}
S.~E. {Boggs}, F.~A. {Harrison}, H.~{Miyasaka}, B.~W. {Grefenstette},
  A.~{Zoglauer}, {\it et al.},
  {$^{44}$Ti gamma-ray emission lines from SN1987A reveal an asymmetric
  explosion}, Science 348 (2015) 670--671.

\bibitem{Seitenzahl:2014}
I.~R. {Seitenzahl}, F.~X. {Timmes}, G.~{Magkotsios}, {The Light Curve of SN
  1987A Revisited: Constraining Production Masses of Radioactive Nuclides},
  \apj 792~(1) (2014) 10.

\bibitem{Jerkstrand:2020}
A.~{Jerkstrand}, A.~{Wongwathanarat}, H.~T. {Janka}, M.~{Gabler}, D.~{Alp},
 {\it et al.},
  {Properties of gamma-ray decay lines in 3D core-collapse supernova models,
  with application to SN 1987A and Cas A}, \mnras 494~(2) (2020) 2471--2497.

\bibitem{Bouchet:1991}
L.~{Bouchet}, P.~{Mandrou}, J.~P. {Roques}, G.~{Vedrenne}, B.~{Cordier},
  A.~{Goldwurm}, F.~{Lebrun}, J.~{Paul}, R.~{Sunyaev}, E.~{Churazov},
  M.~{Gilfanov}, M.~{Pavlinsky}, S.~{Grebenev}, G.~{Babalyan}, I.~{Dekhanov},
  N.~{Khavenson}, {Sigma discovery of variable e(+)-e(-) annihilation radiation
  from the near Galactic center variable compact source 1E 1740.7 - 2942},
  \apjl 383 (1991) L45--L48.

\bibitem{Wooden:1993}
D.~H. {Wooden}, D.~M. {Rank}, J.~D. {Bregman}, F.~C. {Witteborn}, A.~G.~G.~M.
  {Tielens}, M.~{Cohen}, P.~A. {Pinto}, T.~S. {Axelrod}, {Airborne
  Spectrophotometry of SN 1987A from 1.7 to 12.6 Microns: Time History of the
  Dust Continuum and Line Emission}, \apjs 88 (1993) 477.

\bibitem{Lucy:1991}
L.~B. {Lucy}, I.~J. {Danziger}, C.~{Gouiffes}, P.~{Bouchet}, {Dust Condensation
  in the Ejecta of Supernova 1987A - Part Two}, in: S.~E. {Woosley} (Ed.),
  {Supernovae}, (1991) 82.

\bibitem{Jerkstrand:2011}
A.~{Jerkstrand}, C.~{Fransson}, C.~{Kozma}, {The $^{44}$Ti-powered spectrum of
  SN 1987A}, \aap 530 (2011) A45.

\bibitem{Morris:2007}
T.~{Morris}, P.~{Podsiadlowski}, {The Triple-Ring Nebula Around SN 1987A:
  Fingerprint of a Binary Merger}, Science 315~(5815) (2007) 1103.

\bibitem{Podsiadlowski:2017}
P.~{Podsiadlowski}, {The Progenitor of SN 1987A}, in: A.~W. {Alsabti},
  P.~{Murdin} (Eds.), Handbook of Supernovae (2017) 635.

\bibitem{Leibundgut:2003}
B.~{Leibundgut}, N.~B. {Suntzeff}, {Optical Light Curves of Supernovae}, in:
  K.~{Weiler} (Ed.), Supernovae and Gamma-Ray Bursters, Vol. 598 (2003)
  77--90.

\bibitem{Kangas:2022}
T.~{Kangas}, C.~{Fransson}, J.~{Larsson}, K.~{France}, R.~A. {Chevalier}, R.~P.
  {Kirshner}, P.~{Lundqvist}, S.~{Mattila}, J.~{Sollerman}, V.~P. {Utrobin},
  {The morphology of the ejecta of SN 1987A at 31 years from 1150 to 10000
  {\r{A}}}, \mnras  511, 2 (2022) 2977. 

\bibitem{Larsson:2011}
J.~{Larsson}, C.~{Fransson}, G.~{{\"O}stlin}, P.~{Gr{\"o}ningsson},
  A.~{Jerkstrand}, {\it et al.},
  {X-ray illumination
  of the ejecta of supernova 1987A}, \nat 474 (2011) 484--486.

\bibitem{Larsson:2016}
J.~{Larsson}, C.~{Fransson}, J.~{Spyromilio}, B.~{Leibundgut}, P.~{Challis},
 {\it et al.},
 {Three-dimensional Distribution
  of Ejecta in Supernova 1987A at 10,000 Days}, \apj 833~(2) (2016) 147.

\bibitem{Grefenstette:2014}
B.~W. {Grefenstette}, F.~A. {Harrison}, S.~E. {Boggs}, S.~P. {Reynolds}, C.~L.
  {Fryer}, {\it et al.}, {Asymmetries in core-collapse supernovae from
  maps of radioactive $^{44}$Ti in CassiopeiaA}, \nat 506 (2014) 339--342.

\bibitem{Fesen:2006}
R.~A. {Fesen}, M.~C. {Hammell}, J.~{Morse}, R.~A. {Chevalier}, K.~J.
  {Borkowski}, M.~A. {Dopita}, C.~L. {Gerardy}, S.~S. {Lawrence}, J.~C.
  {Raymond}, S.~{van den Bergh}, {The Expansion Asymmetry and Age of the
  Cassiopeia A Supernova Remnant}, \apj 645~(1) (2006) 283--292.

\bibitem{Iyudin:1994}
A.~F. {Iyudin}, R.~{Diehl}, H.~{Bloemen}, W.~{Hermsen}, G.~G. {Lichti},
  {\it et al.}, {COMPTEL observations of Ti-44 gamma-ray line
  emission from CAS A}, \aap 284 (1994) L1--L4.

\bibitem{The:1996}
L.-S. {The}, M.~D. {Leising}, J.~D. {Kurfess}, W.~N. {Johnson}, D.~H.
  {Hartmann}, N.~{Gehrels}, J.~E. {Grove}, W.~R. {Purcell}, {CGRO/OSSE
  observations of the Cassiopeia A SNR.}, \aaps 120 (1996) C357.

\bibitem{Rothschild:1999}
R.~E. {Rothschild}, R.~E. {Lingenfelter}, P.~R. {Blanco}, D.~E. {Gruber}, W.~A.
  {Heindl}, {\it et al.},
  {RXTE observations of Cas A}, Nuclear Physics B Proceedings Supplements 69
  (1999) 68--73.

\bibitem{Vink:2000}
J.~{Vink}, J.~S. {Kaastra}, J.~A.~M. {Bleeker}, H.~{Bloemen}, {The Hard X-Ray
  Emission and 44ti Emission Of Cas A}, Advances in Space Research 25 (2000)
  689--694.

\bibitem{Siegert:2015}
T.~{Siegert}, R.~{Diehl}, M.~G.~H. {Krause}, J.~{Greiner}, {Revisiting
  INTEGRAL/SPI observations of $^{44}$Ti from Cassiopeia A}, \aap 579 (2015)
  A124.

\bibitem{Harrison:2013}
F.~A. {Harrison}, W.~W. {Craig}, F.~E. {Christensen}, C.~J. {Hailey}, W.~W.
  {Zhang}, {\it et al.},
  {The Nuclear Spectroscopic Telescope Array (NuSTAR) High-energy
  X-Ray Mission}, \apj 770 (2013) 103.

\bibitem{Muller:2019}
B.~{M{\"u}ller}, {Neutrino Emission as Diagnostics of Core-Collapse
  Supernovae}, Annual Review of Nuclear and Particle Science 69 (2019)
  253--278.

\bibitem{Logue:2012}
J.~{Logue}, C.~D. {Ott}, I.~S. {Heng}, P.~{Kalmus}, J.~H.~C. {Scargill},
  {Inferring core-collapse supernova physics with gravitational waves}, \prd
  86~(4) (2012) 044023.

\bibitem{Bisnovatyi-Kogan:1973}
G.~S. {Bisnovatyi-Kogan}, A.~A. {Ruzmaikin}, R.~A. {Syunyaev}, {Star
  Contraction and Magnetic-Field Generation in Protogalaxies.}, \sovast 17
  (1973) 137.

\bibitem{Bisnovatyi-Kogan:2018}
G.~S. {Bisnovatyi-Kogan}, S.~G. {Moiseenko}, N.~V. {Ardelyan},
  {Magnetorotational Mechanism of the Explosion of Core-Collapse Supernovae},
  Physics of Atomic Nuclei 81~(2) (2018) 266--278.

\bibitem{Wallner:2021}
A.~{Wallner}, M.~B. {Froehlich}, M.~A.~C. {Hotchkis}, N.~{Kinoshita},
  M.~{Paul}, {\it et al.}, {$^{60}$Fe and
  $^{244}$Pu deposited on Earth constrain the r-process yields of recent nearby
  supernovae}, Science 372~(6543) (2021) 742--745.

\bibitem{Thielemann:2020}
F.-K. {Thielemann}, B.~{Wehmeyer}, M.-R. {Wu}, {r-Process Sites, their Ejecta
  Composition, and their Imprint in Galactic Chemical Evolution}, in: Journal
  of Physics Conference Series, Vol. 1668 of Journal of Physics Conference
  Series (2020) 012044.

\bibitem{Paul:2001}
M.~{Paul}, A.~{Valenta}, I.~{Ahmad}, D.~{Berkovits}, C.~{Bordeanu},
  S.~{Ghelberg},{\it et al.}, {Experimental Limit to Interstellar $^{244}$Pu Abundance},
  \apjl 558~(2) (2001) L133--L135.

\bibitem{Wallner:2004}
C.~{Wallner}, T.~{Faestermann}, U.~{Gerstmann}, K.~{Knie}, G.~{Korschinek},
  C.~{Lierse}, G.~{Rugel}, {Supernova produced and anthropogenic $^{244}$Pu in
  deep sea manganese encrustations}, \nar 48~(1-4) (2004) 145--150.

\bibitem{Paul:2009}
M.~{Paul}, {Counting Rare Atoms for Nuclear Astrophysics}, Acta Physica
  Polonica B 40~(3) (2009) 685.

\bibitem{Raisbeck:2007}
G.~{Raisbeck}, T.~{Tran}, D.~{Lunney}, C.~{Gaillard}, S.~{Goriely},
  C.~{Waelbroeck}, F.~{Yiou}, {A search for supernova produced $^{244}$Pu in a
  marine sediment}, Nuclear Instruments and Methods in Physics Research B
  259~(1) (2007) 673--676.

\bibitem{Wallner:2015}
A.~{Wallner}, T.~{Faestermann}, J.~{Feige}, C.~{Feldstein}, K.~{Knie},
  {\it et al.}, {Abundance of live $^{244}$Pu in deep-sea reservoirs
  on Earth points to rarity of actinide nucleosynthesis}, Nature Communications
  6 (2015) 5956.

\bibitem{Hotokezaka:2015}
K.~{Hotokezaka}, T.~{Piran}, M.~{Paul}, {Short-lived $^{244}$Pu points to
  compact binary mergers as sites for heavy r-process nucleosynthesis}, Nature
  Physics 11 (2015) 1042.

\bibitem{Seitenzahl:2017}
I.~R. {Seitenzahl}, D.~M. {Townsley}, {Nucleosynthesis in Thermonuclear
  Supernovae}, Springer: Berlin, Heidelberg (2017) 1955.

\bibitem{Jha:2019}
S.~W. {Jha}, K.~{Maguire}, M.~{Sullivan}, {Observational properties of
  thermonuclear supernovae}, Nature Astronomy 3 (2019) 706--716.

\bibitem{Maguire:2017}
K.~{Maguire}, {Type Ia Supernovae}, in: A.~W. {Alsabti}, P.~{Murdin} (Eds.),
  Handbook of Supernovae (2017) 293.

\bibitem{Jha:2017}
S.~W. {Jha}, {Type Iax Supernovae}, in: A.~W. {Alsabti}, P.~{Murdin} (Eds.),
  Handbook of Supernovae (2017) 375.

\bibitem{Taubenberger:2017}
S.~{Taubenberger}, {The Extremes of Thermonuclear Supernovae} (2017) 317.

\bibitem{Colgate:1969}
S.~A. {Colgate}, C.~{McKee}, {Early Supernova Luminosity}, \apj 157 (1969) 623.

\bibitem{Clayton:1974}
D.~D. {Clayton}, S.~E. {Woosley}, {Thermonuclear astrophysics}, Reviews of
  Modern Physics 46 (1974) 755--771.

\bibitem{Nadyozhin:1994}
D.~K. {Nadyozhin}, {The Properties of NI CO Fe Decay}, \apjs 92 (1994) 527.

\bibitem{Arnett:1982}
W.~D. {Arnett}, {Type I supernovae. I - Analytic solutions for the early part
  of the light curve}, \apj 253 (1982) 785--797.

\bibitem{Brown:2015}
P.~J. {Brown}, P.~W.~A. {Roming}, P.~A. {Milne}, {The first ten years of Swift
  supernovae}, Journal of High Energy Astrophysics 7 (2015) 111--116.

\bibitem{Kuchner:1994}
M.~J. {Kuchner}, R.~P. {Kirshner}, P.~A. {Pinto}, B.~{Leibundgut}, {Evidence
  for 56Ni 56Co 56Fe Decay in Type IA Supernovae}, \apjl 426 (1994) L89.

\bibitem{Contardo:2000}
G.~{Contardo}, B.~{Leibundgut}, W.~D. {Vacca}, {Epochs of maximum light and
  bolometric light curves of type Ia supernovae}, \aap 359 (2000) 876--886.

\bibitem{Jeffery:1999}
D.~J. {Jeffery}, {Radioactive Decay Energy Deposition in Supernovae and the
  Exponential/Quasi-Exponential Behavior of Late-Time Supernova Light Curves},
 arXiv e-prints (1999) astro--ph/9907015

\bibitem{Stritzinger:2006}
M.~{Stritzinger}, B.~{Leibundgut}, S.~{Walch}, G.~{Contardo}, {Constraints on
  the progenitor systems of type Ia supernovae}, \aap 450 (2006) 241--251.

\bibitem{Scalzo:2014a}
R.~A. {Scalzo}, A.~J. {Ruiter}, S.~A. {Sim}, {The ejected mass distribution of
  Type Ia supernovae: a significant rate of non-Chandrasekhar-mass
  progenitors}, \mnras 445~(3) (2014) 2535--2544.

\bibitem{Kasen:2006}
D.~{Kasen}, {Secondary Maximum in the Near-Infrared Light Curves of Type Ia
  Supernovae}, \apj 649~(2) (2006) 939--953.

\bibitem{Dhawan:2016}
S.~{Dhawan}, B.~{Leibundgut}, J.~{Spyromilio}, S.~{Blondin}, {A reddening-free
  method to estimate the $^{56}$Ni mass of Type Ia supernovae}, \aap 588 (2016)
  A84.

\bibitem{Churazov:2014}
E.~{Churazov}, R.~{Sunyaev}, J.~{Isern}, J.~{Kn{\"o}dlseder}, P.~{Jean},
{\it et al.}:
  {Cobalt-56 {$\gamma$}-ray emission lines from the type Ia
  supernova 2014J}, \nat 512 (2014) 406--408.

\bibitem{Diehl:2015}
R.~{Diehl}, T.~{Siegert}, W.~{Hillebrandt}, M.~{Krause}, J.~{Greiner},
  K.~{Maeda}, F.~K. {R{\"o}pke}, S.~A. {Sim}, W.~{Wang}, X.~{Zhang}, {SN2014J
  gamma rays from the $^{56}$Ni decay chain}, \aap 574 (2015) A72.

\bibitem{Isern:2016}
J.~{Isern}, P.~{Jean}, E.~{Bravo}, J.~{Kn{\"o}dlseder}, F.~{Lebrun},
  {\it et al.}, {Gamma-ray emission from SN2014J near
  maximum optical light}, \aap 588 (2016) A67.

\bibitem{The:2014}
L.-S. {The}, A.~{Burrows}, {Expectations for the Hard X-Ray Continuum and
  Gamma-Ray Line Fluxes from the Type Ia Supernova SN 2014J in M82}, \apj 786
  (2014) 141.

\bibitem{Diehl:2014}
R.~{Diehl}, T.~{Siegert}, W.~{Hillebrandt}, S.~A. {Grebenev}, J.~{Greiner},
  M.~{Krause}, M.~{Kromer}, K.~{Maeda}, F.~{R{\"o}pke}, S.~{Taubenberger}:
  {Early $^{56}$Ni decay gamma rays from SN2014J suggest an unusual explosion},
  Science 345~(6201) (2014) 1162--1165.

\bibitem{Brachwitz:2000}
F.~{Brachwitz}, D.~J. {Dean}, W.~R. {Hix}, K.~{Iwamoto}, K.~{Langanke},
  G.~{Mart{\'\i}nez-Pinedo}, K.~{Nomoto}, M.~R. {Strayer}, F.-K. {Thielemann},
  H.~{Umeda}: {The Role of Electron Captures in Chandrasekhar-Mass Models for
  Type IA Supernovae}, \apj 536~(2) (2000) 934--947.

\bibitem{Dhawan:2018}
S.~{Dhawan}, A.~{Fl{\"o}rs}, B.~{Leibundgut}, K.~{Maguire}, W.~{Kerzendorf},
  S.~{Taubenberger}, M.~H. {Van Kerkwijk}, J.~{Spyromilio}: {Nebular
  spectroscopy of SN 2014J: Detection of stable nickel in near-infrared
  spectra}, \aap 619 (2018) A102.

\bibitem{Floers:2018}
A.~{Fl{\"o}rs}, J.~{Spyromilio}, K.~{Maguire}, S.~{Taubenberger}, W.~E.
  {Kerzendorf}, S.~{Dhawan}: {Limits on stable iron in Type Ia supernovae from
  near-infrared spectroscopy}, \aap 620 (2018) A200.

\bibitem{Floers:2020}
A.~{Fl{\"o}rs}, J.~{Spyromilio}, S.~{Taubenberger}, S.~{Blondin}, R.~{Cartier},
  B.~{Leibundgut}, L.~{Dessart}, S.~{Dhawan}, W.~{Hillebrandt},
  {Sub-Chandrasekhar progenitors favoured for Type Ia supernovae: evidence from
  late-time spectroscopy}, \mnras 491~(2) (2020) 2902--2918.

\bibitem{Noebauer:2017}
U.~M. {Noebauer}, M.~{Kromer}, S.~{Taubenberger}, P.~{Baklanov},
  S.~{Blinnikov}, E.~{Sorokina}, W.~{Hillebrandt}, {Early light curves for Type
  Ia supernova explosion models}, \mnras 472~(3) (2017) 2787--2799.

\bibitem{Reynolds:2008}
S.~P. {Reynolds}: {Supernova remnants at high energy.}, \araa 46 (2008)
  89--126.

\bibitem{Reynolds:2017}
S.~P. {Reynolds}, {Dynamical Evolution and Radiative Processes of Supernova
  Remnants}, in: A.~W. {Alsabti}, P.~{Murdin} (Eds.), Handbook of Supernovae
  (2017) 1981.

\bibitem{Seitenzahl:2019}
I.~R. {Seitenzahl}, P.~{Ghavamian}, J.~M. {Laming}, F.~P.~A. {Vogt}, {Optical
  Tomography of Chemical Elements Synthesized in Type Ia Supernovae}, \prl
  123~(4) (2019) 041101.

\bibitem{Vink:2012}
J.~{Vink}, {Supernova remnants: the X-ray perspective}, \aapr 20 (2012) 49.

\bibitem{Seitenzahl:2013}
I.~R. {Seitenzahl}, G.~{Cescutti}, F.~K. {R{\"o}pke}, A.~J. {Ruiter},
  R.~{Pakmor}, {Solar abundance of manganese: a case for near
  Chandrasekhar-mass Type Ia supernova progenitors}, \aap 559 (2013) L5.

\bibitem{Seitenzahl:2015}
I.~R. {Seitenzahl}, A.~{Summa}, F.~{Krau{\ss}}, S.~A. {Sim}, R.~{Diehl},
 {\it et al.},
  {5.9-keV Mn K-shell X-ray luminosity from the decay of $^{55}$Fe in Type Ia
  supernova models}, \mnras 447~(2) (2015) 1484--1490.

\bibitem{Mori:2018}
K.~{Mori}, M.~A. {Famiano}, T.~{Kajino}, T.~{Suzuki}, P.~M. {Garnavich}, G.~J.
  {Mathews}, R.~{Diehl}, S.-C. {Leung}, K.~{Nomoto}, {Nucleosynthesis
  Constraints on the Explosion Mechanism for Type Ia Supernovae}, \apj 863~(2)
  (2018) 176.

\bibitem{Lach:2020}
F.~{Lach}, F.~K. {R{\"o}pke}, I.~R. {Seitenzahl}, B.~{Cot{\'e}}, S.~{Gronow},
  A.~J. {Ruiter}, {Nucleosynthesis imprints from different Type Ia supernova
  explosion scenarios and implications for galactic chemical evolution}, \aap
  644 (2020) A118.

\bibitem{Gronow:2021a}
S.~{Gronow}, B.~{C{\^o}t{\'e}}, F.~{Lach}, I.~R. {Seitenzahl}, C.~E. {Collins},
  S.~A. {Sim}, F.~K. {R{\"o}pke}, {Metallicity-dependent nucleosynthetic yields
  of Type Ia supernovae originating from double detonations of sub-M$_{Ch}$
  white dwarfs}, \aap 656 (2021) A94.

\bibitem{Patnaude:2015}
D.~J. {Patnaude}, S.-H. {Lee}, P.~O. {Slane}, C.~{Badenes}, A.~{Heger}, D.~C.
  {Ellison}, S.~{Nagataki}, {Are Models for Core-collapse Supernova Progenitors
  Consistent with the Properties of Supernova Remnants?}, \apj 803~(2) (2015)
  101.

\bibitem{Weinberger:2021}
C.~{Weinberger}, Supernova diagnostics from gamma-ray lines in the young
  remnant phase, Ph.D. thesis, TU Munich (2021).

\bibitem{Yamaguchi:2014}
H.~{Yamaguchi}, C.~{Badenes}, R.~{Petre}, T.~{Nakano}, D.~{Castro}, {\it et al.},
{Discriminating the Progenitor
  Type of Supernova Remnants with Iron K-shell Emission}, \apjl 785~(2) (2014)
  L27.

\bibitem{Della-Valle:2020}
M.~{Della Valle}, L.~{Izzo}, {Observations of galactic and extragalactic
  novae}, \aapr 28~(1) (2020) 3.

\bibitem{Chomiuk:2021}
L.~{Chomiuk}, B.~D. {Metzger}, K.~J. {Shen}, {New Insights into Classical
  Novae}, \araa 59 (2021) 391.

\bibitem{Gehrz:1998}
R.~D. {Gehrz}, J.~W. {Truran}, R.~E. {Williams}, S.~{Starrfield},
  {Nucleosynthesis in Classical Novae and Its Contribution to the Interstellar
  Medium}, \pasp 110 (1998) 3--26.

\bibitem{Jose:1998}
J.~{Jose}, M.~{Hernanz}, {Nucleosynthesis in Classical Novae: CO versus ONe
  White Dwarfs}, \apj 494 (1998) 680.

\bibitem{Hernanz:2002}
M.~{Hernanz}, J.~{G{\'o}mez-Gomar}, J.~{Jos{\'e}}, {The prompt gamma-ray
  emission of novae}, New Astronomy Review 46 (2002) 559--563.

\bibitem{Jose:2007}
J.~{Jos{\'e}}, M.~{Hernanz}, {TOPICAL REVIEW: Nucleosynthesis in classical nova
  explosions}, Journal of Physics G Nuclear Physics 34~(12) (2007) R431--R458.

\bibitem{Molaro:2016}
P.~{Molaro}, L.~{Izzo}, E.~{Mason}, P.~{Bonifacio}, M.~{Della Valle}, {Highly
  enriched $^{7}$Be in the ejecta of Nova Sagittarii 2015 No. 2 (V5668 Sgr) and
  the Galactic $^{7}$Li origin}, \mnras 463 (2016) L117--L121.

\bibitem{Molaro:2022}
P.~{Molaro}, L.~{Izzo}, V.~{D'Odorico}, E.~{Aydi}, P.~{Bonifacio},
  G.~{Cescutti}, E.~J. {Harvey}, M.~{Hernanz}, P.~{Selvelli}, M.~{della Valle},
  {$^{7}$Be in the outburst of the ONe nova V6595 Sgr}, \mnras 509~(3) (2022)
  3258--3267.

\bibitem{Siegert:2018}
T.~{Siegert}, A.~{Coc}, L.~{Delgado}, R.~{Diehl}, J.~{Greiner}, {\it et al.},
{Gamma-ray
  observations of Nova Sgr 2015 No. 2 with INTEGRAL}, \aap 615 (2018) A107.

\bibitem{Siegert:2021}
T.~{Siegert}, S.~{Ghosh}, K.~{Mathur}, E.~{Spraggon}, A.~{Yeddanapudi},
  {Nucleosynthesis constraints through {\ensuremath{\gamma}}-ray line
  measurements from classical novae. Hierarchical model for the ejecta of
  $^{22}$Na and $^{7}$Be}, \aap 650 (2021) A187.

\bibitem{Jose:2016b}
J.~{Jos{\'e}}, G.~M. {Halabi}, M.~F. {El Eid}, {Synthesis of C-rich dust in CO
  nova outbursts}, \aap 593 (2016) A54.

\bibitem{Gal-Yam:2019}
A.~{Gal-Yam}, {The Most Luminous Supernovae}, \araa 57 (2019) 305--333.

\bibitem{Nomoto:2010a}
K.~{Nomoto}, T.~{Moriya}, N.~{Tominaga}, {Explosive Nucleosynthesis in
  Supernovae and Hypernovae}, in: {H.~Susa, M.~Arnould, S.~Gales,
  T.~Motobayashi, C.~Scheidenberger, \& H.~Utsunomiya} (Ed.), American
  Institute of Physics Conference Series, Vol. 1238 of American Institute of
  Physics Conference Series (2010) 9--17.

\bibitem{Abbott:2017}
B.~P. {Abbott}, R.~{Abbott}, T.~D. {Abbott}, F.~{Acernese}, K.~{Ackley},
  {\it et al.},
  {Multi-messenger Observations of a Binary Neutron Star Merger}, \apjl 848
  (2017) L12.

\bibitem{Woosley:2010}
S.~E. {Woosley}, {Bright Supernovae from Magnetar Birth}, \apjl 719~(2) (2010)
  L204--L207.

\bibitem{Mukhopadhyay:2000}
B.~{Mukhopadhyay}, S.~K. {Chakrabarti}, {Nucleosynthesis in accretion flows
  around black holes}, \aap 353 (2000) 1029--1043.

\bibitem{Pruet:2003}
J.~{Pruet}, S.~E. {Woosley}, R.~D. {Hoffman}, {Nucleosynthesis in Gamma-Ray
  Burst Accretion Disks}, \apj 586~(2) (2003) 1254--1261.

\bibitem{Surman:2004}
R.~{Surman}, G.~C. {McLaughlin}, {Neutrinos and Nucleosynthesis in Gamma-Ray
  Burst Accretion Disks}, \apj 603~(2) (2004) 611--623.

\bibitem{Wanajo:2012}
S.~{Wanajo}, H.-T. {Janka}, {The r-process in the Neutrino-driven Wind from a
  Black-hole Torus}, \apj 746~(2) (2012) 180.

\bibitem{Janiuk:2014}
A.~{Janiuk}, {Nucleosynthesis of elements in gamma-ray burst engines}, \aap 568
  (2014) A105.

\bibitem{Sollerman:2000a}
J.~{Sollerman}, C.~{Kozma}, C.~{Fransson}, B.~{Leibundgut}, P.~{Lundqvist},
  F.~{Ryde}, P.~{Woudt}, {SN 1998bw at Late Phases}, \apjl 537~(2) (2000)
  L127--L130.

\bibitem{Woosley:1999}
S.~E. {Woosley}, R.~G. {Eastman}, B.~P. {Schmidt}, {Gamma-Ray Bursts and Type
  IC Supernova SN 1998BW}, \apj 516~(2) (1999) 788--796.

\bibitem{Surman:2014}
R.~{Surman}, O.~L. {Caballero}, G.~C. {McLaughlin}, O.~{Just}, H.~T. {Janka},
  {Production of $^{56}$Ni in black hole-neutron star merger accretion disc
  outflows}, Journal of Physics G Nuclear Physics 41~(4) (2014) 044006.

\bibitem{Woosley:2006b}
S.~E. {Woosley}, J.~S. {Bloom}, {The Supernova Gamma-Ray Burst Connection},
  \araa 44~(1) (2006) 507--556.

\bibitem{Nomoto:2010}
K.~{Nomoto}, M.~{Tanaka}, N.~{Tominaga}, K.~{Maeda}, {Hypernovae, gamma-ray
  bursts, and first stars}, \nar 54 (2010) 191--200.

\bibitem{Pian:2017}
E.~{Pian}, P.~{D'Avanzo}, S.~{Benetti}, M.~{Branchesi}, E.~{Brocato},
{\it et al.}, {Spectroscopic identification of r-process nucleosynthesis in a
  double neutron-star merger}, \nat 551~(7678) (2017) 67--70.

\bibitem{Rosswog:1999}
S.~{Rosswog}, M.~{Liebend{\"o}rfer}, F.-K. {Thielemann}, M.~B. {Davies},
  W.~{Benz}, T.~{Piran}, {Mass ejection in neutron star mergers}, \aap 341
  (1999) 499--526.

\bibitem{Chruslinska:2018}
M.~{Chruslinska}, K.~{Belczynski}, J.~{Klencki}, M.~{Benacquista}, {Double
  neutron stars: merger rates revisited}, \mnras 474~(3) (2018) 2937--2958.

\bibitem{Woosley:2007}
S.~E. {Woosley}, S.~{Blinnikov}, A.~{Heger}, {Pulsational pair instability as
  an explanation for the most luminous supernovae}, \nat 450 (2007) 390--392.

\bibitem{Arcavi:2017}
I.~{Arcavi}, D.~A. {Howell}, D.~{Kasen}, L.~{Bildsten}, G.~{Hosseinzadeh},
 {\it et al.}, {Energetic eruptions leading to a peculiar
  hydrogen-rich explosion of a massive star}, \nat 551~(7679) (2017) 210--213.

\bibitem{Woosley:2018}
S.~E. {Woosley}, {Models for the Unusual Supernova iPTF14hls}, \apj 863~(1)
  (2018) 105.

\bibitem{Sollerman:2019}
J.~{Sollerman}, F.~{Taddia}, I.~{Arcavi}, C.~{Fremling}, C.~{Fransson},
 {\it et al.}, {Late-time observations of the extraordinary Type II
  supernova iPTF14hls}, \aap 621 (2019) A30.

\bibitem{Gomez:2019}
S.~{Gomez}, E.~{Berger}, M.~{Nicholl}, P.~K. {Blanchard}, V.~A. {Villar},
  L.~{Patton}, R.~{Chornock}, J.~{Leja}, G.~{Hosseinzadeh}, P.~S.
  {Cowperthwaite}, {SN 2016iet: The Pulsational or Pair Instability Explosion
  of a Low-metallicity Massive CO Core Embedded in a Dense Hydrogen-poor
  Circumstellar Medium}, \apj 881~(2) (2019) 87.

\bibitem{Watson:2019}
D.~{Watson}, C.~J. {Hansen}, J.~{Selsing}, A.~{Koch}, D.~B. {Malesani}, 
{\it et al.}:
  {Identification of strontium in the merger
  of two neutron stars}, \nat 574~(7779) (2019) 497--500.

\bibitem{Kaminski:2018}
T.~{Kami{\'n}ski}, R.~{Tylenda}, K.~M. {Menten}, A.~{Karakas}, J.~M. {Winters},
  A.~A. {Breier}, K.~T. {Wong}, T.~F. {Giesen}, N.~A. {Patel}: {Astronomical
  detection of radioactive molecule $^{26}$AlF in the remnant of an ancient
  explosion}, Nature Astronomy 2 (2018) 778--783.

\bibitem{Franco:2021}
M.~{Franco}, K.~E.~K. {Coppin}, J.~E. {Geach}, C.~{Kobayashi}, S.~C. {Chapman},
 {\it et al.}, {The ramp-up of interstellar medium enrichment at z > 4},
  Nature Astronomy 5 (2021) 1240--1246.

\bibitem{Iben:1984}
I.~{Iben}, Jr., A.~V. {Tutukov}, {Supernovae of type I as end products of the
  evolution of binaries with components of moderate initial mass (M not greater
  than about 9 solar masses)}, \apjs 54 (1984) 335--372.

\bibitem{Maoz:2014}
D.~{Maoz}, F.~{Mannucci}, G.~{Nelemans}, {Observational Clues to the
  Progenitors of Type Ia Supernovae}, \araa 52 (2014) 107--170.

\bibitem{Prantzos:2018}
N.~{Prantzos}, C.~{Abia}, M.~{Limongi}, A.~{Chieffi}, S.~{Cristallo}, {Chemical
  evolution with rotating massive star yields - I. The solar neighbourhood and
  the s-process elements}, \mnras 476~(3) (2018) 3432--3459.

\bibitem{Fuhrmann:2004}
K.~{Fuhrmann}, {Nearby stars of the Galactic disk and halo. III.},
  Astronomische Nachrichten 325~(1) (2004) 3--80.

\bibitem{Reddy:2006}
B.~E. {Reddy}, D.~L. {Lambert}, C.~{Allende Prieto}, {Elemental abundance
  survey of the Galactic thick disc}, \mnras 367~(4) (2006) 1329--1366.

\bibitem{Adibekyan:2013}
V.~Z. {Adibekyan}, P.~{Figueira}, N.~C. {Santos}, A.~A. {Hakobyan}, S.~G.
  {Sousa}, 
  {\it et al.}:
 {Kinematics and chemical properties of the
  Galactic stellar populations. The HARPS FGK dwarfs sample}, \aap 554 (2013)
  A44.

\bibitem{Bensby:2014}
T.~{Bensby}, S.~{Feltzing}, M.~S. {Oey}, {Exploring the Milky Way stellar disk.
  A detailed elemental abundance study of 714 F and G dwarf stars in the solar
  neighbourhood}, \aap 562 (2014) A71.

\bibitem{Gilmore:1989}
G.~{Gilmore}, R.~F.~G. {Wyse}, K.~{Kuijken}, {Kinematics, chemistry, and
  structure of the Galaxy.}, \araa 27 (1989) 555--627.

\bibitem{Tinsley:1980}
B.~M. {Tinsley}, {Evolution of the Stars and Gas in Galaxies}, Fundamentals of
  Cosmic Physics 5 (1980) 287--388.

\bibitem{Magrini:2018}
L.~{Magrini}, F.~{Vincenzo}, S.~{Randich}, E.~{Pancino}, G.~{Casali},
 {\it et al.},
 {The Gaia-ESO Survey: The N/O abundance ratio in the Milky Way},
  \aap 618 (2018) A102.

\bibitem{Tinsley:1972}
B.~M. {Tinsley}, {Galactic Evolution}, \aap 20 (1972) 383.

\bibitem{Tinsley:1974}
B.~M. {Tinsley}, {Constraints on models for chemical evolution in the solar
  neighborhood}, \apj 192 (1974) 629--641.

\bibitem{Matteucci:1989a}
F.~{Matteucci}, P.~{Francois}, {Galactic chemical evolution - Abundance
  gradients of individual elements}, \mnras 239 (1989) 885--904.

\bibitem{Boissier:1999}
S.~{Boissier}, N.~{Prantzos}, {Chemo-Spectral Evolution of the Milky way and of
  Spiral Disks}, \apss 265 (1999) 409--410.

\bibitem{Chiappini:2003}
C.~{Chiappini}, F.~{Matteucci}, G.~{Meynet}, {Stellar yields with rotation and
  their effect on chemical evolution models}, \aap 410 (2003) 257--267.

\bibitem{Gibson:2003}
B.~K. {Gibson}, Y.~{Fenner}, A.~{Renda}, D.~{Kawata}, H.-c. {Lee}, {Galactic
  Chemical Evolution}, \pasa 20~(4) (2003) 401--415.

\bibitem{Kobayashi:2011}
C.~{Kobayashi}, N.~{Nakasato}, {Chemodynamical Simulations of the Milky Way
  Galaxy}, \apj 729 (2011) 16.

\bibitem{Cote:2019a}
B.~{C{\^o}t{\'e}}, M.~{Lugaro}, R.~{Reifarth}, M.~{Pignatari},
  B.~{Vil{\'a}gos}, A.~{Yag{\"u}e}, B.~K. {Gibson}, {Galactic Chemical
  Evolution of Radioactive Isotopes}, \apj 878~(2) (2019) 156.

\bibitem{Kobayashi:2020a}
C.~{Kobayashi}, A.~I. {Karakas}, M.~{Lugaro}, {The Origin of Elements from
  Carbon to Uranium}, \apj 900~(2) (2020) 179.

\bibitem{Minchev:2013}
I.~{Minchev}, C.~{Chiappini}, M.~{Martig}, {Chemodynamical evolution of the
  Milky Way disk. I. The solar vicinity}, \aap 558 (2013) A9.

\bibitem{Izzard:2013}
R.~G. {Izzard}, {Chemical Evolution of Binary Stars}, in: K.~{Pavlovski},
  A.~{Tkachenko}, G.~{Torres} (Eds.), EAS Publications Series, Vol.~64 of EAS
  Publications Series (2013) 13--20.

\bibitem{Mennekens:2013}
N.~{Mennekens}, D.~{Vanbeveren}, J.~P. {De Greve}, {The effect of intermediate
  mass close binaries on the chemical evolution of globular clusters}, \memsai
  84 (2013) 153.

\bibitem{Mandel:2016}
I.~{Mandel}, S.~E. {de Mink}, {Merging binary black holes formed through
  chemically homogeneous evolution in short-period stellar binaries}, \mnras
  458~(3) (2016) 2634--2647.

\bibitem{Song:2016}
H.~F. {Song}, G.~{Meynet}, A.~{Maeder}, S.~{Ekstr{\"o}m}, P.~{Eggenberger},
  {Massive star evolution in close binaries. Conditions for homogeneous
  chemical evolution}, \aap 585 (2016) A120.

\bibitem{Babusiaux:2018}
{Gaia Collaboration}, C.~{Babusiaux}, F.~{van Leeuwen}, M.~A. {Barstow},
  C.~{Jordi}, A.~{Vallenari}, {\it et al.},
  {Gaia Data Release 2. Observational Hertzsprung-Russell
  diagrams}, \aap 616 (2018) A10.

\bibitem{Vincenzo:2019}
F.~{Vincenzo}, E.~{Spitoni}, F.~{Calura}, F.~{Matteucci}, V.~{Silva Aguirre},
  A.~{Miglio}, G.~{Cescutti}, {The Fall of a Giant. Chemical evolution of
  Enceladus, alias the Gaia Sausage}, \mnras 487~(1) (2019) L47--L52.

\bibitem{Ibata:2021}
R.~{Ibata}, K.~{Malhan}, N.~{Martin}, D.~{Aubert}, B.~{Famaey}, {\it et al.}, {Charting the Galactic Acceleration Field. I. A
  Search for Stellar Streams with Gaia DR2 and EDR3 with Follow-up from
  ESPaDOnS and UVES}, \apj 914~(2) (2021) 123.

\bibitem{Helmi:2018}
A.~{Helmi}, C.~{Babusiaux}, H.~H. {Koppelman}, D.~{Massari}, J.~{Veljanoski},
  A.~G.~A. {Brown}, {The merger that led to the formation of the Milky Way's
  inner stellar halo and thick disk}, \nat 563~(7729) (2018) 85--88.

\bibitem{Gratton:2019}
R.~{Gratton}, A.~{Bragaglia}, E.~{Carretta}, V.~{D'Orazi}, S.~{Lucatello},
  A.~{Sollima}, {What is a globular cluster? An observational perspective},
  \aapr 27~(1) (2019) 8.

\bibitem{Milone:2017}
A.~P. {Milone}, G.~{Piotto}, A.~{Renzini}, A.~F. {Marino}, L.~R. {Bedin},
 {\it et al.}, {The Hubble Space Telescope UV Legacy Survey of Galactic globular
  clusters - IX. The Atlas of multiple stellar populations}, \mnras 464~(3)
  (2017) 3636--3656.

\bibitem{Gratton:2001}
R.~G. {Gratton}, P.~{Bonifacio}, A.~{Bragaglia}, E.~{Carretta},
  V.~{Castellani}, {\it et al.}, {The O-Na and Mg-Al anticorrelations in turn-off and early
  subgiants in globular clusters}, \aap 369 (2001) 87--98.

\bibitem{Bastian:2018}
N.~{Bastian}, C.~{Lardo}, {Multiple Stellar Populations in Globular Clusters},
  \araa 56 (2018) 83--136.

\bibitem{Yong:2003}
D.~{Yong}, F.~{Grundahl}, D.~L. {Lambert}, P.~E. {Nissen}, M.~D. {Shetrone},
  {Mg isotopic ratios in giant stars of the globular cluster NGC 6752}, \aap
  402 (2003) 985--1001.

\bibitem{Spitoni:2019}
E.~{Spitoni}, V.~{Silva Aguirre}, F.~{Matteucci}, F.~{Calura}, V.~{Grisoni},
  {Galactic Archaeology with asteroseismic ages: Evidence for delayed gas
  infall in the formation of the Milky Way disc}, \aap 623 (2019) A60.

\bibitem{Spitoni:2020}
E.~{Spitoni}, K.~{Verma}, V.~{Silva Aguirre}, F.~{Calura}, {Galactic
  archaeology with asteroseismic ages. II. Confirmation of a delayed gas infall
  using Bayesian analysis based on MCMC methods}, \aap 635 (2020) A58.

\bibitem{Silva-Aguirre:2018}
V.~{Silva Aguirre}, M.~{Bojsen-Hansen}, D.~{Slumstrup}, L.~{Casagrande},
  D.~{Kawata}, {\it et al.}, {Confirming chemical clocks: asteroseismic age
  dissection of the Milky Way disc(s)}, \mnras 475~(4) (2018) 5487--5500.

\bibitem{Crowther:2007}
P.~A. {Crowther}, {Physical Properties of Wolf-Rayet Stars}, \araa 45 (2007)
  177--219.

\bibitem{Garcia-Hernandez:2016}
D.~A. {Garc{\'\i}a-Hern{\'a}ndez}, P.~{Ventura}, G.~{Delgado-Inglada},
  F.~{Dell'Agli}, M.~{Di Criscienzo}, A.~{Yag{\"u}e}, {Galactic planetary
  nebulae with precise nebular abundances as a tool to understand the evolution
  of asymptotic giant branch stars}, \mnras 461~(1) (2016) 542--551.

\bibitem{Stanghellini:2018}
L.~{Stanghellini}, M.~{Haywood}, {Galactic Planetary Nebulae as Probes of
  Radial Metallicity Gradients and Other Abundance Patterns}, \apj 862~(1)
  (2018) 45.

\bibitem{Gesicki:2003}
K.~{Gesicki}, A.~{Acker}, A.~A. {Zijlstra}, {Kinematics, turbulence and
  evolution of planetary nebulae}, \aap 400 (2003) 957--969.

\bibitem{Kingsburgh:1994}
R.~L. {Kingsburgh}, M.~J. {Barlow}, {Elemental abundances for a sample of
  southern galactic planetary nebulae.}, \mnras 271 (1994) 257--299.

\bibitem{Garcia-Rojas:2022}
J.~{Garc{\'\i}a-Rojas}, C.~{Morisset}, D.~{Jones}, R.~{Wesson}, H.~M.~J.
  {Boffin}, H.~{Monteiro}, R.~L.~M. {Corradi}, P.~{Rodr{\'\i}guez-Gil}, {MUSE
  spectroscopy of planetary nebulae with high abundance discrepancies}, \mnras
  510~(4) (2022) 5444--5463.

\bibitem{Tinsley:1975}
B.~M. {Tinsley}, {Nucleochronology and Chemical Evolution}, \apj 198 (1975)
  145--150.

\bibitem{Meyer:2000}
B.~S. {Meyer}, J.~W. {Truran}, {Nucleocosmochronology}, \physrep 333 (2000)
  1--11.

\bibitem{Cowan:2006}
J.~J. {Cowan}, C.~{Sneden}, {Heavy element synthesis in the oldest stars and
  the early Universe}, \nat 440~(7088) (2006) 1151--1156.

\bibitem{Thielemann:2002b}
F.~{Thielemann}, P.~{Hauser}, E.~{Kolbe}, G.~{Martinez-Pinedo}, I.~{Panov},
  T.~{Rauscher}, K.~{Kratz}, B.~{Pfeiffer}, S.~{Rosswog},
  M.~{Liebend{\"o}rfer}, A.~{Mezzacappa}, {Heavy Elements and Age
  Determinations}, Space Science Reviews 100 (2002) 277--296.

\bibitem{nissen20}
P.~E. {Nissen}, J.~{Christensen-Dalsgaard}, J.~R. {Mosumgaard}, V.~{Silva
  Aguirre}, E.~{Spitoni}, K.~{Verma}, {High-precision abundances of elements in
  solar-type stars. Evidence of two distinct sequences in abundance-age
  relations}, \aap 640 (2020) A81.

\bibitem{Kulkarni:2007a}
V.~P. {Kulkarni}, {The Chemical Composition of Damped Lyman-alpha Galaxies},
  in: J.~{Afonso}, H.~C. {Ferguson}, B.~{Mobasher}, R.~{Norris} (Eds.), Deepest
  Astronomical Surveys, Vol. 380 of Astronomical Society of the Pacific
  Conference Series (2007) 455.

\bibitem{Peroux:2020}
C.~{P{\'e}roux}, J.~C. {Howk}, {The Cosmic Baryon and Metal Cycles}, \araa 58
  (2020) 363--406.

\bibitem{Tolstoy:2009}
E.~{Tolstoy}, V.~{Hill}, M.~{Tosi}, {Star-Formation Histories, Abundances, and
  Kinematics of Dwarf Galaxies in the Local Group}, \araa 47~(1) (2009)
  371--425.

\bibitem{Chiappini:2003a}
C.~{Chiappini}, D.~{Romano}, F.~{Matteucci}, {CNO evolution: Milky way, dwarf
  galaxies and DLAs}, \apss 284 (2003) 771--774.

\bibitem{Venn:2004}
K.~A. {Venn}, M.~{Irwin}, M.~D. {Shetrone}, C.~A. {Tout}, V.~{Hill},
  E.~{Tolstoy}, {Stellar Chemical Signatures and Hierarchical Galaxy
  Formation}, \aj 128~(3) (2004) 1177--1195.

\bibitem{Venn:2007}
K.~A. {Venn}, {Chemical Abundance Constraints on Galaxy Formation}, in: Island
  Universes, Vol.~3 of Astrophysics and Space Science Proceedings, 2007, p.
  245.

\bibitem{Annibali:2022}
F.~{Annibali}, M.~{Tosi}, {Chemical and stellar properties of star-forming
  dwarf galaxies}, Nature Astronomy 6 (2022) 48--58.

\bibitem{Ji:2016a}
A.~P. {Ji}, A.~{Frebel}, A.~{Chiti}, J.~D. {Simon}, {R-process enrichment from
  a single event in an ancient dwarf galaxy}, \nat 531~(7596) (2016) 610--613.

\bibitem{Ji:2018}
A.~P. {Ji}, A.~{Frebel}, {From Actinides to Zinc: Using the Full Abundance
  Pattern of the Brightest Star in Reticulum II to Distinguish between
  Different r-process Sites}, \apj 856~(2) (2018) 138.

\bibitem{Frebel:2005}
A.~{Frebel}, W.~{Aoki}, N.~{Christlieb}, H.~{Ando}, M.~{Asplund}, P.~S.
 {\it et al.},
 {Nucleosynthetic
  signatures of the first stars}, \nat 434 (2005) 871--873.

\bibitem{Skuladottir:2021}
{\'A}.~{Sk{\'u}lad{\'o}ttir}, S.~{Salvadori}, A.~M. {Amarsi}, E.~{Tolstoy},
  M.~J. {Irwin}, {\it et al.}, {Zero-metallicity Hypernova Uncovered
  by an Ultra-metal-poor Star in the Sculptor Dwarf Spheroidal Galaxy}, \apjl
  915~(2) (2021) L30.

\bibitem{Asplund:2005a}
M.~{Asplund}, {New Light on Stellar Abundance Analyses: Departures from LTE and
  Homogeneity}, \araa 43 (2005) 481--530.

\bibitem{Nordlander:2019}
T.~{Nordlander}, M.~S. {Bessell}, G.~S. {Da Costa}, A.~D. {Mackey},
  M.~{Asplund}, {\it et al.}, {The lowest detected stellar Fe abundance: the halo star SMSS
  J160540.18-144323.1}, \mnras 488~(1) (2019) L109--L113.

\bibitem{Caffau:2017}
E.~{Caffau}, P.~{Bonifacio}, E.~{Starkenburg}, N.~{Martin}, K.~{Youakim}, A.~A.
  {Henden}, {\it et al.}, {The Pristine survey II: A sample of
  bright stars observed with FEROS}, Astronomische Nachrichten 338~(6) (2017)
  686--695.

\bibitem{Vanzella:2020}
E.~{Vanzella}, M.~{Meneghetti}, G.~B. {Caminha}, M.~{Castellano}, F.~{Calura},
  {\it et al.}, {Candidate Population III stellar
  complex at z = 6.629 in the MUSE Deep Lensed Field}, \mnras 494~(1) (2020)
  L81--L85.

\bibitem{Nordlander:2017}
T.~{Nordlander}, A.~M. {Amarsi}, K.~{Lind}, M.~{Asplund}, P.~S. {Barklem},
  A.~R. {Casey}, R.~{Collet}, J.~{Leenaarts}, {3D NLTE analysis of the most
  iron-deficient star, SMSS0313-6708}, \aap 597 (2017) A6.

\bibitem{Yong:2021}
D.~{Yong}, C.~{Kobayashi}, G.~S. {Da Costa}, M.~S. {Bessell}, A.~{Chiti},
  A.~{Frebel}, K.~{Lind}, A.~D. {Mackey}, T.~{Nordlander}, M.~{Asplund}, A.~R.
  {Casey}, A.~F. {Marino}, S.~J. {Murphy}, B.~P. {Schmidt}, {r-Process elements
  from magnetorotational hypernovae}, \nat 595~(7866) (2021) 223--226.

\bibitem{Cote:2019}
B.~{C{\^o}t{\'e}}, M.~{Eichler}, A.~{Arcones}, C.~J. {Hansen}, P.~{Simonetti},
  {\it et al.}, {Neutron Star Mergers Might Not Be the Only Source of
  r-process Elements in the Milky Way}, \apj 875~(2) (2019) 106.

\bibitem{Pluschke:2001c}
S.~{Pl{\"u}schke}, R.~{Diehl}, V.~{Sch{\"o}nfelder}, H.~{Bloemen},
  W.~{Hermsen}, {\it et al.}, {The COMPTEL 1.809 MeV survey}, in:
  A.~{Gimenez}, V.~{Reglero}, C.~{Winkler} (Eds.), Exploring the Gamma-Ray
  Universe, Vol. 459 of ESA Special Publication (2001) 55--58.

\bibitem{Kretschmer:2013}
K.~{Kretschmer}, R.~{Diehl}, M.~{Krause}, A.~{Burkert}, K.~{Fierlinger},
  O.~{Gerhard}, J.~{Greiner}, W.~{Wang}, {Kinematics of massive star ejecta in
  the Milky Way as traced by $^{26}$Al}, \aap 559 (2013) A99.

\bibitem{Naya:1996}
J.~E. {Naya}, S.~D. {Barthelmy}, L.~M. {Bartlett}, N.~{Gehrels},
  M.~{Leventhal}, A.~{Parsons}, B.~J. {Teegarden}, J.~{Tueller}, {Detection of
  high-velocity $^{26}$Al towards the Galactic Centre}, \nat 384 (1996) 44--46.

\bibitem{Diehl:2006d}
R.~{Diehl}, H.~{Halloin}, K.~{Kretschmer}, G.~G. {Lichti},
  V.~{Sch{\"o}nfelder}, {\it et al.},
  {Radioactive $^{26}$Al from
  massive stars in the Galaxy}, \nat 439 (2006) 45--47.

\bibitem{Pleintinger:2020}
M.~M.~M. {Pleintinger}, Star groups and their nucleosynthesis, Ph.D. thesis,
  Technische Universit{\"a}t M{\"u}nchen (2020).

\bibitem{Krumholz:2018}
M.~R. {Krumholz}, B.~{Burkhart}, J.~C. {Forbes}, R.~M. {Crocker}, {A unified
  model for galactic discs: star formation, turbulence driving, and mass
  transport}, \mnras 477~(2) (2018) 2716--2740.

\bibitem{Koo:2020}
B.-C. {Koo}, C.-G. {Kim}, S.~{Park}, E.~C. {Ostriker}, {Radiative Supernova
  Remnants and Supernova Feedback}, \apj 905~(1) (2020) 35.

\bibitem{Diehl:2018}
R.~{Diehl}, T.~{Siegert}, J.~{Greiner}, M.~{Krause}, K.~{Kretschmer},
  M.~{Lang}, M.~{Pleintinger}, A.~W. {Strong}, C.~{Weinberger}, X.~{Zhang},
  {INTEGRAL/SPI {$\gamma$}-ray line spectroscopy. Response and background
  characteristics}, \aap 611 (2018) A12.

\bibitem{Green:2019}
D.~A. {Green}, {A revised catalogue of 294 Galactic supernova remnants},
  Journal of Astrophysics and Astronomy 40~(4) (2019) 36.

\bibitem{Chen:1997}
W.~{Chen}, R.~{Diehl}, N.~{Gehrels}, D.~{Hartmann}, M.~{Leising}, J.~E. {Naya},
  N.~{Prantzos}, J.~{Tueller}, P.~{von Ballmoos}: {Implications of the Broad
  $^{26}$Al 1809 keV Line Observed by GRIS}, in: C.~{Winkler}, T.~J.-L.
  {Courvoisier}, P.~{Durouchoux} (Eds.), The Transparent Universe, Vol. 382 of
  ESA Special Publication (1997) 105.

\bibitem{Sturner:1999}
S.~J. {Sturner}, J.~E. {Naya}, {On the Nature of the High-Velocity $^{26}$Al
  near the Galactic Center}, \apj 526 (1999) 200--206.

\bibitem{Krause:2015}
M.~G.~H. {Krause}, R.~{Diehl}, Y.~{Bagetakos}, E.~{Brinks}, A.~{Burkert},
  O.~{Gerhard}, J.~{Greiner}, K.~{Kretschmer}, T.~{Siegert}: {$^{26}$Al
  kinematics: superbubbles following the spiral arms?. Constraints from the
  statistics of star clusters and HI supershells}, \aap 578 (2015) A113.

\bibitem{Schinnerer:2019}
E.~{Schinnerer}, A.~{Hughes}, A.~{Leroy}, B.~{Groves}, G.~A. {Blanc},
 {\it et al.}, {The Gas-Star Formation Cycle
  in Nearby Star-forming Galaxies. I. Assessment of Multi-scale Variations},
  \apj 887~(1) (2019) 49.

\bibitem{Rodgers-Lee:2019}
D.~{Rodgers-Lee}, M.~G.~H. {Krause}, J.~{Dale}, R.~{Diehl}, {Synthetic
  $^{26}$Al emission from galactic-scale superbubble simulations}, \mnras
  490~(2) (2019) 1894--1912.

\bibitem{Krause:2021}
M.~G.~H. {Krause}, D.~{Rodgers-Lee}, J.~E. {Dale}, R.~{Diehl}, C.~{Kobayashi},
  {Galactic $^{26}$Al traces metal loss through hot chimneys}, \mnras 501~(1)
  (2021) 210--218.

\bibitem{Nath:2020}
B.~B. {Nath}, P.~{Das}, M.~S. {Oey}, {Size distribution of superbubbles},
  \mnras 493~(1) (2020) 1034--1043.

\bibitem{Krause:2020}
M.~G.~H. {Krause}, S.~S.~R. {Offner}, C.~{Charbonnel}, M.~{Gieles}, R.~S.
  {Klessen}, E.~{V{\'a}zquez-Semadeni}, J.~{Ballesteros-Paredes},
  P.~{Girichidis}, J.~M. {Diederik Kruijssen}, J.~L. {Ward}, H.~{Zinnecker},
  {The Physics of Star Cluster Formation and Evolution}, \ssr 216~(4) (2020)
  64.

\bibitem{Chevance:2022}
M.~{Chevance}, J.~M.~D. {Kruijssen}, M.~R. {Krumholz}, B.~{Groves}, B.~W.
  {Keller}, {\it et al.}, {Pre-supernova feedback mechanisms drive
  the destruction of molecular clouds in nearby star-forming disc galaxies},
  \mnras 509~(1) (2022) 272--288.

\bibitem{Burrows:1993}
D.~N. {Burrows}, K.~P. {Singh}, J.~A. {Nousek}, G.~P. {Garmire}, J.~{Good}, {A
  multiwavelength study of the Eridanus soft X-ray enhancement}, \apj 406
  (1993) 97--111.

\bibitem{Heiles:1999}
C.~{Heiles}, L.~M. {Haffner}, R.~J. {Reynolds}, {The Eridanus Superbubble in
  its Multiwavelength Glory}, in: {A.~R.~Taylor, T.~L.~Landecker, \& G.~Joncas}
  (Ed.), New Perspectives on the Interstellar Medium, Vol. 168 of Astronomical
  Society of the Pacific Conference Series (1999) 211.

\bibitem{Siegert:2017a}
T.~{Siegert}, R.~{Diehl}, {The $^{26}$Al Gamma-ray Line from Massive-Star
  Regions}, in: S.~{Kubono}, T.~{Kajino}, S.~{Nishimura}, T.~{Isobe},
  S.~{Nagataki}, T.~{Shima}, Y.~{Takeda} (Eds.), 14th International Symposium
  on Nuclei in the Cosmos (NIC2016) (2017) 020305.

\bibitem{Fierlinger:2016}
K.~M. {Fierlinger}, A.~{Burkert}, E.~{Ntormousi}, P.~{Fierlinger},
  M.~{Schartmann}, A.~{Ballone}, M.~G.~H. {Krause}, R.~{Diehl}, {Stellar
  feedback efficiencies: supernovae versus stellar winds}, \mnras 456 (2016)
  710--730.

\bibitem{Tatischeff:2018b}
V.~{Tatischeff}, S.~{Gabici}, {Particle Acceleration by Supernova Shocks and
  Spallogenic Nucleosynthesis of Light Elements}, Annual Review of Nuclear and
  Particle Science 68~(1) (2018) 377--404.

\bibitem{Berezhko:1996}
E.~G. {Berezhko}, V.~K. {Elshin}, L.~T. {Ksenofontov}, {Cosmic ray acceleration
  in supernova remnants.}, Soviet Journal of Experimental and Theoretical
  Physics 82 (1996) 1--21.

\bibitem{Bykov:2001}
A.~M. {Bykov}, {Particle Acceleration and Nonthermal Phenomena in
  Superbubbles}, \ssr 99 (2001) 317--326.

\bibitem{Ellison:2007}
D.~C. {Ellison}, D.~J. {Patnaude}, P.~{Slane}, P.~{Blasi}, S.~{Gabici},
  {Particle Acceleration in Supernova Remnants and the Production of Thermal
  and Nonthermal Radiation}, \apj 661 (2007) 879--891.

\bibitem{Parizot:2004}
E.~{Parizot}, A.~{Marcowith}, E.~{van der Swaluw}, A.~M. {Bykov},
  V.~{Tatischeff}, {Superbubbles and energetic particles in the Galaxy. I.
  Collective effects of particle acceleration}, \aap 424 (2004) 747--760.

\bibitem{Biermann:2010}
P.~L. {Biermann}, J.~K. {Becker}, J.~{Dreyer}, A.~{Meli}, E.-S. {Seo},
  T.~{Stanev}, {The Origin of Cosmic Rays: Explosions of Massive Stars with
  Magnetic Winds and Their Supernova Mechanism}, \apj 725 (2010) 184--187.

\bibitem{Binns:2001}
W.~R. {Binns}, M.~E. {Wiedenbeck}, E.~R. {Christian}, A.~C. {Cummings}, J.~S.
  {George},
  {\it et al.}:
 {GCR neon isotopic abundances:
  Comparison with wolf-rayet star models and meteoritic abundances}, in: R.~F.
  {Wimmer-Schweingruber} (Ed.), Joint SOHO/ACE workshop ``Solar and Galactic
  Composition'', Vol. 598 of American Institute of Physics Conference Series
  (2001) 257--262.

\bibitem{Wiedenbeck:1999}
M.~E. {Wiedenbeck}, W.~R. {Binns}, E.~R. {Christian}, A.~C. {Cummings}, B.~L.
  {Dougherty}, 
  {\it et al.}:
 {Constraints on the Time Delay between Nucleosynthesis and
  Cosmic-Ray Acceleration from Observations of $^{59}$Ni and $^{59}$Co}, \apjl
  523~(1) (1999) L61--L64.

\bibitem{Biermann:2018}
P.~L. {Biermann}, J.~{Becker Tjus}, W.~{de Boer}, L.~I. {Caramete},
  A.~{Chieffi}, {\it et al.}, {Supernova
  explosions of massive stars and cosmic rays}, Advances in Space Research
  62~(10) (2018) 2773--2816.

\bibitem{Yanasak:2001}
N.~E. {Yanasak}, M.~E. {Wiedenbeck}, R.~A. {Mewaldt}, A.~J. {Davis}, A.~C.
  {Cummings}, 
  {\it et al.}:
  {Measurement of the Secondary Radionuclides $^{10}$Be, $^{26}$Al, $^{36}$Cl,
  $^{54}$Mn, and $^{14}$C and Implications for the Galactic Cosmic-Ray Age},
  \apj 563 (2001) 768--792.

\bibitem{Wiedenbeck:2007}
M.~E. {Wiedenbeck}, W.~R. {Binns}, A.~C. {Cummings}, A.~J. {Davis}, G.~A. {de
  Nolfo}, 
  {\it et al.}:
{An Overview of the Origin of Galactic Cosmic Rays as
  Inferred from Observations of Heavy Ion Composition and Spectra}, \ssr
  130~(1-4) (2007) 415--429.

\bibitem{Binns:2016}
W.~R. {Binns}, M.~H. {Israel}, E.~R. {Christian}, A.~C. {Cummings}, G.~A. {de
  Nolfo}, {\it et al.}, {Observation of the $^{60}$Fe
  nucleosynthesis-clock isotope in galactic cosmic rays}, Science 352 (2016)
  677--680.

\bibitem{Woosley:2007b}
S.~E. {Woosley}, A.~{Heger}, {Nucleosynthesis and remnants in massive stars of
  solar metallicity}, \physrep 442 (2007) 269--283.

\bibitem{Sukhbold:2016}
T.~{Sukhbold}, T.~{Ertl}, S.~E. {Woosley}, J.~M. {Brown}, H.-T. {Janka},
  {Core-collapse Supernovae from 9 to 120 Solar Masses Based on
  Neutrino-powered Explosions}, \apj 821 (2016) 38.

\bibitem{Limongi:2018}
M.~{Limongi}, A.~{Chieffi}, {Presupernova Evolution and Explosive
  Nucleosynthesis of Rotating Massive Stars in the Metallicity Range -3
  {\ensuremath{\leq}} [Fe/H] {\ensuremath{\leq}} 0}, \apjs 237~(1) (2018) 13.

\bibitem{Morlino:2020}
G.~{Morlino}, E.~{Amato}, {Impact of transport modeling on the $^{60}$Fe
  abundance inside Galactic cosmic ray sources}, \prd 101~(8) (2020) 083017.

\bibitem{Morlino:2021}
G.~{Morlino}, P.~{Blasi}, E.~{Peretti}, P.~{Cristofari}, {Particle acceleration
  in winds of star clusters}, \mnras 504~(4) (2021) 6096--6105.

\bibitem{Tatischeff:2018}
V.~{Tatischeff}, A.~{De Angelis}, M.~{Tavani}, I.~{Grenier}, U.~{Oberlack},
 {\it et al.}, {The
  e-ASTROGAM gamma-ray space observatory for the multimessenger astronomy of
  the 2030s}, in: \procspie, Vol. 10699 of Society of Photo-Optical
  Instrumentation Engineers (SPIE) Conference Series (2018) 106992J.

\bibitem{Nath:2020a}
B.~B. {Nath}, D.~{Eichler}, {Diffuse galactic Gamma-rays from star clusters},
  \mnras 499~(1) (2020) L1--L5.

\bibitem{Nittler:2018}
L.~R. {Nittler}, C.~M. {O'D. Alexander}, N.~{Liu}, J.~{Wang}, {Extremely
  $^{54}$Cr- and $^{50}$Ti-rich Presolar Oxide Grains in a Primitive Meteorite:
  Formation in Rare Types of Supernovae and Implications for the Astrophysical
  Context of Solar System Birth}, \apjl 856~(2) (2018) L24.

\bibitem{denhartogh22}
J.~{den Hartogh}, M.~K. {Pet{\"o}}, T.~{Lawson}, A.~{Sieverding},
  H.~{Brinkman}, M.~{Pignatari}, M.~{Lugaro}, {Comparison between core-collapse
  supernova nucleosynthesis and meteoric stardust grains: investigating
  magnesium, aluminium, and chromium}, arXiv e-prints (2022)
  arXiv:2201.04692

\bibitem{stephan21}
T.~{Stephan}, A.~M. {Davis}, {Molybdenum Isotope Dichotomy in Meteorites Caused
  by s-Process Variability}, \apj 909~(1) (2021) 8.

\bibitem{kleine20}
T.~{Kleine}, G.~{Budde}, C.~{Burkhardt}, T.~S. {Kruijer}, E.~A. {Worsham},
  A.~{Morbidelli}, F.~{Nimmo}, {The Non-carbonaceous-Carbonaceous Meteorite
  Dichotomy}, \ssr 216~(4) (2020) 55.

\bibitem{warren11}
P.~H. {Warren}, {Stable isotopes and the noncarbonaceous derivation of
  ureilites, in common with nearly all differentiated planetary materials},
  \gca 75~(22) (2011) 6912--6926.

\bibitem{nanne19}
J.~A.~M. {Nanne}, F.~{Nimmo}, J.~N. {Cuzzi}, T.~{Kleine}, {Origin of the
  non-carbonaceous-carbonaceous meteorite dichotomy}, {\it Earth Planet. Sci.
  Lett.} 511 (2019) 44--54.

\bibitem{brasser20}
R.~{Brasser}, S.~J. {Mojzsis}, {The partitioning of the inner and outer Solar
  System by a structured protoplanetary disk}, {Nature Astronomy} (2020)
  8.

\bibitem{Dauphas:2011}
N.~{Dauphas}, M.~{Chaussidon}, {A Perspective from Extinct Radionuclides on a
  Young Stellar Object: The Sun and Its Accretion Disk}, Annual Review of Earth
  and Planetary Sciences 39 (2011) 351--386.

\bibitem{Lugaro:2018}
M.~{Lugaro}, U.~{Ott}, {\'A}.~{Kereszturi}, {Radioactive nuclei from
  cosmochronology to habitability}, Progress in Particle and Nuclear Physics
  102 (2018) 1--47.

\bibitem{lichtenberg16b}
T.~{Lichtenberg}, G.~J. {Golabek}, T.~V. {Gerya}, M.~R. {Meyer}, {The effects
  of short-lived radionuclides and porosity on the early thermo-mechanical
  evolution of planetesimals}, {\it Icarus} 274 (2016) 350--365.

\bibitem{Cote:2021}
B.~{C{\^o}t{\'e}}, M.~{Eichler}, A.~{Yag{\"u}e L{\'o}pez}, N.~{Vassh}, M.~R.
  {Mumpower}, {\it et al.}, {$^{129}$I and $^{247}$Cm in meteorites constrain
  the last astrophysical source of solar r-process elements}, Science
  371 (6532) (2021) 945--948.

\bibitem{adams10}
F.~C. {Adams}, {The Birth Environment of the Solar System}, \araa 48 (2010)
  47--85.

\bibitem{ciesla15}
F.~J. {Ciesla}, G.~D. {Mulders}, I.~{Pascucci}, D.~{Apai}, {Volatile Delivery
  to Planets from Water-rich Planetesimals around Low Mass Stars}, \apj 804
  (2015) 9.

\bibitem{Lichtenberg:2019}
T.~{Lichtenberg}, G.~J. {Golabek}, R.~{Burn}, M.~{Meyer}, Y.~{Alibert},
  T.~{Gerya}, C.~{Mordasini}, {A water budget dichotomy of rocky protoplanets
  from $^{26}$Al-heating}, in: AAS/Division for Extreme Solar Systems
  Abstracts, Vol.~51 of AAS/Division for Extreme Solar Systems Abstracts (2019) 311.01.

\bibitem{Lugaro:2022}
M.~{Lugaro}, B.~{C{\^o}t{\'e}}, M.~{Pignatari}, A.~{Yag{\"u}e L{\'o}pez},
  H.~{Brinkman}, B.~{Cseh}, J.~{Den Hartogh}, C.~L. {Doherty}, A.~I. {Karakas},
  C.~{Kobayashi}, T.~{Lawson}, M.~{Pet{\H{o}}}, B.~{So{\'o}s}, T.~{Trueman},
  B.~{Vil{\'a}gos}, {The RADIOSTAR Project}, arXiv e-prints (2022)
  arXiv:2202.08144

\bibitem{Fields:1970}
P.~R. {Fields}, H.~{Diamond}, D.~N. {Metta}, C.~M. {Stevens}, D.~J. {Rokop},
  P.~E. {Moreland}, {Isotopic Abundances of Actinide Elements in Lunar
  Material:}, Science 167~(3918) (1970) 499.

\bibitem{Cowan:1972}
G.~{Cowan}, in: Proc. Symp. on Cosmochemistry, MA, USA, (1972) 12.

\bibitem{Sakamoto:1974}
K.~{Sakamoto}, {Possible cosmic dust origin of terrestrial plutonium-244}, \nat
  248~(5444) (1974) 130--132.

\bibitem{Ellis:1996}
J.~{Ellis}, B.~D. {Fields}, D.~N. {Schramm}, {Geological Isotope Anomalies as
  Signatures of Nearby Supernovae}, \apj 470 (1996) 1227.

\bibitem{Fields:2005}
B.~D. {Fields}, K.~A. {Hochmuth}, J.~{Ellis}, {Deep-Ocean Crusts as Telescopes:
  Using Live Radioisotopes to Probe Supernova Nucleosynthesis}, \apj 621 (2005)
  902--907.

\bibitem{Athanassiadou:2011}
T.~{Athanassiadou}, B.~D. {Fields}, {Penetration of nearby supernova dust in
  the inner solar system}, \na 16~(4) (2011) 229--241.

\bibitem{Fry:2016}
B.~J. {Fry}, B.~D. {Fields}, J.~R. {Ellis}, {Radioactive Iron Rain:
  Transporting $^{60}$Fe in Supernova Dust to the Ocean Floor}, \apj 827 (2016)
  48.

\bibitem{Fry:2020}
B.~J. {Fry}, B.~D. {Fields}, J.~R. {Ellis}, {Magnetic Imprisonment of Dusty
  Pinballs by a Supernova Remnant}, \apj 894~(2) (2020) 109.

\bibitem{Altobelli:2003}
N.~{Altobelli}, S.~{Kempf}, M.~{Landgraf}, R.~{Srama}, V.~{Dikarev},
  H.~{Kr{\"u}ger}, G.~{Moragas-Klostermeyer}, E.~{Gr{\"u}n}, {Cassini between
  Venus and Earth: Detection of interstellar dust}, Journal of Geophysical
  Research (Space Physics) 108~(A10) (2003) 8032.

\bibitem{Altobelli:2005}
N.~{Altobelli}, S.~{Kempf}, H.~{Kr{\"u}ger}, M.~{Landgraf}, M.~{Roy},
  E.~{Gr{\"u}n}, {Interstellar dust flux measurements by the Galileo dust
  instrument between the orbits of Venus and Mars}, Journal of Geophysical
  Research (Space Physics) 110~(A7) (2005) A07102.

\bibitem{Grun:1993}
E.~{Gr{\"u}n}, {Dust in the planetary system}, Advances in Space Research
  13~(10) (1993) 139--151.

\bibitem{Frisch:2009}
P.~C. {Frisch}, M.~{Bzowski}, E.~{Gr{\"u}n}, V.~{Izmodenov}, H.~{Kr{\"u}ger},
 {\it et al.}, {The Galactic
  Environment of the Sun: Interstellar Material Inside and Outside of the
  Heliosphere}, Space Science Reviews 146 (2009) 235--273.

\bibitem{Mann:2010}
I.~{Mann}, {Interstellar Dust in the Solar System}, \araa 48 (2010) 173--203.

\bibitem{Korschinek:1996}
G.~{Korschinek}, T.~{Faestermann}, K.~{Knie}, C.~{Schmidt}, 60fe, a promising
  ams isotope for many applications, Radiocarbon 38 (1996) 68.

\bibitem{Grun:1993a}
E.~{Grun}, H.~A. {Zook}, M.~{Baguhl}, A.~{Balogh}, S.~J. {Bame}, {\it et al.}, {Discovery of Jovian dust streams and interstellar
  grains by the Ulysses spacecraft}, \nat 362~(6419) (1993) 428--430.

\bibitem{Segl:1984}
M.~{Segl}, A.~{Mangini}, G.~{Bonani}, H.~J. {Hofmann}, M.~{Nessi}, 
{\it et al.}:
{$^{10}$Be-dating of a manganese crust from Central North Pacific
  and implications for ocean palaeocirculation}, \nat 309~(5968) (1984)
  540--543.

\bibitem{Lachner:2020}
J.~{Lachner}, M.~{Ploner}, P.~{Steier}, A.~{Sakaguchi}, A.~{Usui},
  {Accumulation of ferromanganese crusts derived from carrier-free
  $^{10}$Be/$^{9}$Be}, Nuclear Instruments and Methods in Physics Research B
  467 (2020) 146--151.

\bibitem{Hoffman:1971}
D.~C. {Hoffman}, F.~O. {Lawrence}, J.~L. {Mewherter}, F.~M. {Rourke},
  {Detection of Plutonium-244 in Nature}, \nat 234~(5325) (1971) 132--134.

\bibitem{Knie:1999}
K.~{Knie}, G.~{Korschinek}, T.~{Faestermann}, C.~{Wallner}, J.~{Scholten},
  W.~{Hillebrandt}: {Indication for Supernova Produced $^{60}$Fe Activity on
  Earth}, \prl 83~(1) (1999) 18--21.

\bibitem{Fitoussi:2008}
C.~{Fitoussi}, G.~M. {Raisbeck}, K.~{Knie}, G.~{Korschinek}, T.~{Faestermann},
 {\it et al.},
  {Search for Supernova-Produced Fe60 in a Marine Sediment},
  Physical Review Letters 101~(12) (2008) 121101.

\bibitem{Ludwig:2016}
P.~{Ludwig}, S.~{Bishop}, R.~{Egli}, V.~{Chernenko}, B.~{Deneva},
 {\it et al.},
  {Time-resolved 2-million-year-old supernova activity discovered in Earth's
  microfossil record}, Proceedings of the National Academy of Science 113
  (2016) 9232--9237.

\bibitem{Korschinek:2020}
G.~{Korschinek}, T.~{Faestermann}, M.~{Poutivtsev}, A.~{Arazi}, K.~{Knie},
  G.~{Rugel}, A.~{Wallner}, {Supernova-Produced $^{53}$Mn on Earth}, \prl
  125~(3) (2020) 031101.

\bibitem{Wallner:2020}
A.~{Wallner}, J.~{Feige}, L.~K. {Fifield}, M.~B. {Froehlich}, R.~{Golser},
  {\it et al.}, {60Fe deposition
  during the late Pleistocene and the Holocene echoes past supernova activity},
  Proceedings of the National Academy of Science 117~(36) (2020) 21873--21879.

\bibitem{Koll:2020}
D.~{Koll}, T.~{Faestermann}, G.~{Korschinek}, A.~{Wallner}, {Origin of Recent
  Interstellar $^{60}$Fe on Earth}, in: European Physical Journal Web of
  Conferences, Vol. 232 of European Physical Journal Web of Conferences (2020)
  02001.

\bibitem{Fimiani:2016}
L.~{Fimiani}, D.~L. {Cook}, T.~{Faestermann}, J.~M. {G{\'o}mez-Guzm{\'a}n},
  K.~{Hain}, G.~{Herzog}, K.~{Knie}, G.~{Korschinek}, P.~{Ludwig}, J.~{Park},
  R.~C. {Reedy}, G.~{Rugel}, {Interstellar $^{60}$Fe on the Surface of the
  Moon}, Physical Review Letters 116~(15) (2016) 151104.

\bibitem{Koll:2019}
D.~{Koll}, G.~{Korschinek}, T.~{Faestermann}, J.~M. {G{\'o}mez-Guzm{\'a}n},
  S.~{Kipfstuhl}, S.~{Merchel}, J.~M. {Welch}, {Interstellar $^{60}$Fe in
  Antarctica}, \prl 123~(7) (2019) 072701.

\bibitem{Frisch:2017}
P.~{Frisch}, V.~V. {Dwarkadas}, {Effect of Supernovae on the Local Interstellar
  Material}, in: A.~W. {Alsabti}, P.~{Murdin} (Eds.), Handbook of Supernovae
  (2017) 2253.

\bibitem{Cox:1982}
D.~P. {Cox}, P.~R. {Anderson}, {Extended adiabatic blast waves and a model of
  the soft X-ray background}, \apj 253 (1982) 268--289.

\bibitem{Smith:2001}
R.~K. {Smith}, D.~P. {Cox}, {Multiple Supernova Remnant Models of the Local
  Bubble and the Soft X-Ray Background}, \apjs 134~(2) (2001) 283--309.

\bibitem{Breitschwerdt:2016}
D.~{Breitschwerdt}, J.~{Feige}, M.~M. {Schulreich}, M.~A.~D. {Avillez},
  C.~{Dettbarn}, B.~{Fuchs}, {The locations of recent supernovae near the Sun
  from modelling $^{60}$Fe transport}, \nat 532 (2016) 73--76.

\bibitem{Zucker:2022}
C.~{Zucker}, A.~A. {Goodman}, J.~{Alves}, S.~{Bialy}, M.~{Foley}, {\it et al.}, {Star formation near the Sun is
  driven by expansion of the Local Bubble}, \nat 601~(7893) (2022) 334--337.

\bibitem{Payne:1928}
C.~H. {Payne}, {On the Contours of Stellar Absorption Lines, and the
  Composition of Stellar Atmospheres}, Proceedings of the National Academy of
  Science 14~(5) (1928) 399--406.

\bibitem{Caffau:2014}
E.~{Caffau}, M.~{Steffen}, P.~{Bonifacio}, H.~G. {Ludwig}, L.~{Monaco}, G.~{Lo
  Curto}, I.~{Kamp}, {Isotope spectroscopy}, Astronomische Nachrichten 335~(1)
  (2014) 59.

\bibitem{Buder:2021}
S.~{Buder}, S.~{Sharma}, J.~{Kos}, A.~M. {Amarsi}, T.~{Nordlander}, {\it et al.}, {GALAH Collaboration}, {The GALAH+ survey: Third data
  release}, \mnras 506~(1) (2021) 150--201.

\bibitem{Pittard:2010}
J.~M. {Pittard}, E.~R. {Parkin}, {3D models of radiatively driven colliding
  winds in massive O + O star binaries - III. Thermal X-ray emission}, \mnras
  403~(4) (2010) 1657--1683.

\bibitem{Smith:2010}
R.~L. {Smith}, K.~M. {Pontoppidan}, E.~D. {Young}, M.~R. {Morris}, {New
  12CO/13CO Observations in Young Stellar Objects and Molecular Clouds:
  Implications for 12C/13C in the Early Solar Nebula}, Meteoritics and
  Planetary Science Supplement 73 (2010) 5381.

\bibitem{Tang:2019}
X.~D. {Tang}, C.~{Henkel}, K.~M. {Menten}, Y.~{Gong}, S.~{Mart{\'\i}n},
{\it et al.}:
 {ALMA view of
  the $^{12}$C/$^{13}$C isotopic ratio in starburst galaxies}, \aap 629 (2019)
  A6.

\bibitem{Diehl:2013}
R.~{Diehl}, {Nuclear astrophysics lessons from INTEGRAL}, \rpp 76~(2) (2013)
  026301.

\bibitem{Krause:2014}
M.~G.~H. {Krause}, R.~{Diehl}, {Dynamics and Energy Loss in Superbubbles},
  \apjl 794 (2014) L21.

\bibitem{Wiedenbeck:2001a}
M.~E. {Wiedenbeck}, N.~E. {Yanasak}, A.~C. {Cummings}, A.~J. {Davis}, J.~S.
  {George}, {\it et al.}, {The
  Origin of Primary Cosmic Rays: Constraints from ACE Elemental and Isotopic
  Composition Observations}, \ssr 99 (2001) 15--26.

\bibitem{Clayton:2004}
D.~D. {Clayton}, L.~R. {Nittler}, {Astrophysics with Presolar Stardust}, \araa
  42 (2004) 39--78.

\bibitem{Westphal:2014}
A.~J. {Westphal}, H.~A. {Bechtel}, F.~E. {Brenker}, A.~L. {Butterworth},
  G.~{Flynn}, {\it et al.}, {Final reports of the Stardust Interstellar Preliminary
  Examination}, \maps 49~(9) (2014) 1720--1733.

\bibitem{Leung:2018}
S.-C. {Leung}, K.~{Nomoto}, {Explosive Nucleosynthesis in
  Near-Chandrasekhar-mass White Dwarf Models for Type Ia Supernovae: Dependence
  on Model Parameters}, \apj 861~(2) (2018) 143.

\bibitem{Thielemann:2018b}
F.-K. {Thielemann}, J.~{Isern}, A.~{Perego}, P.~{von Ballmoos},
  {Nucleosynthesis in Supernovae}, \ssr 214~(3) (2018) 62.

\bibitem{Kleine:2020}
T.~{Kleine}, G.~{Budde}, C.~{Burkhardt}, T.~S. {Kruijer}, E.~A. {Worsham},
  A.~{Morbidelli}, F.~{Nimmo}, {The Non-carbonaceous-Carbonaceous Meteorite
  Dichotomy}, \ssr 216~(4) (2020) 55.

\bibitem{Thielemann:2017}
F.~K. {Thielemann}, M.~{Eichler}, I.~V. {Panov}, B.~{Wehmeyer}, {Neutron Star
  Mergers and Nucleosynthesis of Heavy Elements}, Annual Review of Nuclear and
  Particle Science 67 (2017) 253--274.


\end{thebibliography}



\end{document}